# Roadmap on Attosecond Science

Rocio Borrego Varillas[1,♦,*], Pierre Agostini[2,3], Fernando Ardana-Lamas[4], Cord L. Arnold[5], David Ayuso[6], Maurizio Reduzzi[7], Jakub Benda[8], Jens Biegert[4,9], Charles Bourassin-Bouchet[11], Thomas Brabec[12], Christian Brahms[13], Andrew C. Brown[14], David Busto[5], Jérémie Caillat[15], Francesca Calegari[16,17], Carlo Callegari[18], Stefanos Carlström[19], Zenghu Chang[12], Ming-Chang Chen[20], Anna G. Ciriolo[22], Paul Corkum[23], Gabriele Crippa[15,24], Rafael de Q. Garcia[16,17], Louis DiMauro[2,3], Nirit Dudovich[25], Per Eng-Johnsson[5], Davide Faccialà[22], Philip Flores[19], Titouan Gadeyne[24], Gianluca Aldo Geloni[26], Chase Geirger[12], Shima Gholam-Mirzaei[23], Jimena D. Gorfinkiel[27], Eleftherios Goulielmakis[28], Mohammed Hassan[29,30], Carlos Hernández-García[31,32], Phay Ho[33], Dandan Hui[34,35], Lynda R. Hutcheson[36], Misha Ivanov[19,21,37], Subhendu Kahaly[38,39], Henry Kapteyn[40], Nicholas Karpowicz[41], Franz X. Kärtner[16,17], Matthias Kling[42,43], Omer Kneller[44], Dong Hyuk Ko[23], Peter M. Kraus[45,46], Maximilian Kubullek[16,17], Stephen R. Leone[47,48,49], Franck Lépine[50], Anne L'Huillier[5], Chen-Ting Liao[51], Thomas Linker[42], Alexander Gabriel Lohr[19,52], Matteo Lucchini[1,7], Lars Bojer Madsen[10], Roland E. Mainz[16,17], Balázs Major[38,53], Jon P. Marangos[54], David Marco[31,32], Hugo Marroux[24], Sean Marshallsay[14], Rebeca Martínez Vázquez[22], Rodrigo Martín-Hernández[31,32], Zdeněk Mašín[8], Michael Meyer[26], Felipe Morales Moreno[19], Margaret Murnane[40], Daniel M. Neumark[47,48], Mauro Nisoli[1,7], Marcus Ossiander[55], Sreelakshmi Palakkal[50], Serguei Patchkovskii[19], Zekun Pi[28], Luis Plaja[31,32], Julita Poborska[4], Miguel A. Porras[56], Kevin C. Prince[18,57], David N. Purschke[23,58], Nicolette G. Puskar[47,48], Giulio Maria Rossi[16,17], Jérémy R. Rouxel[33], Thierry Ruchon[24], Patrick Rupprecht[47,48], Pascal Salières[24], Giuseppe Sansone[59], Fabian Scheiba[16,17], Martin Schultze[55], Bernd Schütte[60], Svitozar Serkez[26], Miguel A. Silva-Toledo[16,17], Olga Smirnova[19,21,52], Salvatore Stagira[7,22], Andrea Trabattoni[16,61,62], John C. Travers[13], Igor Tyulnev[4], Morgane Vacher[63], Giulio Vampa[23], Hugo W. van der Hart[14], Katalin Varjú[38,53], Anne-Lise Viotti[5], Vartika Vishnoi[24], Marc Vrakking[60], Vincent Wanie[16], Stefan Witte[64], Fei Xu[12], Vladislav S. Yakovlev[41,65], Linda Young[33], Diling Zhu[42] and Caterina Vozzi[22,♦,*]

1    Istituto di Fotonica e Nanotecnologie, Consiglio Nazionale delle Ricerche, Milano, Italy
2    Department of Physics, The Ohio State University, Columbus, OH, 43210 USA
3    Institute for Optical Science, The Ohio State University, Columbus, OH 43210 USA
4    ICFO - Institut de Ciencies Fotoniques, The Barcelona Institute of Science and Technology, Spain
5    Department of Physics, Lund University, P.O. Box 118, SE-22100 Lund, Sweden
6    Department of Chemistry, Imperial College, London, UK
7    Dipartimento di Fisica, Politecnico di Milano, Milano, Italy
8    Institute of Theoretical Physics, Faculty of Mathematics and Physics, Charles University, Czech Republic
9    ICREA, Pg. Lluís Companys 23, 08010 Barcelona, Spain
10   Department of Physics and Astronomy, Aarhus University, 8000 Aarhus, Denmark
11   Université Paris-Saclay, CNRS, Institut d'Optique Graduate School, Laboratoire Charles Fabry, France
12   Department of Physics, University of Ottawa, Ottawa K1N 6N5, Ontario, Canada
13   School of Engineering and Physical Sciences, Heriot-Watt University, Edinburgh, EH14 4AS, UK
14   Centre for Light-Matter Interactions, Queen's University Belfast, Belfast, Northern Ireland, UK
15   Sorbonne Université, CNRS, Laboratoire de Chimie Physique-Matière et Rayonnement, France
16   Center for Free-Electron Laser Science CFEL, Deutsches Elektronen-Synchrotron DESY, Germany
17   Physics Department and The Hamburg Centre for Ultrafast Imaging, University of Hamburg, Germany
18   Elettra-Sincrotrone Trieste S.C.p.A., Basovizza (Trieste), 34149, Italy
19   Theory Department, Max-Born Institute, Berlin, Germany
20   Institute of Photonics Technologies, National Tsing Hua University, Hsinchu 30013, Taiwan
21   Technion – Israeli Institute of Technology, Haifa, 32000, Israel

[22]   Istituto di Fotonica e Nanotecnologie, Consiglio Nazionale delle Ricerche, Milano, Italy

[23]   Joint Attosecond Science Laboratory, University of Ottawa and National Research Council of Canada, Ottawa, Canada

[24]   Université Paris-Saclay, CEA, LIDYL, Gif-sur-Yvette 91191, France

[25]   Department of Complex Systems, Weizmann Institute of Science, Rehovot, Israel

[26]   European XFEL, Schenefeld, Germany

[27]   School of Physical Sciences, The Open University, Milton Keynes, United Kingdom

[28]   Institute of Physics, University of Rostock, 18059 Rostock, Germany

[29]   Department of Physics, University of Arizona, Tucson, 85721, USA.

[30]   James C. Wyant College of Optical Sciences, University of Arizona, Tucson, 85721, USA

[31]   Grupo de Investigación en Aplicaciones del Láser y Fotónica, Departamento de Física Aplicada, Universidad de Salamanca, Salamanca, Spain

[32]   Unidad de Excelencia en Luz y Materia Estructuradas (LUMES), Universidad de Salamanca, Spain

[33]   Chemical Sciences and Engineering Division, Argonne National Laboratory, Lemont, IL, USA

[34]   State Key Laboratory of Ultrafast Optical Science and Technology, Shaanxi, China

[35]   Xi'an Institute of Optics and Precision Mechanics, (XIOPM-CAS), Shaanxi, China

[36]   School of Physics, University College Dublin, Ireland

[37]   Humboldt University Berlin, Berlin, Germany

[38]   ELI ALPS, ELI-HU Non-Profit Ltd., Szeged, Hungary

[39]   Institute of Physics, University of Szeged, Hungary

[40]   JILA and Department of Physics, University of Colorado and NIST, Boulder, CO, USA

[41]   Max-Planck-Institut für Quantenoptik, Garching, Germany

[42]   SLAC National Accelerator Laboratory, Menlo Park, CA, USA

[43]   Department of Applied Physics, Stanford, CA 94305, USA

[44]   Department of Physics, University of Regensburg, Regensburg, Germany

[45]   Advanced Research Center for Nanolithography (ARCNL), Amsterdam, The Netherlands

[46]   Department of Physics and Astronomy, Vrije Universiteit, Amsterdam, The Netherlands

[47]   Department of Chemistry, University of California, Berkeley, Berkeley, USA

[48]   Chemical Sciences Division, Lawrence Berkeley National Laboratory, Berkeley, USA

[49]   Department of Physics, University of California, Berkeley, Berkeley, USA

[50]   Université Claude Bernard Lyon 1, CNRS, Institut Lumière Matière, Villeurbanne, France

[51]   Department of Physics, Indiana University, Bloomington, IN, USA

[52]   Institute of Physics, Technical University, Berlin, Germany

[53]   Department of Optics and Quantum Electronics, University of Szeged, Hungary

[54]   Department of Physics - Faculty of Natural Sciences, Imperial College, London, UK

[55]   Institute of Experimental Physics, Graz University of Technology, Graz, Austria

[56]   Complex Systems Group, ETSIME, Universidad Politécnica de Madrid, Spain

[57]   Department of Surface and Plasma Science, Charles University, Czech Republic

[58]   Laboratory for Laser Energetics, University of Rochester, New York, USA

[59]   Institute of Physics, University of Freiburg, Germany

[60]   Max-Born-Institut, Berlin, Germany

[61]   Institute of Quantum Optics, Leibniz Universität Hannover, Hannover, Germany

[62]   Cluster of Excellence PhoenixD (Photonics, Optics, and Engineering-Innovation Across Disciplines), Hannover, Germany

[63]   Nantes Université, CNRS, CEISAM, UMR 6230, F-44000 Nantes, France

[64]   Imaging Physics Department, Faculty of Applied Sciences, Delft University of Technology, The Netherlands

[65]   Ludwig-Maximilians-Universität München, Garching, Germany

♦ Guest Editors of the Roadmap.

\* Author to whom any correspondence should be addressed.

E-mails: rocio.borregovarillas@cnr.it, caterina.vozzi@ifn.cnr.it

## Abstract

Twenty-five years have passed since the first experimental demonstration of attosecond pulses, marking the advent of our ability to resolve and control electron motion in real time. What began as a technological breakthrough - generating the shortest flashes ever produced - has evolved into a powerful approach for probing and steering electronic dynamics in atoms, molecules, and solids. This roadmap, authored by leading experts in the field, surveys the recent rapid progress in the generation and characterization of attosecond pulses, emerging attosecond measurement and control techniques, and their expanding range of applications. It reviews current and future developments in attosecond light sources, including novel laser technologies, waveform synthesizers, new schemes for high-order harmonic generation, attosecond pulse generation at free-electron lasers, and structured light. Advances in attosecond measurement methodologies are also discussed, encompassing all-attosecond pump–probe spectroscopy, attosecond four-wave mixing, attosecond microscopy, spectroscopy with light transients, and attosecond interferometry. Furthermore, the roadmap addresses applications of attosecond spectroscopy to reveal electron dynamics in molecules and condensed matter systems from both theoretical and experimental perspectives, and highlights emerging directions at the interface with quantum optics and quantum entanglement. Overall, this work aims to serve as a comprehensive resource for navigating the evolving landscape of attosecond science.

## Contents

# *Introduction: from light bursts to electron motion*

**Rocío Borrego-Varillas[1], Jon P. Marangos[2,*], and Caterina Vozzi[1]**

[1] CNR-IFN, Piazza Leonardo da Vinci 32, 20133 Milano (Italy)
[2] Blackett Laboratory, Imperial College London, Prince Consort Road, London SW7 2AZ, UK

j.marangos@imperial.ac.uk

25 years have now passed since the first experimental demonstration of attosecond pulses [1,2], marking the beginning of our ability to resolve and control electron motion in real time. These ultrashort pulses were generated through high-harmonic generation (HHG), a nonlinear process in which intense femtosecond laser fields drive electrons in atoms or molecules to emit coherent radiation at multiples of the driving frequency [3]. By precisely shaping the driving laser waveform and controlling the interaction conditions, HHG can produce isolated attosecond pulses or pulse trains, with durations now reaching the tens of attoseconds regime [4].

The award of the 2023 Nobel Prize in Physics for "Attosecond Physics and Technology" to Anne L'Huillier, Pierre Agostini, and Ferenc Krausz underscores the importance of these achievements. It can be seen that the ideas that have been unleashed by these early technological developments in HHG go far beyond the early promise and have unlocked a new way of looking at, and thinking about, the ultrafast world: with resolution and control of electronic motion in matter and sub-cycle laser field sculpting being among the new frontiers that are now open to investigation. Moreover, new technological options based on X-ray free electron lasers and highly stable high rep-rate optical lasers are revolutionizing the tools available. As an example, attosecond pulses from XFELs (e.g. European XFEL and LCLS) give access to the information available from measuring core to valence transitions that adds atomically localised probing to attosecond resolution, giving the possibility to fully map the charge motion in matter. These multiple technological advances are taking the field well beyond the realm of AMO physics, and indeed HHG, where it started, and allowing broad application to complex and challenging problems in materials science, condensed matter, molecular, and even biomolecular science.

What started as a technological feat - compressing light into the shortest flashes ever produced - has evolved into a powerful means of exploring and steering electronic dynamics in atoms, molecules, and solids. Attosecond pulses were first employed for the investigation of electron dynamics in atoms reporting important applications as the measurement of Auger relaxation and the real time observation of valence electron motion [5,6], the characterization of wave packets in helium [7], the measurement of delays in photoemission [8,9], the reconstruction of a correlated two-electron wavepacket [10], the buildup of Fano resonances [11,12] or the analysis of the electron tunnelling in noble gas atoms [13–15]. The first application of attosecond pump-probe spectroscopy in molecules appeared in 2006, when the control of bound electron motion (in $D_2$) by an infrared (IR) electric field was demonstrated [16]; only a few years later, attosecond pulses were combined with IR fields to control the localisation of the electronic charge distribution in $H_2$ and $D_2$ molecules [17]. Utilising the inherent pump-probe attributes of the HHG process itself leads to attosecond resolution of molecular cations in a strong field [18], offering another window into molecular attosecond dynamics. These pioneering works opened the way to investigate many-electron diatomic and small polyatomic molecules, leading to important achievements such as the observation of charge migration [19,20] or the characterisation of the molecular stereo Wigner time delays [21].

The attosecond frontier now extends from fundamental studies of correlated electron behaviour to applications in photonics, materials science, and chemistry. As we look ahead, technological advances in high-order harmonic generation, waveform control, and free electron lasers continue to push attosecond science from observing to actively shaping the microscopic world, transforming our understanding of how light and matter interact on their natural timescale.

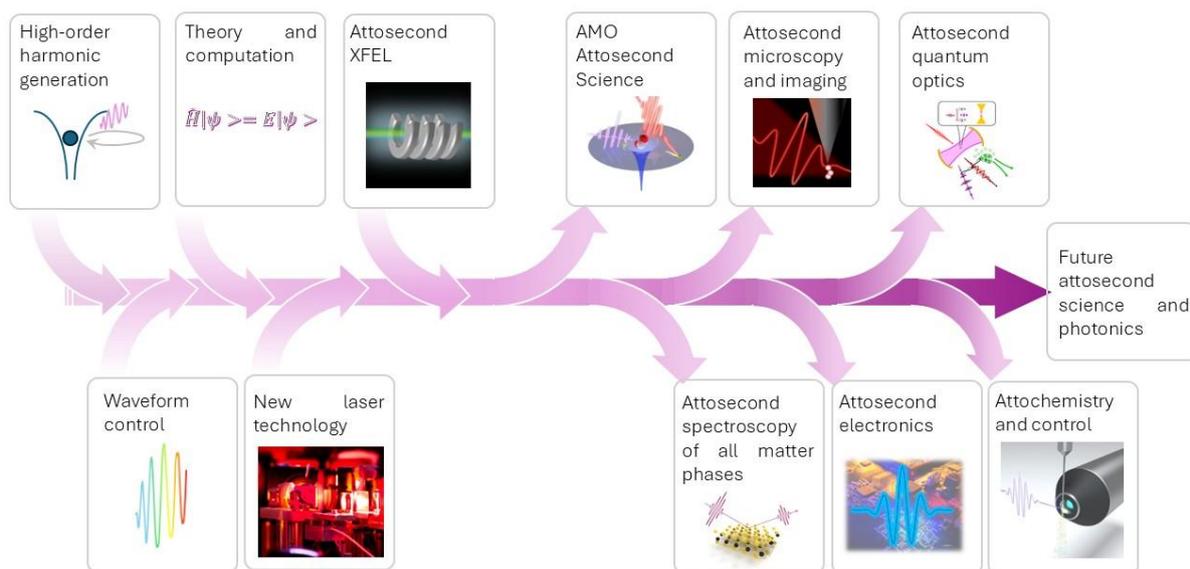

*Figure 1: Overview of the technological advances that are pushing attosecond science beyond simple observation, and the emerging applications of attosecond science.*

This roadmap aims to capture the recent rapid advances in the generation and characterisation of attosecond pulses, new attosecond measurement and control technology, as well as their emerging applications. The chapters are authored by recognised experts and leaders in the field and are written from a range of perspectives to help to illustrate the wide diversity of ideas and techniques in the field as it currently stands.

This roadmap is organised as follows. In the first section, we start by examining the latest and future technological advances towards attosecond light sources (chapters 1 to 12), which include novel laser sources, synthesizers, new schemes for high-order harmonic generation, generation of attosecond pulses at free electron lasers, and structured light. Ongoing progress and perspectives of attosecond measurement techniques and their applications are reviewed in section 2 (chapters 13 to 24). These include all-attosecond pump-probe spectroscopy, attosecond four-wave mixing, attosecond microscopy, attosecond spectroscopy with light transients or attosecond interferometry. Section 3 (chapters 25 to 33) is devoted to attosecond phenomena in condensed phase and molecules, from both theoretical and experimental viewpoints. Finally, section 4 (chapters 34 to 38) deal with the newest perspectives in quantum optics and quantum entanglement.

The full range of contributions represents a valuable survey of the attosecond landscape. We are very grateful to all the authors for their efforts in putting together this roadmap. As an aid to the reader we will briefly highlight a selection of the contributions to point towards important emerging directions.

An important task for the future, already highlighted in many contributions, is to turn attosecond technology into a useable tool that can then widely impact across all of science where fast electronic dynamics are involved i.e. in many natural systems and technological settings. An exciting technical direction, heading towards more accessible and robust attosecond light sources, is outlined in "6. Direct Generation of Bright Isolated Attosecond Pulses from Post-Compressed Yb Lasers" by Ming-Chang Chen. This discusses how highly reliable Yb laser technology offering stable, high power, high repetition rate can be leveraged in generating isolated attosecond pulses through direct pulse compression. The possibilities of upconverting light of novel properties to generate unique attosecond light fields is discussed in "8. Structuring Attosecond Light and High Harmonics in Space and Time" by Rodrigo Martín-Hernández et al". This direction, along with attosecond vortex light (Chapter 7) using upconversion of orbital angular momentum fields is heading towards synthesis of attosecond fields of arbitrary topology with applications in the dynamic imaging of chiral, magnetic and quantum materials.

A developing direction in the measurement of attosecond dynamics in matter of all phases is exemplified by "13. All-attosecond pump-probe spectroscopy" by Bernd Schütte and Marc Vrakking where an implementation in the XUV is presented. These methodologies are now possible in the X-ray range, leveraging the atomic site specificity of x-ray spectroscopy, using the now available attosecond pulses at the LCLS and European XFELs. This latter theme is developed further in "17. Attosecond imaging, radiolysis and chiral dynamics" by Phay Ho, Jérémy Rouxel, and Linda Young where early measurements using these XFEL capabilities are presented and longer term prospects discussed. An exciting direction with XFEL based attosecond pulses due to their > GW peak powers is the possibility to drive non-linear X-ray interactions that opens a whole new vista of measurement possibilities of electron dynamics in matter as touched upon by "12. Generation and application of sub-femtosecond soft X-ray pulses at the European XFEL" M. Meyer, S. Serkez and G. Geloni. The unique opportunities afforded by seeded XUV/Soft X-ray XFELs such as FERMI in Trieste are developed in "19. Attosecond-resolved coherent control" by Carlo Callegari, Kevin C. Prince and Giuseppe Sansone. These are sources of high power attosecond pulse trains where the harmonics can be exquisitely controlled in relative phase, amplitude and polarisation offering new opportunities in measuring photon-ionisation delays and coherent control. Along with the isolated attosecond pulses and pulse pairs available from other XFELs we are seeing huge advances in the control of the coherence and temporal properties of X-rays extending from soft to hard x-rays (see also "11. Attosecond Hard X-ray Pulses for Material Science" by Thomas Linker, Diling Zhu, and Matthias F. Kling).

The new technological direction of controllable and useable attosecond electronics clocked by optical fields is explored in "32. Attosecond Electronics" by Marcus Ossiander and Martin Schultze where emerging tools and ideas are discussed. In a similar and equally exciting direction are the prospects to combine attosecond technology with electron and scanning-tunnelling microscopy to create a capacity to image electron dynamics at their native atomic spatial and attosecond temporal scales in "33. Attomicroscopy: Attosecond Electron Motion Imaging in Real Time and Space" by Dandan Hui and Mohammed Hassan.

The control of the quantum properties of light is now being extended into the attosecond regime at short wavelengths opening up new potential opportunities in metrology and imaging.

The possibilities are discussed in a number of chapters including "34. Attosecond Quantum Optics" Giulio Vampa, David N. Purschke, Thomas Brabec.

Hopefully, this roadmap will prove a useful resource in orientating the researcher in the landscape of attosecond science. This landscape, and the routes available in it, are constantly changing and amongst the most exciting opportunities are in exploring the *terra incognita* at the current edges of the map that may lead into entirely new territory. We hope, therefore, that this roadmap inspires as well as informs.

## References

[1] Paul P M 2001 Observation of a Train of Attosecond Pulses from High Harmonic Generation *Science (80-. ).* **292** 1689–92

[2] Hentschel M, Kienberger R, Spielmann C, Reider G, Milosevic N, Brabec T, Corkum P B, Heinzmann U, Drescher M and Krausz F 2001 Attosecond Meteorology *Nature* **414** 509–13

[3] Ferray M, L'Huillier A, Li X F, Lompre L A, Mainfray G and Manus C 1988 Multiple-harmonic conversion of 1064 nm radiation in rare gases *J. Phys. B At. Mol. Opt. Phys.* **21** L31–5

[4] Gaumnitz T, Jain A, Pertot Y, Huppert M, Jordan I, Ardana-Lamas F and Wörner H J 2017 Streaking of 43-attosecond soft-X-ray pulses generated by a passively CEP-stable mid-infrared driver *Opt. Express* **25** 27506–18

[5] Drescher M, Hentschel M, Kienberger R, Uiberacker M, Yakovlev V, Scrinzi A, Westerwalbesloh T, Kleineberg U, Heinzmann U and Krausz F 2002 Time-resolved atomic inner-shell spectroscopy *Nature* **419** 803–7

[6] Goulielmakis E, Loh Z H, Wirth A, Santra R, Rohringer N, Yakovlev V S, Zherebtsov S, Pfeifer T, Azzeer A M, Kling M F, Leone S R and Krausz F 2010 Real-time observation of valence electron motion *Nature* **466** 739–43

[7] Mauritsson J, Remetter T, Swoboda M, Klünder K, L'Huillier A, Schafer K J, Ghafur O, Kelkensberg F, Siu W, Johnsson P, Vrakking M J J, Znakovskaya I, Zherebtsov S, Kling M F, Lépine F, Benedetti E, Ferrari F, Sansone G and Nisoli M 2010 Attosecond electron spectroscopy using a novel interferometric pump-probe technique *Phys. Rev. Lett.* **105** 053001

[8] Schultze M, Fiess M, Karpowicz N, Gagnon J, Korbman M, Hofstetter M, Neppl S, Cavalieri A L, Komninos Y, Mercouris T, Nicolaides C A, Pazourek R, Nagele S, Feist J, Burgdorfer J, Azzeer A M, Ernstorfer R, Kienberger R, Kleineberg U, Goulielmakis E, Krausz F and Yakovlev V S 2010 Delay in Photoemission *Science (80-. ).* **328** 1658–62

[9] Klünder K, Dahlström J M, Gisselbrecht M, Fordell T, Swoboda M, Guénot D, Johnsson P, Caillat J, Mauritsson J, Maquet A, Taïeb R and L'Huillier A 2011 Probing Single-Photon Ionization on the Attosecond Time Scale *Phys. Rev. Lett.* **106** 143002

[10] Ott C, Kaldun A, Argenti L, Raith P, Meyer K, Laux M, Zhang Y, Blättermann A, Hagstotz S, Ding T, Heck R, Madroñero J, Martín F and Pfeifer T 2014 Reconstruction and control of a time-dependent two-electron wave packet *Nature* **516** 374–8

[11] Gruson V, Barreau L, Jiménez-Galan A, Risoud F, Caillat J, Maquet A, Carré B, Lepetit F, Hergott J F, Ruchon T, Argenti L, Taïeb R, Martín F and Salières P 2016 Attosecond dynamics through a Fano resonance: Monitoring the birth of a photoelectron *Science (80-. ).* **354** 734–8

[12] Kaldun A, Blättermann A, Stooß V, Donsa S, Wei H, Pazourek R, Nagele S, Ott C, Lin C D, Burgdörfer J and Pfeifer T 2016 Observing the ultrafast buildup of a Fano resonance in the time domain *Science (80-. ).* **354** 738–41

[13] Dudovich N, Smirnova O, Levesque J, Mairesse Y, Ivanov M Y, Villeneuve D M and Corkum P B 2006 Measuring and controlling the birth of attosecond XUV pulses *Nat. Phys.* **2** 781–6

[14] Uiberacker M, Uphues T, Schultze M, Verhoef A J, Yakovlev V, Kling M F, Rauschenberger J, Kabachnik N M, Schröder H, Lezius M, Kompa K L, Muller H G, Vrakking M J J, Hendel S, Kleineberg U, Heinzmann U, Drescher M and Krausz F 2007 Attosecond real-time observation of electron tunnelling in atoms *Nature* **446** 627–632

[15] Pfeiffer A N, Cirelli C, Smolarski M, Dimitrovski D, Abu-samha M, Madsen L B and Keller U 2012 Attoclock reveals natural coordinates of the laser-induced tunnelling current flow in atoms *Nat. Phys.* **8** 76–80

[16] Kling M F, Siedschlag C, Verhoef A J, Khan J I, Schultze M, Uphues T, Ni Y, Uiberacker M, Drescher M, Krausz F and Vrakking M J J 2006 Control of electron localization in molecular dissociation *Science (80-. ).* **312** 246–8

[17] Sansone G, Kelkensberg F, Pérez-Torres J F, Morales F, Kling M F, Siu W, Ghafur O, Johnsson P, Swoboda M, Benedetti E, Ferrari F, Lépine F, Sanz-Vicario J L, Zherebtsov S, Znakovskaya I, Lhuillier A, Ivanov M Y, Nisoli M, Martín F and Vrakking M J J 2010 Electron localization following attosecond molecular photoionization *Nature* **465** 763–6

[18] Baker S, Robinson J S, Haworth C A, Teng H, Smith R A, Chirilă C C, Lein M, Tisch J W G and Marangos J P 2006 Probing Proton Dynamics in Molecules on an Attosecond Time Scale *Science (80-. ).* **312** 424–7

[19] Calegari F, Ayuso D, Trabattoni A, Belshaw L, De Camillis S, Anumula S, Frassetto F, Poletto L, Palacios A, Decleva P, Greenwood J B, Martín F, Nisoli M, Camillis S De, Anumula S, Frassetto F, Poletto L, Palacios A,

Decleva P, Greenwood J B and Nisoli M 2014 Ultrafast electron dynamics in phenylalanine initiated by attosecond pulses *Science (80-. ).* **346** 336–9

[20] Kraus P M, Mignolet B, Baykusheva D, Rupenyan A, Horný L, Penka E F, Grassi G, Tolstikhin O I, Schneider J, Jensen F, Madsen L B, Bandrauk A D, Remacle F and Wörner H J 2015 Measurement and laser control of attosecond charge migration in ionized iodoacetylene *Science (80-. ).* **350** 790–5

[21] Vos J, Cattaneo L, Patchkovskii S, Zimmermann T, Cirelli C, Lucchini M, Kheifets A, Landsman A S and Keller U 2018 Orientation-dependent stereo Wigner time delay and electron localization in a small molecule *Science (80-. ).* **360** 1326–30

## *1. Multidimensional optimization of high-order harmonic generation*

**A.-L. Viotti[1], P. Eng-Johnsson[1], A. L'Huillier[1] and C. L. Arnold[1]\***

[1] Department of Physics, Lund University, P.O. Box 118, SE-22100 Lund, Sweden

cord.arnold@fysik.lu.se

**Status**

Coherent radiation spanning the extreme-ultraviolet (XUV) to soft X-ray spectral range, produced through high-order harmonic generation (HHG) in gases, has emerged as a versatile tool across a wide array of applications. Gas HHG requires a driving femtosecond (fs) laser pulse focused to high intensity in the generation medium. Today, HHG sources are operated with a large variety of driving femtosecond laser technologies, including Titanium-Sapphire and Ytterbium chirped pulse amplifiers (CPAs), possibly with post-compression of the pulses, and optical parametric chirped pulse amplifiers (OPCPAs), with wavelengths from the ultraviolet to the mid-infrared. Moreover, there is a large diversity of generation geometries, like short and dense gas jets, long and dilute gas cells, semi-infinite cells, and gas-filled capillaries.

Depending on the specific application, such as probing ultrafast electronic dynamics on the attosecond timescale or performing high-resolution coherent imaging at the nanoscale [1-3], the optimization of the XUV source may focus on various attributes, such as overall conversion efficiency (CE), temporal structure, coherence properties, or the ability to tightly refocus the beam to reach high peak intensity. The experimental conditions for HHG, such as laser wavelength, pulse duration, beam profile, wavefront shape, gas medium density, atomic species, and the position of the generation medium relative to the laser focus, play a pivotal role in determining the characteristics of the resulting radiation. Consequently, precise understanding, characterization and control of these parameters are essential for optimizing HHG, not only to maximize the photon flux or efficiency but also to ensure high spatial and temporal quality of the emitted radiation.

HHG in gases involves the response of a single atom to a strong laser field (strong-field atomic physics) and the coherent buildup of the emission in the finite medium (phase-matching). In the following, we distinguish three levels of source optimization related to (1) the single-atom response, (2) the spatial properties of the emitted radiation, which directly affect the spatiotemporal structure of the attosecond pulses and (3) phase-matching. Those optimization levels, involving different parameters of the light-matter interaction, operate approximately in zero, two and three dimensions, which is why we call this contribution to the roadmap *Multidimensional optimization of high-order harmonic generation*.

**Current and future challenges**

**Challenge 1: Single-atom response**

The single-atom response is well described by an intuitive three-step model, where an electron tunnels through the atomic potential modified by the laser field and is further driven by it. When it returns to the parent ion, it may recombine back to the ground state, emitting a high-energy photon. Figure 1 (a) shows the short- and long electron trajectories calculated in argon for a laser wavelength of 1 μm and an intensity of $10^{14}$ W/cm$^2$. The colours indicate the electron return energies, with the brightest colour indicating the highest energy, called the cutoff. Figure 1 (b) represents the return energy as a function of return time, thus linking time to energy for a given family of trajectories, short or long. The optimization of the photon energy range and/or the efficiency by manipulating the strong laser-atom interaction and specifically the electron trajectories has been pursued for several decades [4,5]. A key challenge is to determine which driving waveform enhances the ionization and recombination steps.

### Challenge 2: Spatial properties

The spatial properties of the harmonics follow those of the driving field: to obtain high quality XUV beams requires a high degree of control of the laser. The wavefront of the harmonics is not simply equal to that of the fundamental, as in perturbative nonlinear optics, but includes an intrinsic, chromatic contribution induced by the spatial intensity profile of the driving laser pulse. For conventional drivers with Gaussian spatial shape, the size and position of the waist of the harmonics depend on the harmonic order and the position of the medium in relation to the geometrical focus, as shown in Figure 1 (e) [6-8]. This affects the refocusing, the achievable intensity and spatiotemporal properties of the attosecond pulse train. In addition, hardly detectable amounts of astigmatism or ellipticity in the fundamental, can lead to significantly distorted XUV wavefronts, often manifested as cross-shaped far-field spatial profiles [9]. Such aberrations pose serious challenges for applications that require high spatial beam quality, such as nanoscale imaging, or intense field strengths, as in table-top XUV pump–probe experiments. While most theoretical models traditionally assume idealized driving fields, the impact of more realistic beam conditions has only recently begun to receive attention.

### Challenge 3: Phase-matching

Phase-matching requires that the phase velocities of the driving and generated fields remain equal over the length of the macroscopic medium. It depends on the dispersion in the neutral medium, the presence of free electrons, the focusing conditions and the intensity-dependent electron trajectories. Figure 1 (f) presents simulations of the $23^{rd}$ harmonic CE in argon as a function of medium pressure and length for an 800 nm laser wavelength, a pulse duration of 22 fs and an intensity of $2\times10^{14}$ W/cm². The axes are normalized according to a geometrical scaling, which predicts that the CE is invariant, if pulse energy, gas density, target length and focusing conditions are adjusted accordingly [10]. The figure shows that the optimal CE, limited by absorption [11], is achieved on a "pressure-length" hyperbola [12], encompassing high pressure, short media (vertical branch) and long, dilute cells (horizontal branch). Despite 30 years of research, optimizing HHG remains a very open problem since it depends on many macroscopic and microscopic parameters, as well as on the interplay between nonlinear propagation in a partially ionized medium and strong-field, ultrafast light-matter interaction.

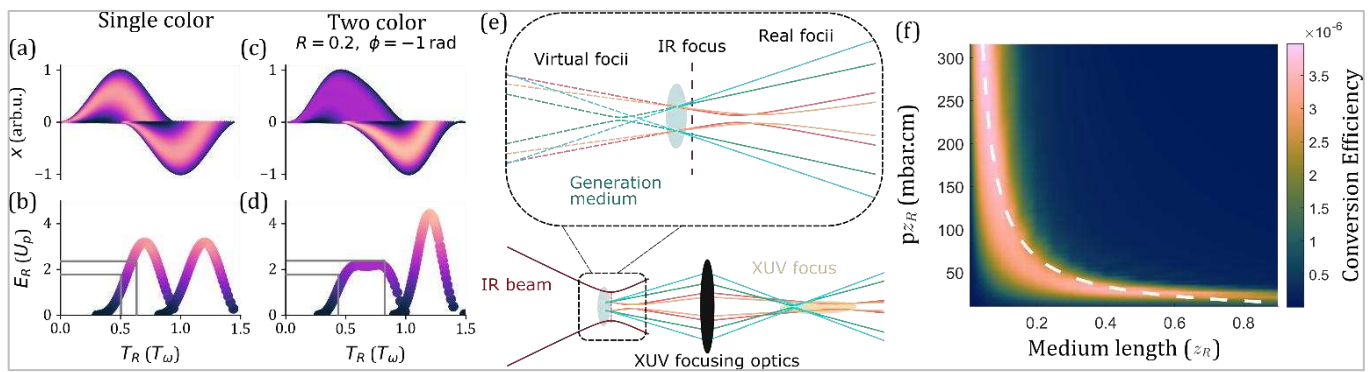

*Figure 1.* *(a) and (c): Electron trajectories calculated from the semiclassical three-step model. (b) and (d): Electron return energy, in unit of the ponderomotive energy $U_p$, as a function of return time, in unit of the period of the fundamental optical cycle. (a) and (b) correspond to the single-colour case and (c) and (d) to the two-colour (fundamental and second harmonic) case. We assume a 1 μm driving wavelength, $10^{14}$ W/cm2 intensity, generation in argon, and, for the two-colour case, a relative intensity ratio R=0.2 and a relative phase ϕ =-1 rad. The electron trajectories are modified from the single-color case (a) to the two-colour case (c), leading to a higher cutoff energy (second half-cycle in (d)) and to a yield enhancement (first half-cycle), coupled to the flattening of the time-energy curve. (e) Illustration of spatiotemporal couplings of attosecond pulses upon refocusing. The location of the gas target with respect to the laser geometrical focus determines the wavefront and divergence of individual harmonics, which originate from real or virtual source points depending on the order. (f) Numerical simulations of the CE for harmonic 23 in argon for different pressures p and medium lengths at a laser intensity of $2\times10^{14}$ W/cm2 and wavelength of 800 nm ($z_R$ is the Rayleigh length). The medium is centred around the laser geometrical focus. The analytical hyperbole, based on a 1D phase-matching model (dashed line), predicts configurations of high CE with good accuracy.*

**Advances in science and technology to meet challenges**

Experimentally, the challenges mentioned above have started to be addressed partially thanks to the technological transition towards robust Ytterbium laser systems operating at high repetition rates and average powers. Regarding challenge 1, systematic studies of HHG driven by a fundamental and its second harmonic have been performed as a function of the relative intensity ratio and phase between both fields over a large parameter space. As shown in Figure 1 (c) and (d), the trajectories can be manipulated to, e.g., extend the cut-off or enhance a certain spectral range. The results show that a given relative phase leads to a significant enhancement of the harmonic yield, independently of the order. The optimum intensity ratio, which depends on the process order, can be predicted using a simple formula [13]. This optimization study can be extended to more general waveforms, including several colour driving fields and non-collinear geometries [14]. Cross-polarized fields can also be advantageously employed, e.g., to obtain elliptically polarized harmonics [15].

Challenge 2 can be addressed by correcting and/or manipulating the spatial properties of the fundamental field, using adaptive optics, e.g., spatial light modulators or deformable mirrors. For instance, flat-top [16], hollow Gaussian beams or beams carrying orbital angular momentum [17] can be used to generate harmonics in a focal plane where the driver's intensity is spatially uniform, thus avoiding the intrinsic divergent contribution from a Gaussian fundamental beam. Such investigations require advanced spatiotemporal diagnostics of the emitted harmonics [18], which are only recently emerging in the community.

Finally, to tackle challenge 3, more versatile simulation tools, which include reshaping and few-cycle effects, as well as an accurate description of the light-matter interaction by solving the time-dependent Schrödinger equation, are needed. Recent experimental results show that the highest CE is obtained for combinations of pulse durations and intensities, where phase-matching is universally achieved for an optimum density of free electrons [19]. Shorter pulse durations generally allow for higher intensities and result in higher CEs. HHG using long driving wavelengths leads to high photon energies in the soft x-ray range, of interest for several applications. Unfortunately, the CE is drastically reduced, which motivates further theoretical and experimental work to understand and optimize HHG in these conditions [20]. The optimization of the harmonics/attosecond pulse refocused intensity could benefit from different strategies, for instance "out-of-focus" generation where the harmonics are generated several Rayleigh lengths away from the laser focus [1] leading to a small XUV focus, or via clever spatial shaping of the driving field.

**Concluding remarks**

We have discussed HHG optimization strategies related to the single-atom response, the spatial properties of the emitted radiation, and phase-matching. To optimize HHG for a specific application remains a formidable challenge, which continuously grows due to the new possibilities offered by laser technology, in terms of available wavelengths and multidimensional shaping capabilities. On the other hand, our understanding of the physics of the HHG process, as well as our ability to quantitatively simulate it, has dramatically advanced over the last decade. In addition, the experimental possibilities to study and control HHG have exploded with increasing repetition rates, available spectral ranges, flexible geometries and novel diagnostics. The envisioned developments will be essential for practical and industrial implementation of HHG sources to meet societal needs.

## Acknowledgements

The authors acknowledge the financial support from the Swedish Research Council (Grant no. 2017-04106, 2021-04691, 2021-05992, 2022-03519, 2023-04603, 2023-04684), the European Research Council (advanced grant QPAP, Grant No. 884900), The Knut and Alice Wallenberg Foundation (Grant no. KAW 2020.0111), the Crafoord Foundation, the Lund Laser Centre and the Swedish Foundation for Strategic Research (Grant no. FFL24-0144). ALH is partly supported by the Wallenberg Center for Quantum Technology (WACQT), funded by the Knut and Alice Wallenberg Foundation.

We thank P. Smorenburg, D. O'Dwyer, S. Roscam Abbing, A. Ross, Y. Tao and S. Edward from ASML for theoretical and experimental support. We thank M. Redon and R. Weissenbilder for their assistance with the figures.

## References

[1] Kretschmar M, Svirplys E, Volkov M, Witting T, Nagy T, Vrakking M J J and Schütte B, 2024, Compact realization of all-attosecond pump-probe spectroscopy, *Sci. Adv.* **10**, eadk9605

[2] Seaberg M D, Adams D E, Townsend E L, Raymondson D A, Schlotter W F, Liu Y, Menoni C S, Rong L, Miao J, Kapteyn H C and Murnane M M, 2011, Ultrahigh 22 nm resolution coherent diffractive imaging using a desktop 13 nm high harmonic source, *Opt. Express* **19**, 22470-22479

[3] Rothhardt J, Tadese G K, Eschen W and Limpert J, 2018, Table-top nanoscale coherent imaging with XUV light, *J. Opt.* **20**, 113001

[4] Chipperfield L E, Robinson J S, Tisch J W G and Marangos J P, 2009, Ideal waveform to generate the maximum possible electron recollision energy for any given oscillation period, *Phys. Rev. Lett.* **102**, 063003

[5] Ratz O, Pedatzur O, Bruner B D and Dudovich N, 2012, Spectral caustics in attosecond science, *Nat. Photon.* **6**, 170-173

[6] Frumker E, Paulus G G, Niikura H, Naumov A, Villeneuve D M and Corkum P B, 2012, Order-dependent structure of high harmonic wavefronts, *Opt. Express* **20**, 13870-13877

[7] Quintard L, Strelkov V, Vabek J, Dubrouil A, Descamps D, Burgy F, Pejot C, Mevel E, Catoire F and Constant E, 2019, Optics-less focusing of XUV high-order harmonics, *Sci. Adv.* **5**, eaau7175

[8] Wikmark H, Guo C, Vogelsang J, Smorenburg P, Coudert-Alteirac H, Lahl J, Peschel J, Rudawski P, Dacasa H, Carlström S, Maclot S, Gaarde M B, Eng-Johnsson P, Arnold C L and L'Huillier A, 2019, Spatiotemporal coupling of attosecond pulses, *Proc. Natl. Acad. Sci. U.S.A.* **116**, 4779-4787

[9] Plach M, Vismarra F, Appi E, Poulain V, Peschel J, Smorenburg P, O'Dwyer D, Edward S, Tao Y, Borrego-Varillas R, Nisoli M, Arnold C L, L'Huillier A and Eng-Johnsson P, 2024, Spatial aberrations in high-order harmonic generation, *Ultrafast Sci.* **4**, 0054

[10] Heyl C M, Coudert-Alteirac H, Miranda M, Louisy M, Kovacs K, Tosa V, Balogh E, Varju K, L'Huillier A, Couairon A and Arnold C L, 2016, Scale-invariant nonlinear optics in gases, *Optica* **3**, 75-81

[11] Constant E, Garzella D, Breger P, Mevel E, Dorrer Ch, Le Blanc C, Salin F and Agostini P, 1999, Optimizing high harmonic generation in absorbing gases: model and experiment, *Phys. Rev. Lett.* **82**, 1668

[12] Weissenbilder R, Carlström S, Rego L, Guo C, Heyl C M, Smorenburg P, Constant E, Arnold C L and L'Huillier A, 2022, How to optimize high-order harmonic generation in gases, *Nat. Rev. Phys.* **4**, 713-722

[13] Raab A-K, Redon M, Roscam Abbing S, Fang Y, Guo C, Smorenburg P, Mauritsson J, Viotti A-L, L'Huillier A and Arnold C L, 2025, XUV yield optimization of two-color high-order harmonic generation in gases, Nanophotonics, 0579

[14] Chappuis C, Bresteau D, Auguste T, Gobert O and Ruchon T, 2019, High-order harmonic generation in an active grating, *Phys. Rev. A* **99**, 033806

[15] Kfir O, Grychtol P, Turgut E, Knut R, Zusin D, Popmintchev D, Popmintchev T, Nembach H, Shaw J M, Fleischer A, Kapteyn H C, Murnane M M and Cohen O, 2015, Generation of bright phase-matched circularly-polarized extreme ultraviolet high harmonics, *Nat. Photon.* **9**, 99-105

[16] Strelkov V, Mevel E and Constant E, 2009, Isolated attosecond pulse generated by spatial shaping of femtosecond laser beam, *Eur. Phys. J. Spec. Top.* **175**, 15

[17]   Hernandez-Garcia C, Picon A, San Roman J and Plaja L, 2013, Attosecond extreme ultraviolet vortices from high-order harmonic generation, *Phys. Rev. Lett.* **111**, 083602

[18]   Liu X, Pelekanidis A, Du M, Zhang F, Eikema K S E and Witte S, 2023, Observation of chromatic effects in high-order harmonic generation, *Phys. Rev. Res.* **5**, 043100

[19]   Westerberg S, Redon M, Raab A-K, Beaufort G, Arias Velasco M, Guo C, Sytcevich I, Weissenbilder R, O'Dwyer D, Smorenburg P, Arnold C L, L'Huillier A and Viotti A-L, 2025, Influence of the laser pulse duration in high-order harmonic generation, *APL Photon.* **10**, 096103

[20]   Klas R, Gebhardt M, Rothhardt J and Limpert J, 2025, Unleashing HHG efficiency: the role of driving pulse duration, arXiv:2510.04259

## 2. New routes to efficient table-top sources of coherent XUV attosecond pulses exploiting microfluidic devices

**Anna G. Ciriolo[1]\*, Davide Faccialà[1], Rebeca Martínez Vázquez[1], Salvatore Stagira[1,2], and Caterina Vozzi[1]\***

[1] Istituto di Fotonica e Nanotecnologie, Consiglio Nazionale delle Ricerche, Milano, Italy
[2] Dipartimento di Fisica, Politecnico di Milano, Milano, Italy

annagabriella.ciriolo@cnr.it, caterina.vozzi@cnr.it

### Status

X-ray radiation has been essential for probing the structural, elemental, and electronic properties of matter. Yet, understanding and ultimately controlling physical processes at the level of atoms and electrons requires access to ultrafast timescales. The development of coherent eXtreme Ultraviolet (XUV) and soft X-ray (SXR) sources driven by femtosecond laser via High-order Harmonic Generation (HHG) has opened a new pathway for tracking electronic dynamics in real time, by enabling the generation of XUV/SXR light pulses with attosecond duration [1,2]. However, the low conversion efficiency of the process remains a critical limitation. Generating bright, broadband attosecond pulses in the SXR persists as an ongoing challenge. Conventional HHG systems, based on the first-generation Ti:Sapphire laser technology, typically generate photon energies up to 150 eV, which is insufficient for probing key absorption edges in biological and chemical systems, particularly within the SXR *water window* (282–533 eV). Recent efforts using mid-infrared (MIR) laser drivers have extended HHG into this spectral region, enabling attosecond access to carbon, nitrogen, and oxygen K-edges [3-8]. Recent advances in solid-state and fiber-laser technologies have enabled the combination of high peak power with high repetition rates, boosting the photon flux of HHG sources by 1–3 orders of magnitude compared to earlier systems [9]. Currently, optical parametric chirped pulse amplification (OPCPA) enables the generation of high-energy, ultrashort pulses in the MIR, effectively compensating for the inherently low efficiency of HHG in the SXR regime [10].

In parallel, a persistent challenge in attosecond science remains the optimization of HHG efficiency, both at the microscopic (single atom) level and through macroscopic phase matching. To address this, a wide range of strategies have been developed. On the one hand, advanced optical approaches, such as the optical shaping of spatial and polarization properties of the driving fields and the use of multi-colour waveforms, enabled precise control over electron trajectories [11]. On the other hand, the implementation of extended gas media with tailored laser-gas interaction schemes pushed the up-conversion efficiency of HHG closer to its fundamental physical limit [12].

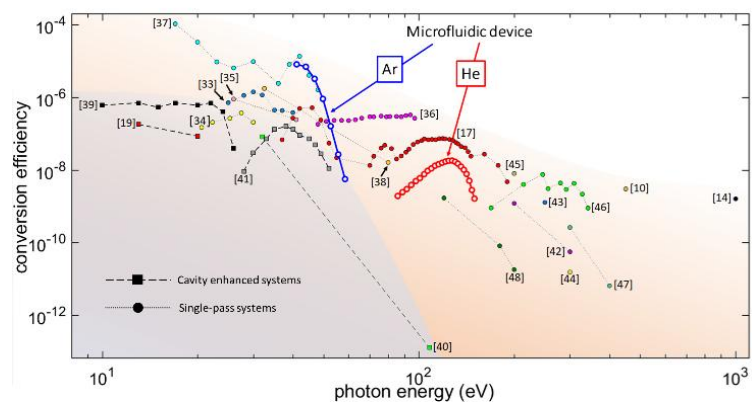

*Figure 1:* Comparison of HHG source efficiencies reported in the literature. The performance of the microfluidic source in He and Ar is highlighted. Figure from ref. [14], see ref [14] for detailed references.

In this context, microfluidic devices emerged as a new platform to precisely regulate the gas flow and control the propagation of the driving laser field, thereby enabling fine-tuning of the HHG conditions [13, 14]. These systems have already demonstrated enhanced HHG flux (see fig. 1) and offer promising strategies for both sample delivery within a micro-controlled environment and optical flexibility in terms of waveguiding and manipulation of the fundamental laser beam [15].

**Current and future challenges**

A central goal for attosecond sources in the coming years is to meet the technical requirements for performing both fundamental and applied experiments using HHG-based SXR sources. This includes achieving enhancement and precise control over the photon flux, as well as the temporal and spatial characteristics of the beam. Particular attention is focused on the water window spectral region, which holds exceptional potential for the investigation of biologically relevant systems with atomic specificity.

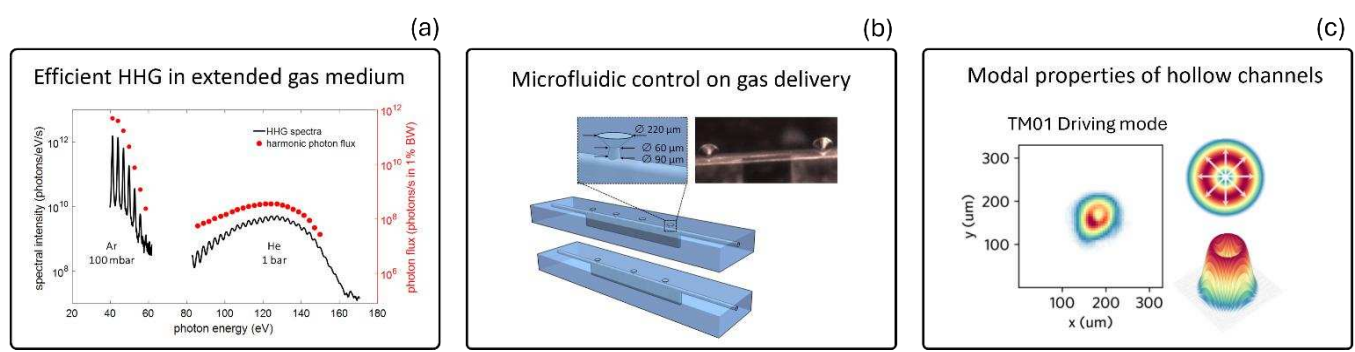

*Figure 2:* *Key benefits of the microfluidic approach, including (a) efficient generation in extended gas media [14], (b) precise control of gas delivery and distribution within the active region [14], and (c) driving and XUV beam manipulation through modal properties [16].*

Significant research efforts have led to the demonstration of HHG schemes operating in the water window and beyond, primarily using femtosecond drivers in the near- to mid-infrared. However, generating high-brightness isolated attosecond pulses (IAPs) in the SXR remains challenging. A major limitation is the intrinsic *attochirp* due to the frequency-dependent emission timing that stretches the pulse duration. Current dispersion compensation techniques are ineffective in the SXR range, where conventional materials provide insufficient control. This calls for the development of advanced attochirp compensation methods, such as tailored phase shaping of the driving field.

Furthermore, the absence of conventional optical components in the XUV and X-ray regimes calls for alternative approaches to manipulate the polarization properties of the HHG beam. The introduction of multi-colour driving fields, optical vortices, and non-collinear interfering driving fields has opened new possibilities, including the demonstration of XUV radiation with three-fold symmetry and the control of the orbital angular momentum [17] and the spin angular momentum [18] of the generated XUV waveform. However, detrimental phase-matching effects in extended gas media result in critical bottlenecks in SXR driven by multi-fold polarization and spatiotemporal symmetry.

Phase matching is critical for the efficient coherent buildup of the harmonic field. Its disruption generally leads to longer pulse duration. However, certain regimes can exploit these dynamics advantageously. At high laser intensities, transient phase-matching conditions can confine efficient HHG to specific temporal windows, as in the ionization gating [19,20]. Highly nonlinear propagation effects like self-focusing and defocusing can further reshape the driving field creating dynamic phase matching conditions that confine HHG to sub-optical-cycle durations. An example is the overdriven regime, which can be used to generate isolated attosecond pulses (IAPs) in the soft X-ray range [6].

Quasi-phase-matching strategies, involving gas pressure gradients, optical gradients within waveguides, or multi-color drivers, show promise but are not yet fully implemented in the SXR regime.

Engineering the spatial and temporal profiles of the driving pulse, using waveform synthesis, adaptive optics, and diffractive elements, makes it possible to finely control emission timing, spectral content, and phase-matching conditions, thus opening new pathways for optimizing SXR pulse generation. However, this approach remains underexploited, primarily due to the limited availability of suitable optical components operating effectively in the near-to mid-infrared spectral region and supporting a large spectral bandwidth.

**Advances in science and technology to meet challenges**

Microfluidic technology holds significant potential for XUV and SXR science. The HHG process within gas or liquid media demands precise environmental control to ensure both vacuum compatibility and tailored fluid distribution. In this framework, the key advantage of microfluidic platforms lies in the ability to precisely control and manipulate gas distributions at the micrometer scale within the generation volume, offering unprecedented accuracy and flexibility. This level of control is essential for tailoring the generation conditions with high precision, potentially enabling quasi-phase-matching in specific spectral regions through custom-designed gas injection modules (see fig.2(b)). Moreover, the technology supports high-density gas delivery and is scalable to accommodate molecular samples, offering flexibility and accuracy in the delivery of samples with demanding environmental conditions through integration with temperature and flow regulation systems.

An additional advantage lies in the ability to exploit the modal properties of hollow microfluidic channels to manipulate the propagation of the driving laser fields, thereby opening new possibilities for all-optical laser field tailoring (see fig.2(c)). By selectively exciting specific waveguide modes, the spatial distribution of the laser's intensity and polarization can be precisely shaped, thus enabling HHG driven by complex and structured spatial field distributions. Moreover, by leveraging total internal reflection on glass interfaces, hollow channels can also serve as waveguides for light in the XUV and soft X-ray regimes, providing an additional degree of control over the propagation of the generated radiation.

These systems have already demonstrated enhanced HHG flux (see fig 2(a)) and represent a promising platform for integration with high-repetition-rate laser sources. In such regimes, collective effects arising from long-lived plasma dynamics become significant, necessitating flexible and precise control over the interaction conditions. Therefore, microfluidics offers a powerful route for improving both the efficiency and scalability of XUV and SXR sources.

Ultimately, the microfluidic approach addresses future challenges by offering a miniaturized, chip-based alternative to large-scale and technologically demanding workstations for X-ray science. This technology provides a first demonstration of the potential for extending the Lab-on-a-Chip (LOC) concept to x-ray beamlines, offering a path toward a new generation of attosecond spectroscopy and x-ray-based technologies. This could encompass the development of miniaturized microfluidic experimental stations capable of supporting increasingly complex optical functionalities, including interaction with either gas or liquid samples, and the handling of multiple laser beams by incorporating components like pulse splitting, delay lines, and interferometers in one device. Future iterations of this technology may also feature the possibility of integrating vacuum-compatible micro-electrodes, micro-actuators, or optical switches for remote control and dynamic reconfiguration of the operation conditions within a single monolithic device.

**Concluding remarks**

The advancements in attosecond technologies are closely tied to progress in laser sources, with a focus on achieving higher intensities and greater average power to compensate for the inherently low efficiency of the HHG process, especially in the SXR spectral region. Simultaneously, increasingly sophisticated techniques, such as optical shaping of the driving laser field, are being explored to enhance the single-atom response, as well as to tailor the polarization and spatiotemporal characteristics of the generated XUV/SXR radiation. These approaches aim to maximize performance within conventional HHG setups.

The emerging microfluidic solutions that we proposed and demonstrated may soon become competitive by combining waveguide-based manipulation of optical fields with micrometer-scale control of the gas medium. Such innovations position these compact, efficient microfluidic-based HHG sources as key enablers for next-generation SXR attosecond science and ultrafast spectroscopy. Moreover, microfluidic platforms provide a unique opportunity to integrate source generation with sample preparation and delivery, paving the way for miniaturized, all-in-one experimental environments for transformative studies in physics, chemistry, and biology.

**Acknowledgements**

The reported research has received funding from the European Union's Horizon 2020 Research and Innovation Program under Grant Agreement No. 964588 (X-PIC), the European COST Action Grant No. 22148 (NEXT), the European Union's NextGenerationEU Programme with the I-PHOQS Infrastructure [CUP B53C22001750006] and the PRIN Project HAPPY P20224AWLB [CUP B53D23025210001].

**References**

[1]  Linda Young, Kiyoshi Ueda, Markus Gühr, Philip H Bucksbaum, Marc Simon, Shaul Mukamel, Nina Rohringer, Kevin C Prince, Claudio Masciovecchio, Michael Meyer, Artem Rudenko, Daniel Rolles, Christoph Bostedt, Matthias Fuchs, David A Reis, Robin Santra, Henry Kapteyn, Margaret Murnane, Heide Ibrahim, François Légaré, Marc Vrakking, Marcus Isinger, David Kroon, Mathieu Gisselbrecht, Anne L'Huillier, Hans Jakob Wörner and Stephen R Leone, Roadmap of ultrafast x-ray atomic and molecular physics, 2018 J. Phys. B: At. Mol. Opt. Phys. 51 032003

[2]  Jie Li, Xiaoming Ren, Yanchun Yin, Kun Zhao, Andrew Chew, Yan Cheng, Eric Cunningham, Yang Wang, Shuyuan Hu, Yi Wu, Michael Chini & Zenghu Chang, 53-attosecond X-ray pulses reach the carbon K-edge. Nat. Commun. 8, 186 (2017).

[3]  Tenio Popmintchev 1, Ming-Chang Chen, Dimitar Popmintchev, Paul Arpin, Susanna Brown, Skirmantas Alisauskas, Giedrius Andriukaitis, Tadas Balciunas, Oliver D Mücke, Audrius Pugzlys, Andrius Baltuska, Bonggu Shim, Samuel E Schrauth, Alexander Gaeta, Carlos Hernández-García, Luis Plaja, Andreas Becker, Agnieszka Jaron-Becker, Margaret M Murnane, Henry C Kapteyn, Bright coherent ultrahigh harmonics in the keV x-ray regime from mid-infrared femtosecond lasers, SCIENCE 336, 1287 (2012)

[4]  V Cardin, B E Schmidt, N Thiré, S Beaulieu, V Wanie, M Negro, C Vozzi, V Tosa and F Légaré, Self-channelled high harmonic generation of water window soft x-rays, J. Phys. B: At. Mol. Opt. Phys. 51 174004 (2018)

[5]  Gregory J Stein, Phillip D Keathley, Peter Krogen, Houkun Liang, Jonathas P Siqueira, Chun-Lin Chang, Chien-Jen Lai, Kyung-Han Hong, Guillaume M Laurent and Franz X Kärtner, Water-window soft x-ray high-harmonic generation up to the nitrogen K-edge driven by a kHz, 2.1 µm OPCPA source, J. Phys. B: At. Mol. Opt. Phys. 49 (2016) 155601

[6] Allan S. Johnson, Dane R. Austin, David A. Wood, Christian Brahms, Andrew Gregory, Konstantin B. Holzner, Sebastian Jarosch, Esben W. Larsen, Susan Parker, Christian S. Strüber, Peng Ye, John W. G. Tisch, and Jon P. Marangos, High-flux soft x-ray harmonic generation from ionization-shaped few-cycle laser pulses, Sci. Adv. 2018;4: eaar3761

[7] J. Pupeikis, P.-A. Chevreuil, N. Bigler, L. Gallmann, C. R. Phillips, and U. Keller, Water window soft x-ray source enabled by a 25 W few-cycle 2.2 μm OPCPA at 100 kHz, Optica 7, 168 (2020)

[8] Yuxi Fu, Kotaro Nishimura, Renzhi Shao, Akira Suda, Katsumi Midorikawa, Pengfei Lan & Eiji J. Takahashi, High efficiency ultrafast water-window harmonic generation for single-shot soft X-ray spectroscopy, Communications Physics volume 3, Article number: 92 (2020)

[9] Robert Klas, Alexander Kirsche, Martin Gebhardt, Joachim Buldt, Henning Stark, Steffen Hädrich, Jan Rothhardt & Jens Limpert, Ultra-short-pulse high-average-power megahertz-repetition-rate coherent extreme-ultraviolet light source, PhotoniX volume 2, Article number: 4 (2021)

[10] Daniel Walke, Azize Koç, Florian Gores, Minjie Zhan, Nicolas Forget, Raman Maksimenka, and Iain Wilkinson, High-average-power, few-cycle, 2.1 μm OPCPA laser driver for soft-X-ray high-harmonic generation, Optics Express Vol. 33, 10006-10019 (2025)

[11] Barry D. Bruner, Michael Krüger, Oren Pedatzur, Gal Orenstein, Doron Azoury, and Nirit Dudovich, Robust enhancement of high harmonic generation via attosecond control of ionization, Optics Express Vol. 26, 9310-9322 (2018)

[12] Zongyuan Fu, Yudong Chen, Sainan Peng, Bingbing Zhu, Baochang Li, Rodrigo Martín-Hernández, Guangyu Fan, Yihua Wang, Carlos Hernández-García, Cheng Jin, Margaret Murnane, Henry Kapteyn, and Zhensheng Tao, Extension of the bright high-harmonic photon energy range via nonadiabatic critical phase matching, Science Advances 2022 Vol 8, Issue 51

[13] A. G. Ciriolo, R. M. Vázquez, V. Tosa, A. Frezzotti, G. Crippa, M. Devetta, D. Faccialá, F. Frassetto, L. Poletto, A. Pusala, C. Vozzi, R. Osellame, and S. Stagira, "High-order harmonic generation in a microfluidic glass device," J. Phys.: Photonics 2, 024005 (2020).

[14] A. G. Ciriolo, R. Martínez Vázquez, G. Crippa, M. Devetta, D. Faccialà, P. Barbato, F. Frassetto, M. Negro, F. Bariselli, L. Poletto, V. Tosa, A. Frezzotti, C. Vozzi, R. Osellame, and S. Stagira, "Microfluidic devices for quasi-phase-matching in high-order harmonic generation," APL Photonics 7, 110801 (2022).

[15] R. Martínez Vázquez, A. G. Ciriolo, G. Crippa, V. Tosa, F. Sala, M. Devetta, C. Vozzi, S. Stagira, and R. Osellame, "Femtosecond laser micromachining of integrated glass devices for high-order harmonic generation," Int. J. Appl. Glass Sci. 13, 162–170 (2022).

[16] Riccardo Piccoli, Marco Bardellini, Stavroula Vovla, Linda Oberti, Kamal A. A. Abedin, Anna G. Ciriolo, Rebeca Martínez Vázquez, Roberto Osellame, Luca Poletto, Fabio Frassetto, Davide Faccialá, Michele Devetta, Caterina Vozzi, Salvatore Stagira, Synthesizing extreme-ultraviolet vector beams in a chip, arXiv:2403.11006 [physics.optics].

[17] D. Gauthier, P. Rebernik Ribič, G. Adhikary, A. Camper, C. Chappuis, R. Cucini, L. F. DiMauro, G. Dovillaire, F. Frassetto, R. Géneaux, P. Miotti, L. Poletto, B. Ressel, C. Spezzani, M. Stupar, T. Ruchon & G. De Ninno, Tunable orbital angular momentum in high-harmonic generation, Nat Commun 8, 14971 (2017).

[18] Avner Fleischer, Ofer Kfir, Tzvi Diskin, Pavel Sidorenko & Oren Cohen, Spin angular momentum and tunable polarization in high-harmonic generation. Nature Photon 8, 543–549 (2014).

[19] F. Ferrari, F. Calegari, M. Lucchini, C. Vozzi, S. Stagira, G. Sansone & M. Nisoli, High-energy isolated attosecond pulses generated by above saturation few-cycle fields, Nature Photonics volume 4, pages 875–879 (2010)

[20] Ming-Chang Chen, Christopher Mancuso, Carlos Hernández-García, Franklin Dollar, Ben Galloway, Dimitar Popmintchev, Pei-Chi Huang, Barry Walker, Luis Plaja, Agnieszka A. Jaroń-Becker, Andreas Becker, Margaret M. Murnane, Henry C. Kapteyn, and Tenio Popmintchev, Generation of bright isolated attosecond soft X-ray pulses driven by multicycle midinfrared lasers. Proc. Natl Acad. Sci. USA 111, E2361–E2367 (2014)

## 3. *The optical route of attosecond physics*

Zekun Pi[1] and Eleftherios Goulielmakis[1]*

[1] Institute of Physics, University of Rostock, 18059 Rostock, Germany

e.goulielmakis@uni-rostock.de

**Status**

Attosecond technology [1] has progressed along two principal pathways. The first and historically dominant route is the generation of extreme-ultraviolet (XUV) attosecond pulses via high-harmonic generation [2-4]. This approach has shaped the field for years and was long considered the only viable strategy for accessing the attosecond regime. Synthesizing attosecond pulses in the visible and adjacent spectral regions was typically considered challenging if not impossible, based on oversimplified arguments: a light pulse cannot be shorter than one optical cycle, and that a Gaussian pulse in the visible whose bandwidth is broad enough to support attosecond confinement of its waveform would inevitably have DC spectral components incompatible with Maxwell's equations for propagating waves.

Yet, the duration of practical electromagnetic signals is not characterized by the temporal extend of a waveform but by quantities that reflect their resolution in experiments such as full width at half maximum (FWHM) of their intensity profile. In this sense, an optical attosecond pulse is defined as a waveform with sub-femtosecond concentration of a significant portion of its energy within its FWHM and not a waveform that builds up and disappears within a period or less. Moreover, once the Gaussian model describing temporally and spectrally a pulse is put aside, furth perceived limitations vanish [5]: a quasi-uniform in spectral intensity pulse in the visible—achievable via nonlinear broadening in hollow-core capillaries [6,7]—can make synthesis of attosecond pulses accessible. For instance, a pulse whose intensity envelope is ~950 as in its FWHM requires a spectrum spanning about two optical octaves (1100 nm–230 nm), fully compatible with modern continuum generation.

The above realizations have enabled a second path in attosecond technology: The synthesis and sub-optical cycle control of ultrashort pulses [8-13] and their confinement to a fraction of their carrier period. In a broad class of approaches taken, few-tens-of-femtosecond pulses from commercial amplifiers are broadened in gas-filled hollow-core fibers (HCFs) to generate a multi-octave continuum [10]. This continuum is divided into spectral channels—near-IR, visible, visible-UV, and deep-UV—using broadband, dispersive dichroic optics [14]. Each channel is compressed with dispersive mirrors, and all channels are recombined on an interferometrically stable platform to synthesize the final waveform. While individual channels can be characterized with standard few-fs techniques, accessing the full two-octave synthesized waveform required attosecond streaking [15], which enabled sampling of light waves with attosecond precision. A state-of-the-art example of light field synthesis of an optical attosecond pulse is summarized in Fig.1.

The synthesis of optical attosecond pulses has opened new horizons in attosecond metrology and ultrafast control. Applications already include probing the delay nonlinear response of electrons gasses [10], enabling isolated half-cycle EUV emission and identifying intra-band transport as the dominant mechanism high harmonics in solids [16], establishing attosecond pump–probe spectroscopy with soft-x rays [17], and enabling attosecond-resolved field emission from nanotips [18].

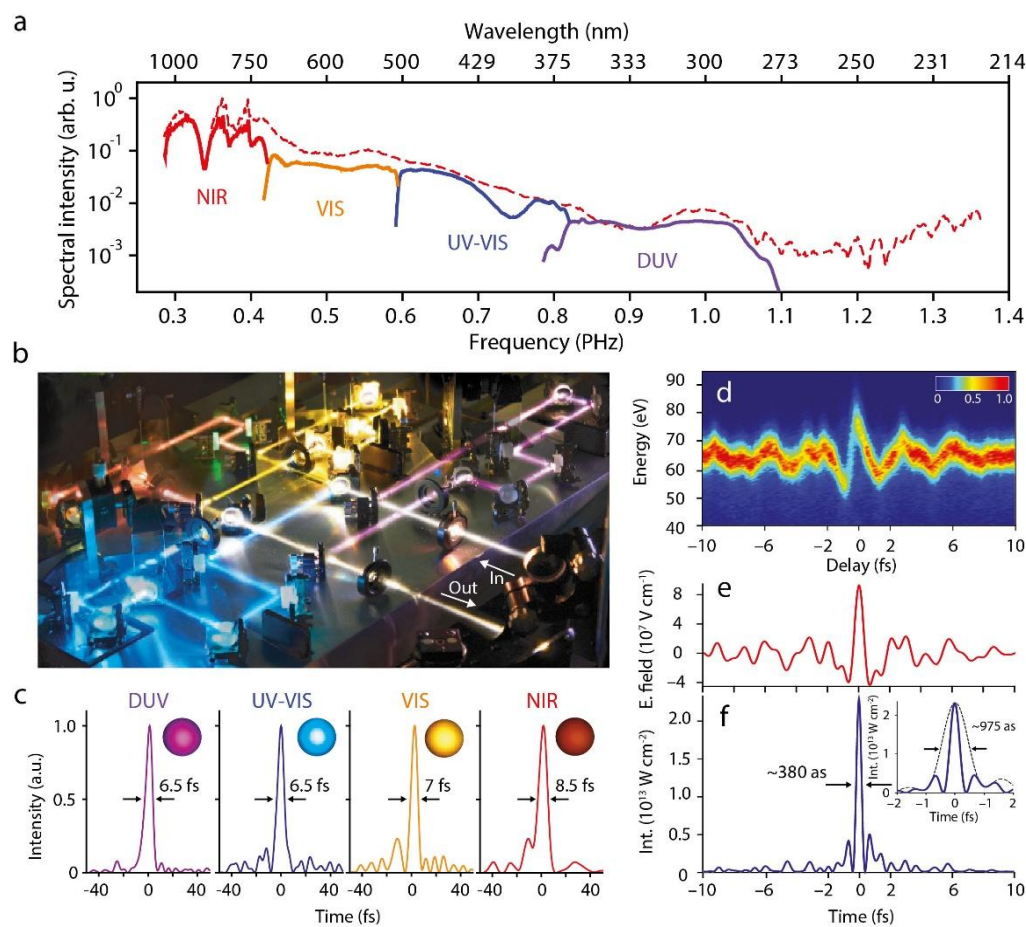

***Figure 1.*** *(a) A multi-octave supercontinuum from Ne-filled HCF capillary post-compressor, seeded by 22 fs pulse with central wavelength of 790 nm and an energy of ~ 1 mJ. (b) Photograph of an attosecond light field synthesizer. The pulse in (a) is divided into four nearly equal-width spectral bands by dichroic beam splitters. The pulse in each spectral band is compressed to a duration of several femtoseconds, as shown in the temporal intensity distribution in figure (c). And insets in each panel shows a typical beam profile in the far field. These beams are then spatiotemporally superimposed to form a single pulse at the device exit. (d) Attosecond streaking spectrogram of a synthesized optical attosecond pulse. (e) Evaluated electric field (red line) and (f) instantaneous intensity profile (blue line) of the pulse. The intensity profile has a FWHM duration of approximately 380 as. The inset of (f) shows a close-up of this instantaneous intensity profile (blue line) under the intensity envelope (dashed black line), with a FWHM duration of approximately 975 as. Adapted from ref [10].*

## Current and future challenges

A major challenge for the next phase of optical attosecond synthesis is its integration with emerging Yb:KGW laser technology. These systems, while offering high average power and repetition rates, deliver significantly longer pulses. Bringing them into a regime where multi-octave continua can be efficiently produced and compressed to the single-cycle limit is essential.

A second challenge concerns the spectral extension of light field synthesis toward both the ultraviolet and the infrared. Access to the deep ultraviolet is essential for pushing optical attosecond pulses to ever shorter durations, yet dispersion control, dispersive optics coating performance and material transparency in this range remain technically challenging. Extending synthesis toward the infrared, on the other hand, is critical for attosecond-scale studies in semiconductors and other low-bandgap materials. Achieving simultaneous UV and IR access within a single synthesizer architecture—while preserving energy—constitutes a demanding physics and engineering problem.

A third challenge is the realization of fully optical attosecond pump–probe spectroscopies. Optical attosecond pulses such as those presented in Fig.1 can, in principle, drive highly nonlinear processes with intrinsic attosecond resolution; however, this capability is compromised when used for linear or excitation of materials by the presence of satellite pulses and pulse pedestals that result in insufficient contrast.

Finally, a key challenge is the development of light field characterization techniques that do not rely on XUV pulses. Traditional attosecond streaking has been so far indispensable but requires XUV sources and therewith high power and more complex infrastructure.

**Advances in science and technology to meet challenges**

Meeting the emerging challenges of optical attosecond synthesis and spectroscopy will require coordinated advances in laser technology, dispersion engineering and attosecond metrology. A priority is the development of highly efficient multioctave spectral-broadening schemes compatible with Yb:KGW sources. These lasers must be guided into a regime where their long pulses and modest nonlinearities can still generate the multi-octave continua necessary for attosecond confinement. Novel gas-based broadening mechanisms based on pressurized high ionization potential, noble gasses seem to offer a viable path [19].

A second area of needed progress is the development of optical components that support a broader usable spectral range. In the deep ultraviolet, advances in multilayer mirror coatings, and a new generation of dispersion-controlled reflective elements will be essential. Extending synthesis to longer infrared wavelengths will require low-loss broadband IR optics, but it is admittedly less challenging than these in the deep ultraviolet range. Ultimately, achieving seamless UV-to-IR synthesis in a single platform will depend on progress in both material science and thin-film engineering.

For all-optical attosecond pump–probe experiments, technology must advance toward waveforms with far higher contrast. This includes improved amplitude- and phase-shaping dispersive optics capable of selectively enhancing the dominant half-cycle while suppressing unwanted satellite pulse components.

Equally important is the development of fully optical, EUV-independent diagnostics. Homochromatic attosecond streaking [18,20] represents a promising foundation, but several technological advances are needed before it becomes a universal tool. For instance, harnessing the full potential of visible light transients calls also for techniques that can capture not only the field waveform by also the complete vectorial characterization of light transients of complex evolution of their polarization vector. This will be an essential condition for extending the synthesis of light to include sculpting of the complete polarization vector of a light transient.

Together, these advances will establish a robust technological ecosystem for enabling more advanced application of the optical route of attosecond physics.

**Concluding remarks**

The optical route to attosecond science has matured into a powerful and versatile platform for controlling and probing matter on its natural timescales. By moving beyond conventional assumptions about pulse duration and by exploiting multi-octave continua, optical attosecond synthesis has opened a powerful pathway to the XUV-based approach. The ability to sculpt light-wave transients with sub-cycle precision has enabled attosecond-resolved studies of electron dynamics in solids, isolated-cycle extreme nonlinear optics, soft-x-ray pump–probe spectroscopy, and the generation and complete characterization of attosecond electron pulses from nanostructures.

Looking ahead, major advances will come from extending synthesis to new laser technologies such as Yb:KGW systems, broadening the accessible bandwidth toward both the deep-UV and the IR, enabling all-optical attosecond pump–probe methodologies, and establishing EUV-independent optical field diagnostics. These developments will extend the reach of optical attosecond methods well beyond the state of the art.

## Acknowledgements

Deutsche Forschungsgemeinschaft (441234705, 437567992); European Research Council (101098243).

## References

[1]   Krausz F, Ivanov M 2009 Attosecond physics *Rev. Mod. Phys*. **81** 163–234
[2]   Li X F, l'Huillier A, Ferray M, Lompré L A and Mainfray G 1989 Multiple-harmonic generation in rare gases at high laser intensity *Phys. Rev. A* **39** 5751-5761
[3]   Hentschel M, Kienberger R, Spielmann C, Reider G A, Milosevic N, Brabec T, Corkum P, Heinzmann U, Drescher M and Krausz F 2001 Attosecond metrology *Nature* **414** 509–513
[4]   Paul P M, Toma E S, Breger P, Mullot G, Augé F, Balcou P, Muller H G and Agostini P 2001 Observation of a train of attosecond pulses from high harmonic generation *Science* **292** 1689–1692
[5]   Brabec T and Krausz F 1997 Nonlinear optical pulse propagation in the single-cycle regime *Phys. Rev. Lett.* **78** 3282-3285
[6]   Nisoli M, De Silvestri S and Svelto O 1996 Generation of high-energy 10 fs pulses by a new pulse compression technique *Appl. Phys. Lett.* **68** 2793–2795
[7]   Nisoli M, De Silvestri S, Svelto O, Szipöcs R, Ferencz K, Spielmann C, Sartania S and Krausz F 1997 Compression of high-energy laser pulses below 5 fs *Opt. Lett.* **22** 522–524
[8]   Rausch S, Binhammer T, Harth A, Kärtner F X and Morgner U 2008 Few-cycle femtosecond field synthesizer *Opt. Express* **16** 17410–17419
[9]   Krauss G, Lohss S, Hanke T, Sell A, Eggert S, Huber R and Leitenstorfer A 2010 Synthesis of a single cycle of light with compact erbium-doped fibre technology *Nat. Photonics* **4** 33–36
[10]  Hassan M T, Luu T T, Moulet A, Raskazovskaya O, Zhokhov P, Garg M, Karpowicz N, Zheltikov A M, Pervak V, Krausz F and Goulielmakis E 2016 Optical attosecond pulses and tracking the nonlinear response of bound electrons *Nature* **530** 66–70
[11]  Wirth A et al 2011 Synthesized light transients *Science* **334** 195–200
[12]  Liang H et al 2017 High-energy mid-infrared sub-cycle pulse synthesis from a parametric amplifier *Nat. Commun.* **8** 141
[13]  Alqattan H, Hui D, Pervak V and Hassan M T 2022 Attosecond light field synthesis *APL Photonics* **7** 041301
[14]  Razskazovskaya O, Krausz F and Pervak V 2017 Multilayer coatings for femto-and attosecond technology *Optica* **4** 129–138
[15]  Goulielmakis E, Uiberacker M, Kienberger R, Baltuška A, Yakovlev V, Scrinzi A, Westerwalbesloh T, Kleineberg U, Heinzmann U, Drescher M and Krausz F 2004 Direct measurement of light waves *Science* **305** 1267-1269
[16]  Garg M, Zhan M, Luu T T, Lakhotia H, Klostermann T, Guggenmos A and Goulielmakis E 2016 Multi-petahertz electronic metrology *Nature* **538** 359–363
[17]  Moulet A, Bertrand J B, Klostermann T, Guggenmos A, Karpowicz N and Goulielmakis E 2017 Soft x-ray excitonics *Science* **357** 1134–1138
[18]  Kim H Y, Garg M, Mandal S, Seiffert L, Fennel T, Goulielmakis E 2023 Attosecond field emission *Nature* **613** 662-666
[19]  Pi Z, Kim H Y and Goulielmakis E 2022 Petahertz-scale spectral broadening and few-cycle compression of Yb:KGW laser pulses in a pressurized, gas-filled hollow-core fiber *Opt. Lett.* **47** 5865–5868
[20]  Pi Z, Kim H Y and Goulielmakis E 2025 Synthesis of single-cycle pulses based on a Yb:KGW laser amplifier *Optica* **12** 296-301

## *4. Isolated attosecond pulses driven by tailored waveforms*

**Rafael de Q. Garcia[1,2]\*, Miguel A. Silva-Toledo[1], Fabian Scheiba[1,2], Maximilian Kubullek[1,2], Roland E. Mainz[1], Franz X. Kärtner[1,2], and Giulio Maria Rossi[1,2]**

[1] Center for Free-Electron Laser Science, Deutsches Elektronen-Synchrotron DESY, Hamburg, Germany
[2] Physics Department and The Hamburg Centre for Ultrafast Imaging, University of Hamburg, Hamburg, Germany

rafael.garcia@desy.de

**Status**

Laser pulses with spectra spanning more than one octave and intensity envelopes lasting from few down to less than a single optical cycle (sub-cycle pulses) are becoming more widely available. These technological advancements have made it possible to synthesize sub-femtosecond light transients in the UV-VIS range. They have also opened up opportunities for the coherent control of isolated attosecond pulses (IAPs) generation in the XUV and soft X-ray regimes via high harmonic generation (HHG) [1,2]. With a single to sub-cycle HHG driver, the emission can be confined to one single dominant attosecond bust. If phase control is available across different spectral regions, the full field dependence of HHG at the single-atom level can be harnessed. This allows tailoring the field to an optimized non-sinusoidal shape that can enhance the flux and/or the cutoff. Boosting the efficiency is particularly critical in the soft X-ray spectral range, where HHG driven by longer-wavelength infrared fields suffers from the unfavourable $\lambda^{-6}$ scaling of conversion efficiency. Phase matching further requires higher gas pressures ($P \propto \lambda^2$), which increases waveform distortions. In addition, it tolerates progressively lower plasma fractions ($\eta \propto \lambda^{-2}$), which limits the usable peak intensity and reduces the number of contributing emitters. Parametric waveform synthesis (PWS) remains the main energy-scalable approach producing stable, few to sub-cycle tailored pulses [3,4]. Recent advances in hollow core fiber (HCF) compression technology have also demonstrated sub-cycle pulses with mJ-level energies with good waveform stability for driving IAPs [5]. HCF systems, however, need to be combined with a synthesis frontend to achieve a more systematic control over the waveforms produced, as demonstrated in the past [1].

The search of an "ideal waveform" for enhancing certain HHG characteristics has been pioneered by several groups [6–8]. Although most of the studies are not necessarily aiming at optimizing IAPs, when these tailored ideal waveforms have approximately only one cycle, the consequence is the direct control of IAP properties. Specific waveforms, such as those in Figure 1(a) have demonstrated cutoff extension [6,7,9]. So far, this was done only with multi-cycle pulses, which only provide IAPs when filtering the harmonic emission relatively close to its cutoff. On the other hand, dramatic yield enhancements have been predicted for $\omega + 3\omega$ or $\omega + 2\omega + 3\omega$ synthesis by up to 100-1000 times for waveforms such as in Figure 1(b) [8,10]. These tailored waveforms simultaneously optimize recombination of short trajectories, decreasing the overall HHG beam divergence. The addition of the $2\omega$ field plays a role on the temporal gating of these pulses, consequently providing broadband and high contrast IAPs. Other parameters of waveform tailoring, including chirp and spectral shaping, have also been explored extensively in simulation studies [11].

**Current and future challenges**

The different tailoring schemes studied so far provide useful guiding principles for experimentalists seeking optimal control of IAPs. Yet with the growing availability of sub-cycle pulses with pulse energies capable to drive IAPs even in the water window (~ 285 - 535 eV) [12], waveform optimization at the single-atom level alone is no longer sufficient. The combination of high intensity, multi-octave bandwidth, and high gas pressures in the nonlinear medium makes optimization increasingly dependent on macroscopic propagation effects, underscoring the need for systematic studies that bridge waveform design with realistic experimental conditions.

Aspects of macroscopic propagation like the use of waveguides or free-space gas targets and the multiparameter space spanned by beam waists, Rayleigh ranges ($z_R$) and focus positions of synthesized pulses remains insufficiently explored [10,13]. In addition, more attention is needed on tight ($z_R < L$) and intermediate ($z_R \sim L$) focusing geometries, with L being the medium`s length. In these cases, the geometric phase variation becomes more pronounced and intensities are higher in comparison to most of the prior work, that has addressed loose focusing geometries ($z_R \gg L$). Simulating multi-bar pressure media leading to extreme propagation effects on the waveforms is still challenging because of the lack of practical metrology tools for characterizing the pressure distribution in common gas cell geometries. Beyond simple beam parameters (e.g., $z_R$), more complex spatial control using spatial light modulators and deformable mirrors is already feasible and promising, especially for multi-cycle two-color setups [14]. Whether such approaches can also bring significant benefits in the few to sub-cycle regime is still unclear, given the more stringent demands for broadband operation.

On the experimental side, field synthesis now has the chance to demonstrate its unique capability of producing high-flux, water-window IAPs with high temporal contrast, potentially spanning multiple K-edges of light elements such as carbon, nitrogen, and oxygen, realizing its full potential for time-resolved spectroscopy studies. Achieving this will require a new generation of high-power PWS systems to tackle the still unexplored potential of waveform synthesis. For example, the 100-1000 times yield enhancement of the three-color synthesis case of Fig. 1(b) has still not been explored down to the single-cycle pulse regime. Beyond the water window, the optimal synthesis scheme to more efficiently generate keV-order harmonics remains an open question. The extremely low photon fluxes achieved at this spectral region have been so far prohibitive for spectroscopy applications. Exploring alternative synthesis schemes (e.g., Fig. 1(c)) and experimentally achieving phase matching of multi-color fields will be key to advancing attosecond science [15].

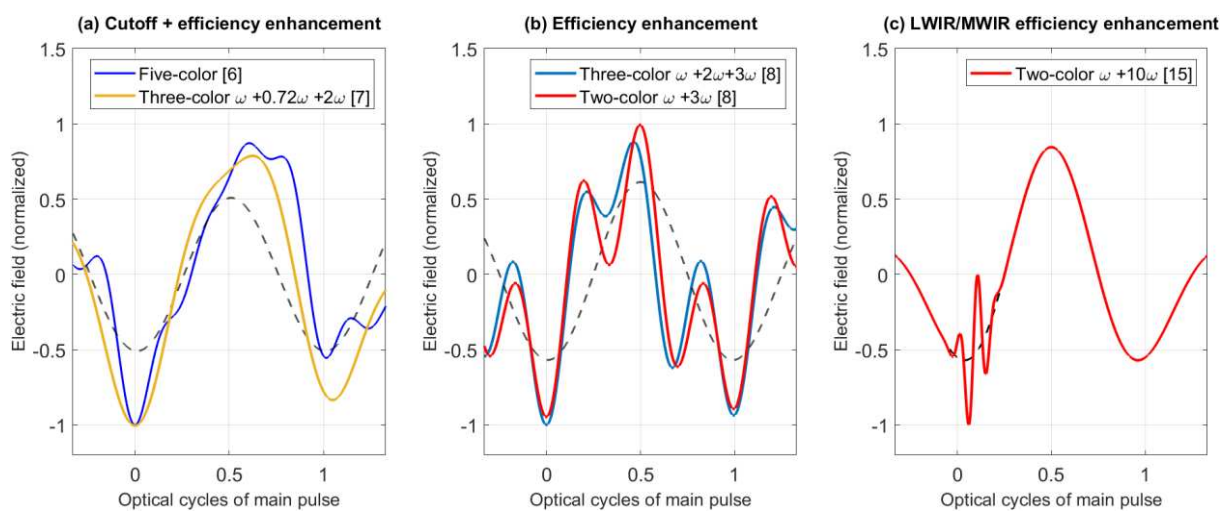

**Figure 1.** *Three tailoring schemes focusing on cutoff (a) and efficiency (b,c) enhancement are depicted above. The horizontal axis is normalized based on the period of the main pulse (grey dashed line) used in the synthesis, with each tailoring method having different limits of wavelength scalability depending on the propagation effects. LWIR = Long wave infrared, MWIR= Mid wave infrared.*

**Advances in science and technology to meet challenges**

Further advances are necessary in the analytical and simulation toolbox to investigate the scaling laws and optimization of HHG when transitioning from few-cycle to sub-cycle pulses driving IAPs [16,17]. In fact, pulses shorter than a single cycle can become "too short" for efficiently driving HHG at high photon energies. Due to the shorter envelope duration, sub-cycle pulses require higher peak intensities to achieve the same ionization yield and cutoff in comparison to a single or few-cycle pulse [16]. Sources capable of producing transform-limited, sub-cycle pulses often need to be operated out of their fully compressed state if tailoring schemes such as in Fig. 1 are to be implemented. In practice, one still needs to determine the most viable combination of pulses and synthesis parameters that allow the electric field within a single cycle to be shaped for optimal generation. The more pulses are used in synthesis, the longer they can be individually and thus easier to generate and compress. However, this also implies more synthesis parameters and higher complexity in the waveform control. Hence, more developments in this direction will shape the design criteria of next generation PWSs and HCF systems with tailoring capabilities.

Considering the challenges of simulations to reproduce propagation effects, there are recent works that try to simulate tailoring experiments as closely as possible [18]. This close feedback between experiment and theory achieves a balance between simplifying assumptions and predictive accuracy, providing indeed new insights on how to use synthesized fields experimentally. Still, faster simulation tools are needed to explore the larger parameter spaces pertained to waveform optimization. At the same time, improved ionization rate models with broader validity ranges are essential for single-cycle level tailored fields, with promising developments already emerging [19].

Advances in the experimental side are also needed to provide simulations with better constraints-such as accurate measurements of the gas density distribution in commonly used HHG media. Techniques for characterizing the synthesized transients are advancing considerably, being able to provide in-situ and spatio-temporal characterizations of few to sub-cycle pulses [20,21]. This precise knowledge of the waveform is critical to uncover the underlying phase matching mechanisms of HHG. In addition, there is a pressing need for IAP characterization techniques beyond attosecond streaking capable of handling the hundreds-of-eV bandwidths of soft X-ray pulses and the additional drawback of low photoionization cross sections in this spectral range [9].

**Concluding remarks**

With the progress on waveform tailoring during the last few years, different avenues have been opened up for exploration. First, we expect record yield enhancements to be achieved with synthesis schemes as in Fig. 1 b-c. The highest expectations lie on the enhancement at the water window region and eventually up to keV energies with long wavelength drivers. This includes not only Ti:Sa IR OPCPA technology but also longer wavelength drivers such as Yb, MWIR and LWIR lasers. Recent developments of more efficient few-cycle UV sources can also make optical + UV synthesis viable for generating bright EUV IAPs. In addition, more accurate propagation simulations and waveform characterization techniques will narrow the boundaries between experiment and simulation, informing future design choices for a new generation of sub-cycle sources and tailoring schemes.

## Acknowledgements

Funding: This work was supported by the Helmholtz Association via the Program MML-Matter, by the Cluster of Excellence "CUI:Advanced Imaging of Matter" of the Deutsche Forschungsgemeinschaft (DFG) (EXC 2056 - project ID 390715994), and by PIER, the partnership of Universität Hamburg and DESY (grant ID PIF-2022-07). M. Kubullek acknowledges support by the Max Planck School of Photonics

## References

[1]    Wirth A, Hassan M Th, Grguraš I, Gagnon J, Moulet A, Luu T T, Pabst S, Santra R, Alahmed Z A, Azzeer A M, Yakovlev V S, Pervak V, Krausz F and Goulielmakis E 2011 Synthesized Light Transients *Science* **334** 195–200

[2]    Yang Y, Mainz R E, Rossi G M, Scheiba F, Silva-Toledo M A, Keathley P D, Cirmi G and Kärtner F X 2021 Strong-field coherent control of isolated attosecond pulse generation *Nat Commun* **12** 6641

[3]    Rossi G M, Mainz R E, Yang Y, Scheiba F, Silva-Toledo M A, Chia S-H, Keathley P D, Fang S, Mücke O D, Manzoni C, Cerullo G, Cirmi G and Kärtner F X 2020 Sub-cycle millijoule-level parametric waveform synthesizer for attosecond science *Nat. Photonics* **14** 629–35

[4]    Veisz L, Fischer P, Vardast S, Schnur F, Muschet A, De Andres A, Kaniyeri S, Li H, Salh R, Ferencz K, Nagy G N and Kahaly S 2025 Waveform-controlled field synthesis of sub-two-cycle pulses at the 100 TW peak power level *Nat. Photon.* **19** 1013–9

[5]    Pi Z, Kim H Y and Goulielmakis E 2025 Synthesis of single-cycle pulses based on a Yb:KGW laser amplifier *Optica, OPTICA* **12** 296–301

[6]    Chipperfield L E, Robinson J S, Tisch J W G and Marangos J P 2009 Ideal Waveform to Generate the Maximum Possible Electron Recollision Energy for Any Given Oscillation Period *Phys. Rev. Lett.* **102** 063003

[7]    Haessler S, Balčiūnas T, Fan G, Chipperfield L E and Baltuška A 2015 Enhanced multi-colour gating for the generation of high-power isolated attosecond pulses *Sci Rep* **5** 10084

[8]    Jin C, Wang G, Wei H, Le A-T and Lin C D 2014 Waveforms for optimal sub-keV high-order harmonics with synthesized two- or three-colour laser fields *Nat Commun* **5** 4003

[9]    Dong D, Wang H, Xue B, Imasaka K, Kanda N, Fu Y, Nabekawa Y and Takahashi E J 2025 Perturbed three-channel waveform synthesizer for efficient isolated attosecond pulse generation and characterization *Opt. Lett., OL* **50** 1461–4

[10] Li B, Tang X, Wang K, Zhang C, Guan Z, Wang B, Lin C D and Jin C 2022 Generation of Intense Low-Divergence Isolated Soft-X-Ray Attosecond Pulses in a Gas-Filled Waveguide Using Three-Color Synthesized Laser Pulses *Phys. Rev. Appl.* **18** 034048

[11] He L, Yuan G, Wang K, Hua W, Yu C and Jin C 2019 Optimization of temporal gate by two-color chirped lasers for the generation of isolated attosecond pulse in soft X rays *Photon. Res., PRJ* **7** 1407–15

[12] Scheiba F, Mainz R E, Rossi G M, Silva-Toledo M A, Kubullek M and Kärtner F X 2023 Soft X-ray continua generation via HHG with sub-cycle synthesized laser fields *Ultrafast Optics 2023 - UFOXIII (2023), paper W2.4* (Optica Publishing Group)

[13] Tang X, Li B, Wang K, Yin Z, Zhang C, Guan Z, Wang B, Lin C D and Jin C 2023 Role of the Porras factor in phase matching of high-order harmonic generation driven by focused few-cycle laser pulses *Opt. Lett., OL* **48** 3673–6

[14] Raab A-K, Schmoll M, Simpson E R, Redon M, Fang Y, Guo C, Viotti A-L, Arnold C L, L'Huillier A and Mauritsson J 2024 Highly versatile, two-color setup for high-order harmonic generation using spatial light modulators *Review of Scientific Instruments* **95** 073002

[15] Shim B and Chang Z 2025 Attosecond X-ray pulse generation by waveform control of mid-infrared laser *Opt. Express, OE* **33** 31465–81

[16] Rajpoot R and Takahashi E J 2025 Systematic analysis of an attosecond pulse generation by a subcycle laser field *Phys. Rev. Res.* **7**

[17] Bódi B, Balogh E, Tosa V, Goulielmakis E, Varjú K and Dombi P 2016 Attosecond pulse generation with an optimization loop in a light-field-synthesizer *Opt. Express, OE* **24** 21957–62

[18] Zhang C, Tang X, Li B, Yin Z, You J, Wang B, Li X, Lin C-D and Jin C 2025 Probing Spatiotemporal Reshaping of Three-Color Laser Waveforms in a Gas Medium via High-Order Harmonic Generation Spectroscopy *ACS Photonics*

[19] Yakovlev V S and Agarwal M 2025 Unveiling Photoinjection Dynamics 2025 *Conference on Lasers and Electro-Optics Europe & European Quantum Electronics Conference (CLEO/Europe-EQEC)*

[20] Kubullek M, Silva-Toledo M A, Mainz R E, Scheiba F, de Q. Garcia R, Ritzkowsky F, Rossi G M and Kärtner F X 2025 Complete Electric Field Characterization of Ultrashort Multicolor Pulses *Ultrafast Science* **5** 0081

[21] Mamaikin M, Ridente E, Krausz F and Karpowicz N 2024 Spatiotemporal electric-field characterization of synthesized light transients *Optica, OPTICA* **11** 88–93

## 5. Optical attosecond pulse generation through soliton self-compression in hollow-core fibres

**John C. Travers[1]\*, Christian Brahms[1]**

[1] School of Engineering and Physical Sciences, Heriot-Watt University, Edinburgh, EH14 4AS, UK

j.travers@hw.ac.uk

**Status**

The ability to generate and shape sub-femtosecond electric-field transients in the ultraviolet to infrared region has unlocked several transformative research directions. Because they enable attosecond driving of field-sensitive processes at non-ionising photon energies, a primary application of such *optical attosecond pulses* is in cutting-edge ultrafast science [1]. For instance, tailored sub-femtosecond driving fields can enable petahertz-speed electronics through all-optical switching of currents in dielectrics and semiconductors [2]. The fully temporally compressed supercontinuum which optical attosecond pulses represent can also be used as a spectroscopic probe rather than a driving field, as demonstrated in the study of ultrafast phase transitions [3].

The first route to optical attosecond pulse generation was light-field synthesis [4], which consists of extreme nonlinear spectral broadening followed by separate phase compensation in multiple spectral channels. This approach overcomes the bandwidth limits of chirped mirrors and enables tailoring of the waveform. A second approach has recently emerged: soliton-effect self-compression in gas-filled hollow-core fibres [5]. This process combines nonlinear spectral broadening with continuous phase compensation from the anomalous dispersion of the waveguide. The dispersion is smooth and broadband from the vacuum ultraviolet to the mid-infrared, directly supporting multiple octaves of bandwidth and enabling a simpler and direct route to the generation of optical attosecond pulses. Soliton self-compression is often accompanied by resonant dispersive-wave (RDW) emission, which provides high-efficiency generation of wavelength-tuneable few-cycle pulses in the vacuum and deep-ultraviolet spectral region.

Optical soliton self-compression was pioneered in solid-core optical fibres [6]. However, the large higher-order dispersion and lack of ultraviolet guidance in such fibres inherently prohibit the generation of sub-femtosecond pulses. Power scaling was also limited due to optical damage. Gas-filled hollow-core fibres overcome these limitations and have been widely used for few-cycle pulse compression [7]. Soliton dynamics were demonstrated in hollow photonic band-gap fibres [8] and anti-resonant guiding optical fibres [9,10]. Recently, soliton self-compression in hollow-capillary fibres enabled the generation of sub-femtosecond electric field transients with 40-gigawatt peak power [5], rivalling those that could be produced in light-field synthesizers for the first time, along with tuneable high-energy RDW emission spanning from 110 nm to 350 nm. Since then, considerable effort has been made to further extend this approach, and several applications have been demonstrated [3,11]. This technology is set to enable a new class of experiments in ultrafast science, pushing the frontiers of our understanding and control of the ultrafast structure and dynamics of matter.

**Current and future challenges**

In the initial demonstration, 1.2 fs pulse envelopes were generated. This corresponded to a 412 attosecond field transient. However, the pulse characterisation relied on frequency-resolved optical gating and back-propagation. Recently, several experiments have addressed the major challenge of field-resolved characterisation. Optical field transients as short as 352 attoseconds have been directly measured [12] when driving with a Ti:sapphire laser and field-resolved measurements of pulses shorter than a single-cycle of the drive wavelength have also been obtained when driving with an optical parametric amplifier [13]. Both results were phase stable. Single-femtosecond-scale pulses with more than 1 mJ of energy have recently been directly measured indicating that the peak power was at the terawatt scale [14]. There have also been major advances in the characterisation of the few-femtosecond ultraviolet RDW pulses [15,16]. However, none of these approaches provide single-shot and/or spatio-temporal measurements of the pulse—performing such a characterisation of a multi-octave sub-femtosecond pulse is a major challenge.

The result in Ref. [14] represent significant energy up-scaling, opening new applications towards relativistic nonlinear optics with transients. Further upscaling with facility-class lasers is an outstanding challenge which could significantly broaden access to such unique regimes. Down-scaling to much lower energy is also attractive as a means to shrink and simplify sources of optical attosecond transients and few-cycle ultraviolet pulses. Apart from promoting much wider access by reducing space, cost and complexity, this will also enable frequency-comb operation. We can envisage frequency combs in the deep and vacuum ultraviolet, along with phase-stable, high repetition-rate sub-femtosecond multi-octave supercontinua.

High pulse repetition rates (hundreds of kHz to MHz) are crucial for many applications, including frequency combs and especially spectroscopy applications that are fluence-limited but require improved accumulated signal strengths. High-energy soliton self-compression sources in hollow capillaries have reached the 10's to 100's of kHz level by moving to ytterbium-doped pump lasers [17]. In anti-resonant fibres, multiple MHz have been achieved for compression to single cycle pulses [18] (but not yet attosecond transients) and for deep ultraviolet generation [19]. Apart from the usual damage and thermal issues to be resolved at high average powers, a major ongoing challenge is due to ionisation of the gas inside the fibre, which can lead to heating and long-lived thermal and gas density depletion effects which accumulate over successive pulses [20]. The resulting refractive-index changes reach a quasi-steady-state and alter the waveguide modes and dispersion, disrupting the soliton dynamics.

Most work to date has been based on linearly polarised, fundamental mode, and simple spatio-temporal fields. Interesting possibilities arise if attosecond optical transients can be transformed into more complex spatial structures and polarisation states. This could open new avenues in chiral-sensitive spectroscopy and strong-field physics and has not been widely investigated.

**Advances in science and technology to meet challenges**

Addressing these challenges effectively requires targeted advances in both fundamental scientific understanding and the development of enabling technologies.

Many advanced applications, such as attosecond spectroscopy, demand exceptional stability in pulse energy, interferometric timing, and the carrier-envelope phase (CEP). Ensuring that these properties are preserved throughout the highly nonlinear compression process is a crucial requirement for transitioning soliton-based sources from specialized physics demonstrations to workhorse scientific instruments. Beyond these specific parameters, the overall system stability is paramount. Attosecond-scale timing necessitates nanometre-scale optical path length stability over many hours, a demanding requirement that is challenged by thermal drifts and mechanical vibrations across the entire beamline. When combined with the peculiarities of soliton-based light sources, such as direct-to-vacuum delivery from a hollow-core fibre and extreme spectral bandwidths, the demands of attosecond science become all the more challenging. Much more detailed characterization and mitigation of such instabilities are required before these sources can be routinely used.

Attosecond field transients at optical frequencies inherently correspond to multi-octave supercontinua; this is one of their outstanding and most useful features. However, it also introduces challenges, as managing such broad spectra with minimal dispersion is currently beyond all but the simplest optical coatings, such as thinly coated aluminium. While a perfect optic for such sources is likely unattainable, considerable progress is needed on more realistic goals: developing more broadband, low-dispersion ultraviolet optics, improving damage resistance for ultrabroadband metal optics, and creating lower-dispersion polarization manipulation components. Advances in meta-optics [21] and machine learning [22] may help achieve these goals.

There are numerous challenges related to pulse quality and contrast. Soliton self-compression inherently results in a femtosecond-scale pre-pulse pedestal due to self-steepening. Furthermore, some strong-field applications require excellent contrast on the picosecond to nanosecond timescale. Therefore, the development of contrast-enhancing soliton self-compression techniques is important. More generally, spatio-temporal control of attosecond transients, including the role of polarization, could open new applications. While low-dispersion, multi-octave polarization optics are not yet feasible, they are also unnecessary, as the soliton dynamics preserve simple polarization states. Circularly [23] and radially polarized self-compression has already been demonstrated. A bigger challenge is to leverage this capability to generate more advanced polarization states.

For repetition-rate scaling, a deeper theoretical analysis of longitudinal fibre effects and long-term dynamics is required, alongside testing possible mitigations beyond current techniques, such as gas gradients and gas mixtures. This could include running systems in burst-mode operation or employing optical cooling techniques.

**Concluding remarks**

Soliton self-compression in gas-filled hollow-core fibres has transitioned from a promising concept to a demonstrated reality, offering a robust and accessible route to optical attosecond pulses. With the successful generation and characterisation of sub-femtosecond transients and high-energy dispersive waves now realized in the laboratory, these sources are effectively ready for deployment in transformative applications ranging from petahertz electronics to time-resolved spectroscopy. However, this maturity does not signal the end of fundamental development; rather, it highlights a new set of intriguing challenges that are research frontiers in their own right.

Addressing the limitations of repetition-rate scaling, for instance, requires a deeper understanding of cumulative plasma and thermal dynamics within the waveguide. Similarly, mastering spatio-temporal and polarization control opens the door to novel chiral-sensitive interactions previously inaccessible to attosecond science. Furthermore, the push for extreme stability and single-shot field-resolved characterization demands innovation in both optical engineering and measurement methodology. Solving these problems will not only refine these sources into reliable workhorse instruments but also uncover new regimes of nonlinear optics. Ultimately, the continued evolution of this technology promises to democratize access to the attosecond timescale, driving the next generation of discoveries in the structure and dynamics of matter.

## Acknowledgements

JCT is funded by the European Research Council (ERC) under the European Union's Horizon 2020 research and innovation programme Consolidator Grant agreement XSOL no. 101001534, the Institution of Engineering and Technology (IET) through the IET A F Harvey Engineering Research Prize, and the Royal Academy of Engineering through a Chair in Emerging Technologies. CB is funded by the European Research Council (ERC) under the European Union's Horizon Europe research and innovation programme Starting Grant agreement FASTER no. 101161675 and by a Royal Academy of Engineering Research Fellowship (RF/202122/21/133).

## References

[1]  Calegari F and Martin F 2023 Open questions in attochemistry *Commun Chem* **6** 1–5

[2]  Hui D, Alqattan H, Zhang S, Pervak V, Chowdhury E and Hassan M Th 2023 Ultrafast optical switching and data encoding on synthesized light fields *Sci. Adv.* **9** eadf1015

[3]  Brahms C, Zhang L, Shen X, Bhattacharya U, Recasens M, Osmond J, Grass T, Chhajlany R W, Hallman K A, Haglund R F, Pantelides S T, Lewenstein M, Travers J C and Johnson A S 2025 Decoupled few-femtosecond phase transitions in vanadium dioxide *Nat Commun* **16** 3714

[4]  Wirth A, Hassan M T, Grguraš I, Gagnon J, Moulet A, Luu T T, Pabst S, Santra R, Alahmed Z A, Azzeer A M, Yakovlev V S, Pervak V, Krausz F and Goulielmakis E 2011 Synthesized Light Transients *Science* **334** 195–200

[5]  Travers J C, Grigorova T F, Brahms C and Belli F 2019 High-energy pulse self-compression and ultraviolet generation through soliton dynamics in hollow capillary fibres *Nat. Photonics* **13** 547–54

[6]  Mollenauer L F, Stolen R H, Gordon J P and Tomlinson W J 1983 Extreme picosecond pulse narrowing by means of soliton effect in single-mode optical fibers *Opt. Lett.* **8** 289–91

[7]  Nisoli M, De Silvestri S and Svelto O 1996 Generation of high energy 10 fs pulses by a new pulse compression technique *Appl. Phys. Lett.* **68** 2793

[8]  Ouzounov D G, Ahmad F R, Müller D, Venkataraman N, Gallagher M T, Thomas M G, Silcox J, Koch K W and Gaeta A L 2003 Generation of Megawatt Optical Solitons in Hollow-Core Photonic Band-Gap Fibers *Science* **301** 1702–4

[9]  Im S-J, Husakou A and Herrmann J 2010 High-power soliton-induced supercontinuum generation and tunable sub-10-fs VUV pulses from kagome-lattice HC-PCFs *Opt. Express* **18** 5367–74

[10]  Joly N Y, Nold J, Chang W, Hölzer P, Nazarkin A, Wong G K L, Biancalana F and Russell P St J 2011 Bright Spatially Coherent Wavelength-Tunable Deep-UV Laser Source Using an Ar-Filled Photonic Crystal Fiber *Phys. Rev. Lett.* **106** 203901

[11] Jackson S L, Prentice A W, Bertram L, Hutton I, Kotsina N, Brahms C, Sparling C, Travers J C, Kirrander A, Paterson M J and Townsend D 2025 Decoupling structural molecular dynamics from excited state lifetimes using few-femtosecond ultraviolet resonant dispersive waves *Nature Communications* **16** 9986

[12] Heinzerling A M, Tani F, Agarwal M, Yakovlev V S, Krausz F and Karpowicz N 2025 Field-resolved attosecond solitons *Nat. Photon.* 1–6

[13] Utrio Lanfaloni V, Vismarra F, Ardali E, Monahan N, Wiese J, Kopp T, Ardana-Lamas F, Fazio G, Redaelli L, Pertot Y, Zinchenko K, Balčiūnas T and Wörner H J 2025 Self-compressed waveform-stable light transients enabling water-window attosecond spectroscopy *Nat. Photon.*

[14] Kotsina N, Heynck M, Nordmann J, Gebhardt M, Grigorova T, Brahms C and Travers J C 2025 Extreme Soliton Dynamics for Terawatt-Scale Optical Attosecond Pulses and 30 GW-Scale Sub-3 fs Far-ultraviolet Pulses *Conference on Lasers and Electro-Optics/Europe (CLEO/Europe 2025) and European Quantum Electronics Conference (EQEC 2025) (2025), paper jpd_2_3* (Optica Publishing Group)

[15] Reduzzi M, Pini M, Mai L, Cappenberg F, Colaizzi L, Vismarra F, Crego A, Lucchini M, Brahms C, Travers J C, Borrego-Varillas R and Nisoli M 2023 Direct temporal characterization of sub-3-fs deep UV pulses generated by resonant dispersive wave emission *Opt. Express, OE* **31** 26854–64

[16] Andrade J R C, Kretschmar M, Danylo R, Carlström S, Witting T, Mermillod-Blondin A, Patchkovskii S, Ivanov M Y, Vrakking M J J, Rouzée A and Nagy T 2025 Temporal characterization of tunable few-cycle vacuum ultraviolet pulses *Nature Photonics* **19** 1240–6

[17] Brahms C and Travers J C 2023 Efficient and compact source of tuneable ultrafast deep ultraviolet laser pulses at 50 kHz repetition rate *Opt. Lett., OL* **48** 151–4

[18] Köttig F, Schade D, Koehler J R, Russell P S J and Tani F 2020 Efficient single-cycle pulse compression of an ytterbium fiber laser at 10 MHz repetition rate *Opt. Express, OE* **28** 9099–110

[19] Köttig F, Tani F, Biersach C M, Travers J C and Russell P S J 2017 Generation of microjoule pulses in the deep ultraviolet at megahertz repetition rates *Optica, OPTICA* **4** 1272–6

[20] Koehler J R, Köttig F, Trabold B M, Tani F and Russell P St J 2018 Long-Lived Refractive-Index Changes Induced by Femtosecond Ionization in Gas-Filled Single-Ring Photonic-Crystal Fibers *Phys. Rev. Applied* **10** 064020

[21] Zhang C, Chen L, Lin Z, Song J, Wang D, Li M, Koksal O, Wang Z, Spektor G, Carlson D, Lezec H J, Zhu W, Papp S and Agrawal A 2024 Tantalum pentoxide: a new material platform for high-performance dielectric metasurface optics in the ultraviolet and visible region *Light: Science & Applications* **13** 23

[22] Chattopadhyay U, Carstens F, Steinecke M, Kellermann T, Wienke A, Hartl I, Ay N, Heyl C M and Tünnermann H 2025 Efficient optical coating design using an autoencoder-based neural network model *J. Phys. Photonics* **8** 015007

[23] Lekosiotis A, Brahms C, Belli F, Grigorova T F and Travers J C 2021 Ultrafast circularly polarized pulses tunable from the vacuum to deep ultraviolet *Opt. Lett., OL* **46** 4057–60

## 6. Direct Generation of Bright Isolated Attosecond Pulses from Post-Compressed Yb Lasers

### Ming-Chang Chen

Institute of Photonics Technologies, National Tsing Hua University, Hsinchu 30013, Taiwan

mingchang@mx.nthu.edu.tw

**Status**

Since the first demonstration of isolated attosecond pulses (IAPs) via high-order harmonic generation (HHG) in 2001 [1], attosecond science has advanced rapidly, offering unprecedented tools to probe electron motion on its natural timescale. Early developments were driven by Ti:sapphire lasers operating around 800 nm, which enabled few-cycle, carrier-envelope-phase (CEP) stabilized driving pulses and established IAP generation as a practical reality [2–4] (see Fig. 1). Temporal gating methods such as polarization gating, double optical gating, and later the attosecond lighthouse effect became key to confining the HHG emission to a single half-cycle, producing pulses as short as 67 as with photon energies reaching beyond 100 eV [5]. These sources have been central to breakthrough studies of photoionization delays, charge migration, and correlated electron dynamics across atoms, molecules, and solids.

Around 2010, a shift occurred toward mid-infrared (MIR) optical parametric amplifiers (OPAs), which leveraged the $\lambda^2$ scaling of the HHG cutoff to reach the water-window spectral region and beyond 300 eV [6–8]. This advance enabled the generation of attosecond pulses below 50 as, with intensities exceeding a gigawatt and bandwidths spanning several hundred eV. However, conversion efficiency drops steeply with increasing wavelength, and the photon flux of IAPs remains a bottleneck for demanding applications such as attosecond pump–probe spectroscopy.

To date, most demonstrations of IAPs have relied on either Ti:sapphire or OPA-based MIR drivers, each with inherent trade-offs. Ti:sapphire lasers are mature and reliable but limited in repetition rate and scalability, whereas MIR OPAs extend the spectral reach at the expense of efficiency and system complexity. This landscape motivates the growing attention toward ytterbium (Yb)-based lasers. Their combination of excellent stability, high average power, scalability to high pulse energies, and intrinsic compatibility with high-repetition-rate operation makes them highly attractive candidates for the next generation of driving sources. Yb-based femtosecond amplifiers, including thin-disk and fiber architectures, can already deliver hundreds of watts of average power with MHz repetition rates, providing the stability and photon flux essential for translating attosecond methods from specialized laboratories to widespread scientific use. However, Yb-based femtosecond lasers typically produce longer pulse durations (several hundred femtoseconds) due to the gain bandwidth limitations. The current and future challenge lies in bridging the gap between these Yb platforms and robust IAP generation—through advanced post-compression technology, dispersion management, and CEP stabilization—paving the way for attosecond science at unprecedented flux, stability, and accessibility.

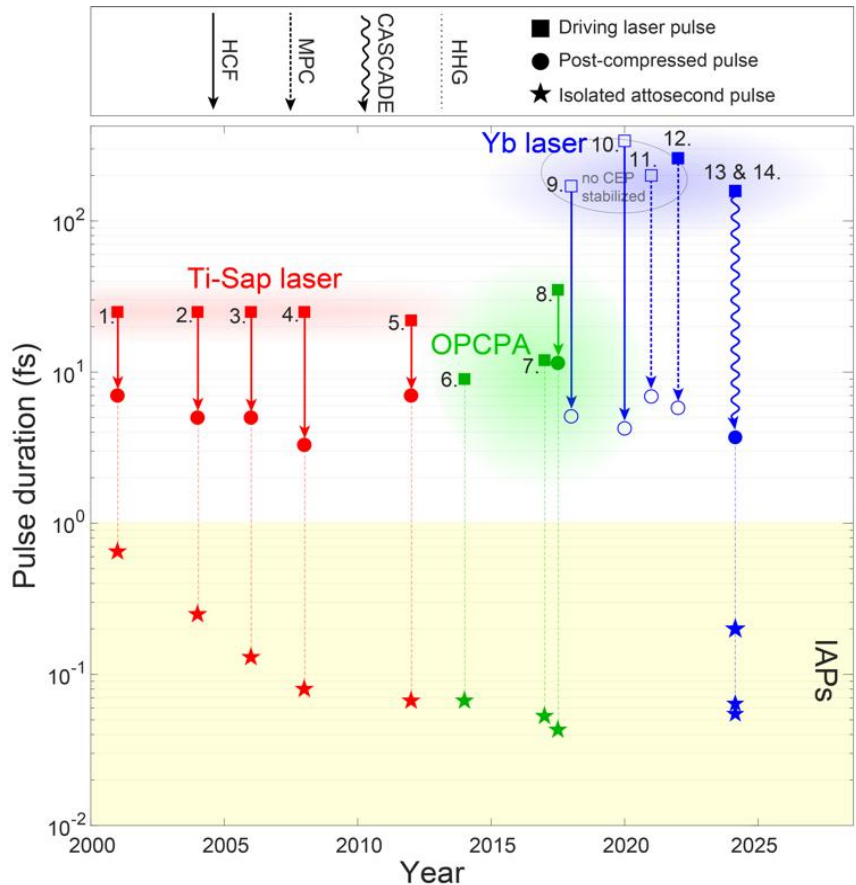

**Figure 1.** *Laser sources and post-compression techniques demonstrated for IAP generation include three types of drivers: (1) CEP-stabilized Ti:Sapphire lasers (red), (2) CEP-stabilized OPA/OPCPA driven by Ti:Sapphire or Yb lasers (green), and (3) direct CEP-stabilized Yb lasers (blue). Hollow-core fibers (HCF) have been the primary post-compression method for Ti:Sapphire and OPA/OPCPA, while multi-pass cells (MPC) and cascaded focus and compression (CASCADE) have recently emerged. For Yb lasers, the long initial pulse duration (>150 fs) necessitates >40× compression to reach few-cycle pulses; approaches such as HCF, MPC, and CASCADE have therefore been implemented. References are listed chronologically: [1–14].*

## Current and future challenges

The post-compression process is achieved through self-phase modulation (SPM) followed by dispersion correction. When an intense pulse propagates through a medium, the refractive index becomes intensity dependent, $n = n_0 + n_2 I(t,x,y)$, where $n_0$ is the refractive index, $n_2$ is the third-order nonlinear optical coefficient, and $I(t,x,y)$ is the spatiotemporal intensity distribution of the pulse with $t$ denoting time and $x,y$ the transverse coordinates. The accumulated nonlinear phase along the optical axis $(x,y=0)$ is $\Delta\phi_{NL}(t) = (2\pi/\lambda)n_2 I(t)L_{eff}$, which results in an instantaneous frequency shift, $d\Delta\phi_{NL}(t)/dt = -(2\pi/\lambda)n_2(dI(t)/dt)L_{eff}$, where $\lambda$ is the central wavelength, and $L_{eff}$ denotes the effective interaction length over which the nonlinear process accumulates. Since $n_2 > 0$ in transparent media (i.e., off-resonance), the SPM produces a positive chirp: red-shift on the leading edge and blue-shift on the trailing edge of the pulse. Temporal compression is achieved by compensating this positive chirp with the negative chirp provided by dispersive optics such as prisms, gratings, or chirped mirrors. Since the spectral broadening factor scales approximately linearly with the maximum nonlinear phase shift, it is characterized by the B-integral, $\Delta\phi_{NL,max} = B = (2\pi/\lambda)n_2 I_0 L_{eff}$, where $I_0$ is the peak intensity. For instance, compressing a 200-fs pulse (~60 cycles at 1 μm) to <30 fs requires a total accumulated B-integral of ~2.5π, while further reduction to <5 fs (~1.5 cycles, suitable for IAP generation) corresponds to total accumulated B-integrals of ~9π.

In the spatial domain, spectral broadening results in wavefront modulation and distortion. In a Gaussian beam profile at focus, a B-integral of 2π corresponds to a peak-intensity–induced nonlinear phase shift equivalent to one optical cycle at 1 μm (~3.3 fs) at the spatiotemporal center of the pulse. Achieving a total accumulated nonlinear phase shift of ~9π, as required for compressing to <5 fs, corresponds to an on-axis delay of ~15 fs (equivalent to ~4.5 μm of optical path). Accumulating such large phase modulations in a single focus gives rise to highly complex spatiotemporal distortions of the wavefront that cannot be corrected with conventional spherical optics (e.g., lenses or concave mirrors).

In the temporal domain, a large B-integral introduces high-order phase components that standard chirped mirrors, typically designed to compensate GVD and TOD, cannot fully correct, producing residual satellite pulses, reducing the fraction of energy in the main peak, and lowering the achievable peak power. At extreme intensities, plasma generation adds further nonlinear phase shifts, complicating the temporal profile.

**Advances in science and technology to meet challenges**

Reaching few-cycle durations from long Yb-based pulses requires the accumulation of very large B-integrals, yet maintaining high beam quality and compressibility is nontrivial. Overcoming spatiotemporal distortion, and coherence loss has driven the following three post-compression technologies.

Hollow-core fiber (HCF) [9,10]: Gas-filled HCFs provide long interaction lengths where the nonlinear phase accumulates gradually. Wavefront distortions remain confined to the guided mode, preserving spatial and temporal profiles. An ~1–1.5 m HCF typically accumulates and saturates at ~3π B-integral, as temporal spreading lowers the peak intensity and limits further broadening, while still supporting compression factors of ~6. To post-compress Yb lasers into the few-cycle regime, two stages are usually required, since each HCF stage must be followed by dispersion compensation to re-establish peak intensity before further accumulating nonlinear phase shift (spectral broadening). HCFs are extremely sensitive to beam pointing and alignment, requiring careful stabilization.

Multi-pass cell (MPC) [11,12]: In Herriott-type cavities, pulses undergo multiple passes through a gas medium, encountering tens of foci. A single MPC typically allows the accumulation of a B-integral of ~3π—distributed over ~20 foci (~0.15π each)—which provides sufficient spectral broadening for compression factors of ~6. The weak wavefront perturbation introduced at each focus is naturally compensated by the cavity geometry. As with HCFs, this saturation results from spatiotemporal spreading that lowers the peak intensity and limits further broadening. For Yb lasers, reaching the few-cycle regime generally requires two stages. Achieving this also depends critically on high-performance mirror coatings to preserve both reflectivity and bandwidth. Compared with HCFs, MPCs exhibit greater tolerance to beam pointing and allow more straightforward intensity scaling.

Cascaded focus and compression (CASCADE) [13]: Each stage of CASCADE consists of a long focus in a gas cell followed by dispersion compensation, typically with chirped mirrors, before the next stage. A long focus is chosen to accumulate a B-integral of ~2π while keeping the Gaussian-driven nonlinear phase shift close to a spherical wavefront, which can be corrected with standard lenses or concave mirrors. Compression to near–transform-limited after each step maximizes the efficiency of frequency generation, since the instantaneous frequency shift is proportional to the temporal intensity gradient, $dI(t)/dt$. This staged approach allows very large B-integrals to be accumulated while preserving beam quality. In practice, four such stages can provide a total accumulated nonlinear phase of ~9π, sufficient to compress >150 fs Yb pulses to ~3.5 fs (one-cycle regime), corresponding to compression factors of >40. The technique has demonstrated high efficiency (>70%) and has already enabled the generation of bright, stable IAPs when driven by a CEP-stabilized Yb:KGW laser [14] (see Fig. 2).

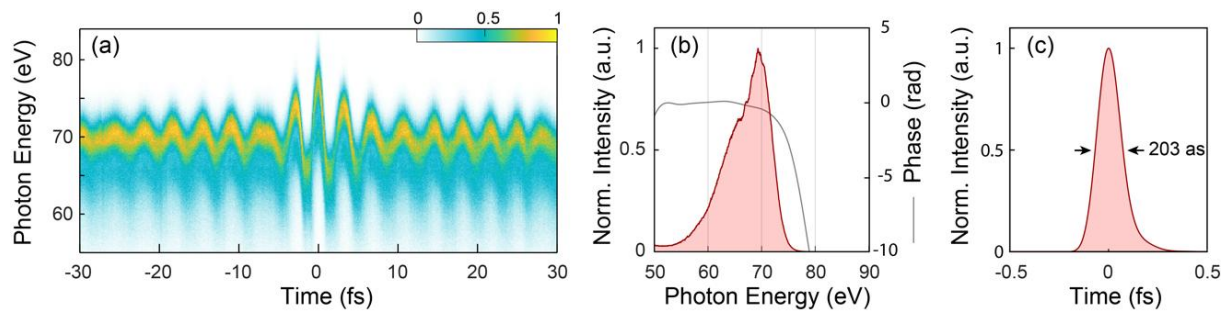

***Figure 2.*** *Attosecond pulse generation in argon driven by a post-compressed Yb laser. (a) Streaking spectrogram, (b) retrieved EUV spectrum and spectral phase, and (c) temporal profile of an isolated attosecond pulse in Ar, with a retrieved duration of about 200 as. The pulse is slightly positively chirped. Figure adapted from [14].*

## Concluding remarks

Yb lasers operating near ~1 µm have become key drivers for pumping secondary sources such as OPA or OPCPA [15], which provide long-wavelength drivers (1.5–2 µm) that extend the HHG cutoff into the water-window region, but these secondary systems typically achieve only ~10–30% conversion efficiency. In contrast, post-compression of Yb pulses is highly efficient and has demonstrated compression down to one-cycle operation, enabling direct driving of IAPs via HHG. Looking ahead, exploiting gas species with very large nonlinear coefficients ($n_2$) may red-shift the spectrum to generate long-wavelength pulses that extend the IAP cutoff into the water-window region [16,17]. Moreover, with repetition rates scalable to the MHz regime [10] and power compatibility reaching kilowatt-class Yb lasers [11], post-compression offers a promising route to high-average-power, table-top sources of IAPs. Interestingly, post-compression itself has the potential to provide a direct attosecond light source without HHG, as demonstrated by hollow-core fibers with resonant dispersive wave emission [10,18] and multi-stage plasma-based self-compression [19], which are capable of approaching—and even surpassing—the sub-femtosecond regime. Taken together, these advances position post-compression of Yb laser a robust and versatile platform for the next generation of attosecond pump–probe spectroscopy, coherent soft X-ray science, ultrafast imaging, and precision metrology.

## Acknowledgements

M.-C. C. acknowledges the support from the National Science and Technology Council, Taiwan, grant no. 113-2112-M-007-042-MY3.

## References

[1] Hentschel M, Kienberger R, Spielmann C, Reider G A, Milosevic N, Brabec T, Corkum P, Heinzmann U, Drescher M and Krausz F 2001 Attosecond metrology *Nature* **414** 509–13

[2]   Kienberger R, Goulielmakis E, Uiberacker M, Baltuska A, Yakovlev V, Bammer F, Scrinzi A, Westerwalbesloh T, Kleineberg U and Heinzmann U 2004 Atomic transient recorder *Nature* **427** 817–21

[3]   Sansone G, Benedetti E, Calegari F, Vozzi C, Avaldi L, Flammini R, Poletto L, Villoresi P, Altucci C, Velotta R, Stagira S, De Silvestri S and Nisoli M 2006 Isolated Single-Cycle Attosecond Pulses *Science* **314** 443–6

[4]   Goulielmakis E, Schultze M, Hofstetter M, Yakovlev V S, Gagnon J, Uiberacker M, Aquila A L, Gullikson E M, Attwood D T, Kienberger R, Krausz F and Kleineberg U 2008 Single-Cycle Nonlinear Optics *Science* **320** 1614–7

[5]   Zhao K, Zhang Q, Chini M, Wu Y, Wang X and Chang Z 2012 Tailoring a 67 attosecond pulse through advantageous phase-mismatch *Optics letters* **37** 3891–3

[6]   Ishii N, Kaneshima K, Kitano K, Kanai T, Watanabe S and Itatani J 2014 Carrier-envelope phase-dependent high harmonic generation in the water window using few-cycle infrared pulses *Nature Communications* **5** 3331

[7]   Li J, Ren X, Yin Y, Zhao K, Chew A, Cheng Y, Cunningham E, Wang Y, Hu S and Wu Y 2017 53-attosecond X-ray pulses reach the carbon K-edge *Nature communications* **8** 186

[8]   Gaumnitz T, Jain A, Pertot Y, Huppert M, Jordan I, Ardana-Lamas F and Wörner H J 2017 Streaking of 43-attosecond soft-X-ray pulses generated by a passively CEP-stable mid-infrared driver *Optics express* **25** 27506–18

[9]   Jeong Y-G, Piccoli R, Ferachou D, Cardin V, Chini M, Hädrich S, Limpert J, Morandotti R, Légaré F and Schmidt B E 2018 Direct compression of 170-fs 50-cycle pulses down to 1.5 cycles with 70% transmission *Scientific reports* **8** 1–6

[10]  Köttig F, Schade D, Koehler J R, Russell P S J and Tani F 2020 Efficient single-cycle pulse compression of an ytterbium fiber laser at 10 MHz repetition rate *Optics Express* **28** 9099–110

[11]  Müller M, Buldt J, Stark H, Grebing C and Limpert J 2021 Multipass cell for high-power few-cycle compression *Optics letters* **46** 2678–81

[12]  Hädrich S, Shestaev E, Tschernajew M, Stutzki F, Walther N, Just F, Kienel M, Seres I, Jójárt P and Bengery Z 2022 Carrier-envelope phase stable few-cycle laser system delivering more than 100 W, 1 mJ, sub-2-cycle pulses *Optics Letters* **47** 1537–40

[13]  Tsai M-S, Liang A-Y, Tsai C-L, Lai P-W, Lin M-W and Chen M-C 2022 Nonlinear compression toward high-energy single-cycle pulses by cascaded focus and compression *Sci. Adv.* **8** eabo1945

[14]  Chien Y-E, Tsai M-S, Liang A-Y and Chen M-C 2024 Isolated 64-attosecond pulses driven by a postcompressed Yb-laser *High Intensity Lasers and High Field Phenomena* (Optica Publishing Group) pp HTh4B-1

[15]  Kretschmar M, Tuemmler J, Schütte B, Hoffmann A, Senfftleben B, Mero M, Sauppe M, Rupp D, Vrakking M J and Will I 2020 Thin-disk laser-pumped OPCPA system delivering 4.4 TW few-cycle pulses *Optics Express* **28** 34574–85

[16]  Beetar J E, Nrisimhamurty M, Truong T-C, Nagar G C, Liu Y, Nesper J, Suarez O, Rivas F, Wu Y, Shim B and Chini M 2020 Multioctave supercontinuum generation and frequency conversion based on rotational nonlinearity *Sci. Adv.* **6** eabb5375

[17]  Dorner-Kirchner M, Shumakova V, Coccia G, Kaksis E, Schmidt B E, Pervak V, Pugzlys A, Baltuška A, Kitzler-Zeiler M and Carpeggiani P A 2023 HHG at the Carbon K-Edge Directly Driven by SRS Red-Shifted Pulses from an Ytterbium Amplifier *ACS Photonics* **10** 84–91

[18]  Kotsina N, Heynck M, Nordmann J, Gebhardt M, Grigorova T, Brahms C and Travers J C 2025 Extreme Soliton Dynamics for Terawatt-Scale Optical Attosecond Pulses and 30 GW-Scale Sub-3 fs Far-ultraviolet Pulses *European Quantum Electronics Conference* (Optica Publishing Group) p jpd_2_3

[19]  Chien Y-E, Hsieh C-W, Tsai M-S and Chen M-C 2025 Single-Cycle 2.6 fs Pulses via Two-Stage Plasma Self-Compression *The European Conference on Lasers and Electro-Optics* (Optica Publishing Group) p cf_8_3

## 7. Attosecond vortex light

**Titouan Gadeyne[1], Vartika Vishnoi[1] and Thierry Ruchon[1]***

[1] Université Paris-Saclay, CEA, LIDYL, 91191 Gif-sur-Yvette, France

thierry.ruchon@cea.fr

**Status**

Over the last three decades, vortex beams of light have earned their place among the fundamental tools of modern optics. These beams are characterized by a transverse ring-like energy distribution, centered on a singular line around which their phase winds by an integer number $\ell$ of $2\pi$ cycles, known as their topological charge. As a consequence of this helical phase structure, they transport orbital angular momentum (OAM) oriented along their propagation axis, in an amount corresponding to $\hbar\,\ell$ per photon. This property has sparked their application in tasks as diverse as optical manipulation, information encoding or high-dimensional quantum experiments.

While long restricted to the visible or infrared domains, extreme-ultraviolet (XUV) or X-ray vortex beams have recently entered the scene as new tools for spectroscopy.

Pioneering experiments have demonstrated vortex X-ray light to work as a probe of the handedness of chiral molecules [15] or of mesoscopic magnetic structures [3,11], thus identifying so-called helical dichroisms (HD) wherein a sample responds differently to light with opposite OAM sign. Furthermore, *ultrashort* XUV vortices are just starting to showcase their potential for time-resolved studies: magnetic HD measurements recently revealed picosecond dynamics in magnetic vortices 4] and HD has been proposed as a probe of femtosecond evolutions of chirality [8] or electronic coherences [19] in molecules. These developments motivate the extension of short-wavelength vortex sources down to the attosecond scale.

While femtosecond XUV pulses with OAM can now be accessed using diffractive optics or advanced undulator setups at free-electron laser facilities, high-harmonic generation (HHG) remains a most promising route towards attosecond vortex pulses. Owing to the coherent nature of the process, the phase structure of the driving laser gets imprinted onto the XUV beams: HHG driven by a vortex beam of charge $\ell$ thus results in vortex harmonics with charges $q\ell$ varying linearly with the harmonic order q. In this case, each single harmonic forms a vortex pulse with well-defined topological charge, but their coherent superposition was shown to form a so-called *light spring* [7,5]. This structure is shown in fig. 1(a) and can be viewed as an azimuthally-delayed attosecond pulse train. Alternatively, trains of successive vortex pulses of attosecond duration Fig.1(b) were recently devised using a two-beam HHG setup, in which the two drivers have opposite circular polarization states as well as opposite topological charges [2].

**Current and future challenges**

While the metrology of femtosecond optical pulses has tremendously improved over the past decades [1], attosecond vortex pulses (and more generally, ultrashort XUV structured light) currently face a lack of precise space-time characterization methods in this regime. Indeed, the staple pulse measurement methods of attosecond science (such as RABBITT or streaking) typically integrate signal originating from the whole beam focus, and are thereby largely blind to imperfections like space-time couplings or astigmatism. As a result, it remains unclear how close experimental XUV vortices really get to the idealized structures of Fig. 1. To our knowledge, the sole experimental attempt at attosecond metrology of a vortex structure is the characterization of a light spring in [5]: in this study a RABBITT measurement was performed, in which interferences could only be observed when the dressing beam itself carried OAM, confirming the q$\ell$ topological charge scaling of the harmonics. The emitted electron wavepacket could then be reconstructed (Fig. 2(a)), assuming ideal azimuthal phases for the harmonics. This trailblazing measurement remains far from a precise three-dimensional measurement of the pulse.

In fact, wavefront-sensing measurements [12] later revealed the highly multimodal nature of the vortex beams produced by HHG: as shown in (Fig. 2(b)), these are more accurately described as a sum of many vortex beams, with a broad distribution of topological charges around q$\ell$. Recent ptychographic reconstructions of the intensity and phase of high-harmonic vortices [13], have also emphasized the importance of aberrations on the generated beams after refocusing.

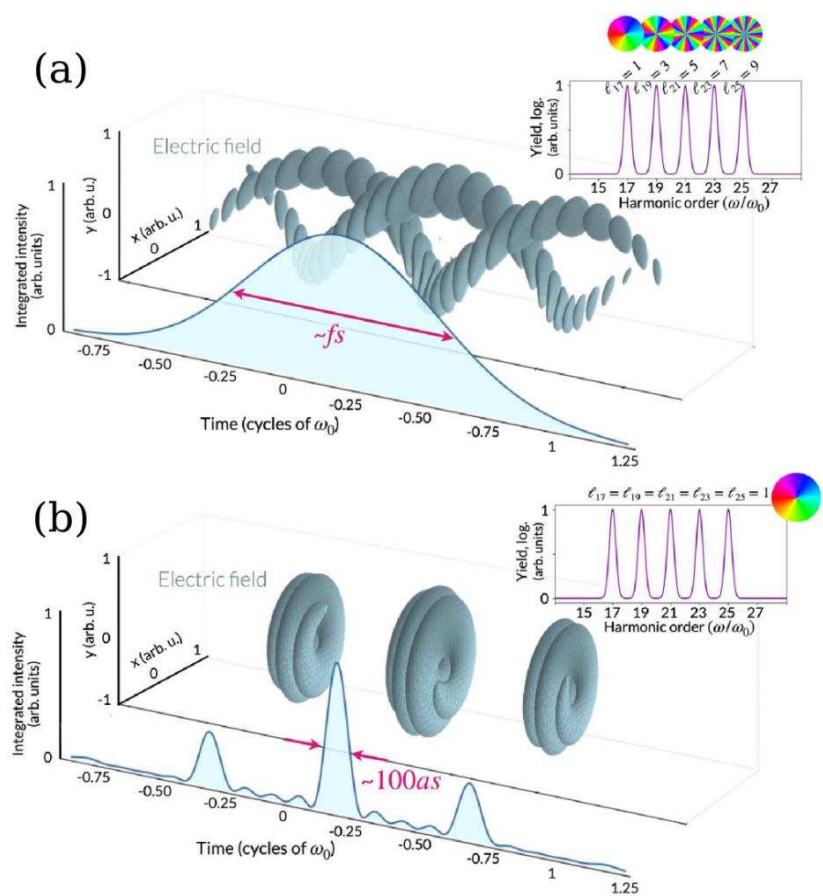

**Figure 1.** *Volumes enclosing positive electric field values for two ultrashort vortex structures. (a) attosecond light spring resulting from the interference of five harmonics with increasing topological charges (see inset). (b) Train of attosecond vortex pulses from the interference of harmonics with the same charge (see inset). Adapted from [2].*

Working towards more accurate characterization standards will be essential to future HD experiments, which ideally require pure OAM modes, and identical intensity profiles for the $\ell$ and -$\ell$ beams. From a practical perspective, a notable challenge lies in the reproducible switching of the OAM value, which typically involves changing a phase plate and realigning the beam and sample. Another difficulty is the need for tight focusing of the vortices: for a HD signal to emerge, phase gradients must be felt over a scale comparable to that of the samples of interest, which can range from sub-micron magnetic textures like skyrmions down to molecules. This requirement only reinforces the challenge surrounding near-field characterization of these pulses.

**Advances in science and technology to meet challenges**

Coupling wavefront-sensing or ptychography techniques to experiments sensitive to the harmonic spectral phase, full space-time measurements of attosecond vortex structures could be envisioned, extending recent efforts towards three-dimensional metrology of attosecond pulses [18] to cases with more complex topologies. As these approaches rely on scanning the position of diffractive elements and time delays, devising less time-consuming or even single-shot imaging methods currently stands as a major challenge, especially if those are to be used to optimize the generated pulses in real time. To this end, the development of compact, high-resolution XUV cameras or wavefront sensors able to be moved inside a vacuum chamber may also revolutionize in-situ beam diagnostics. As of today, the feasibility of spatially-resolved RABBITT or streaking-like measurements remains elusive, and calls for creative protocols.

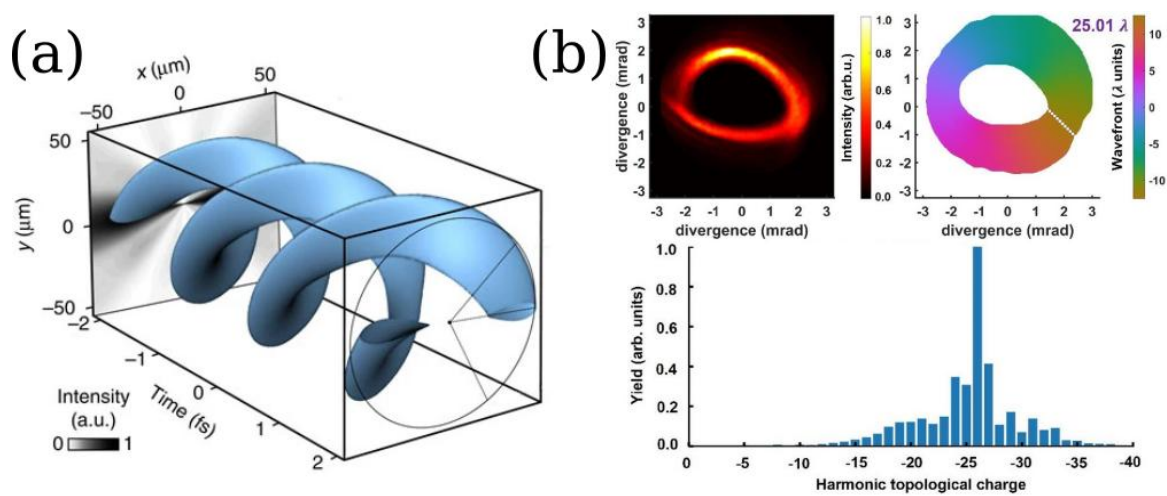

**Figure 2.** *(a) Iso-intensity contour of an electron wavepacket ionized by an attosecond light spring, reconstructed from experimental intensity profiles and spectral phases, assuming an ideal $q\ell$ vortex phase for each harmonic order. Reproduced from [5]. (b) Experimental intensity and phase front of the 25th harmonic of a $\ell$ =-1 driving laser, measured with a Hartmann wavefront sensor. The bar graph shows its decomposition on vortex modes of different topological charges. Adapted from [16].*

Research in several directions could expand the toolkit for attosecond vortex pulse synthesis. First, it is believed that the scheme of [2] could produce *isolated* vortex atto pulses if driven by a sufficiently short laser source. Direct post-compression of a vortex laser pulse in a hollow-core fiber is one viable route to such intense, few-cycle vortex pulses [6]. Alternatively, innovative post-shaping methods can be envisioned to shape plain attosecond pulses into vortex modes. Indeed, the diffractive optics used thus far such as spiral zone plates typically have low conversion efficiency and are highly chromatic. As an alternative, spiral phase mirrors with a helically-shaped surface have been theoretically proposed [9]. Meanwhile, magnetic metamaterials are opening a promising route to reconfigurable X-ray optical elements, and have already been used to produce an on/off OAM switch activated by a temperature or magnetic field change [17].

Last but not least, theoretical investigations of the interaction of XUV vortices with matter have remained few and far between, and the possible targets and protocols have likely not been exhausted. Beyond fundamental considerations, it is desirable that future theoretical predictions of HD take into account experimental subtleties which can importantly affect the magnitude of the expected signals: these include the averaging effect from having a sample of emitters distributed over the transverse plane of the vortex, as well as the sensitivity of HD signals to unavoidable modal imperfections and asymmetries between the $\ell$ and -$\ell$ pulses.

**Concluding remarks**

The ability to control the OAM of short-wavelength light is expanding the possibilities of ultrafast spectroscopy. In order for this tool to reach into the attosecond regime, specific metrology methods and improved generation schemes are urgently needed. These advances will be elemental to the application of ever more finely structured wavepackets, such as *spatiotemporal* optical vortices (see Chapter 8. Structuring Attosecond Light and High Harmonics in Space and Time), but also self-torqued pulses [14], toroidal vortices or optical hopfions [10].

**Acknowledgements**

This work is part of the TORNADO and ULTRA-FAST projects of PEPR LUMA and was supported by the French National Research Agency, as a part of the France 2030 program, under grant ANR-23-EXLU-0004 and ANR-23-EXLU-0002. This work was also supported by the Indo-French Centre for the Promotion of Advanced Research – CEFIPRA under grand Project No. 7104-1.

**References**

[1]  Benjamín Alonso, Andreas Döpp, and Spencer W. Jolly. Space–time characterization of ultrashort laser pulses: A perspective. *APL Photonics*, 9(7), July 2024.

[2]  Alba de las Heras, David Schmidt, Julio San Román, Javier Serrano, Jonathan Barolak, Bojana Ivanic, Cameron Clarke, Nathaniel Westlake, Daniel E. Adams, Luis Plaja, Charles G. Durfee, and Carlos Hernández-García. Attosecond vortex pulse trains. *Optica*, 11(8):1085, August 2024.

[3]  Mauro Fanciulli, Matteo Pancaldi, Emanuele Pedersoli, Mekha Vimal, David Bresteau, Martin Luttmann, Dario De Angelis, Primož Rebernik Ribic, Benedikt Rösner, Christian David, Carlo Spezzani, Michele Manfredda, Ricardo Sousa, Ioan-Lucian Prejbeanu, Laurent Vila, Bernard Dieny, Giovanni De Ninno, Flavio Capotondi, Maurizio Sacchi, and Thierry Ruchon. Observation of magnetic helicoidal dichroism with extreme ultraviolet light vortices. *Physical Review Letters*, 128(7):077401, February 2022.

[4]  Mauro Fanciulli, Matteo Pancaldi, Anda-Elena Stanciu, Matthieu Guer, Emanuele Pedersoli, Dario De Angelis, Primož Rebernik Ribic, David Bresteau, Martin Luttmann, Pietro Carrara, Arun Ravindran, Benedikt Rösner, Christian David, Carlo Spezzani, Michele Manfredda, Ricardo Sousa, Laurent Vila, Ioan Lucian Prejbeanu, Liliana D. Buda-Prejbeanu, Bernard Dieny, Giovanni De Ninno, Flavio Capotondi, Thierry Ruchon, and Maurizio Sacchi. Magnetic vortex dynamics probed by time-resolved magnetic helicoidal dichroism. *Physical Review Letters*, 134(15):156701, April 2025.

[5]  R. Géneaux, A. Camper, T. Auguste, O. Gobert, J. Caillat, R. Taïeb, and T. Ruchon. Synthesis and characterization of attosecond light vortices in the extreme ultraviolet. *Nature Communications*, 7:12583, Aug 2016.

[6]  Matthieu Guer, Martin Luttmann, Jean-François Hergott, Fabien Lepetit, Olivier Tcherbakoff, Thierry Ruchon, and Romain Géneaux. Few-cycle optical vortices for strong-field physics. *Optics Letters*, 49(1):93, December 2023.

[7]  Carlos Hernández-Garc, Antonio Picón, Julio San Román, and Luis Plaja. Attosecond extreme ultraviolet vortices from high-order harmonic generation. *Phys. Rev. Lett.*, 111:083602, Aug 2013.

[8]  Xiang Jiang, Yeonsig Nam, Jérémy R. Rouxel, Haiwang Yong, and Shaul Mukamel. Time-resolved enantiomer-exchange probed by using the orbital angular momentum of x-ray light. *Chemical Science*, 14(40):11067–11075, 2023.

[9]  Sunwoo Lee, Dong Uk Kim, Ji Yong Bae, Ilkyu Han, Sangwon Hyun, Hwan Hur, Kye-Sung Lee, Ki Soo Chang, Woo-Jong Yeo, Minwoo Jeon, Hwan-Jin Choi, Mincheol Kim, Jangwoo Kim, Il Woo Choi, Soojong Pak, and I Jong Kim. Generation of wavelength- and orbital angular momentum-tunable extreme-ultraviolet vortex beams using a spiral phase mirror. *Optics Communications*, 570:130909, November 2024.

[10] Zijian Lyu, Yiqi Fang, and Yunquan Liu. Formation and controlling of optical hopfions in high harmonic generation. *Physical Review Letters*, 133(13):133801, September 2024.

[11] Margaret R. McCarter, Ahmad I. U. Saleheen, Arnab Singh, Ryan Tumbleson, Justin S. Woods, Anton S. Tremsin, Andreas Scholl, Lance E. De Long, J. Todd Hastings, Sophie A. Morley, and Sujoy Roy. Antiferromagnetic real-space configuration probed by dichroism in scattered x-ray beams with orbital angular momentum. *Physical Review B*, 107(6):l060407, February 2023.

[12] Alok Kumar Pandey, Alba de las Heras, Tanguy Larrieu, Julio San Román, Javier Serrano, Luis Plaja, Elsa Baynard, Moana Pittman, Guillaume Dovillaire, Sophie Kazamias, Carlos Hernández-Garcá, and Olivier Guilbaud. Characterization of extreme ultraviolet vortex beams with a very high topological charge. *ACS Photonics*, 9(3):944–951, February 2022.

[13] Antonios Pelekanidis, Fengling Zhang, Kjeld S. E. Eikema, and Stefan Witte. Generation dynamics of broadband extreme ultraviolet vortex beams. *ACS Photonics*, 12(3):1638–1649, February 2025.

[14] Laura Rego, Kevin M. Dorney, Nathan J. Brooks, Quynh L. Nguyen, Chen-Ting Liao, Julio San Román, David E. Couch, Allison Liu, Emilio Pisanty, Maciej Lewenstein, Luis Plaja, Henry C. Kapteyn, Margaret M. Murnane, and Carlos Hernández-Garcá. Generation of extreme-ultraviolet beams with time-varying orbital angular momentum. *Science*, 364(6447):eaaw9486, jun 2019.

[15] Jérémy R. Rouxel, Benedikt Rösner, Dmitry Karpov, Camila Bacellar, Giulia F. Mancini, Francesco Zinna, Dominik Kinschel, Oliviero Cannelli, Malte Oppermann, Cris Svetina, Ana Diaz, Jérôme Lacour, Christian David, and Majed Chergui. Hard x-ray helical dichroism of disordered molecular media. *Nature Photonics*, 16(8):570–574, July 2022.

[16] F. Sanson, A. K. Pandey, I. Papagiannouli, F. Harms, G. Dovillaire, E. Baynard, J. Demailly, O. Guilbaud, B. Lucas, O. Neveu, M. Pittman, D. Ros, M. Richardson, E. Johnson, W. Li, Ph. Balcou, and S. Kazamias. Highly multimodal structure of high topological charge extreme ultraviolet vortex beams. *Optics Letters*, 45(17):4790, aug 2020.

[17] Justin S. Woods, Xiaoqian M. Chen, Rajesh V. Chopdekar, Barry Farmer, Claudio Mazzoli, Roland Koch, Anton S. Tremsin, Wen Hu, Andreas Scholl, Steve Kevan, Stuart Wilkins, Wai-Kwong Kwok, Lance E. De Long, Sujoy Roy, and J. Todd Hastings. Switchable x-ray orbital angular momentum from an artificial spin ice. *Physical Review Letters*, 126(11), mar 2021.

[18] Mingdong Yan, Yaodan Hu, Zijuan Wei, and Zhengyan Li. 3d characterization of spatiotemporally coupled high harmonic attosecond pulses. *Laser & Photonics Reviews*, 19(9), February 2025.

[19] Haiwang Yong, Jérémy R. Rouxel, Daniel Keefer, and Shaul Mukamel. Direct monitoring of conical intersection passage via electronic coherences in twisted x-ray diffraction. *Physical Review Letters*, 129(10):103001, August 2022.

## 8. Structuring Attosecond Light and High Harmonics in Space and Time

**Rodrigo Martín-Hernández[1,2]\*, David Marco[1,2], Chen-Ting Liao[3], Luis Plaja[1,2], Henry Kapteyn[4], Margaret Murnane[4], Miguel A. Porras[5], Carlos Hernández-García[1,2]**

[1] Grupo de Investigación en Aplicaciones del Láser y Fotónica, Departamento de Física Aplicada, Universidad de Salamanca, 37008, Salamanca, Spain
[2] Unidad de Excelencia en Luz y Materia Estructuradas (LUMES), Universidad de Salamanca, Salamanca, Spain
[3] Department of Physics, Indiana University, Bloomington, IN 47405, United States of America
[4] JILA and Department of Physics, University of Colorado and NIST, Boulder, CO 80309, United States of America
[5] Complex Systems Group, ETSIME, Universidad Politécnica de Madrid, Ríos Rosas 21, 28003 Madrid, Spain

rodrigomh@usal.es

**Status**

High-order harmonic generation (HHG) is a robust tool for producing broadband, ultrashort attosecond pulses through the nonlinear up-conversion of intense visible or infrared (IR) pulses into the extreme-ultraviolet (EUV) and soft x-rays. Traditionally, efforts have focused on improving the temporal—towards shorter durations— and spectral —towards higher photon energies—properties of the high frequency emission. In terms of spatial property, the goal was to obtain clean, homogeneous modes that could be focused to maximize the intensity at the target. Recently, however, attention has broadened to engineering attosecond pulses with non-trivial structures in their intensity, phase and polarization, simultaneously in space and time. This trend has been boosted by the development of structured light fields across a broad range of spectral regimes and applications [1-2]. Structured light in attosecond science has focused on two key directions: the generation of structured beams at the attosecond timescale [3-7], enabling applications such as imaging, probing quantum materials or magnetic dichroism [8-10]; and the enhanced control of attosecond beam properties [11, 12].

Generating structured spatiotemporal EUV fields opens new degrees of freedom for light-matter interactions at the attosecond timescale. This includes transferring topological properties—such as the topological charge (associated to optical vortices), the Poincaré index (vector beams) or the Skyrme number (skyrmionic light textures)—into the EUV/attosecond regime. Yet, while structured light fields are well developed in the IR/visible regimes, their extension to higher photon energies is challenging, as standard optical elements are inefficient. To circumvent this limitation, HHG offers the possibility to transfer the spatiotemporal properties and topology of structured IR/visible drivers into EUV harmonics. However, this mapping is not straightforward. Indeed, polarization states up-convert differently, with efficiency dropping drastically with the ellipticity of the driving field. This limitation was circumvented by assigning separate conserved quantities—such as photon energy [13], linear momentum [14], or orbital angular momentum (OAM) [15]—to the right- and left- circular components of the driving field, enabling the generation of circularly polarized far-field harmonics.

A paradigmatic example is the up-conversion of linearly polarized IR/visible longitudinal optical vortices (LOV), whose OAM is characterized through the topological charge, $\ell$. When HHG is driven by a LOV with $\ell_0$, the topological charge of the q-th-order harmonic follows the scaling law: $\ell_q = q\ell_0$ [2]. This rule opened the way to the up-conversion of other structured beams, such as harmonic vector beams [16], harmonic vector-vortex beams [17] or harmonics with fractional OAM [18,19], among others. A similar scenario has been found recently in the up-conversion of spatiotemporal optical vortices (STOVs) that carry transverse OAM, whose topological charge, $\ell_0^{st}$, describes a phase winding coupled in one spatial and the temporal dimension. Despite not being eigenmodes of propagation, focused STOV driving beams carrying $\ell_0^{st}$ up-convert into high-order harmonic STOVs carrying $q\ell_0^{st}$ a result theoretically predicted in [20] and corroborated experimentally [21] (Fig. 1a).

**Current and future challenges:**

    Main current challenges lie in the generation and full temporal characterization of EUV harmonic structured pulses. While significant progress has been made in up-converting structured beams into the EUV, controlling their spatiotemporal structure remains non-trivial. This difficulty is a natural outcome of the angular momentum conservation laws. For instance, the linear scaling law in LOV-driven HHG, related to OAM conservation, prevents the direct synthesis of attosecond vortex pulses via the superposition of several harmonic orders. Such superposition actually leads to attosecond light springs [3, 4]. True attosecond vortex pulse trains can be recovered only if all harmonics share the same topological charge, ensuring wavefront-locking across different orders. Recent experiments have shown that by assigning linear momentum (in a non-collinear geometry) and opposite OAM to the right- and left- handed components of the driving field, circularly polarized LOV harmonics with identical topological charges can be produced, enabling the synthesis of attosecond vortex pulse trains [6] (see Chapter 7 'Attosecond vortex light' T. Ruchon).

    A similar scenario is found in the up-conversion of STOVs. The scaling law in STOV-driven HHG, $\ell_q^{st} = q\ell_0^{st}$, hinders the direct synthesis of attosecond STOV pulses. However, recent theoretical work predicts that harmonic STOVs with non-scaling, order independent topological charge can be generated when HHG is driven by their spectral counterpart, spatio-spectral optical vortex fields (SSOV) [7] (Fig. 1b). This result reveals that the topological charge and angular momentum can scale differently in HHG driven by topological spatiotemporal fields, in which case a detailed analysis in terms of transverse- and longitudinal-OAM is required [7, 22].

    Looking forward, an exciting prospect is the generation of custom polarization textures in both the spectral and temporal domains of EUV/attosecond pulses, tailored specifically for applications. A paradigmatic example is provided by paraxial skyrmionics beams, light fields whose transverse profiles cover all possible polarization states on the Poincaré sphere. Their topology is quantified by the Skyme number, $N_s$, which counts the number of times that the sphere is covered. Recent theoretical work has predicted that high-order harmonics with identical Skyrme number can be generated by assigning appropriate fractional OAM to the right- and left- handed components of the driving field, leading to the emission of attosecond skyrmion pulse trains [23] (Fig. 1c).

        Attosecond pulses structured in both space and time promise new opportunities for spatially-resolved ultrafast spectroscopy and imaging applications in advanced materials. Extending these schemes into the soft x-ray regime is particularly appealing, as it would cover the core-level absorption edges of most relevant elements in quantum materials. In this context, extending attosecond structured fields to the soft x-rays through HHG requires the need of further development of mid-IR driving sources with advanced spatiotemporal structuring and optimal phase matching conditions.

a) HHG driven by STOVs: high harmonic STOVs with scaling topological charge

b) HHG driven by SSOVs: attosecond STOV pulses

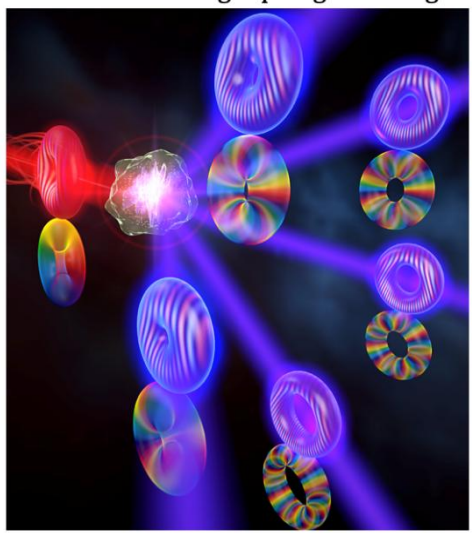

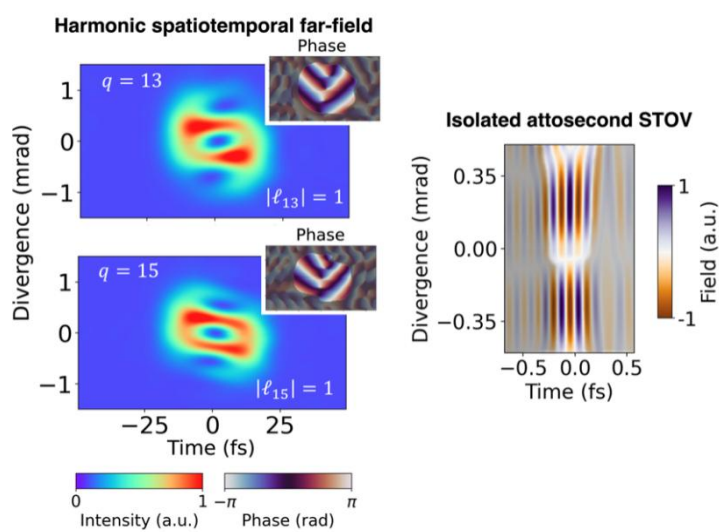

c) Attosecond skyrmion pulse

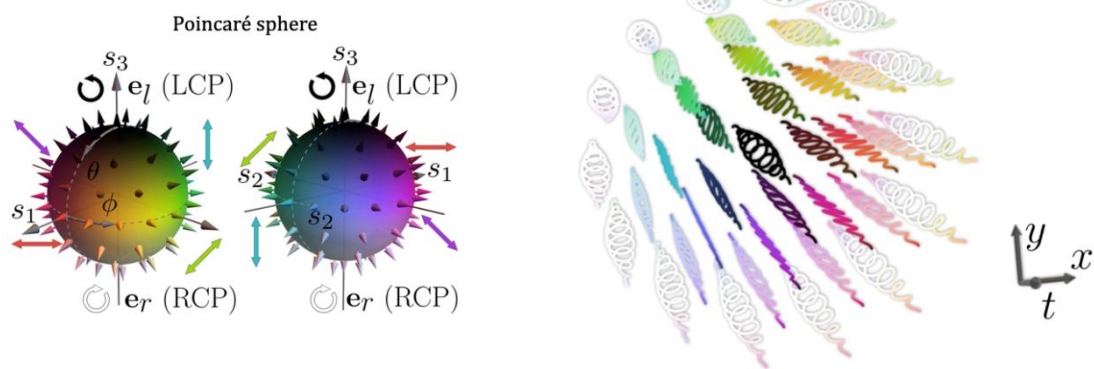

*Figure 1.* Attosecond structured pulses. a) High harmonic STOVs with scaling topological charge resulting from HHG driven by STOV pulses [23] (Credit, Steve Burrows and KM group, JILA). b) Attosecond STOV pulses, obtained from the coherent superposition of SSOV-driven harmonics [7]. d) Attosecond skyrmion pulses, obtained from the superposition of harmonic polarization textures that cover one Poincaré sphere, ie, with Skyrme number one [23].

## Advances in science and technology to meet challenges

The generation of structured light fields at the attosecond timescale still faces a critical bottleneck for unlocking their full potential: achieving complete spatiotemporal characterization of attosecond pulses. Most advanced diagnostic techniques with full polarization and spatiotemporal reconstruction have been developed for the IR/visible/femtosecond regime, and their extension to the EUV and soft x-ray domains remains highly challenging. Current EUV diagnostic tools provide only partial access to the properties of structured EUV radiation.

On the spatial side, wavefront sensors [17] can retrieve intensity and phase spectral distributions of EUV beams, while coherent imaging methods such as ptychography have enabled high-resolution reconstructions of EUV harmonics [9]. On the temporal side, attosecond pump–probe techniques, such as attosecond streaking [24] or RABBITT (reconstruction of attosecond beating by interference of two-photon transitions) [25], provide temporal information by recording photoemission delays induced by an EUV attosecond pump pulse (or train) and a phase-locked IR probe. However, these methods lack spatial resolution of the EUV beam itself. A notable exception is RABBITT applied to the characterization of LOV-driven HHG, where careful wavefront matching between the EUV vortex pump and the IR probe enabled the reconstruction of attosecond light springs [4]. Nevertheless, achieving full spatiotemporal characterization of structured EUV or soft-x-ray pulses remains an open challenge. Addressing it will likely require the development of new diagnostic tools and methods capable of combining nanometer-scale spatial resolution with attosecond temporal precision.

From the perspective of applications, fundamental questions remain regarding the interaction of spatiotemporally structured attosecond pulses with matter. In particular, the interaction of novel EUV structured beams—such as STOV/SSOV fields, or skyrmions—with dynamic excitations or transport in magnetic, chiral, and quantum materials is largely unexplored. A deeper understanding of these interactions could accelerate applications of helicity-sensitive dichroism and ultrafast spin and orbital manipulation, or enable new approaches for probing non-trivial dynamic topological phases in condensed-matter systems such as magnetic skyrmions or spin ices.

Progress in these areas will require advances in several fronts: (i) next-generation of structured mid-IR driving sources for optimized harmonic phase-matching at high photon energies, (ii) bright soft X-ray high harmonic sources with tailored properties, and (iii) innovative metrology approaches capable of disentangling the intertwined spatial, temporal, and polarization degrees of freedom in structured attosecond beams. Ultimately, these advances will not only address outstanding scientific questions but also enable transformative applications in ultrafast spectroscopy, metrology, and imaging—where the spatiotemporal topology of up-converted EUV and x-ray radiation is expected to play a significant role.

**Concluding remarks**

Structured EUV/soft-x-ray beams in the attosecond regime are opening transformative opportunities for spectroscopy, metrology, and ultrafast science. By coherently transferring and tailoring spatial, temporal, and polarization properties through HHG, it becomes possible to synthesize attosecond light fields with complex topologies—such as LOVs, STOVs, SSOVs or even skyrmionic textures—offering unprecedented control over the structure of ultrashort pulses in both space and time.

Yet, realizing their full potential requires addressing several central challenges: the development of advanced EUV diagnostic tools capable of complete spatiotemporal characterization at the nanometer and attosecond scales, the generation of bright soft x-ray structured sources, and a deeper understanding of the interaction of structured attosecond pulses with magnetic, chiral, and quantum materials. Progress in these areas will not only deepen our insight into spatially-resolved ultrafast light–matter interactions, but also enable new opportunities in attosecond science, ultimately positioning structured attosecond pulses as a cornerstone for next-generation of ultrafast photonics.

**Acknowledgements**

R.M.-H., D.M., L. P. and C.H.-G. acknowledge funding from the European Research Council (ERC) under the European Union's Horizon 2020 research and innovation programme (grant agreement No 851201), from the Department of Education of the Junta de Castilla y León and FEDER Funds (Escalera de Excelencia CLU-2023-1-02 and grant No. SA108P24), and from Ministerio de Ciencia e Innovacion (Grant PID2022-142340NB-I00). M.A.P. acknowledges support from the Spanish Ministry of Science and Innovation, Gobierno de España, under Contract No. PID2021-122711NB-C21. M.M.M. and H.C.K. gratefully acknowledge support from the Department of Energy Basic Energy Sciences Award No. DE-FG02-99ER14982 for the experiments done at JILA. C.-T.L. acknowledges support from the U.S. Air Force Office of Scientific Research (AFOSR), award no. FA9550-23-1-0234.

## References

[1] Bliokh KY, et al. 2023 Roadmap on structured waves. *Journal of Optics* **25**, 103001
[2] Shen Y, et al. 2023 Roadmap on spatiotemporal light fields, *Journal of Optics* **25**, 093001
[3] Hernández-García C, Picón A, San Román J and Plaja L 2013 Attosecond extreme ultraviolet vortices from high-order harmonic generation *Phys. Rev. Lett.* **111** 083602
[4] Geneaux R et al. 2016 Synthesis and characterization of attosecond light vortices in the extreme ultraviolet *Nature Communications* **7**, 12583
[5] Rego L et al 2019 Generation of extreme-ultraviolet beams with time-varying orbital angular momentum *Science* **364**, eaaw9486
[6] De las Heras A, Schmidt D et al. 2024 Attosecond vortex pulse trains, *Optica* **11**, 1085-1093
[7] Martín-Hernández R, Plaja L, Hernández-García C, Porras M A 2025 Isolated attosecond spatio-temporal optical vortices: Interplay between the topological charge and orbital angular momentum scaling in high harmonic generation *Phys. Rev. Lett. **22** 223801*.
[8] Shi X et al 2020 *Journal of Physics B: Atomic, Molecular and Optical Physics* **53**, 184008
[9] Wang B, et al. 2023 High-fidelity ptychographic imaging of highly periodic structures enabled by vortex high harmonic beams, *Optica* **10**, 1245-1252.
[10] Fanciulli M et al 2022 Observation of magnetic helicoidal dichroism with extreme ultraviolet light vortices *Phys. Rev. Lett.* **128**, 07740
[11] Rego L et al. 2022 Necklace-structured high-harmonic generation for low-divergence, soft x-ray harmonic combs with tunable line spacing *Science Advances* **8**, eabj7380
[12] Martín-Hernández R et al. 2025 Compact, intense attosecond sources driven by hollow Gaussian beams *arXiv:2507.04550*
[13] Fleischer et al. 2014 Spin angular momentum and tunable polarization in high-harmonic generation *Nature Photonics* **8**, 543–549
[14] Hickstein D et al. 2015 Non-collinear generation of angularly isolated circularly polarized high harmonics. *Nature Photonics* **9**, 743–750
[15] Brooks N et al. 2024 Circularly Polarized Attosecond Pulses Enabled by an Azimuthal Phase and Polarization Grating *ACS Photonics* **12**, 495-504
[16] Hernandez-Garcia C et al 2017 Extreme ultraviolet vector beams driven by infrared lasers *Optica* **4**, 520-526
[17] De Las Heras A et al 2022 Extreme-ultraviolet vector-vortex beams from high harmonic generation *Optica* **9**, 71
[18] Turpin A et al Extreme ultraviolet fractional orbital angular momentum beams from high harmonic generation 2017 *Scientific Reports* **7**, 43888
[19] Luttmann et al. 2024 Nonlinear up-conversion of a polarization Möbius strip with half-integer optical angular momentum *Science Advances* **9**, eadf3486
[20] Fang Y et al 2021 Controlling Photon Transverse Orbital Angular Momentum in High Harmonic Generation *Phys. Rev. Lett.* **127**, 273901
[21] Martín-Hernández R et al. 2025 Extreme-ultraviolet spatiotemporal vortices via high harmonic generation. *Nat. Photon.* **19**, 817–824.
[22] Gadeyne T et al. 2025 Energy and photon centroids of spatiotemporal light pulses and consequences for their intrinsic orbital angular momentum, *Phys. Rev. A.* (https://doi.org/10.1103/1wsz-48br)
[23] Marco D, Plaja L, Hernández-García C, 2025 Attosecond Light Skyrmion Pulses via High Harmonic Generation, *arXiv:2509.19113*.
[24] Constant E, Taranukhin V D, Stolow A, Corkum P B 1997 Methods for the measurement of the duration of high-harmonic pulses. *Physical Review A* **56**, 3870.
[25] Véniard, V, Taïeb R, Maquet A. 1996. Phase dependence of (N+ 1)-color (N> 1) ir-uv photoionization of atoms with higher harmonics. Physical Review A 54, 721.

## *9. ELI ALPS – a user facility opening new horizons in ultrafast science*

**Balázs Major[1,2]\*, Subhendu Kahaly[1,3]\* and Katalin Varjú[1,2]\***

[1] ELI ALPS, ELI-HU Non-Profit Ltd., Wolfgang Sandner utca 3., Szeged, H-6728, Hungary
[2] Department of Optics and Quantum Electronics, University of Szeged, Dóm tér 9., Szeged, H-6720, Hungary
[3] Institute of Physics, University of Szeged, Dóm tér 9., Szeged, H-6720, Hungary

balazs.major@eli-alps.hu, subhendu.kahaly@eli-alps.hu, katalin.varju@eli-alps.hu

**Status**

The concept of establishing a user facility dedicated to ultrafast laser-driven research was conceived decades ago, gradually refined over the years, and has recently culminated in opening the Extreme Light Infrastructure Attosecond Light Pulse Source (ELI ALPS) facility to the user community, now accessible through the ELI ERIC User Program. Notably, this milestone coincided with the award of the Nobel Prize in Physics "for experimental methods that generate attosecond pulses of light for the study of electron dynamics in matter". The core instrumentation at ELI ALPS was inspired by the pioneering work of the 2023 Nobel laureates [1], and the facility now offers six advanced beamlines: four based on high-harmonic generation (HHG) in gases, and two relying on HHG from surface plasma oscillations [2]. In response to the evolving landscape of attoscience, the facility has also introduced HHG from semiconductors [3] to expand its capabilities (see Table 1 for the driving mid-infrared (MIR) laser parameters). Furthermore, development is underway for an additional attosecond source aimed at accessing the water-window regime.

While state-of-the-art instrumentation is essential for advancing cutting edge science – and in this regard, ELI ALPS provides ample contribution to laser-based research – it is equally crucial to continuously explore new scientific frontiers. This is particularly evident in the emergence of multi- and interdisciplinary applications stemming from recent theoretical and experimental breakthroughs in attosecond physics. These developments have fostered intersections with other rapidly evolving fields, such as quantum science and technology [4] along with artificial intelligence and machine learning [5]. Moreover, attosecond research is increasingly addressing acute societal challenges. For example, in medicine by making femtosecond fieldoscopy and molecular fingerprinting an existing tool [6], or by allowing for real-life attosecond science applications in quantum physics, chemistry, and biology using attosecond electron microscopy [7].

Building on the previous discussion, this section highlights key advancements enabled by the research instrumentation at ELI ALPS. As a global hub for collaboration, the facility is opening new horizons in ultrafast science, empowering researchers from around the world to push the boundaries of attosecond technology. Through this overview, we aim to present a comprehensive – though not exhaustive – summary of the transformative potential attosecond science holds for the near future, and to illustrate how actively the ELI ALPS user facility is involved in addressing the challenges poised to reshape everyday life.

**Current and future challenges**

Ultrafast dynamics is a rapidly evolving field, marked by ever-increasing precision and a steady stream of novel concepts. Figure 1 illustrates the wide spectrum of research areas explored across all three ELI facilities in Szeged (Hungary), Dolní Břežany (Czechia) and Măgurele (Romania). Investigations into sub-femtosecond processes in various phases of matter are made possible through HHG, a cornerstone technique in attoscience. Among these endeavours, HHG in gases has reached a level of technological maturity that allows it to serve as a reliable attosecond light source, even in user-oriented environments like ELI ALPS [2].

In contrast, HHG from bulk solids, first demonstrated in 2011 [8], remains in a developmental phase. Recent advances – such as the temporal metrology of vacuum ultraviolet attosecond pulses from semiconductors [3] – represent a critical step toward enabling attosecond spectroscopy using solid-state sources. This interaction is also being explored as a diagnostic tool of few-cycle laser pulses, potentially paving the way for a compact, all-solid-state system for laser pulse characterization [9] with further enhancements possible using machine learning techniques [5].

Photoelectron holography offers another frontier, enabling the visualization of quantum phenomena on the attosecond timescale [10]. By precisely shaping extreme ultraviolet pulses, one can gain control over the wave packet of ionized electrons, thereby tuning interference and holographic reconstruction in a desirable manner. This control is exemplified by techniques such as laser-assisted dynamical interference of chirped photoelectrons [11]. Notably, foundational work synthesizing quantum optics with attosecond science has been pioneered at ELI ALPS [12], revealing quantum optical signatures in a strong laser pulse interacting with semiconductors.

Alongside these compelling application opportunities, there is continuous effort to develop the HHG-based attosecond light sources further. To this end, ELI ALPS has hosted a user experiment series focused on optimizing phase-matching regimes for increased HHG efficiency [13], based on the theoretical work of the group, widely applied in the HHG community.

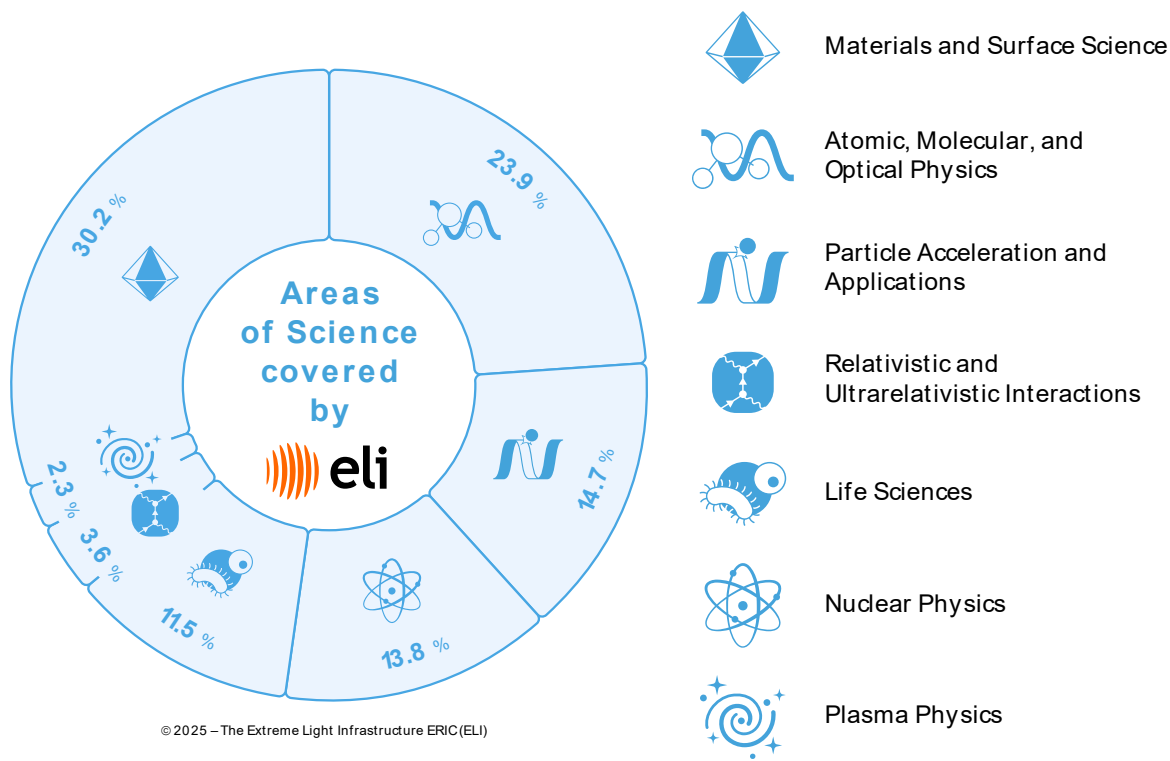

*Figure 1* The distribution of the 369 user experiments to which beamtime was granted up until summer 2025 at the ELI facilities. Source: 2024–2025 edition of the ELI Annual Report, https://www.eli-laser.eu/media/3012/eli_annual_report_2024-2025_digital.pdf

**Advances in science and technology to meet challenges**

HHG became accessible in the late 1980s, following the development of ultrashort pulsed laser sources. The continued advancement of HHG-based sources is closely tied to progress in ultrafast laser technologies. In this regard ELI (ALPS) has played a pivotal role by investing in next-generation laser developments. The SYLOS project, for example, has developed cutting-edge (non-collinear) optical parametric chirped pulse amplification ((N)OPCPA) based laser sources [14], providing few-cycle pulses with terawatt peak powers to the broad scientific community. The most advanced of these devices, the SYLOS3 laser, is housed at ELI ALPS (see main specifications in Table 1).

Another major contribution from ELI ALPS and its collaborators is the advancement of ytterbium fiber chirped pulse amplifier-based systems. These are now commonly paired with multi-pass cell (MPC) post-compression stages to provide few-cycle pulses based on commercially available, often industrial grade lasers. The HR2 laser system of ELI ALPS is the first kW-class few-cycle laser system (see the main specifications in Table 1) [15] of its kind.

Laser-plasma HHG sources require high peak power (100 TW to PW) drivers, traditionally achieved using Titanium:sapphire-based and chirped pulse amplification (CPA) systems. However, the front ends of these instruments are often replaced by ytterbium oscillators due to the multiple advantages they offer, e.g. in the HF laser system of ELI ALPS (see the main specifications in Table 1), or the sub-two-cycle 100-TW laser at Umea University, which reached ultrarelativistic – $10^{21}$ W cm$^{-2}$ – intensities with contributions from our facility team [16]. In combination with replenishable liquid-based targets, ELI ALPS is uniquely positioned for high repetition rate relativistic interactions.

Beyond primary laser sources, the attosecond beamlines require technological innovations to enable pump–probe experiments on the attosecond timescale across the spatial dimensions of ELI ALPS, which pose extra stability requirements. Recently, we found a solution that provides access to ultralong acquisition times in attosecond pump–probe beamlines driven by high-power lasers [17]. Additionally, the optimization of phase matching through tailored gas density distributions has proven critical for enhancing HHG flux in gas-based sources [18].

Emerging scientific domains also demand innovative measurement techniques and novel diagnostics. One of the great examples of this aspect is ultrafast ellipsometry and polarimetry [19], which allows, for example, testing Kramer-Kronig relations and causality on the fastest timescales currently observable experimentally [20], or studying novel chiro-optical phenomena.

**Table 1.** Main parameters of the most recently finished and cutting-edge lasers of the ELI ALPS facility.

| Laser name | Technology | Pulse duration | Pulse energy | Repetition rate | Central wavelength |
|---|---|---|---|---|---|
| SYLOS3 | NOPCPA [a] | 8 fs | 120 mJ | 1 kHz | 825 nm |
| HR2 | Fiber amplifier, MPC [a] post-compression | 6.2 fs | 4 mJ | 100 kHz | 1030 nm |
| HF | CPA [a] | 24 fs | 10 J | 2.5 Hz | 800 nm |
| MIR | OPA [a] | 20 fs | 70 uJ | 100 kHz | 3.2 um |

[a] Abbreviations in the table: CPA – chirped pulse amplification, MPC – multi-pass cell, NOPCPA -non-collinear optical parametric chirped pulse amplification, OPA – optical parametric amplification.

**Concluding remarks**

Although HHG and attosecond pulse technologies have reached a mature stage in certain implementations, their full potential – particularly in areas of societal relevance – is only beginning to unfold. Accordingly, there is a huge void to fill, which requires the engagement of many more researchers in this field and an increasing number of scientific platforms available for them. The current trend in international scientific cooperation materializes in the form of user facilities. ELI ALPS is the first of its kind that operates by this principle. In addition to the achievements listed in this paper, to tackle future challenges it is key to leverage artificial intelligence and machine learning, and to build a scalable research ecosystem with concerted international effort. We look forward to the advancements of attosecond science and we are keen on contributing to it on joint ground, by providing a synergistic integration of experimental apparatus, advanced simulation tools, and data-driven approaches for the future of our research field.

**Acknowledgements**

The ELI ALPS project (GINOP-2.3.6-15-2015-00001) is supported by the European Union and co-financed by the European Regional Development Fund. The work of B. M. was supported by the János Bolyai Research Scholarship of the Hungarian Academy of Sciences.

**References**

[1]   O. Alexander, D. Ayuso, M. Matthews, L. Rego, J. W. G. Tisch, B. Weaver, J. P. Marangos; Attosecond physics and technology, *Appl. Phys. Lett.* 126, 170501 (2025). https://doi.org/10.1063/5.0251524

[2]   M. Shirozhan, S. Mondal, T. Grósz, B. Nagyillés, B. Farkas, A. Nayak, N. Ahmed. I. Dey, S. Choudhary De Marco, K. Nelissen, M. Kiss, L. Gulyás Oldal, T. Csizmadia, Z. Filus, M. De Marco, S. Madas, M. Upadhyay Kahaly, D. Charalambidis, P. Tzallas, E. Appi, R. Weissenbilder, P. Eng-Johnsson, A. L'Huillier, Z. Diveki, B. Major, K. Varjú, S. Kahaly; High-Repetition-Rate Attosecond Extreme Ultraviolet Beamlines at ELI ALPS for Studying Ultrafast Phenomena, *Ultrafast Sci.* 4, 0067 (2024). https://doi.org/10.34133/ultrafastscience.0067

[3]   A. Nayak, D. Rajak, B. Farkas, C. Granados, P. Stammer, J. Rivera-Dean, T. Lamprou, K. Varju, Y. Mairesse, M. F. Ciappina, M. Lewenstein. P. Tzallas;  Attosecond metrology of vacuum-ultraviolet high-order harmonics generated in semiconductors via laser-dressed photoionization of alkali metals, *Nat. Commun.* 16, 1428 (2025). https://doi.org/10.1038/s41467-025-56759-0

[4]   M. Lewenstein, M. F. Ciappina, E. Pisanty, J. Rivera-Dean, P. Stammer, Th. Lamprou, P. Tzallas; Generation of optical Schrödinger cat states in intense laser–matter interactions, *Nature Phys.* 17,  1104–1108 (2021). https://doi.org/10.1038/s41567-021-01317-w

[5]   B. Nagyillés, G. N. Nagy, B. Kiss, E. Cormier, P. Földi, K. Varjú, S. Kahaly, M. Upadhyay Kahaly, Z. Diveki; MIR laser CEP estimation using machine learning concepts in bulk high harmonic generation, *Opt. Express* 32, 46500-46510 (2024). https://doi.org/10.1364/OE.537172

[6]   A. Srivastava, A. Herbst, M. M. Bidhendi, M. Kieker, F. Tani, H. Fattahi; Near-petahertz fieldoscopy of liquid, *Nat. Photon.* 18, 1320–1326 (2024). https://doi.org/10.1038/s41566-024-01548-2

[7]   D. Hui, H. Alqattan, M. Sennary, N. V. Golubev, M Th. Hassan; Attosecond electron microscopy and diffraction, *Sci. Adv.* 10, eadp5805 (2024). https://doi.org/10.1126/sciadv.adp5805

[8]   S. Ghimire, A. D. DiChiara, E. Sistrunk, P. Agostini, L. F. DiMauro, D. A. Reis; Observation of high-order harmonic generation in a bulk crystal, *Nature Phys.* 7, 138–141 (2011). https://doi.org/10.1038/nphys1847

[9]   M. Awad, A. Manna, S. Hell, B. Ying, L. Ábrók, Z. Divéki, E. Cormier, B. Kiss, J. Böhmer, C. Ronning, S. Heon Han, A. George, A. Turchanin, A. N. Pfeiffer, M. Kübel; Few-cycle laser pulse characterization on-target using high-harmonic generation from nano-scale solids, *Opt. Express* 32, 1325-1333 (2024). https://doi.org/10.1364/OE.508062

[10]    G. Porat, G. Alon, S. Rozen, O. Pedatzur, M. Krüger, D. Azoury, A. Natan, G. Orenstein, B. D. Bruner, M. J. J. Vrakking, N. Dudovich; Attosecond time-resolved photoelectron holography, *Nat. Commun.* 9, 2805 (2018). https://doi.org/10.1038/s41467-018-05185-6

[11]    F. Vismarra, M. Bertolino, E. Appi, M. Plach, L. Gulyas Oldal, T. Grosz, G. L. Dolso, V. Poulain, D. Mocci, G. Inzani, C. Biswas, M. De Marco, G. Zeni, F. Frassetto, L. Poletto, M. Reduzzi, R. Borrego-Varillas, H. J. Worner, Z.

Filus, I. Seres, P. Jojart, B. Major, T. Csizmadia, M. Nisoli, P. Eng-Johnsson, J. M. Dahlstrom, M. Lucchini; Dynamic Interference of Chirped Photoelectrons, *Phys. Rev. Lett.* **135**, 033202 (2025). https://doi.org/10.1103/73tl-w87y

[12]     N. Tsatrafyllis, S. Kühn, M. Dumergue, P. Foldi, S. Kahaly, E. Cormier, I. A. Gonoskov, B. Kiss, K. Varju, S. Varro, P. Tzallas; Quantum Optical Signatures in a Strong Laser Pulse after Interaction with Semiconductors, *Phys. Rev. Lett.* **122**, 193602 (2019). https://doi.org/10.1103/PhysRevLett.122.193602

[13]     E. Appi, R. Weissenbilder, B. Nagyillés, Z. Diveki, J. Peschel, B. Farkas, M. Plach, F. Vismarra, V. Poulain, N. Weber, C. L. Arnold, K. Varjú, S. Kahaly, P. Eng-Johnsson, and A. L'Huillier; Two phase-matching regimes in high-order harmonic generation,  *Opt. Express* **31**, 31687-31697 (2023). https://doi.org/10.1364/OE.488298

[14]     S. Toth, T. Stanislauskas, I. Balciunas, R. Budriunas, J. Adamonis, R. Danilevicius, K. Viskontas, D. Lengvinas, G. Veitas, D. Gadonas, A. Varanavičius, J. Csontos, T. Somoskoi, L. Toth, A. Borzsonyi, K. Osvay; SYLOS lasers – the frontier of few-cycle, multi-TW, kHz lasers for ultrafast applications at extreme light infrastructure attosecond light pulse source, *J. Phys. Photonics* **2**, 045003 (2020).  https://doi.org/10.1088/2515-7647/ab9fe1

[15]     I. Seres et al., manuscript in prep., (2025)

[16]     L. Veisz, P. Fischer, S. Vardast, F. Schnur, A. Muschet, A. De Andres, S. Kaniyeri, H. Li, R. Salh, K. Ferencz, G. N. Nagy, S. Kahaly; Waveform-controlled field synthesis of sub-two-cycle pulses at the 100 TW peak power level, *Nat. Photon.* **19**, 1013–1019 (2025). https://doi.org/10.1038/s41566-025-01720-2

[17]     T. Csizmadia, L. Gulyás Oldal, B. Gilicze, D. Kiss, T. Bartyik, K. Varjú, S. Kahaly, B. Major; Active stabilization for ultralong acquisitions in an attosecond pump–probe beamline, *APL Photonics* **10**, 080803 (2025). https://doi.org/10.1063/5.0273558

[18]     Z. Filus, T. Grósz, C. Biswas, L. Gulyás Oldal, T. Bartyik, B. Gilicze, S. Kahaly, K. Varjú, B. Major; High-harmonic generation yield enhancement with tailored gas target design, in preparation, (2025).

[19]     L. Gulyás Oldal, B. Gilicze, T. Bartyik, D. Kiss, M. Devetta, G. Zeni, F. Frassetto, L. Poletto, T. Csizmadia, B. Major, submitted to *J. Phys. Photonics* (2025).

[20]     V. Leshchenko, S. J. Hageman, C. Cariker, G. Smith, A. Camper, B. K. Talbert, P. Agostini, L. Argenti, L. F. DiMauro; Kramers–Kronig relation in attosecond transient absorption spectroscopy, *Optica* **10**, 142-146 (2023). https://doi.org/10.1364/OPTICA.474960

## 10. Status and future of driving lasers for attosecond X-ray sources beyond water window

### Chase Geiger, Fei Xu and Zenghu Chang*

Department of Physics, University of Ottawa, Ottawa, Canada

zchang@uottawa.ca

### Status

The first generation of attosecond light sources was based on Ti:Sapphire chirped-pulse amplification (CPA) systems operating near 800 nm, delivering millijoule-level pulse energies at kiloherz repetition rates. When combined with hollow-core fiber (HCF) pulse compression, these systems evolved into a mature and commercially available technology capable of generating few-cycle, carrier-envelope phase (CEP) stable pulses. These lasers are now widely employed for producing isolated attosecond pulses (IAPs) in the extreme ultraviolet (XUV) range (20–150 eV) [1]. More recently, ytterbium (Yb) fiber and thin-disk CPA systems operating near 1 μm have achieved sufficient repetition rate and average power for high-flux XUV generation. Commercial and laboratory Yb femtosecond laser systems are often combined with multi-pass compressors or HCFs and now routinely operate at repetition rates up to 100 kHz.

Gas-phase high harmonic generation (HHG) spectra display a high-energy photon limit called the harmonic cutoff, which scales favorably with increasing wavelength of the driving laser [2]. Early extensions of the harmonic cutoff therefore employed frequency conversion of Ti:Sapphire lasers to wavelengths between 1.1 and 2.4 μm using optical parametric amplifiers (OPAs) [3] and optical parametric chirped-pulse amplifiers (OPCPAs). Using such drivers, record-setting IAPs of 53 as at the carbon K-edge (282 eV) were achieved [4], enabling attosecond transient absorption spectroscopy in the water-window spectral region (282–533 eV) [5]. Progress towards similar-wavelength sources with less complex setups have been enabled by robust Yb-based pump lasers for OPCPA systems, typically employing PPLN, BBO, or LBO crystals and operating at high repetition rates, leading to demonstrations of HHG extending beyond 0.6 keV at 100 kHz [6]. Alternatively, CPA systems in the ~2 μm range, based on the Ho/Tm laser family, have also emerged as strong candidates for attosecond drivers. Tm:fiber and Ho:YLF CPAs routinely deliver multi-millijoule, sub-picosecond to picosecond pulses at 2 μm, which can be compressed to sub-100 fs durations via nonlinear methods [7]. These technologies are maturing rapidly, and new broadband gain media such as Ho:CALGO are under active investigation [8], pointing toward turnkey 2-μm platforms for next-generation attosecond light sources.

### Current and future challenges

Up until now, high harmonic generation in gases has enabled attosecond X-ray transient absorption spectroscopy at photon energies below the oxygen K-edge (533 eV) [9]. A key objective moving forward is to extend this technique to generate high-flux attosecond pulses beyond the water window and into the keV regime, thereby enabling core-level spectroscopy of light elements, including 3d transition metals, with attosecond temporal resolution. In addition, pump–probe studies of ultrafast processes such as charge migration and charge transfer in molecules demand higher intensity attosecond ultraviolet and extreme ultraviolet pulses [10-12].

Generating isolated attosecond pulses at keV photon energies with sufficient flux for experimentation necessitates mid-infrared (MIR) drivers (3–8 μm) delivering multi-millijoule, few-cycle pulses with excellent carrier-envelope phase stability at multi-kilohertz repetition rates. Such sources are under active development. However, scaling to longer driving wavelengths introduces unfavorable conditions: the single-atom HHG response diminishes due to quantum diffusion of the electron wavepacket, while phase-matching becomes increasingly stringent. Higher pulse energies at high repetitio rates are therefore needed at longer wavelengths to compensate for the lower HHG efficiency. MIR source development is a significant bottleneck in extending attosecond science beyond the water window, since commercial systems are not yet available.

**Advances in science and technology to meet challenges**

Two gain media, Fe:ZnSe and ZnGeP$_2$, have emerged as particularly attractive candidates for CPA and OPCPA systems operating at 4 μm and beyond. Fe:ZnSe combines a broad gain bandwidth with excellent mid-IR transparency [13]. Its upper-state lifetime of ~50 μs at liquid-nitrogen temperature enables efficient pumping with pulsed Er:YAG lasers at 2.94 μm, originally developed for medical applications [14]. Advances in high-quality crystal growth and OPA seeding have enabled multi-pass CPA operation, achieving multi-millijoule, few-hundred-femtosecond pulses [15]. Further progress in cryogenic CPA has pushed repetition rates and average powers for operation near 4 μm [16]. Both single-crystal and polycrystalline Fe:ZnSe show promising gain for high-energy CPA stages, indicating scalability toward kHz repetition rates and multi-mJ outputs. As pulse compression and CEP stabilization techniques continue to mature [17], Fe:ZnSe CPAs are poised to become direct mid-IR attosecond drivers.

Recent advances in 2-μm Ho-doped, picosecond, high-energy pump lasers have enabled ZnGeP$_2$ (ZGP)-based mid-IR OPCPAs spanning the 3–8 μm range. ZGP offers a large nonlinear coefficient ($d_{36}$ = 75 pm/V), broad transparency (0.72–12 μm), and a high damage threshold (10 GW/cm$^2$ for 5 ps pulses at 2.05 μm). Pumped at ~2 μm, ZGP OPCPAs have already demonstrated millijoule-level pulses at 5 μm and 1 kHz with excellent stability [18]. Combined with 4f Fourier-plane pulse shaping of linear and nonlinear spectral phase [19], compression of broadband MIR pulses to near their Fourier-transform limit is within reach. Continued progress is expected to significantly enhance average power in the near future.

Outlook (2–5 year horizon). High-repetition-rate mid-IR sources: Yb-pumped PPLN OPA front ends are expected to seed ZGP OPCPA chains in the 3–5 μm range, combining Yb reliability with mid-IR scalability. Fe:ZnSe CPA maturation: Cryogenically cooled, CEP-stabilized Fe:ZnSe CPAs, complemented by hollow-core fiber or multipass compression, as well as ZGP-based OPCPAs, are anticipated to deliver multi-kHz, multi-mJ, few-cycle pulses near 4 μm. Toward shorter IAPs: Mid-IR drivers in the 3–5 μm region will support ultrabroad X-ray continua. With innovative compression schemes to mitigate attochirp [20], isolated pulses shorter than one atomic unit of time with enhanced soft X-ray flux should become feasible.

**Concluding remarks**

In summary, although realizing keV attosecond pulses from gas-phase HHG remains a formidable challenge, advances in mid-IR laser technology and HHG target engineering outline a clear and feasible pathway toward compact, laboratory-scale keV attosecond X-ray sources. This spectral range is particularly valuable, offering element-specific sensitivity and nanometer-scale spatial resolution, while the attosecond temporal resolution enables direct observation of electron dynamics in real time.

## Acknowledgements

Canada Excellence Research Chairs, Government of Canada; National Science Foundation (2207674); Air Force Office of Scientific Research (FA9550-18-1-0223).

## References

[1] Chini M, Zhao K and Chang Z 2014 The generation, characterization and applications of broadband isolated attosecond pulses Nat. Photonics 8 178–186

[2] Han S, Li J, Zhu Z, Chew A, Larsen E W, Wu Y, Pang S S and Chang Z 2020 Tabletop attosecond X-rays in the water window Adv. At. Mol. Opt. Phys. 69 1–65

[3] Shan B and Chang Z 2001 Dramatic extension of the high-order harmonic cutoff by using a long-wavelength driving field Phys. Rev. A 65 011804

[4] Li J et al. 2017 53-attosecond X-ray pulses reach the carbon K-edge Nat. Commun. 8 186

[5] Saito N et al. 2021 Attosecond electronic dynamics of core-excited states of $N_2O$ in the soft x-ray region Phys. Rev. Res. 3 043222

[6] Pupeikis J, Chevreuil P-A, Bigler N, Gallmann L, Phillips C R and Keller U 2020 Water window soft x-ray source enabled by a 25 W few-cycle 2.2 µm OPCPA at 100 kHz Optica 7 168–171

[7] Nagy T, von Grafenstein L, Ueberschaer D and Griebner U 2021 Femtosecond multi-10-mJ pulses at 2 µm wavelength by compression in a hollow-core fiber Opt. Lett. 46 3033–3036

[8] Suzuki A, Kassai B, Wang Y, Omar A, Löscher R, Tomilov S, Hoffmann M and Saraceno C J 2025 High-peak-power 2.1 µm femtosecond holmium amplifier at 100 kHz Optica 12 534–537

[9] Buades B, Picón A, Berger E, León I, Di Palo N, Cousin S L, Cocchi C, Pellegrin E, Martin J H, Mañas-Valero S and Coronado E 2021 Attosecond state-resolved carrier motion in quantum materials probed by soft x-ray XANES Appl. Phys. Rev. 8 011408

[10] [10] Golubev N V, Vaniček J and Kuleff A I 2021 Core-valence attosecond transient absorption spectroscopy of polyatomic molecules Phys. Rev. Lett. 127 123001

[11] [11] Ruberti M 2021 Quantum electronic coherences by attosecond transient absorption spectroscopy: ab initio B-spline RCS-ADC study Faraday Discuss. 228 286–311

[12] [12] Kobayashi Y and Leone S R 2022 Characterizing coherences in chemical dynamics with attosecond time-resolved x-ray absorption spectroscopy J. Chem. Phys. 157 184306

[13] [13] Mirov S B, Fedorov V V, Martyshkin D, Moskalev I S, Mirov M and Vasilyev S 2014 Progress in mid-IR lasers based on Cr and Fe-doped II–VI chalcogenides IEEE J. Sel. Top. Quantum Electron. 21 292–310

[14] [14] Evans J W, Harris T R, Reddy B R, Schepler K L and Berry P A 2017 Optical spectroscopy and modeling of $Fe^{2+}$ ions in zinc selenide J. Lumin. 188 541–550

[15] [15] Migal E, Pushkin A, Bravy B, Gordienko V, Minaev N, Sirotkin A and Potemkin F 2019 3.5-mJ 150-fs Fe:ZnSe hybrid mid-IR femtosecond laser at 4.4 µm for driving extreme nonlinear optics Opt. Lett. 44 2550–2553

[16] [16] Marra Z A, Wu Y, Zhou F and Chang Z 2023 Cryogenically cooled Fe:ZnSe-based chirped pulse amplifier at 4.07 µm Opt. Express 31 13447–13454

[17] [17] Marra Z A, Wu Y, Belden N and Chang Z 2024 Few-cycle, mJ-level, mid-wave infrared pulses generated via post-compression of a chirped pulse amplifier Opt. Lett. 49 3170–3173

[18] [18] von Grafenstein L, Bock M, Ueberschaer D, Zawilski K, Schunemann P, Griebner U and Elsaesser T 2017 5 µm few-cycle pulses with multi-gigawatt peak power at a 1 kHz repetition rate Opt. Lett. 42 3796–3799

[19] [19] Nicolai, F., et al., Acousto-optic modulator based dispersion scan for phase characterization and shaping of femtosecond mid-infrared pulses. Optics Express, 2021. 29(13): p. 20970-20980

[20] [20] Du J-X, Wang G-L, Yang Z-Q, Jiao Z-H, Zhao S-F and Zhou X-X 2025 Generation of Fourier-limited ultrashort soft X-ray isolated attosecond pulses by two-color chirped lasers without chirp compensation J. Opt. Soc. Am. B 42 1105–1113

## 11. Attosecond Hard X-ray Pulses for Material Science

**Thomas Linker[1]\* , Diling Zhu[1], and Matthias F. Kling[1,2]**

[1] SLAC National Accelerator Laboratory, Menlo Park, CA 94025, USA
[2] Department of Applied Physics, Stanford, CA 94305, USA

tlinker@slac.stanford.edu

**Status**

Attosecond Hard-X-ray pulses have the potential to be a transformative probe for materials with the capability to be damage free imagers, directly probe electrons dynamics on their natural angstrom–attosecond (Å–as) space–time scales, as well as coherently drive and probe deep core-level electron dynamics. X-ray free-electron lasers (XFELs) have demonstrated isolated attosecond soft-X-ray(≤1keV) pulses [1-5], where first applications have focused on atomic and molecular systems [5-8] though there have been recent extensions into condensed matter [9,10]. Within these frameworks temporal reconstruction of the XFEL pulses as well co-timing with tabletop laser systems through angular streaking has enabled attosecond time-resolved laser pump, soft XFEL probe experiments [1,5,6,8]. Concurrently FEL developments for generating twin attosecond pulses in the soft X-ray regime have enabled attosecond X-ray pump/probe studies [7,9].

Recent FEL advances aim to extend these concepts into the hard-X-ray regime (≥5 keV), with single spike hard XFEL pulses being established as early as 2017 [11], and recently TW-class attosecond hard-X-ray pulses having been demonstrated at EuXFEL [12]. Nano-focused attosecond hard-X-ray beams offer route for driving non-linear coherent core-level dynamics opening new views into ultrafast correlated core level electron motion [13,14]. Most non-linear hard X-spectroscopy techniques are challenging for materials with femtosecond pulses due to cascade processes melting the electronic order within a few femtoseconds. Sub-femtosecond pulses below typical core hole lifetimes can outrun the electronic cascade dynamics offering unprecedent insight into materials electronic structure via spectroscopy as well as diffraction techniques. Overall, incorporation of developed streaking/cross-correlation timing tools into the hard X-ray regime combined with emerging time-domain theory for X-ray excitations and diffraction could drastically enhance our ability to probe sub-fs coherent electron dynamics such as charge migration across atomic lattice sites or interfaces (example dynamics illustrated in Fig. 1a). Such information is critical for linking fundamental electron dynamics to macroscopic functionalities such as switching and energy efficiency in practical device platforms.

Beyond single pulse studies, twin hard X-ray pump–probe can open a new frontier for exploring the physics of coupled electron–lattice interactions that underpin emergent phenomena in solids, including superconductivity, charge/spin ordering, and topological phases. Many of these exotic states depend critically on charge localization processes, such as the formation of polarons or other localized quasiparticles, that govern how electrons interact with defects, phonons, and other collective excitations. When a deep core electron is excited by a hard X-ray, it produces a highly localized point charge within the lattice (see Fig. 1b). How the surrounding valence electrons (and for low Z materials the ions) respond before the core hole decays remains poorly understood, yet this response is a uniquely sensitive probe of electron correlation and screening dynamics. Attosecond hard-X-ray pump–probe techniques could provide unprecedented insight into how localized charges drive correlated behaviour, which is critically relevant for designing quantum materials, tailoring superconductors, and engineering functional oxides and catalyst. There has been recent success in utilizing femtosecond hard-X-ray–core-level excitations to generate localized electronic disturbances that influence subsequent lattice and collective dynamics [15], underscoring the opportunity for attosecond hard-X-ray pulses to selectively initiate and track the ultrafast processes that give rise to correlated and emergent behaviour in solids.

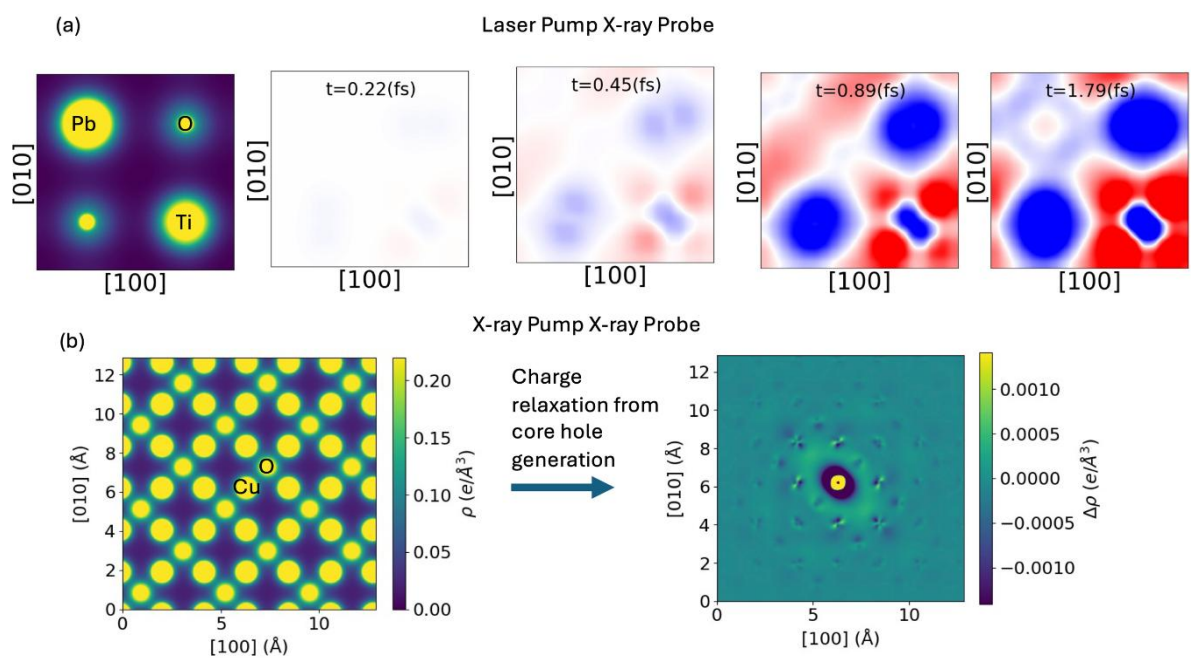

*Figure 1.* Example of materials dynamics for which hard X-ray attosecond pulses can help observe. (a) Simulation of attosecond electron dynamics in photoexcited ferroelectric PbTiO3 resulting in O to Ti charge transfer. The plots are based on simulations described in ref. [16]. Core level X-ray spectroscopy combined with diffraction has potential to directly image such charge transfer dynamics. (b) DFT simulation illustrating charge relaxation around generated 1s core hole in CuO crystal. In the presence of core hole, electrons will relax around it to screen out the coulomb charge. Both experimentally and theoretically the timescale this occurs on and the role of electronic correlation in the screening are not understood. Hard X-ray pump/probe schemes can give insight into such dynamics.

## Current and future challenges

Despite the promise, significant challenges remain. Current hard X-ray attosecond pulse durations have been inferred from spectral measurements and simulations with no direct temporal reconstruction measurements. Promising experiments have recently demonstrated more direct evidence for attosecond pulse duration through amplified spontaneous emission measurements with known intensity threshold requirements [17]; however, such measurements similarly require simulation guidance and do not allow for co-timing with optical laser systems. While today's SASE FEL pulses generally consist of isolated sub-femtosecond spikes, they often come with low stability, random spectral content, and limited tunability. This makes quantitative pump–probe measurements challenging Controllable temporal/spectral content is also essential for non-linear hard X-ray spectroscopies, where multispike nature of SASE pulses can lead to multiple excitation/interference effects [14]. For X-ray pump/probe experiments, specialized designed optics need to be developed for attosecond pulse manipulation. Most current split and delay optics using Bragg reflectors, cannot maintains sub-femtosecond pulse duration. While upgrades to high-rep rate facilities offer tremendous opportunities for attosecond science especially for non-linear signals with low signal to noise/background, average beam heating effects will need mitigation in particular for the study of quantum states at cryogenic temperatures. In addition, for measurements of irreversible processes induced by high x-ray intensity, platforms for higher speed sample refreshment will be needed.

Beyond experimental difficulties, interpreting attosecond hard-X-ray signals in solids requires multiscale theory that bridges electron dynamics localized to core of the lattice with delocalized valence electron and phonon dynamics. While there has been significant progress within the atomic and molecular communities for direct first principle of simulations attosecond core dynamics [18,19], attosecond solid state electronic dynamics are less explored.

Overall, robust attosecond operation at Å-wavelengths, reproducible pulse metrology, precise pulse manipulation, as well as clear observables that can couple with complex first-principles dynamical simulations for core-level spectroscopies and diffraction remain key hurdles before broad materials adoption.

**Advances in science and technology to meet challenges.**

There are current efforts at X-ray free electron laser facilities to better control attosecond pulses, with recent experiments at LCLS demonstrating the ability to controllably shape attosecond X-ray pulses in the soft X-ray regime [20]. Extension of such studies into hard X-ray offers a path forward to better tunability and reliability of hard X-ray attosecond pulses. The development of cavity-based X-ray FEL concepts, and the potential using their stable output to drive further stabilization of cavity-mode and 2nd stage pulse broadening/compression, are paths toward generation of stable attosecond hard x-ray pulse trains [21-24]]. For X-ray pump/probe schemes, development of hard-X-ray attosecond split-delay line using broadband optics that can achieve sub-20 as scanning resolution over ~20 fs delays will significantly advance the field [23]. Experiments at LCLS to utilize such optics for attosecond pulse duration measurements through intensity autocorrelations are currently underway. For co-timing with optical lasers, deployment of hard-X-ray streaking or other metrology techniques to allow for X-ray/optical arrival time monitoring with attosecond precision is the critical bottle neck in extending many studies applied in atomic and molecular domain with soft X-rays to the hard X-ray regime for materials. On the theoretical side, recent work has demonstrated the ability to simulate time resolved RIXS in solids for O K edge and transition metal L edges [25]. Extension of such computational frameworks to hard X-ray excitations as well as incorporation within attosecond diffraction [26] theories will be essential for interpreting new hard X-ray attosecond experiments.

**Concluding remarks**

Attosecond hard-X-ray science is poised to shift materials dynamics research from femtosecond "movies" to sub-fs, site-specific electron dynamics with atomic spatial reach, closing a key gap between spectroscopy, imaging and function. Achieving this requires concurrent progress in source stability and shaping, attosecond-grade metrology at Å wavelengths, and interpretable experiments tightly coupled to predictive, open computational toolchains. With LCLS-II-class facilities enabling high-rep operation and advanced pulse control, and with community investment in theory, and robust diagnostics, the next five years can deliver definitive case studies for applying attosecond X-rays to studying quantum materials and petahertz electronics [27].

## Acknowledgements

The work performed by T.L. and M.F.K. were supported by the Chemical Sciences, Geosciences, and Biosciences Division (CSGB), Basic Energy Sciences, Department of Energy. The Linac Coherent Light Source (LCLS), SLAC National Accelerator Laboratory, is supported by the U.S. Department of Energy (DOE), Office of Science, Office of Basic Energy Sciences (BES) under contract no. DE-AC02-76SF00515.

## References

[1] Li S et al. (2018) Characterizing isolated attosecond pulses with angular streaking. *Opt. Express* **26**, 4531–4547.
[2] Duris J et al. (2020) Tunable isolated attosecond X-ray pulses with gigawatt peak power from a free-electron laser. *Nat. Photonics* **14**, 30–36.
[3] Franz P et al. (2024) Terawatt-scale attosecond X-ray pulses from a cascaded superradiant free-electron laser. *Nat. Photonics* **18**, 698–703.
[4] Prat E et al. (2023) Coherent sub-femtosecond soft X-ray free-electron laser pulses with nonlinear compression. *APL Photonics* **8**, 111302.
[5] Funke L et al. (2024) Capturing nonlinear electron dynamics with fully characterised attosecond X-ray pulses. *arXiv:2408.03858*.
[6] Li S et al. (2022) Attosecond coherent electron motion in Auger–Meitner decay. *Science* **375**, 285–289.
[7] Guo Z et al. (2024) Experimental demonstration of attosecond pump–probe spectroscopy with an X-ray free-electron laser. *Nat. Photonics* (2024), doi:10.1038/s41566-024-01419-w.
[8] Driver T et al. (2024) Attosecond delays in X-ray molecular ionization. *Nature* **632**, 762–767.
[9] Li S et al. (2024) Attosecond-pump attosecond-probe X-ray spectroscopy of liquid water. *Science* **383**, 1118–1122.
[10] Alexander O et al. (2024) Attosecond impulsive stimulated X-ray Raman scattering in liquid water. *Sci. Adv.* **10**, eadp0841.
[11] Huang S et al. (2017) Generating single-spike hard X-ray pulses with nonlinear bunch compression in free-electron lasers. *Phys. Rev. Lett.* **119**, 154801.
[12] Yan J et al. (2024) Terawatt-attosecond hard X-ray free-electron laser at high repetition rate. *Nat. Photonics* **18**, 1293–1298.
[13] Inoue I et al. (2025) Nanofocused attosecond hard X-ray free-electron laser with intensity exceeding $10^{19}$ W/cm$^2$. *Optica* **12**, 309–310.
[14] Linker T M et al. (2025) Attosecond inner-shell lasing at ångström wavelengths. *Nature* **642**, 934–940.
[15] Li H et al. (2025) Nanoscale ultrafast lattice modulation with hard X-ray free-electron laser. *arXiv:2506.03428*.
[16] Linker T et al. (2022) Exploring far-from-equilibrium ultrafast polarization control in ferroelectric oxides with excited-state neural network quantum molecular dynamics. Science Advances 8 (12), eabk2625
[17] Inoue I et al. (2025) Experimental demonstration of attosecond hard X-ray pulses. *arXiv:2506.07968*.
[18] Bruner A, Hernandez S, Mauger F, Abanador P M, LaMaster D J, Gaarde M B, Schafer K J, and Lopata K (2017). Attosecond charge migration with TDDFT: Accurate dynamics from a well-defined initial state. *J. Phys. Chem. Lett.* **8**, 3991–3997.
[19] Chergui M et al. (2023) Progress and prospects in nonlinear extreme-ultraviolet and X-ray optics and spectroscopy. *Nat. Rev. Phys.* 1–19.
[20] Robles R R et al. (2025) Spectrotemporal shaping of attosecond X-ray pulses with a fresh-slice free-electron laser. *Phys. Rev. Lett.* **134**, 115001.
[21] Marcus G et al. (2019) Cavity-based free-electron laser research and development: A joint Argonne National Laboratory and SLAC National Accelerator Laboratory collaboration. *Proc. FEL'19*, TUD04.
[22] Rauer P et al. (2023) Cavity-based X-ray free-electron laser demonstrator at the European X-ray Free Electron Laser facility. *Phys. Rev. Accel. Beams* **26**, 020701.
[23] Sun Y et al. (2025) An ultrastable hard X-ray attosecond split–delay line. *arXiv:2505.06865*.

[24] Li H et al. (2022) Femtosecond-terawatt hard X-ray pulse generation with chirped pulse amplification on a free electron laser. *Phys. Rev. Lett.* **129**, 213901.

[25] Jost D et al. (2025) Time-resolved X-ray spectroscopy from the atomic orbital ground state up. *Phys. Rev. X* **15**, 011012.

[26] Yuan M, and Golube N V (2025) Attosecond diffraction imaging of electron dynamics in solids. *Phys. Rev. Research* **7**, L022042.

[27] Heide C, Keathley P D, and Kling, M F (2025) Petahertz electronics. *Nat. Rev. Phys.* **6**, 648–662.

## *12. Generation and application of sub-femtosecond soft X-ray pulses at the European XFEL*

**M. Meyer*, S. Serkez and G. Geloni**

European XFEL, Schenefeld, Germany

michael.meyer@xfel.eu

**Status**

A strategy paper in 2018 [1] was the starting point of activities related to attosecond science at the European XFEL. Commissioning activities were initiated in parallel to standard user operation of the facility until the first ultrashort soft X-ray pulses could be produced at the SASE3 undulator in 2021 in the high-repetition rate burst mode of the European XFEL (ten bursts per second, each including, typically, several hundreds pulses at multi-MHz intra-burst rate)**.** The characterization of these pulses was based on spectral diagnostics showing single pulses with spectral distributions with widths in the order of 5-6 eV (Figure 1). In the following year the temporal width of the pulses was determined during an experimental campaign at the Small Quantum Systems (SQS) scientific instrument and pulse durations of about 300-500 attoseconds were measured [2] using the electron angular streaking technique [3] (Figure 2).

Both methods, spectral and temporal diagnostic, are now available at the SQS scientific instrument and can be applied in parallel to standard user operation. The availability of these ultra-short soft X-ray pulses with pulse energies of up to 1 mJ opens up a large variety of novel scientific applications in the field of non-linear and time-resolved studies in the short wavelength region. Dynamics in the electron cloud of atoms, molecules and clusters can be investigated making use of site-selective excitations of core electrons introduced by the soft X-ray pulses. New detailed information about charge transfer processes, electronic shielding, Auger relaxation dynamics, plasma formation, electron wavepacket dynamics can be obtained with sub-femtosecond time resolution and non-linear multi-photon processes can be strongly enhanced compared to the generally dominating single-photon excitations.

Particularly important for the application of pump-probe excitation schemes is the capability of the SASE3 undulator to generate two soft X-ray pulses with independently controllable photon energy, pulse energy and temporal delay (Figure 1). At present, a magnetic chicane introduced in the undulator [4] enables X-ray – X-ray pump-probe experiments with temporal delays between almost zero and several hundred femtoseconds and hundreds of micro Joule pulse energy in the individual pulses. Combination with a fresh-bunch technique [5] allows for better independence of the pulse energies and for zero-crossing delay between pump and probe pulses. Most recently, with the additional use of the APPLE X radiator at the SASE3 undulator we demonstrated production of circularly polarized attosecond soft X-ray pulses, bringing dynamical studies of dichroic phenomena into reach.

**Current and future challenges**

The generation of soft X-ray pulses at the European XFEL is based on particular manipulations of the electron bunch: first, owing to special compression settings and to the presence of self-interactions along the accelerator, the electron bunch acquires large energy correlations as a function of the longitudinal position. Second, by tuning the magnetic field in quadrupole magnets in a dispersive area of the machine, we create correlation between electron energy and time-of-flight as well as transverse position in the beam that enters the undulator. The combined effect with the bunch energy chirps results in an extremely short lasing bunch part during the FEL process.

The FEL mode of operation for attosecond pulse generation, however, is still Self Amplified Spontaneous Emission (SASE) [6,7], which results in strong variability of the individual pulses in terms of their spectral and temporal pulse shapes. Consequently, a single-pulse characterization of the pulses is of highest importance to fully exploit the application of these pulses in advanced experimental investigations. At the SQS instrument, we have established simultaneous shot-to-shot measurements of the spectral profile besides analysis of the temporal distribution by angular streaking to characterize the soft X-ray pulses. However, routine operation under changing experimental conditions with respect to photon energy, pulse energy and pulse durations is still highly demanding. In practice, the experimental campaigns are becoming extremely complex, since meaningful data recording requires executing two experiments in parallel, one applying the attosecond pulses in the science-related experiment and another for the precise characterization of these pulses.

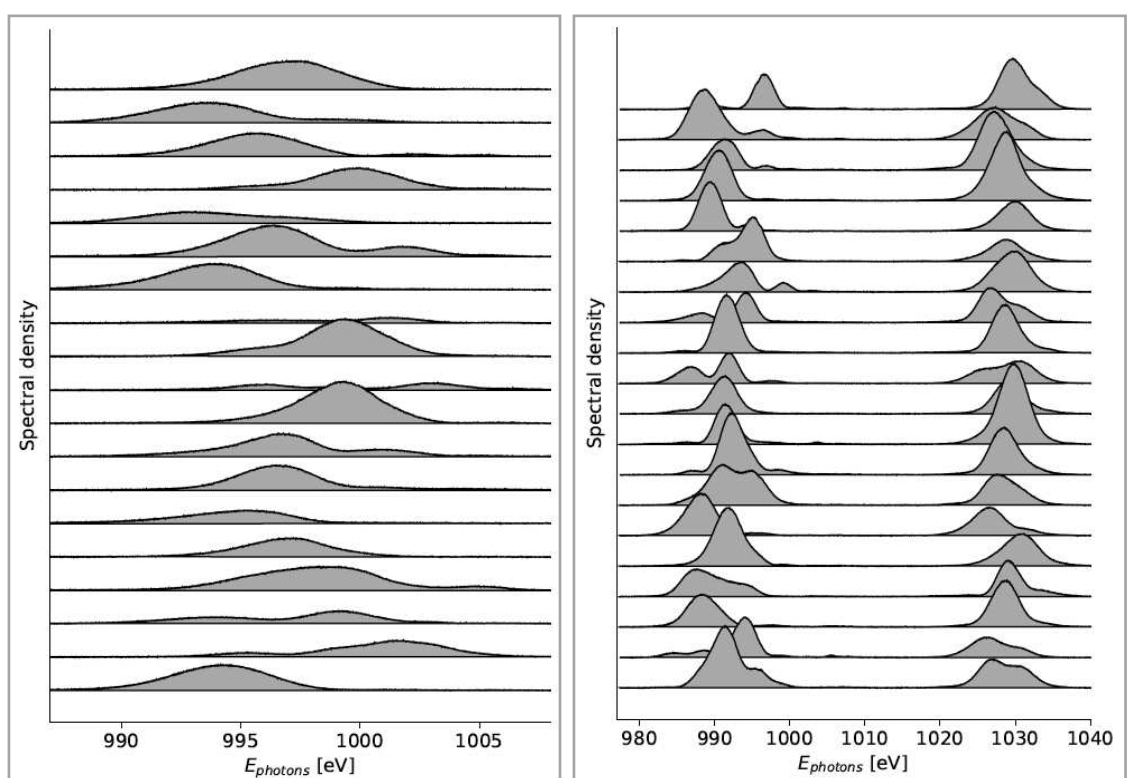

**Figure 1** *(left) Spectral distribution of 20 consecutive single soft X-ray pulses with pulse energies up to 1 mJ and nominal photon energy of 997 eV, measured behind the SASE3 monochromator. (right) Spectral distribution of two ultrashort soft X-ray pulses with pulse energies of 200-300 muJ in each of the pulses and with nominal photon energies of 990 and 1030 eV, generated simultaneously in the two-color operation mode of the SASE3 undulator.*

Particularly challenging is the simultaneous generation of two well-controlled soft X-ray pulses. Generally, the complex electron dynamics in the electron bunches results in strong interaction between the two pulses, especially in the regime of temporally overlapping or nearly overlapping pulses. The precise control of the temporal delay is therefore rather challenging, although recent studies have indicated that even the control of the spectral phase seems to be possible. A specific challenge for experiments at the European XFEL lies in the parallel operation of three dedicated undulators: two for the hard and one for the soft X-ray wavelength [8]. As a consequence, sophisticated manipulation of the electron bunches, as required for the generation of attosecond pulses, is often in conflict with the specific operation mode of another undulator. Therefore, it is rather difficult to fully exploit all capabilities of the machine and to push the parameter space over a certain limit.

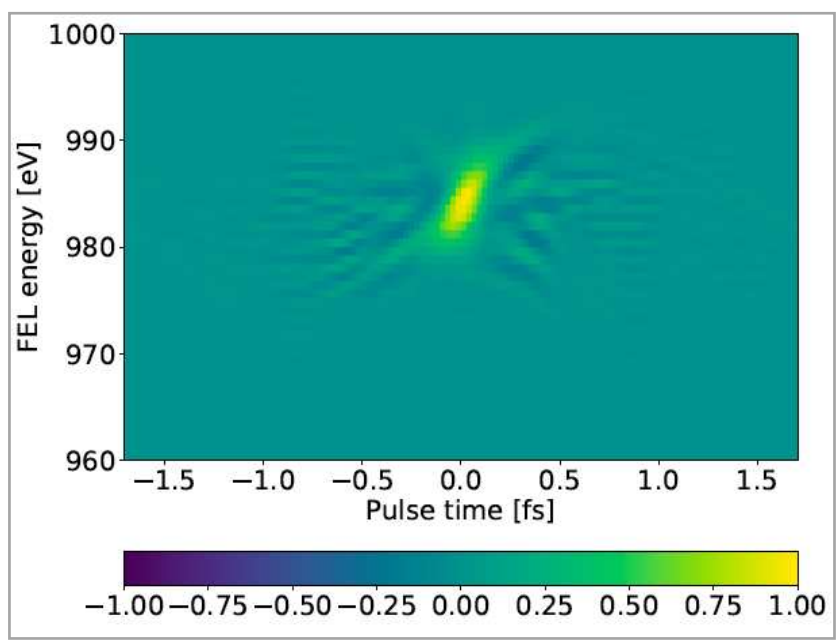

**Figure 2.** *Two-dimensional map of the spectral and temporal distribution of a sub-femtosecond soft X-ray pulse with nominal photon energy of 984 eV reconstructed from the analysis of the electron angular streaking measurements of the Ne 1s photoelectron using an array of 16 time-of-flight electron spectrometers [9].*

**Advances in science and technology to meet challenges**

The understanding of the electron dynamics in the accelerator has been strongly improved during the last years allowing for a more precise control of the different electron parameters. However, a number of steps for improving their control still needs to be undertaken: in fact, electron bunch control in terms of tilts, orbit and dispersion is pivotal to the generation of attosecond pulses with controlled properties.

For the two-color attosecond operation of the SASE3 undulator, an optical delay line [4] was recently inserted into the magnetic chicane. For future applications, this will enable a strong reduction of the mutual perturbation between the two pulses leading thereby to a better control of the zero-crossing, as well as allowing the simple switch-off of the first color.

The photon diagnostics at free-electron laser facilities has also been developed largely and various techniques are at hand to characterize the pulses. However, relatively simple diagnostics devices that may become standard for attosecond pulse characterization still need to be developed. Moreover, when addressing pump-probe experiments with two x-ray pulses or, a more challenging case, with an optical pump used together with an X-ray probe, timing of pump and probe with attosecond-level accuracy becomes of fundamental importance.

Finally, several experiments with attosecond pulses have been performed already at the European XFEL. However, the scientific instruments still need to be optimized for regular attosecond-science investigations. In order to address these challenges, a strategy program focusing on delivering attosecond pulses to users in a well-controlled and well-characterized way is being launched at the European XFEL, including but not limited to the soft X-ray region of the spectrum. It addresses generation of radiation, in particular optimization and control of pulse characteristics and simplified tuning procedure; photon diagnostics, towards characterization of attosecond pulses by means of standardized diagnostics devices; and finally preparing instruments for experiments. This complex program will benefit from a number of synergies and capabilities that will be developed, pending funding availability, independently of it. These include, but are not limited to sub-fs UV laser pulses for optical/X-ray pump-probe, optimized synchronization between external pump laser and X-ray probe, down to the few femtosecond level, electron beam phase-space diagnostics, greatly helping the FEL tuning process.

## Concluding remarks

In summary, the European XFEL is currently featuring attosecond capabilities that span from the generation of attosecond pulses, to the delivery at the experimental stations, including pulse characterization capabilities. There are still a number of steps to be undertaken, in order to deliver attosecond pulses to users in a fully reliable and reproducible way. However, first experiments have already been performed demonstrating the huge potential to develop new scientific investigations. By characterising, on a pulse-by-pulse level, the individual soft X-ray pulses and by applying suitable sorting procedures with respect to pulse energy, pulse duration and photon energy, novel information on the dynamics of ultrafast processes have already been obtained. Future applications will be able to address various unresolved problems related to time-resolved investigations of electron dynamics as well as to new approaches for non-linear studies [1, 2, 10-14].

In parallel to offering the attosecond pulses already for scientific application, the control on the pulse properties and their characterization will gradually be improved. Besides modulation of wavelength, pulse energy and pulse shape, the control of chirp as well as the spectral phase of the soft X-ray pulses represent very challenging but also possible developments in the future at free-electron laser sources.

## Acknowledgements

We acknowledge the European XFEL in Schenefeld, Germany, for the provision of X-ray free-electron laser beam time at the SQS instrument and thank the EuXFEL staff for their assistance. In particular, we are very thankful to the SQS and the FEL R&D teams as well as to all co-authors of reference [2].

## References

[1] Serkez S, Geloni G, Tomin S, Feng G, Gryzlova E V, Grum-Grzhimailo A N and Meyer M 2018 Overview of options for generating high-brightness attosecond x-ray pulses at free-electron lasers and applications at the European XFEL *J. Opt.* **20** 024005

[2] Funke L, Ilchen M, Dingel K, Mazza T, Mullins T, Otto T, Rivas D E, Savio S, Serkez S, Walter P, Wieland N, Wülfing L, Bari S, Boll R, Braune M, Calegari F, De Fanis A, Decking W, Duensing A, Düsterer D, Egun F, Ehresmann A, Erk B, Ferreira de Lima D E, Galler A, Geloni G, Guetg M, Grünert J, Grychtol P, Hans A, Held A, Hindriksson R, Jahnke T, Laksman J, Larsson M, Liu J, Marangos J P, Marder L, Meier D, Meyer M, Mirian N, Ott C, Passow C, Pfeifer T, Rupprecht P, Schletter A, Schmidt P, Scholz F, Schott S, Schneidmiller E, Sick B, Tiedtke K, Usenko S, Wanie V, Wurzer M, Yurkov M, Zhaunerchyk V and Helml W 2025 Capturing Nonlinear Electron Dynamics with Fully Characterised Attosecond X-ray Pulses https://doi.org/10.48550/arXiv.2408.03858

[3] Hartmann N, Hartmann G, Heider R, Wagner M S, Ilchen M, Buck J, Lindahl A O, Benko C, Grünert J, Krzywinski J, Liu J, Lutman A A, Marinelli A, Maxwell T, Miahnahri A A, Moeller S P, Planas M, Robinson J, Kazansky A K, Kabachnik N M, Viefhaus J, Feurer T, Kienberger R, Coffee R N and Helml W 2018 Attosecond time-energy structure of X-ray free-electron laser pulses Nature Photonics **12,** 215

[4] Serkez S, Decking W, Froehlich L, Gerasimova N, Grünert J, Guetg M, Huttula M, Karabekyan S, Koch A, Kocharyan V, Kot Y, Kukk E, Laksman J, Lytaev P, Maltezopoulos T, Mazza T, Meyer M, Saldin E,  Schneidmiller E, Scholz M, Tomin S, Vannoni M, Wohlenberg T, Yurkov M, Zagorodnov I and Geloni G 2020 Opportunities for Two-Color Experiments in the Soft X-ray Regime at the European XFEL Appl. Sci.**10**, 2728

[5] Lutman A,  Maxwell J, MacArthur J, Guetg M, Berrah N, Coffee R, Ding Y, Huang Z, Marinelli A, Moeller S, Zemella J 2016 Fresh-slice multicolour X-ray free-electron lasers, Nature Photonics **10**, 745 http://www.nature.com/doifinder/10.1038/nphoton.2016.201

[6] Kondratenko A and Saldin E 1980 Generating of coherent radiation by a relativistic electron beam in an ondulator Part. Accel. **10**, 207

[7] Bonifacio R, Pellegrini C and Narducci L M 1984 Collective instabilities and high-gain regime in a free electron laser Opt. Commun. **50**, 373

[8] Tschentscher T, Bressler C, Grünert J, Madsen A, Mancuso A P, Meyer M, Scherz A, Sinn H and Zastrau U 2017 Photon Beam Transport and Scientific Instruments at the European XFEL Appl. Sci. **7**, 592

[9] Ferreira de Lima D 2025 private communication

[10]     Calegari F, Sansone G, Stagira S, Vozzi C  and Nisoli M 2016 Advances in attosecond science J. Phys. B: At. Mol. Opt. Phys. **49** 062001

[11]     Kowalewski M, Bennett K, Dorfman K E and Mukamel S 2015 Catching Conical Intersections in the Act: Monitoring Transient Electronic Coherences by Attosecond Stimulated X-Ray Raman Signals Phys. Rev. Lett. **115**, 193003

[12]     Kuleff A I, Kryzhevoi NV, Pernpointner M and Cederbaum L S 2016 Core Ionization Initiates Subfemtosecond Charge Migration in the Valence Shell of Molecules Phys. Rev. Lett. **117**, 093002

[13]     Calegari F and Martin F 2023 Open questions in attochemistry Communications Chemistry **6**, 184

[14]     Alexander O, Egun F, Rego L, Martinez Gutierrez A, Garratt D, Cárdenas G A, Nogueira J J, Lee J P, Zhao K, Ru-Wang R-P, Ayuso D, Barnard J C T, Beauvarlet S, Bucksbaum P H, Cesar D, Coffee R, Duris J, Frasinski L J, Huse N, Kowalczyk K M, Larsen K A, Matthews M, Mukamel S, O'Neal J T, Penfold T, Thierstein E, Tisch J W G, Turner J R, Vogwell J, Driver T, Berrah N, Lin M-F, Dakovski G L, Moeller S P, Cryan J P, Marinelli A, Picón A and Marango J P 2024 Attosecond impulsive stimulated X-ray Raman scattering in liquid water Sci. Adv. 10, eadp0841

## *13. All-attosecond pump-probe spectroscopy*

**Bernd Schütte* and Marc J. J. Vrakking***

Max-Born-Institut, Max-Born-Str. 2A, 12489 Berlin, Germany

Bernd.Schuette@mbi-berlin,de, Marc.Vrakking@mbi-berlin.de

**Status**

Most attosecond experiments so far have combined an attosecond pump or probe pulse—typically in the extreme-ultraviolet (XUV) or soft X-ray range—with a femtosecond infrared pulse. This approach has revealed unprecedented details of ultrafast electron dynamics in atoms, molecules, liquids, and solids, but it also has important limitations. Therefore, since the first demonstration of attosecond pulses [1], a key objective has been to develop attosecond-pump attosecond-probe spectroscopy (APAPS). APAPS offers three major advantages: (i) both pump and probe pulses have attosecond durations, making it possible to track electron dynamics from attosecond to few-femtosecond timescales; (ii) strong near-infrared (NIR) laser fields are avoided, preventing them from changing or masking the dynamics of interest; and (iii) both pump and probe pulses can selectively excite valence and/or core levels.

Progress toward APAPS has been slow, as it requires intense attosecond pulses ($\approx 10^{14}$ W/cm$^2$). Given the low conversion efficiencies of high-harmonic generation (HHG, typically $<10^{-4}$), this has been a major challenge. First proof-of-principle APAPS measurements demonstrated XUV autocorrelation in atoms and molecules, revealing pulse durations between 0.5 and 1.5 fs [2–4]. Recently, APAPS has been demonstrated at the free-electron laser (FEL) LCLS [5,6], extending the method into the soft X-ray regime. HHG- and FEL-based attosecond sources are therefore complementary, providing access to different wavelength ranges.

In a recent breakthrough, we demonstrated table-top all-attosecond transient absorption spectroscopy (AATAS), which has allowed us to observe valence hole motion in rare-gas atoms [7]. This accomplishment was made possible through the following strategy: (i) Instead of scaling the XUV pulse energy, we emphasized optimizing the XUV intensity [8], at the same time reducing the required NIR driving laser pulse energy. (ii) This approach allows the use of turn-key laser systems operating at kHz repetition rates for APAPS, improving experimental statistics [9,10]. (iii) We applied a highly stable and robust cascaded post-compression scheme [10,11], overcoming limitations associated with hollow-core fibers and multipass cells. In Fig. 1, a two-color APAPS experiment in Ar ions is shown, exhibiting an excellent signal-to-noise ratio. Together, these advances have enabled us to obtain AATAS data in various molecules and solids, with results currently in preparation for publication.

**Current and future challenges**

In attochemistry, a major challenge is the investigation of charge migration in biologically relevant molecules. This research is still in its infancy [12–14]. The understanding and control of charge migration may enable charge-directed reactivity. APAPS may contribute to the accomplishment of this goal by studying the pump-pulse induced dynamics in these molecules on attosecond to femtosecond timescales while avoiding strong NIR laser fields. APAPS may moreover be used to study conical intersection dynamics, a major challenge in atto- and femtochemistry [15].

Until now, the investigation of solids using APAPS has received little attention, both experimentally and theoretically. A major technical challenge in these experiments is preventing sample degradation while still obtaining sufficiently strong signals. Recently, we demonstrated that this is possible by performing all-attosecond reflectivity and absorption spectroscopy in the ionic crystal $BaF_2$. The ability to achieve core excitation and element specificity in both the pump and the probe steps is particularly relevant in these experiments. While the full potential of APAPS in solids is yet to be realized, promising applications may include the real-time observation of core-hole decay processes, charge transfer dynamics, dephasing, as well as exciton and trion dynamics.

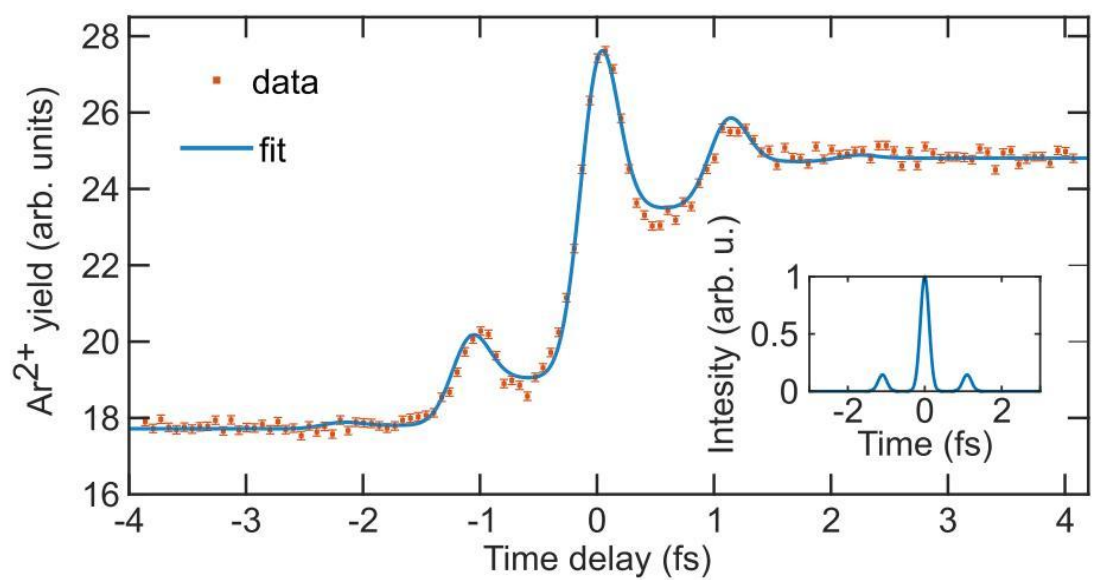

**Figure 1.** *Two-color APAPS performed in Ar. The pump pulse, dominated by contributions below the second ionization potential (27.6 eV), ionizes Ar. When the probe pulse, centered at 31 eV, arrives after the pump pulse (positive time delays), it can further ionize Ar+, leading to an increased Ar2+ signal. Additional maxima appear when the individual attosecond bursts overlap in time. The blue solid line shows a fit based on a simple model, which allowed us to estimate the attosecond pulse structure, as shown in the inset. The main attosecond burst has a width of 270 as, and the pre- and post-pulses have relative intensities of 14 %. Reprinted from Ref. [10].*

**Advances in science and technology to meet challenges**

APAPS will benefit from the development of intense attosecond sources operating at high repetition rates. In this context, ytterbium lasers are expected to play a key role. When combined with cascaded post-compression, they can provide excellent long-term stability [11], enabling long acquisition times and effective noise suppression. A notable drawback, however, is their central wavelength of 1030 nm, which results in lower HHG conversion efficiencies compared to 800 nm driving pulses. To mitigate this, the second harmonic at 515 nm can be employed to drive HHG, reaching conversion efficiencies on the order of $10^{-4}$ [16–18]. However, the obtained pulse durations > 14 fs so far remain too long for generating isolated attosecond pulses with high pulse energies. In the future, cascaded post-compression may provide the few-femtosecond pulses required.

To access inner-shell levels in atoms and molecules, it is desirable to perform HHG-driven APAPS at higher photon energies. Recently, we demonstrated APAPS in the 100 eV region. Given the rapid progress in the field, it is foreseeable that laboratory-scale APAPS will extend into the soft X-ray region, using an XUV attosecond pump pulse in combination with a soft X-ray attosecond probe pulse.

## Concluding remarks

Recent progress in the development of all-attosecond pump–probe spectroscopy indicates that it is on its way to becoming a mature technology, ready to be implemented in other HHG-based laboratories. The ability to investigate ultrafast electron dynamics will be further enhanced by the advancement of attosecond sources operating at higher repetition rates and at higher photon energies in the XUV and X-ray regions.

## Acknowledgements

We acknowledge funding from the DFG projects 456137830 and 471478110 as well as from the Leibniz Collaborative Excellence Programme K612/2024.

## References

[1]  Hentschel M, Kienberger R, Spielmann C, Reider G A, Milosevic N, Brabec T, Corkum P, Heinzmann U, Drescher M and Krausz F 2001 Attosecond metrology *Nature* **414** 509–13

[2]  Sekikawa T, Kosuge A, Kanai T and Watanabe S 2004 Nonlinear optics in the extreme ultraviolet *Nature* **432** 605–8

[3]  Tzallas P, Skantzakis E, Nikolopoulos L a. A, Tsakiris G D and Charalambidis D 2011 Extreme-ultraviolet pump–probe studies of one-femtosecond-scale electron dynamics *Nature Phys* **7** 781–4

[4]  Takahashi E J, Lan P, Mücke O D, Nabekawa Y and Midorikawa K 2013 Attosecond nonlinear optics using gigawatt-scale isolated attosecond pulses *Nature Communications* **4** 2691

[5]  Li S, Lu L, Bhattacharyya S, Pearce C, Li K, Nienhuis E T, Doumy G, Schaller R, Moeller S and Lin M-F 2024 Attosecond-pump attosecond-probe x-ray spectroscopy of liquid water *Science* eadn6059

[6]  Guo Z, Driver T, Beauvarlet S, Cesar D, Duris J, Franz P L, Alexander O, Bohler D, Bostedt C, Averbukh V, Cheng X, DiMauro L F, Doumy G, Forbes R, Gessner O, Glownia J M, Isele E, Kamalov A, Larsen K A, Li S, Li X, Lin M-F, McCracken G A, Obaid R, O'Neal J T, Robles R R, Rolles D, Ruberti M, Rudenko A, Slaughter D S, Sudar N S, Thierstein E, Tuthill D, Ueda K, Wang E, Wang A L, Wang J, Weber T, Wolf T J A, Young L, Zhang Z, Bucksbaum P H, Marangos J P, Kling M F, Huang Z, Walter P, Inhester L, Berrah N, Cryan J P and Marinelli A 2024 Experimental demonstration of attosecond pump–probe spectroscopy with an X-ray free-electron laser *Nat. Photon.* **18** 691–7

[7]  Mikhail Volkov, Evaldas Svirplys et al. in preparation

[8]  Senftleben B, Kretschmar M, Hoffmann A, Sauppe M, Tümmler J, Will I, Nagy T, Vrakking M J J, Rupp D and Schütte B 2020 Highly non-linear ionization of atoms induced by intense high-harmonic pulses *Journal of Physics: Photonics* **2** 034001

[9]  Kretschmar M, Svirplys E, Volkov M, Witting T, Nagy T, Vrakking M J J and Schütte B Compact realization of all-attosecond pump-probe spectroscopy *Science Advances* **10** eadk9605

[10] Sobolev E, Volkov M, Svirplys E, Thomas J, Witting T, Vrakking M J J and Schütte B 2024 Terawatt-level three-stage pulse compression for all-attosecond pump-probe spectroscopy *Opt. Express, OE* **32** 46251–8

[11] Tsai M-S, Liang A-Y, Tsai C-L, Lai P-W, Lin M-W and Chen M-C 2022 Nonlinear compression toward high-energy single-cycle pulses by cascaded focus and compression *Science Advances* **8** eabo1945

[12] Cederbaum L S and Zobeley J 1999 Ultrafast charge migration by electron correlation *Chemical Physics Letters* **307** 205–10

[13] Calegari F, Ayuso D, Trabattoni A, Belshaw L, De Camillis S, Anumula S, Frassetto F, Poletto L, Palacios A, Decleva P, Greenwood J B, Martín F and Nisoli M 2014 Ultrafast electron dynamics in phenylalanine initiated by attosecond pulses *Science* **346** 336–9

[14] Kraus P M, Mignolet B, Baykusheva D, Rupenyan A, Horný L, Penka E F, Grassi G, Tolstikhin O I, Schneider J, Jensen F, Madsen L B, Bandrauk A D, Remacle F and Wörner H J 2015 Measurement and laser control of attosecond charge migration in ionized iodoacetylene *Science* **350** 790–5

[15] Schuurman M S and Stolow A 2018 Dynamics at Conical Intersections *Annual Review of Physical Chemistry* **69** 427–50

[16] Klas R, Kirsche A, Gebhardt M, Buldt J, Stark H, Hädrich S, Rothhardt J and Limpert J 2021 Ultra-short-pulse high-average-power megahertz-repetition-rate coherent extreme-ultraviolet light source *PhotoniX* **2** 4

[17] Hell S, Späthe J, Førre M, Klas R, Rothhardt J, Limpert J, Moshammer R, Ott C, Paulus G G, Fritzsche S and Kübel M 2025 Coincidence measurement of two-photon double ionization of argon through an autoionizing resonance *Phys. Rev. Res.* **7** L032030

[18] Descamps D, Guichard F, Petit S, Beauvarlet S, Comby A, Lavenu L and Zaouter Y 2021 High-power sub-15 fs nonlinear pulse compression at 515 nm of an ultrafast Yb-doped fiber amplifier *Opt. Lett., OL* **46** 1804–7

## 14. State Resolved Dynamics by Attosecond Noncollinear Four-Wave-Mixing

**Patrick Rupprecht[1,2], Nicolette G. Puskar[1,2], Daniel M. Neumark[1,2] and Stephen R. Leone[1,2,3]\***

[1] Department of Chemistry, University of California, Berkeley, Berkeley, USA
[2] Chemical Sciences Division, Lawrence Berkeley National Laboratory, Berkeley, USA
[3] Department of Physics, University of California, Berkeley, Berkeley, USA

srl@berkeley.edu

**Status**

The direct observation of the shortest time dynamics in nature became possible with the generation of attosecond laser pulses [1]. Table-top high harmonic generation (HHG) can create attosecond pulses spanning the extreme ultraviolet (XUV) to x-ray regimes, enabling the study of electronic motion in valence and core-level orbitals. At higher photon energies, photoabsorption is localized to atomic edges, enabling XUV or x-ray absorption experiments to serve as atomically specific probes for electronic processes. Pump-probe techniques such as attosecond transient absorption spectroscopy (ATAS) capitalize on this idea by introducing an XUV or x-ray pulse either before or after a relevant visible (VIS) or ultraviolet (UV) pulse, to time-resolve element-specific electronic dynamics sensitive to spin, oxidation states, chirality, and chemical environment [2].

A particularly powerful means of using attosecond pulses to probe the dynamics of highly excited electronic states is attosecond noncollinear four-wave-mixing (FWM) spectroscopy [3, 4]. In this scheme, one attosecond XUV pulse interacts with two time-delayable, noncollinear NIR pulses in a third-order nonlinear process, generating phase-matched, spatially isolated FWM signals (Fig. 1). The background-free nature of attosecond noncollinear FWM avoids the spectral congestion typical in ATAS spectra, such as Stark effects, hyperbolic sidebands, and multi-photon interferences. Pulse-ordering can be tailored to elucidate the dynamics of either optically dipole-allowed ("bright") or dipole-forbidden ("dark") states relative to the XUV ground state. These geometric and temporal conditions make FWM an exquisitely quantum-path-specific technique.

Over the last decade, table-top attosecond noncollinear FWM has been established as a technique to disentangle the fastest electronic dynamics in atoms, molecules, and materials. In atoms, it has been used to probe ultrafast Auger-Meitner decay of doubly excited electronic states in helium [5] and investigated non-perturbative effects and Rabi cycling in argon [6]. In molecules, it has revealed population dynamics of the $CO_2$ molecule's Henning diffuse Rydberg series contained within the signal for the Henning sharp Rydberg series due to the XUV-induced coherence transfer between multiple FWM pathways [7]. Moreover, the dependence of the quantum-state lifetime on the vibrational sublevel has been investigated with FWM for the 3s Rydberg series of $O_2$ [8]. In materials, it has been used to decipher core-exciton dynamics in NaCl [9]. Attosecond transient grating spectroscopy (ATGS), a variation of noncollinear FWM, has explored ultrafast photoexcited dynamics in Sb semimetal thin films, such as the generation of coherent phonons, lattice and carrier dynamics, and carrier-phonon interactions [10]. ATGS now is a dedicated setup at the FERMI FEL, where pioneering work on the thermal and coherent phonon dynamics in crystalline silicon and amorphous silicon nitride have been conducted [11]. Broadly applicable across chemical elements and species, attosecond noncollinear FWM continues to advance our understanding of fundamental chemical physics at the shortest time scales.

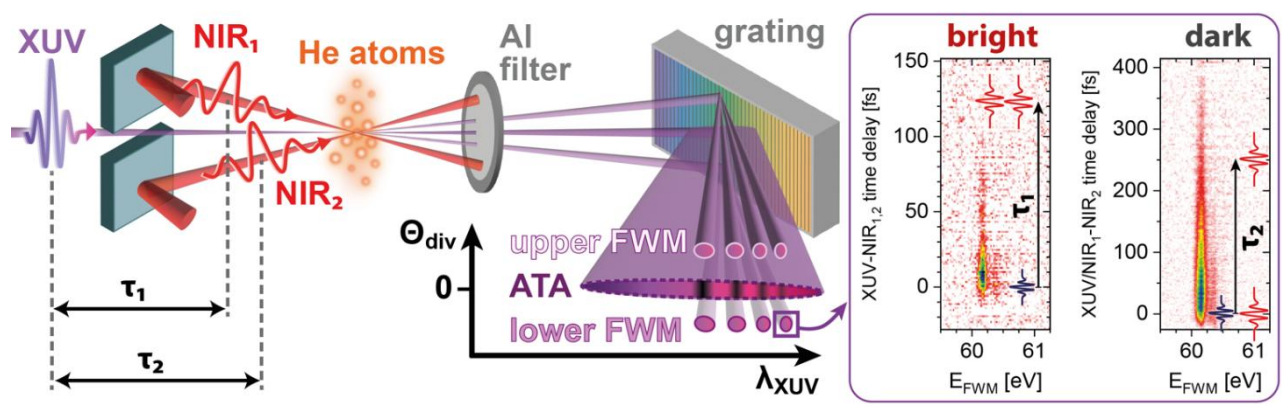

**Figure 1** *Attosecond noncollinear four-wave-mixing scheme applied to doubly excited states in helium. The noncollinear few-femtosecond pulses NIR1 and NIR2 mix with the XUV attosecond pulse in a gas-phase helium target. After cleaning the XUV emission from residual NIR light with a thin aluminium (Al) filter, the background-free FWM signals (here depicted as a Rydberg series) are spectrally dispersed with a reflective grating. Depending on which time delay is scanned, the 17.6 fs lifetime of the dipole-accessible ("bright") state 2s2p can be measured or alternatively the 119 fs lifetime of the 2p2 ("dark") state is obtained, which is only accessible via a two-photon transition from the ground state. Reprinted from [5].*

**Current and future challenges**

Attosecond FWM techniques show great potential to disentangle quantum dynamics in complex systems. Targets include combining atomic-site with quantum-state specificity in chemical dynamics, tracing charge migration in solid-state materials on small length scales, and deciphering population dynamics from vibronic or electronic coherences in photochemistry. To successfully apply FWM in such scientific scenarios, current challenges to be addressed concerning spatial and temporal resolution, quantum-path and energetic selectivity, as well as advanced pulse sequence schemes.

One way to gain more insights into complex quantum dynamics is to observe them from the point of view of a single atom within a molecule. This can be achieved by addressing tightly bound core-level electrons with x-ray pulses. To apply FWM spectroscopy to the dynamics of organic molecules, the carbon K-edge around 280 eV is targeted. While FWM of core-level electrons has recently been demonstrated in the XUV spectral region for xenon atoms [12], soft x-ray attosecond FWM spectroscopy will offer unprecedented insights into dynamics of chemically and biologically relevant molecules. Furthermore, core-level photon spectroscopy is extremely sensitive to coherent motion between nuclei, e.g. vibrations in molecules, down to the femtometer level [13]. Attosecond FWM studies with well-defined impulsive vibrational excitation pulses will be crucial to investigate the interplay between vibrational motion with specific electronic states. If the transient grating itself is generated via intense x-ray pulses at FELs, this provides grating periods in the nanometer scale, ideally suited to investigate and spatially resolve phonon dynamics in solid-state systems [14].

Excitations that include core-hole states, however, tend to have very short, few-femtosecond lifetimes, necessitating a temporal resolution in FWM experiments on this timescale or below. So far, sub-6 fs state lifetimes have been successfully characterized using few-cycle laser pulses [8, 12]. To significantly improve the temporal resolution, sub-cycle transients in the infrared-to-ultraviolet spectral regime are needed. A drawback of such pulses is their large spectral bandwidth, which compromises quantum-path selectivity as multiple dark states can be coupled to the bright states at the same time. One solution to this problem is to employ a spectral pulse-shaper for the optical pulses to selectively block small spectral windows and hence specific dark-state couplings while maintaining the overall laser pulse duration. This approach is also beneficial to characterize the resonance energies of dark states, which are often not precisely measurable with other techniques. Furthermore, recent progress on table-top sources for intense attosecond pulses as well attosecond FEL pulses lead the way to all-attosecond FWM experiments.

**Advances in science and technology to meet challenges**

The successful application of FWM spectroscopy to ever-more complex systems requires a detailed understanding of the underlying nonlinear processes. Recent studies have focused on nonperturbative effects in FWM spectroscopy [6, 12]. The interpretation of these experiments has been enabled by simulations that take the multi-emitter character of FWM into consideration [15,16]. Due to the low XUV pulse intensities and transition-dipole moments, FWM experiments often require intense NIR or VIS pulses to yield FWM signals with an acceptable signal-to-noise ratio (snr). With the introduction of Ytterbium laser technology to attosecond science [17], high-repetition-rate HHG sources for XUV and soft x-ray pulses will produce acceptable snr with perturbative NIR/VIS intensities to guarantee reliable state-specific lifetime extraction for a wide variety of quantum systems.

Recent progress in the generation of few-femtosecond UV pulses via resonant-dispersive-wave generation in a hollow-core fiber [18] opens new opportunities to apply the quantum-state-specificity of FWM spectroscopy to photochemistry. Novel FWM schemes target open questions in nonadiabatic dynamics, e.g.: How long will an electronic coherence induced by a conical intersection (CI; where two photoexcited potential energy surfaces are degenerate) persist and can this coherence be measured experimentally? An electronic Raman scheme, TRUECARS [19], was proposed in 2015 to detect electronic coherences after CIs but has yet to be experimentally demonstrated due to the high technical requirements of phase-stable attosecond hard x-ray pulses. Heterodyned FWM (Hd-FWM) spectroscopy can offer a table-top alternative by interfering a population-encoding FWM signal in the XUV with a coherence-sensitive one [20]. While intense femtosecond NIR/VIS pulses provide the spectral quantum-pathway specificity, a weak attosecond XUV or x-ray pulse enables a sub-femtosecond temporal resolution. Hd-FWM relies on laser pulse parameters that are readily available with current ultrafast laser technology. Applying Hd-FWM to organic molecules might grant unprecedented insights into the role of electronic coherence or its decay/absence in the photodynamics occurring in nature. Hd-FWM may provide the interferometric precision and quantum-pathway specificity required to investigate entanglement of dissociating particles by tomographic means in the future.

A noncollinear geometry can also benefit XUV/x-ray second-harmonic generation (SHG) spectroscopy. SHG is ideally suited to investigate interfaces, e.g., of solids. Combining SHG with the elemental specificity provided by XUV and x-ray pulses results in the capability to temporally resolve and disentangle complex dynamics at surfaces without being dominated by the bulk response [21]. Here, spatially isolating the SHG signal via a noncollinear input geometry, e.g. using masks, helps to separate SHG from other contributions in the same spectral region.

**Concluding remarks**

Noncollinear attosecond four-wave-mixing is becoming a crucial tool to disentangle complex ultrafast dynamics with both atomic site-and quantum-state selectivity. Recent efforts at FELs towards nanometer-period XUV transient gratings specifically target temporally resolving quantum-dynamics in solid-state systems with high spatial resolution. Table-top attosecond XUV spectroscopy, on the other hand, focuses on measuring few-femtosecond population and coherence dynamics in gases and solid-state media with quantum-pathway specificity in the perturbative regime and beyond. New table-top FWM schemes pave the way for quantum-state tomography of nonadiabatic dynamics and entanglement in photochemistry. A noncollinear geometry can also help to unambiguously extract nonlinear signals of surface-selective techniques like XUV and x-ray SHG spectroscopy.

## Acknowledgements

The authors gratefully acknowledge support from the Department of Energy, Office of Science, Basic Energy Sciences (Contract No. DE-AC02-05CH11231 from the Atomic, Molecular and Optical Sciences Program, through Lawrence Berkeley National Laboratory). Additional support is provided by the Air Force Office of Scientific Research (Contract Nos. FA9550-20-1-0334 and FA9550-24-1-0184, SRL), the Alexander von Humboldt Foundation (PR), Soroptimist International of the Americas (NGP), as well as important input from Rafael Quintero-Bermudez and Kevin Xiong.

## References

[1] Paul, P. M., Toma, E. S., Breger, P., Mullot, G., Augé, F., Balcou, Ph., Muller, H. G., and Agostini, P. 2001 Observation of a Train of Attosecond Pulses from High Harmonic Generation *Science* **292** 1689

[2] Geneaux, R., Marroux, H. J. B., Guggenmos, A., Neumark, D. M. and Leone, S. R. 2019 Transient absorption spectroscopy using high harmonic generation: a review of ultrafast X-ray dynamics in molecules and solids *Philos. Trans. R. Soc. A* **2145** 20170463

[3] Puskar, N. G., Lin, Y-C., Gaynor, J. D., Schubert, M. C., Chattopadhyay, S., Marante, C. Fidler, A. P., Keenan, C. L., Argenti, L. Neumark, D. M., and Leone, S. R. 2023 Measuring autoionization decay lifetimes of optically forbidden inner valence excited states in neon atoms with attosecond noncollinear four-wave-mixing spectroscopy *Phys. Rev. A* **107** 033117

[4] Leone, S. R., and Neumark, D. M. 2023 Probing matter with nonlinear spectroscopy *Science* **379** 536-537

[5] Rupprecht, P., Puskar, N. G., Neumark, D. M., and Leone, S. R. 2024 Extracting doubly excited state lifetimes in helium directly in the time domain with attosecond noncollinear four-wave-mixing spectroscopy *Phys. Rev. Research* **6** 043100

[6] Yanez-Pagans, S., Harkema, N., Sandhu, A., Cariker, C., and Argenti, L. 2025 Non-perturbative effects in attosecond four-wave mixing spectra *Phys. Rev. A* **112** 013107

[7] Fidler, A. P., Lin, Y-C., Gaynor, J. D., McCurdy, C. W., Leone, S. R., Lucchese, R. R., and Neumark, D. M. 2022 State-selective probing of $CO_2$ autoionizing inner valence Rydberg states with attosecond extreme ultraviolet four-wave-mixing spectroscopy *Phys. Rev. A* **106** 063525

[8] Lin, Y.-C., Fidler, A. P., Sandhu, A., Lucchese, R. R., McCurdy, C. W., Leone, S. R. and Neumark, D. M. 2021 Coupled nuclear–electronic decay dynamics of $O_2$ inner valence excited states revealed by attosecond XUV wave-mixing spectroscopy *Faraday Discuss.* **228** 537-554

[9] Gaynor, J. D., Fidler, A. P., Lin, Y-C., Chang, H-T., Zuerch, M., Neumark, D. M., and Leone, S. R. 2021 Solid state core-exciton dynamics in NaCl observed by tabletop attosecond four-wave-mixing spectroscopy *Phys. Rev. B* **103** 245140

[10] Quintero-Bermudez, R., Drescher, L., Eggers, V., Xiong, K. G., and Leone, S. R. 2025 Attosecond transient grating spectroscopy with near-infrared grating pulses and an extreme ultraviolet diffracted probe *ACS Photonics* **12** 2097

[11] Bencivenga, F., Mincigrucci, R., Capotondi, F., Foglia, L., Naumenko, D., Maznev, A. A., Pedersoli, E., Simoncig, A., Caporaletti, F., Chiloyan, V., Cucini, R., Dallari, F., Duncan, R. A., Frazer, T. D., Gaio, G., Gessini, A., Giannessi, L., Huberman, S., Kapteyn, H., Knobloch, J., Kurdi, G., Mahne, N., Manfredda, M., Martinelli, A., Murnane, M., Principi, E., Raimondi, L., Spampinati, S., Spezzani, C., Trovò, M., Zangrando, M., Chen, G., Monaco, G., Nelson, K. A., and Masciovecchio, C. 2019 Nanoscale transient gratings excited and probed by extreme ultraviolet femtosecond pulses *Sci. Adv.* **5** eaaw5805

[12] Puskar, N. G. , Rupprecht, P., Dvořák, J., Lin, Y.-C., Greene, A. E., Lucchese, R. R., McCurdy, C. W., Leone, S. R. and Neumark, D. M. 2025 Probing autoionization decay lifetimes of the $4d^{-1}6\ell$ core-excited states in xenon using attosecond noncollinear four-wave-mixing spectroscopy *J. Chem. Phys.* **163** 184302

[13] Rupprecht, P., Aufleger, L., Heinze, S., Magunia, A., Ding, T., Rebholz, M., Amberg, S., Mollov, N., Henrich, F., Haverkort, M. W., Ott, C. and Pfeifer, T. 2023 Resolving vibrations in a polyatomic molecule with femtometer precision via x-ray spectroscopy *Phys. Rev. A* **108** 032816

[14]      Ferrari, E., Ueda, H., Fainozzi, D., Osaka, T., Bencivenga, F., Burian, M., Carrara, P., Vila-Comamala, J., Cucini, R., David, C., Gessini, A., Gerber, S., Goloborodko, A., Hrabec, A., Inoue, I., Inubushi, Y., Diniz Leroy, L. M., Mincigrucci, R., Paris, E., Pedrini, B., Roesner, B., Rouxel, J. R., Serrat, C., Scagnoli, V., Tono, K., Yabashi, M., Yamada, J., Yamamoto, K., Zdora, M. C., Beye, M., Masciovecchio, C., Chergui, M., Staub, U. and Svetina, C. 2025 All hard X-ray transient grating spectroscopy *Commun. Phys.* **8** 257

[15]      Mi, K., Cao, W., Xu, H., Zhang, Q., and Lu, P. 2021 Method for high precision measurement of decaying dynamics using attosecond wave-mixing spectroscopy *Opt. Express* **29** 2798-2808

[16]      Rupprecht, P., Neumark, D. M. and Leone, S. R. 2025 All-optical logic gates for extreme ultraviolet switching via attosecond four-wave mixing *arXiv* 2510.00699

[17]      Truong, T.-C., Khatri, D., Lantigua, C., Kincaid, C. and Chini, M. 2025 Few-cycle Yb-doped laser sources for attosecond science and strong-field physics *APL Photonics* **10** 040902

[18]      Travers, J. C., Grigorova, T. F., Brahms, C. and Belli, F. 2019 High-energy pulse self-compression and ultraviolet generation through soliton dynamics in hollow capillary fibres *Nat. Photonics* **13** 547-554

[19]      Kowalewski, M., Bennett, K., Dorfman, K. E. and Mukamel, S. 2015 Catching conical intersections in the act: Monitoring transient electronic coherences by attosecond stimulated X-ray Raman signals *Phys. Rev. Lett.* **115** 193003

[20]      Rupprecht, P., Montorsi, F., Xu, L., Puskar, N. G., Garavelli, M., Mukamel, S., Govind, N., Neumark, D. M., Keefer, D. and Leone, S. R. 2025 Tracing long-lived atomic coherences generated via molecular conical intersections *Phys. Rev. Lett.* **135** 233201

[21]      Helk, T., Berger, E., Jamnuch, S., Hoffmann, L., Kabacinski, A., Gautier, J., Tissandier, F., Goddet, J.-P., Chang, H. Z., Oh, J.,  Pemmaraju, C. D., Pascal, T. A., Sebban, S., Spielmann, C., and Zuerch, M. 2021 Table-top extreme ultraviolet second harmonic generation *Sci. Adv.* **7** eabe2265

## 15. Attosecond transient interferometry

**Omer Kneller[1]\* and Nirit Dudovich[2]\***

[1] Present address: Department of Physics, University of Regensburg, Regensburg, Germany
[2] Department of Complex Systems, Weizmann Institute of Science, Rehovot, Israel

omer.kneller@ur.de, nirit.dudovich@weizmann.ac.il

**Status**

Attosecond transient absorption spectroscopy (ATAS) captures the subcycle dynamics of light-driven quantum systems by the time-dependent transmission spectra of attosecond pulses, typically in the extreme-ultraviolet (XUV) spectral range. ATAS has become one of the primary metrology schemes in attosecond science, enabling the observation of a broad range of fundamental attosecond-scale phenomena, in atoms, molecules and solids [1,2]. However, the multi-eV bandwidth of the attosecond pulse leads to the inevitable excitation of many quantum paths that coherently contribute to the measured signal. Achieving a complete picture of the underlying dynamics thus requires identifying the role of individual excitation pathways and their coupling to electronic populations and coherences, on subcycle timescales. Establishing new dynamical experimental observables offers a promising route to address this challenge.

As attosecond pulses are transmitted through the target system, the electronic dynamics are imprinted not only in their spectral intensity, but also in their spectral phase and polarization state. Such transient phase information is inherently lost by the intensity measurement at the XUV spectrometer, while the weak transient polarization response remains extremely challenging to resolve. Reconstructing the temporal evolution of the quantum wavefunction from spectral-domain measurements demands access to the full complex information encoded in the attosecond pulses. Retrieving such complex information requires a fundamental step -- advancing from spectral intensity measurements to phase-resolved measurements. Recent advances in attosecond transient interferometry [3–5] have established the transient phase, which reflects the transient dynamics of the XUV refractive index, as a new and exciting observable in attosecond metrology.

Pioneering ATAS experiments have retrieved phase-sensitive information by utilizing the interference mechanisms, intrinsic to the nonlinear light–matter interaction [6-11]. In contrast, attosecond transient interferometry directly measures the transient phase of the transmitted attosecond pulses by interfering them with an external reference XUV source, integrating ATAS with XUV-XUV interferometry [12]. Time-dependent XUV interferometry was first demonstrated in free-induction decay metrology, evolving on femtosecond time scales [13]. More recently, dynamical interferometry has been integrated with ATAS, where attosecond pulses serve as probing fields [3–5]. The pioneering first demonstration [3] followed the multicycle evolution of the transient phase, validating the Kramers–Kronig relations. Later, this concept was implemented to reveal phase dynamics beyond the single-atom response [4]. Resolving the transient phase with attosecond precision revealed sub-cycle phase oscillations and enabled the direct visualization and decomposition of the quantum pathways governing the ultrafast dynamics.

**Current and future challenges**

An important challenge arises from the spectral dynamic range of the measurement. First, attosecond transient interferometry has so far been implemented using attosecond pulse trains (APT), which produce an XUV spectrum consisting of a discrete set of harmonics [3-5]. The typical few-electronvolt spacing between adjacent harmonics constitutes a key constraint, resulting in a partial spectral coverage of the electronic states. Unlocking the full potential of transient interferometry requires spectrally continuous XUV pulses, corresponding to isolated attosecond pulses (IAP). Second, current measurements have focused on the $20 - 30\ eV$ photon energy range, corresponding to single and double electronic excitations in noble atomic gases [3–5], benchmarking the approach relative to model electronic systems. Extending attosecond transient interferometry to molecular and solid-state systems requires a significant broadening of the accessible spectral range. Higher photon energies would access absorption edges associated with core-level transitions, revealing element-specific phase dynamics in core excitons of solids or charge migration in molecules [2]. Notably, this range would require an extremely high phase stability of both sources. Conversely, lower photon energies, towards the vacuum ultraviolet regime, would open the possibility of observing charge-carrier dynamics in wide-bandgap insulators [2] and bound nonadiabatic or chiral dynamics in molecules [14].

In addition, attosecond transient interferometry requires complex experimental geometries, relying on two XUV sources with attosecond relative phase stability; One probes the sample medium together with the driving laser, while the other is spatially decoupled from the sample, serving as an external reference source. Simplifying the experimental geometry may prove instrumental in enhancing the interferometric stability, thereby significantly extending the accessible spectral range. The development of new attosecond interferometry schemes will pave the way for applying this technique to a broader range of systems, enabling the observation of complex quantum phenomena with unprecedented temporal resolution.

Finally, attosecond transient metrology is predominantly scalar in nature, inducing and measuring the dynamics along a single polarization direction. Yet, vectorial dynamics play a key role [14,15], revealing broken symmetries, anomalous currents, and their interplay with the driving field. The ability to detect the subtle vectorial polarization signal is obstructed by the scalar, parallel component of the interaction, which completely overwhelms the measurement. Advancing attosecond transient metrology into the vectorial regime requires the capability to detect the weak orthogonal polarization response, determine the transient phase of the two polarization components, and simultaneously retain the ability to identify and track the underlying quantum pathways.

**Advances in science and technology to meet challenges**

Recent progress has addressed key challenges in attosecond transient interferometry -- extending the spectral dynamic range, simplifying the experimental geometry, and accessing vectorial dynamics.

New schemes for generating phase-locked pairs of isolated attosecond pulses (IAPs) that interact with a target, together with a driving laser [16–18], overcome the limited spectral range of attosecond transient interferometry. A key step in this direction has already been demonstrated through the application of IAP pairs to track femtosecond-scale dynamics of the XUV refractive index [18]. More recently, this scheme has been extended into the attosecond regime [19]. The spectrally resolved transmission of the IAP pair through a laser-driven medium encodes phase-sensitive information, effectively corresponding to two simultaneous ATAS measurements that interfere coherently at the detector. Importantly, while both IAPs interact with the medium, one of them can still act as a simple reference source by temporally separating it from any nonlinear interaction. This technique provides a background-free map of dynamical electronic coherences, with subcycle temporal and millielectronvolt spectral resolution, across the broad bandwidth of the IAP, offering a powerful route to resolve attosecond-scale phase dynamics in complex systems.

New schemes in attosecond transient interferometry offer promising avenues for reducing the complexity of the experimental geometry. The application of an IAP pair, with attosecond relative stability, generated in a single medium [16,17,19], drastically reduces the complexity of the in-vacuum optical setup, becoming comparable to that used by ATAS, while providing phase information. In addition, using split-and-delay mirrors for XUV pulse pair generation and delay control is a promising alternative approach for attosecond transient interferometry [4,18]. Interestingly, using a four-quadrant mirror enables the combination of a laser pulse pair with an attosecond pulse pair, potentially providing a multidimensional view of light–matter interactions with attosecond resolution [4].

Generalizing transient interferometry into the vectorial regime is based on combining a vectorial excitation with a highly sensitive polarimetric detection scheme. Recently, such an approach was demonstrated by integrating a vectorial driving field with vectorial XUV–XUV interferometry [20]. The vectorial field induces a dynamical symmetry breaking between consecutive half-cycles of the fundamental field, while the internal dynamics are mapped into the instantaneous polarization state of a transmitted APT. The interferometric configuration enables the heterodyne detection of the weak vectorial response, enhancing the sensitivity to subtle polarization dynamics, with subcycle temporal resolution.

**Concluding remarks**

Attosecond transient interferometry is emerging as a powerful new scheme for attosecond metrology, accessing the transient dynamics of the refractive index and revealing the underlying electronic coherences in the frequency domain. Its recent extensions into the vectorial regime and to IAPs offer an important step toward future studies, where simplified experimental geometries will enable access to new spectral frontiers and the exploration of molecular and solid-state dynamics. The direct measurement of electronic coherences and their decomposition into two orthogonal polarization components impose a stringent benchmark for theoretical models and approximations. Tracing the evolution of the vectorial complex wavefunction has the potential to reveal processes ranging from geometric effects in topological systems to field-driven chirality in complex molecules, paving the way for uncovering novel quantum phenomena at the forefront of attosecond science.

**Acknowledgements**

N.D. is the incumbent of the Robin Chemers Neustein Professorial Chair. N.D. acknowledges the Minerva Foundation, the Israeli Science Foundation, the Crown Center of Photonics and the European Research Council (ERC) for financial support. O.K. acknowledges the Yad Hanadiv foundation and the Israeli council for higher education for the awards of a Rothschild fellowship and a quantum science and technologies postdoctoral fellowship.

## References

[1] Wu, M., Chen, S., Camp, S., Schafer, K.J. and Gaarde, M.B. (2016) 'Theory of strong-field attosecond transient absorption', *Journal of Physics B: Atomic, Molecular and Optical Physics*, 49, 062003.

[2] Di Palo, N., Inzani, G., Dolso, G.L., Talarico, M., Bonetti, S. and Lucchini, M. (2024) 'Attosecond absorption and reflection spectroscopy of solids', *APL Photonics*, 9(2), 020901.

[3] Leshchenko, V. et al. (2023) 'Kramers–Kronig relation in attosecond transient absorption spectroscopy', *Optica*, 10, pp. 142–146.

[4] Hedewig, L., Kleine, C., He, Y., Wieder, F., Ott, C. and Pfeifer, T. (2025) 'State-resolved femtosecond phase control in dense-gas laser–atom interaction enabled by attosecond XUV interferometry', *Optics Letters*, 50, pp. 3006–3009.

[5] Kneller, O., Mor, C., Klimkin, N.D., et al. (2025) 'Attosecond transient interferometry', *Nature Photonics*, 19, pp. 134–141.

[6] Ott, C. et al. (2013) 'Lorentz meets Fano in spectral line shapes: a universal phase and its laser control', *Science*, 340, pp. 716–720.

[7] Ott, C. et al. (2014) 'Reconstruction and control of a time-dependent two-electron wave packet', *Nature*, 516, pp. 374–378.

[8] Goulielmakis, E., Loh, Z.-H., Wirth, A., et al. (2010) 'Real-time observation of valence electron motion', *Nature*, 466, pp. 739–743.

[9] Kaldun, A. et al. (2014) 'Extracting phase and amplitude modifications of laser-coupled Fano resonances', *Physical Review Letters*, 112, 103001.

[10] Stooß, V. et al. (2018) 'Real-time reconstruction of the strong-field-driven dipole response', *Physical Review Letters*, 121, 173005.

[11] Borisova, G.D. et al. (2024) 'Laser-induced modification of an excited-state vibrational wave packet in neutral $H_2$ observed in a pump-control scheme', *Physical Review Research*, 6, 033326.

[12] Krüger, M. and Dudovich, N. (2024) 'Attosecond interferometry', in Ueda, K. (ed.) *Ultrafast Electronic and Structural Dynamics*. Singapore: Springer. https://doi.org/10.1007/978-981-97-2914-2_2

[13] Beaulieu, S. et al. (2017) 'Phase-resolved two-dimensional spectroscopy of electronic wave packets by laser-induced XUV free induction decay', *Physical Review A*, 95, 041401.

[14] Wanie, V., Bloch, E., Månsson, E.P., et al. (2024) 'Capturing electron-driven chiral dynamics in UV-excited molecules', *Nature*, 630, pp. 109–115.

[15] Siegrist, F., Gessner, J.A., Ossiander, M., et al. (2019) 'Light-wave dynamic control of magnetism', *Nature*, 571, pp. 240–244.

[16] Koll, L.-M. et al. (2022) 'Experimental control of quantum-mechanical entanglement in an attosecond pump–probe experiment', *Physical Review Letters*, 128(4), 043201.

[17] Koll, L.M., Maikowski, L., Drescher, L., Vrakking, M.J. and Witting, T., 2022. Phase-locking of time-delayed attosecond XUV pulse pairs. *Optics Express*, 30(5), pp.7082-7095.

[18] Oshima, A., Mashiko, H., Chen, M.C., et al. (2025) 'Spectral interferometric transient complex refraction spectroscopy with extreme ultraviolet double attosecond pulses', *Communications Physics*, 8, 353.

[19] Kneller, O., Witting, T., Koll, L.-M., et al. (2025) 'Attosecond Fourier transform spectroscopy', *Research Square* [preprint], 4 March. Available at: https://doi.org/10.21203/rs.3.rs-6041019/v1

[20] Yaffe, N., Mor, C., Dudovich, N., et al. (2025) 'Vectorial attosecond transient spectroscopy', *Research Square* [preprint], 5 June. Available at: https://doi.org/10.21203/rs.3.rs-6656668/v1

## 16. Optical-field-resolved attosecond science

**Nicholas Karpowicz[1]\*, Vladislav S. Yakovlev[1,2]\***

[1] Max-Planck-Institut für Quantenoptik, Hans-Kopfermann-Str. 1, Garching 85748, Germany
[2] Ludwig-Maximilians-Universität München, Am Coulombwall 1, Garching 85748, Germany

nicholas.karpowicz@mpq.mpg.de, vladislav.yakovlev@mpq.mpg.de

**Status**

Attosecond science, since its inception, has been intimately connected to the generation and application of the attosecond pulses that emerge from high-harmonic generation (HHG). This has yielded a multitude of insights into the motion of electrons on their "natural" timescale: for example, the "breathing" motion of a hydrogen atom in a superposition of 1s and 2p states has a period of 405 attoseconds. The photons released during HHG are usually energetic, in the extreme ultraviolet and soft x-ray. This places a constraint on the interactions and materials that they can be used to interrogate: resonant interactions with soft x-ray light favour inner-shell electrons, and the low transmission in most materials combined with relatively low photon flux makes the observation of weaker interactions experimentally challenging.

Direct interaction of an attosecond pulse with the system under study is not the only possible experimental approach with attosecond time resolution, however. A different approach, which we call optical-field-resolved measurements, relies on techniques that give direct access to the electric field of light. Such measurements are standard in the terahertz spectral range, but performing them in the optical domain required the development of novel techniques, such as: Nonlinear Photoconductive Sampling (NPS) [1], Tunneling Ionization with a Perturbation for the Time-domain Observation of an Electric field (TIPTOE) [2], and Generalized Heterodyne Optical-Sampling Techniques (GHOSTs) [3]. Unlike attosecond streaking [4], these all-optical techniques do not require attosecond XUV pulses to measure the time-dependent electric field of an optical wave.

It is the underlying principle of every kind of spectroscopy that charge motion that light induces in a medium is encoded in the transmitted, reflected, or otherwise emitted electromagnetic waves. A general depiction of such a measurement in the field-resolved context is shown in Figure 1. Since the electric fields are observable in the context of field-resolved measurements, the time-dependent current density may be inferred. From the time-dependent fields and the electric current they induce, other quantities, such as the sub-cycle-resolved energy transfer, can be calculated [5].

The temporal resolution of such a measurement is not limited by the duration of the pulses, and instead is given by the bandwidth of the measurement technique. This is in contrast to the majority of pump-probe techniques, where the durations of the pump and probe pulses limit the time resolution. These factors make field-resolved measurements well-suited for exploring attosecond dynamics where lower-energy photons are involved, such as multiphoton interband transitions, coherent light-driven electron motion, coherent phonon generation, and electron scattering [6].

**Current and future challenges**

Among the main prerequisites for the future progress in field-resolved attosecond science are advancements in the measurement techniques and data analysis, design of experiments that maximize the amount and quality of information encoded by various physical processes in measured optical waves, as well as in our theoretical understanding of such experiments.

While petahertz-scale bandwidth is easily achievable in NPS and TIPTOE measurements, accurate retrieval of electric fields from measured waveforms remains a challenge. Every field detection technique is associated with a response function. In the ideal case of a linear response, this function is simply convolved with the true field to yield the measurement. Even then, deconvolution requires particular care to avoid dramatic increases of noise at spectral minima of the response. This requires improved modelling and understanding of each step in the data acquisition chain: the microscopic effects that create the observable and how macroscopic effects such as net currents form from the accumulation of these microscopic effects [7]. The key microscopic effect here is multiphoton ionization or photoinjection—NPS and TIPTOE are sensitive to how exactly a strong electric field photoinjects carriers [8]. Another essential experimental challenge here is increasing the signal-to-noise ratio of field detection techniques.

From the theoretical perspective, the primary challenge involves establishing reliable connections between measured macroscopic optical responses and underlying microscopic quantum dynamics. Brute-force *ab initio* simulation of strong-field-driven many-body dynamics remains computationally intractable, which requires development of physically motivated approximation schemes that capture essential physics while remaining computationally feasible. Current theoretical frameworks typically employ oversimplified treatments of scattering and decoherence mechanisms, despite these processes playing key roles in many experiments.

An intriguing future direction involves extending optical-field-resolved methodologies to atomically thin (two-dimensional) materials, strongly correlated electron systems, and other quantum systems, whose unique properties may present new opportunities for controlling electron motion with light.

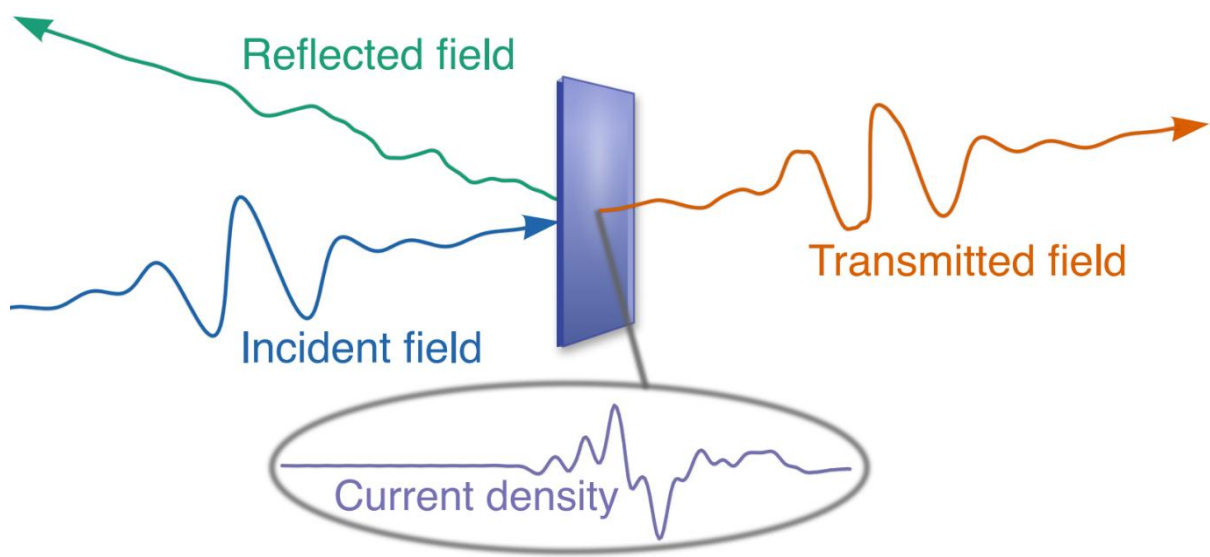

**Figure 1.** *In field-resolved attosecond science, there are four physical quantities that are coupled during the interaction with a material: the incident, reflected, and transmitted fields, and the current density. Field measurement techniques have made these observable quantities: the fields are directly observable, and the relationship between them determines the current.*

**Advances in science and technology to meet challenges**

The experimental challenges associated with field-resolved attosecond science are fundamentally different from those of HHG-based measurements, which are often limited by photon or electron counting statistics associated with the low conversion efficiency of the process. With the electric field as the observable, the bottleneck for waveform measurements is due to laser fluctuations and their effect on the signal-to-noise ratio [9]. This tends to favour higher repetition rates, and provides less incentive to increase laser power, especially when this is associated with increased technical noise. Carrier-envelope-phase (CEP) stability is especially important as CEP noise can strongly reduce the temporal resolution.

Field-resolved measurements benefit from photoinjection confined to a $\sim 1$ fs time interval, which can be achieved using recent advances in soliton self-compression of high-energy pulses in hollow-core fibres, as has been championed by John Travers and colleagues [10]. These pulses can not only compress themselves to sub-femtosecond duration, providing a reliable source of optical attosecond pulses, but also can be accompanied by an ultraviolet resonant dispersive wave carrying significant energy in the form of a short ultraviolet pulse [11]. Combining these sources with field-resolved metrology with sufficient bandwidth to resolve their field oscillations presents a wealth of new possible experimental avenues for field-resolved studies, as well as challenges associated with high-power ultraviolet sources such as degradation of the sample and optics.

Future theoretical developments must address the challenge of bridging timescales—from attosecond electron dynamics to femtosecond decoherence processes and eventual thermalization. Recent advances in time-dependent density functional theory [12] and non-equilibrium many-body theory [13, 14] offer promising avenues. However, the approximations that enable realistic modelling of the relevant physical processes are yet to be established and verified. These approximations should take advantage of the ultrafast time scales and the dominant role of a strong laser field, while still accounting for electron-phonon and electron-electron interactions. In addition, the quest for controlling electron motion with light will drive the transition from studying electron dynamics in periodic crystal lattices to experimenting with simple petahertz-scale devices, which will require novel, multiscale modelling.

**Concluding remarks**

Field-resolved attosecond science is a complementary approach to the established techniques that expands the range of material systems that can be accessed on our current frontier of ultrafast science. With expanded observation of extremely fast electronic processes comes the hope of manipulating them as well, giving access to petahertz (lightwave) electronics [15-19]. At the same time, we should be realistic about its implications: after all, the Auston switch enabled sub-picosecond optoelectronics decades ago [20]. This did not lead to terahertz electronics in the sense of computer processors. However, just like terahertz metrology enabled advances from fundamental spectroscopy to industrial quality control, we hope that field-resolved attosecond science may find a similar range of exciting and useful applications.

**Acknowledgements**

We thank Ferenc Krausz for illuminating discussions.

**References**

[1]   Keiber S, Sederberg S, Schwarz A, Trubetskov M, Pervak V, Krausz F and Karpowicz N 2016 Electro-optic sampling of near-infrared waveforms *Nature Photonics* **10** 159
[2]   Park S B, Kim K, Cho W, Hwang S I, Ivanov I, Nam C H and Kim K T 2018 Direct sampling of a light wave in air *Optica* **5** 402
[3]   Zimin D A, Yakovlev V S and Karpowicz N 2022 Ultra-broadband all-optical sampling of optical waveforms *Sci Adv* **8** eade1029
[4]   Goulielmakis E, Uiberacker M, Kienberger R, Baltuska A, Yakovlev V, Scrinzi A, Westerwalbesloh T, Kleineberg U, Heinzmann U, Drescher M and Krausz F 2004 Direct measurement of light waves *Science* **305** 1267

[5]    Sommer A, Bothschafter E M, Sato S A, Jakubeit C, Latka T, Razskazovskaya O, Fattahi H, Jobst M, Schweinberger W, Shirvanyan V, Yakovlev V S, Kienberger R, Yabana K, Karpowicz N, Schultze M and Krausz F 2016 Attosecond nonlinear polarization and light-matter energy transfer in solids *Nature* **534** 86

[6]    Zimin D A, Karpowicz N, Qasim M, Weidman M, Krausz F and Yakovlev V S 2023 Dynamic optical response of solids following 1-fs-scale photoinjection *Nature* **618** 276

[7]    Schotz J, Maliakkal A, Blochl J, Zimin D, Wang Z, Rosenberger P, Alharbi M, Azzeer A M, Weidman M, Yakovlev V S, Bergues B and Kling M F 2022 The emergence of macroscopic currents in photoconductive sampling of optical fields *Nat Commun* **13** 962

[8]    Agarwal M, Scrinzi A, Krausz F and Yakovlev V S 2023 Theory of Nonlinear Photoconductive Sampling in Atomic Gases *Ann Phys-Berlin* **535** 2300322

[9]    Krausz F and Ivanov M 2009 Attosecond physics *Rev Mod Phys* **81** 163

[10]   Travers J C, Grigorova T F, Brahms C and Belli F 2019 High-energy pulse self-compression and ultraviolet generation through soliton dynamics in hollow capillary fibres *Nature Photonics* **13** 547

[11]   Heinzerling A M, Tani F, Agarwal M, Yakovlev V S, Krausz F and Karpowicz N 2025 Field-resolved attosecond solitons *Nature Photonics* **19** 772

[12]   Sato S A, Hübener H, De Giovannini U and Rubio A 2025 Technical review: Time-dependent density functional theory for attosecond physics ranging from gas-phase to solids *Npj Comput Mater* **11** 233

[13]   Joost J P, Schlünzen N and Bonitz M 2020 G1-G2 scheme: Dramatic acceleration of nonequilibrium Green functions simulations within the Hartree-Fock generalized Kadanoff-Baym ansatz *Physical Review B* **101** 245101

[14]   Perfetto E, Pavlyukh Y and Stefanucci G 2022 Real-Time GW: Toward an Ab Initio Description of the Ultrafast Carrier and Exciton Dynamics in Two-Dimensional Materials *Phys Rev Lett* **128** 016801

[15]   Goulielmakis E, Yakovlev V S, Cavalieri A L, Uiberacker M, Pervak V, Apolonski A, Kienberger R, Kleineberg U and Krausz F 2007 Attosecond control and measurement: lightwave electronics *Science* **317** 769

[16]   Zheltikov A 2019 Multioctave supercontinua and subcycle lightwave electronics [Invited] *Journal of the Optical Society of America B-Optical Physics* **36** A168

[17]   Borsch M, Meierhofer M, Huber R and Kira M 2023 Lightwave electronics in condensed matter *Nat Rev Mater* **8** 668

[18]   Hassan M T 2024 Lightwave Electronics: Attosecond Optical Switching *Acs Photonics* **11** 334

[19]   Heide C, Keathley P D and Kling M F 2024 Petahertz electronics *Nat Rev Phys* **6** 648

[20]   Auston D H 1975 Picosecond Optoelectronic Switching and Gating in Silicon *Appl Phys Lett* **26** 101

## 17. Attosecond imaging, radiolysis and chiral dynamics

**Phay Ho[1], Jérémy R. Rouxel[1], and Linda Young[1]\***

[1] Chemical Sciences and Engineering Division, Argonne National Laboratory, Lemont, Illinois 60439, United States

young@anl.gov

**Status**

Tunable isolated attosecond x-ray pulses became available in 2020 at the Linac Coherent Light Source [1]. The energy available in a single soft x-ray pulse was enhanced over high-harmonic generation (HHG) sources [2] by ∼ 106 to the microjoule level, and the photon energy range was extended to the kilovolt regime. By 2024, the attosecond pulse intensity reached the terawatt-scale [3] and synchronized pulse pairs were produced [4]. In addition, EUV and x-ray attosecond pulses have recently been generated with polarization control capabilities [5, 6, 7]. These powerful, ultrafast x-ray pulses provide new opportunites in attosecond single particle imaging, radiolysis and chiral dynamics as described below.

Single-Particle Imaging (SPI), proposed in 2000, aims to resolve the structure of individual, noncrystalline nanoscale objects, such as biological particles (e.g., viruses, metalloproteins), aerosols, or catalytic nanoparticles—without requiring crystallization [8]. It employs intense, femtosecond XFEL pulses to capture diffraction patterns before the sample is destroyed. Each pulse delivers 1012 photons in a few femtoseconds, enabling single-shot imaging of fragile samples in their native environments. SPI offers strong potential for high-resolution imaging of heterogeneous systems and enables time-resolved stud- ies of ultrafast processes when combined with excitation lasers. However, achieving atomic resolution remains difficult. Radiation damage from high x-ray fluence, due to multiphoton absorption and tran- sient electronic changes, limits the achievable resolution. The current best is 2 nanometers [9], still short of the atomic scale. Ongoing efforts aim to enhance scattering efficiency [10] and reduce damage using attosecond or few-femtosecond pulses [11], pushing SPI closer to atomic-scale imaging.

Attosecond x-ray pulse pairs have enabled the study of ultrafast electron dynamics in condensed phase [12] and have established the powerful technique of all x-ray attosecond transient absorption spectroscopy (AX-ATAS), which allows to capture spectral snapshots at specific atomic centers free from nuclear motion. Such snapshots allow us to probe the origin of reactive species resulting from radiolysis [13]. An understanding of radiolysis, i.e. ionization-induced chemistry, impacts many fields ranging from cosmochemistry, human health, microelectronics to nuclear energy. The attosecond studies rep- resent a marked advance; early pulse radiolysis studies on the microsecond timescale identified the prominent optical signature for the hydrated electron and current installations access the picosecond regime – neither captures the femtosecond timescales of bond-breaking and making. With relatively weak attosecond x-ray pulses from HHG sources one can observe photon-induced processes at specific atomic sites via transient absorption [14], however intense x-ray pulse pairs are required for the study of elementary ionization mechanisms with x-ray pump/x-ray probe methods[12].

Circularly polarized EUV/x-ray pulses offer new avenues to probe molecular chirality with attosec- ond resolution [15, 16, 17]. Molecular chirality originates from the asymmetric configuration of a nuclear geometry.Ultrafast nuclear chiral dynamics are usually in the femtosecond regime and above, but can become even faster when the motion of a single hydrogen atom connected to a chiral center is sufficient to modulate the chiral signal.

Molecular chirality also impacts attosecond electronic dynamics and cir- cularly polarized attosecond pump-probe experiment are promising to enable their study. Beaulieu et al. [15] have used attosecond photoelectron interferometry to demonstrate that direct photoionization and autoionization delays are enantioselective. Molecular electronic wavepacket in achiral molecules can also acquire an asymmetric geometry through interactions with circularly polarized light pulses, as shown by Chen et al.[17]. Han et al.[16] have used circularly polarized EUV pulses to generate electron vortices in Argon and probe them by IR photoionization.

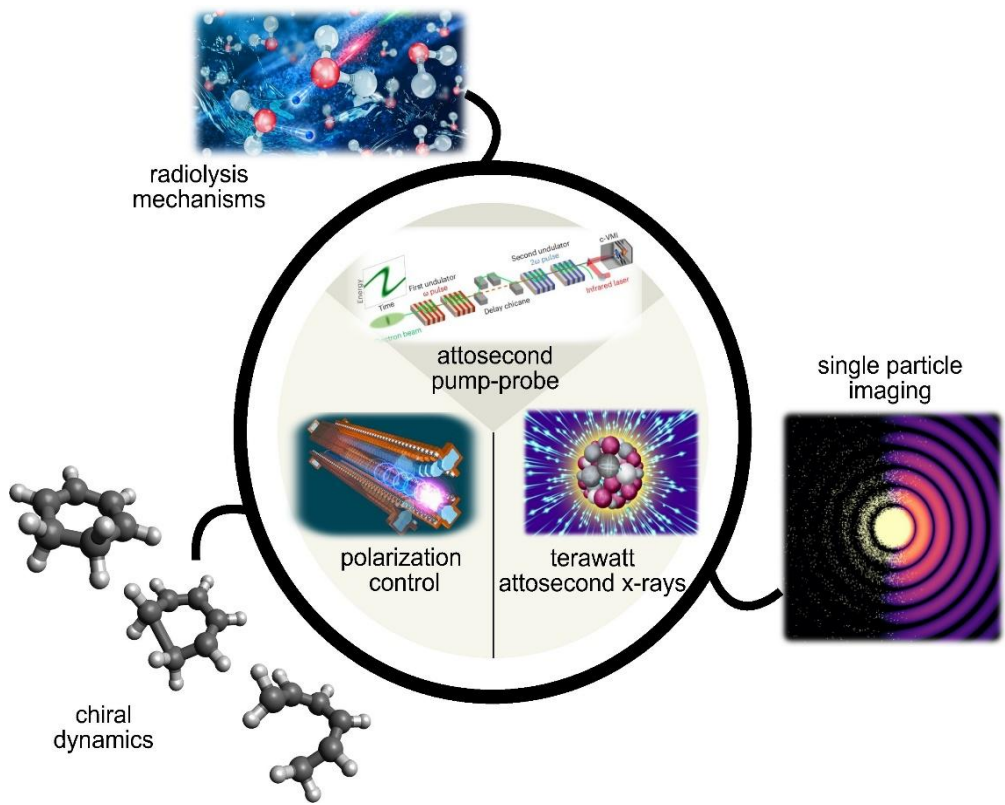

*Figure 1: The advancement of terawatt attosecond x-rays, attosecond pump-probe technique and polarization control capability enable single particle imaging with improved resolution [10], studies of the early-stage of radiolysis mechanisms [12] and ultrafast chiral dynamics [15] respectively.*

## Current and future challenges

Since the early vision for SPI experiments with XFELs, radiation induced damage has been an important topic. Most experimental and theoretical studies suggest that the damage are detrimental to quality of the scattering image. Recent studies reexamine the role of radiation damage in SPI[18]. Intense attosecond pulses allows selectively accessing transient resonance states [10] with enhanced scattering response beyond ground state. These enhancements have been demonstrated in the soft x-ray regime, with several efforts underway to extend this approach into the hard x-ray regime. These experiments exploit nonlinear processes, where the desired transient states are both prepared and probed within the same XFEL pulse. The near-temporal coherence of attosecond pulses also opens the possibility of quantum control of inner-shell electrons via Rabi oscillation for mitigating radiation damage [11] and enhancing scattering signals.

So far, the studies of radiolysis on the attosecond timescale have been limited to pure liquid water and executed with a synchronized pulse pair operating in the ω/2ω pump/probe configuration. These early isolated attosecond x-ray pulse-pair experiments [4, 12] were accomplished at LCLS with a Cu linac running at 120 Hz, vastly limiting statistics and access, prior to an upgrade to a high-repetition rate superconducting linac. Clearly extension to more complex systems with independently tunable x-ray pulse pairs and better statistics is of high scientific interest. More recently, attosecond pulses at mega- hertz repetition rates were realized in the hard x-ray regime with hundreds of microjoule pulse energy at the Euopean XFEL [19] - enabling attosecond scattering experiments to complement the spectroscopy that dominates the soft x-ray regime. The use of high-repetition rate sources for x-ray pump/x-ray probe experiments is still in its infancy - due to accessibility and target replenishment challenges. More- over, the tuning range in photon energy, pulse-pair temporal separation, overall characterization of spa- tial/temporal overlap and rapid pump on-off switching pose significant challenges at the accelerator- based facilities.

In ultrafast chirality measurements, most observables occur as a differential signal between two quantities. Those can be differences between different states of light (e.g. the difference between left and right incident polarization), or differences between measurement geometry (for example the for- ward/backward asymmetry in PECD). Recent progress involving continuum states shows that asym- metries can be large compared to the asymmetries involving only bound states. Nonetheless, well- controlled and stable circular polarization states for ultrashort EUV and X-ray pulses are needed. They are becoming increasingly available in the EUV regime with HHG-based light sources, but capabilities remain scarce in the x-ray regime at facilities. For HHG sources, it remains difficult to obtain isolated circularly polarized attosecond pulses [5].

### Advances in science and technology to meet challenges

Advances required to make progress in attosecond imaging and spectroscopy are both instrumental at FELs and theoretical/computational to handle the data deluge coming from the new high-repetition rate sources. On the instrumental side, since attosecond pulses at FELs are recent, diagnostics and control of the pulses are still developing. For example, in situ characterization of the incident pulse duration and pulse pair separation, by e.g. streaking methods [1, 4], is missing for liquid phase experiments. Moreover, to date, it has been not possible, via accelerator-based manipulations, to have x-ray pump/x- ray probe experiments that smoothly traverse zero delay time and to rapidly switch the pump x-ray beam off – this is in stark contrast to experiments employing optical pump lasers. Additional require- ments for chirality-sensitive attosecond spectrocopies are the capacity to switch between left, right and linear polarizations on-demand, with a high control of the degree of circular polarization. The ability to switch rapidly between opposite circular polarization will also become an advantage to eliminate any long term drift in the experiment. Beyond delivery of attosecond pulses, endstation designs that incorporate x-ray detectors that read out individual frames at high repetition rate (e.g. 100 kHz for a cw superconducting XFEL - analogous to the one-of-a-kind DSSC detector at the pulsed superconducting XFEL ) will eliminate averaging and allow the full capabilities of the facility to be realized as statistical correlation methods can then be applied [20].

High-repetition rate experiments present significant big data challenges due to the sheer volume and velocity of data generated. These experiments produce vast amounts of raw data at rapid intervals, necessitating real-time processing capabilities to ensure timely feedback for experimental adjustments. The exponential growth in data collection rates demands scalable algorithms, high-performing com- puting facilities, and theoretical and computational models capable of real-time handling, analysis and interpretation petascale to exascale data volumes swiftly.

## Concluding remarks

The recent and continuing strides in intensity, polarization and pulse-pair properties of attosecond x-ray pulses from accelerator-based XFELs are enabling new experimental frontiers for single particle imaging, radiolysis and chiral dynamics. These must be coupled with advanced data analysis for the extraction of new scientific insights.

## Acknowledgements

Work was supported by the U.S. Department of Energy (DOE), Office of Science, Basic Energy Science (BES), Chemical Sciences, Geosciences and Biosciences Division (CSGB) under Contract No. DE-AC02- 06CH11357.

## References

[1]  Joseph Duris, Siqi Li, Taran Driver, Elio G Champenois, James P MacArthur, Alberto A Lutman, Zhen Zhang, Philipp Rosenberger, Jeff W Aldrich, Ryan Coffee, et al. Tunable isolated attosecond x-ray pulses with gigawatt peak power from a free-electron laser. Nature Photonics, 14(1):30–36, 2020.

[2]  Stephan M Teichmann, F Silva, SL Cousin, M Hemmer, and J Biegert. 0.5-kev soft x-ray attosecond continua. Nature Communications, 7(1):11493, 2016.

[3]  Paris Franz, Siqi Li, Taran Driver, River R Robles, David Cesar, Erik Isele, Zhaoheng Guo, Jun Wang, Joseph P Duris, Kirk Larsen, et al. Terawatt-scale attosecond x-ray pulses from a cascaded superradiant free-electron laser. Nature Photonics, 18(7):698–703, 2024.

[4]  Zhaoheng Guo, Taran Driver, Sandra Beauvarlet, David Cesar, Joseph Duris, Paris L Franz, Oliver Alexander, Dorian Bohler, Christoph Bostedt, Vitali Averbukh, et al. Experimental demonstra- tion of attosecond pump–probe spectroscopy with an x-ray free-electron laser. Nature Photonics, 18(7):691–697, 2024.

[5]  Pei-Chi Huang, Carlos Herna ́ndez-Garc ́ıa, Jen-Ting Huang, Po-Yao Huang, Chih-Hsuan Lu, Laura Rego, Daniel D Hickstein, Jennifer L Ellis, Agnieszka Jaron-Becker, Andreas Becker, et al. Polariza- tion control of isolated high-harmonic pulses. Nature Photonics, 12(6):349–354, 2018.

[6]  Giovanni Perosa, Jonas Wa ̈tzel, David Garzella, Enrico Allaria, Matteo Bonanomi, Miltcho Boy- anov Danailov, Alexander Brynes, Carlo Callegari, Giovanni De Ninno, Alexander Demidovich, et al. Femtosecond polarization shaping of free-electron laser pulses. Physical Review Letters, 131(4):045001, 2023.

[7]  Markus Ilchen, Enrico Allaria, Primoz ̌ Rebernik Ribic ̌, Heinz-Dieter Nuhn, Alberto Lutman, Evgeny Schneidmiller, Markus Tischer, Mikail Yurkov, Marco Calvi, Eduard Prat, et al. Oppor- tunities for gas-phase science at short-wavelength free-electron lasers with undulator-based polar- ization control. Physical Review Research, 7(1):011001, 2025.

[8]  Richard Neutze, Remco Wouts, David Van der Spoel, Edgar Weckert, and Janos Hajdu. Potential for biomolecular imaging with femtosecond x-ray pulses. Nature, 406(6797):752–757, 2000.

[9]  Hirokatsu Yumoto, Takahisa Koyama, Akihiro Suzuki, Yasumasa Joti, Yoshiya Niida, Kensuke Tono, Yoshitaka Bessho, Makina Yabashi, Yoshinori Nishino, and Haruhiko Ohashi. High-fluence and high-gain multilayer focusing optics to enhance spatial resolution in femtosecond x-ray laser imaging. Nature Communications, 13(1):5300, 2022.

[10] Stephan Kuschel, Phay J Ho, Andre Al Haddad, Felix F Zimmermann, Leonie Flueckiger, Matthew R Ware, Joseph Duris, James P MacArthur, Alberto Lutman, Ming-Fu Lin, et al. Non- linear enhancement of ultrafast x-ray diffraction through transient resonances. Nature Communica- tions, 16(1):847, 2025.

[11] Anatoli Ulmer, Phay Ho, Bruno Langbehn, Stephan Kuschel, Linos Hecht, Razib Obaid, Simon Dold, Taran Driver, Joseph Duris, Ming-Fu Lin, David Cesar, Paris Franz, Zhaoheng Guo, Philip Hart, Andrei Kamalov, Kirk Larsen, Xiang Li, Michael Meyer, Kazutaka Nakahara, and Tais Gorkhover. Nonlinear reversal of photo-excitation on the attosecond time scale improves ultra- fast x-ray diffraction images, 06 2025.

[12] Shuai Li, Lixin Lu, Swarnendu Bhattacharyya, Carolyn Pearce, Kai Li, Emily T Nienhuis, Gilles Doumy, Richard D Schaller, S Moeller, M-F Lin, et al. Attosecond-pump attosecond-probe x-ray spectroscopy of liquid water. Science, 383(6687):1118–1122, 2024.

[13] Z-H Loh, G Doumy, C Arnold, Ludvig Kjellsson, SH Southworth, A Al Haddad, Y Kumagai, M-F Tu, PJ Ho, AM March, et al. Observation of the fastest chemical processes in the radiolysis of water. Science, 367(6474):179–182, 2020.

[14] Romain Geneaux, Hugo JB Marroux, Alexander Guggenmos, Daniel M Neumark, and Stephen R Leone. Transient absorption spectroscopy using high harmonic generation: a review of ultra- fast x-ray dynamics in molecules and solids. Philosophical Transactions of the Royal Society A, 377(2145):20170463, 2019.

[15] Samuel Beaulieu, Antoine Comby, Alex Clergerie, Je´re´mie Caillat, Dominique Descamps, Nirit Dudovich, Baptiste Fabre, Romain Ge´neaux, Franc¸ois Le´gare´, Ste´phane Petit, et al. Attosecond- resolved photoionization of chiral molecules. Science, 358(6368):1288–1294, 2017.

[16] Meng Han, Jia-Bao Ji, Tadas Balcˇiūnas, Kiyoshi Ueda, and Hans Jakob Wörner. Attosecond circular-dichroism chronoscopy of electron vortices. Nature Physics, 19(2):230–236, 2023.

[17] Yunjiao Chen, Dietrich Haase, Jörn Manz, Huihui Wang, and Yonggang Yang. From chiral laser pulses to femto- and attosecond electronic chirality flips in achiral molecules. Nature Communica- tions, 15(1):565, 2024.

[18] Phay J Ho, Benedikt J Daurer, Max F Hantke, Johan Bielecki, Andre Al Haddad, Maximilian Bucher, Gilles Doumy, Ken R Ferguson, Leonie Flückiger, Tais Gorkhover, et al. The role of transient reso- nances for ultra-fast imaging of single sucrose nanoclusters. Nature communications, 11(1):167, 2020.

[19] Jiawei Yan, Weilun Qin, Ye Chen, Winfried Decking, Philipp Dijkstal, Marc Guetg, Ichiro Inoue, Naresh Kujala, Shan Liu, Tianyun Long, et al. Terawatt-attosecond hard x-ray free-electron laser at high repetition rate. Nature Photonics, pages 1–6, 2024.

[20] Kai Li, Christian Ott, Marcus Agåker, Phay J. Ho, Gilles Doumy, Alexander Magunia, Marc Reb- holz, Marc Simon, Tommaso Mazza, Alberto De Fanis, Thomas M. Baumann, Jacobo Montano, Nils Rennhack, Sergey Usenko, Yevheniy Ovcharenko, Kalyani Chordiya, Lan Cheng, Jan-Erik Rubens- son, Michael Meyer, Thomas Pfeifer, Mette B. Gaarde, and Linda Young. Super-resolution stimu- lated x-ray raman spectroscopy. Nature, 643(8072):662–668, 2025.

## 18. All-optical attosecond chiral measurements

**Alexander Gabriel Lohr[1,2], David Ayuso[3], S. Patchkovskii[1], Olga Smirnova[1,2,4]\***

[1] Theory Department, Max-Born Institute, Berlin, Germany
[2] Institute of Physics, Technical University, Berlin, Germany
[3] Department of Chemistry, Imperial College, London, UK
[4] Solid state institute, Technion, Israeli Institute of Technology, Haifa, Israel

olga.smirnova@mbi-berlin.de

**Status**

A chiral molecule cannot be superimposed on its mirror image (enantiomer), making many biological processes inherently enantio-sensitive. Exquisite methods for enantio-sensitive detection using e.g. gas and liquid chromatography are rooted in thermodynamics and kinetics. Operating at equilibrium, these processes are intrinsically slow, constraining their applicability to areas like small-molecule chiral detection. Nonequilibrium dynamics may hold the key to overcoming these limitations.

Attosecond technology provides the fastest means for tracking and manipulating non-equilibrium properties of matter at its electronic timescale. Attosecond measurements rely on either using attosecond light pulses, becoming available in laboratories or at FELs, or on our ability to sculpt light oscillations on a sub-cycle timescale. Both approaches enable exciting and probing electronic currents before they decohere or dissipate, opening unique opportunities for enantio-sensing [1].

First, light pulses where the polarization vector traces a two-dimensional Lissajous figure (e.g. circularly polarised fields) can excite [2] or probe [3] enantio-sensitive electron currents in randomly oriented molecules. The chiral response is encoded in the instantaneous direction of the current [2,3], which is opposite in two enantiomers. Such an ultrafast excitation and detection of chiral electron currents involve only electric-dipole interactions, resulting in a chiral coupling strength that is, in principle (as elaborated below), three orders of magnitude higher [1] than in standard optical methods relying on the interaction with light's magnetic field. Fundamentally, it means that chiral interactions can be defined *locally*, which may seem counterintuitive, since chirality is a geometric property of an extended structure. Yet, out-of-equilibrium matter provides a unique means to encode chirality in the temporal evolution of a local vectorial observable, such as an electron current or induced polarization. Its detection requires resolving the phase of the oscillations of such a vector. Attosecond spectroscopy has been developed to do just that. Here we focus on all-optical chiral detection leaving aside photoelectron spectroscopy [4,5,6], where induced chiral photoelectron currents can be static.

Second, light's polarization can be shaped in 3D to create *locally* chiral light [7,8] by combining several colours and polarization directions, so that the tip of the electric-field vector draws a chiral three-dimensional structure locally, at every point in space. Such *temporally* chiral light can pull the electrons inside the molecule on a sub-cycle timescale to probe and manipulate molecular handedness on the scale of the molecule. One can fully control the shape of this 3D chiral Lissajous figure by adjusting relative phases of multicolour fields. In contrast with circularly polarized light, locally chiral light provides access to scalar enantio-sensitive observables within the electric-dipole approximation, such as total strength of nonlinear emission or the amount of excited or photoionized electrons.

**Current and future challenges**

The discovery of chiral sum-frequency generation (SFG) in the electric-dipole regime by Giordmain [9] marks the beginning of the era of non-linear chiral measurements. It predates the concept of temporal chirality by half a century and effectively serves as its precursor, facing many of the same limitations. The early SFG experiments from the bulk of chiral solutions [10,11] revealed that unlocking the electric-dipole enhancement is nontrivial. The challenges for all-optical detection of chiral electric-dipole signals are threefold.

First, driving macroscopic SFG signals requires that the incident light field carries a longitudinal electric-field component [10,11], which can be achieved via *non-collinear* arrangements of two-colour pulses. However, the strength of the longitudinal field component is weak in the range of opening angles maximising macroscopic signal [11]. Second, the chiral isotropic second-order nonlinearity, relevant for perturbative light-matter interactions, is inherently weak as it lacks the static limit [12]. Third, dispersion in liquid samples leads to rather small coherence lengths, estimated to be 5 microns in early experiments [11]. The combination of these factors results in vanishingly small signals in non-resonant conditions [11]. Indeed, vibrational resonances led to chiral SFG signals about 25 times stronger than the respective non-dipole bulk signal [11]. Thus, the early measurements yielded a 1.4-order-of-magnitude gain in the coupling strength, falling short of the projected three orders.

What does the concept of temporal chirality bring to the era of non-linear chiral measurements? The key factors are 3D polarisation and phase locking between different colours, e.g. a fundamental $\omega$ frequency and its second harmonic $2\omega$, driving a non-linear response in a chiral medium not limited to SFG. It is this phase lock that controls the temporal shape of the resulting field interacting with the medium and exciting ultrafast currents. This sub-cycle control translates the *non-linear optical* set-up into an *attosecond setting*. Stronger driving fields lead to the dressed-states picture, where resonant interactions appear naturally and non-adiabatic transitions between different quasienergy states can be quantified by analysing the Fourier harmonics of the signal wrt the locked phase. In simple terms, one can view such dynamics as coupled Rabi oscillations driven by phase locked two-colour pulses in unison. Stronger fields and resonant interactions lead to stronger signals, while each Fourier harmonic wrt the locked phase gives access to individual quantum pathways involving different number of 2ω photons. Physically, the dependence of the detected signal on the two-color phase encodes the relative phase between chiral and achiral responses, which is enantio-sensitive and molecular specific.

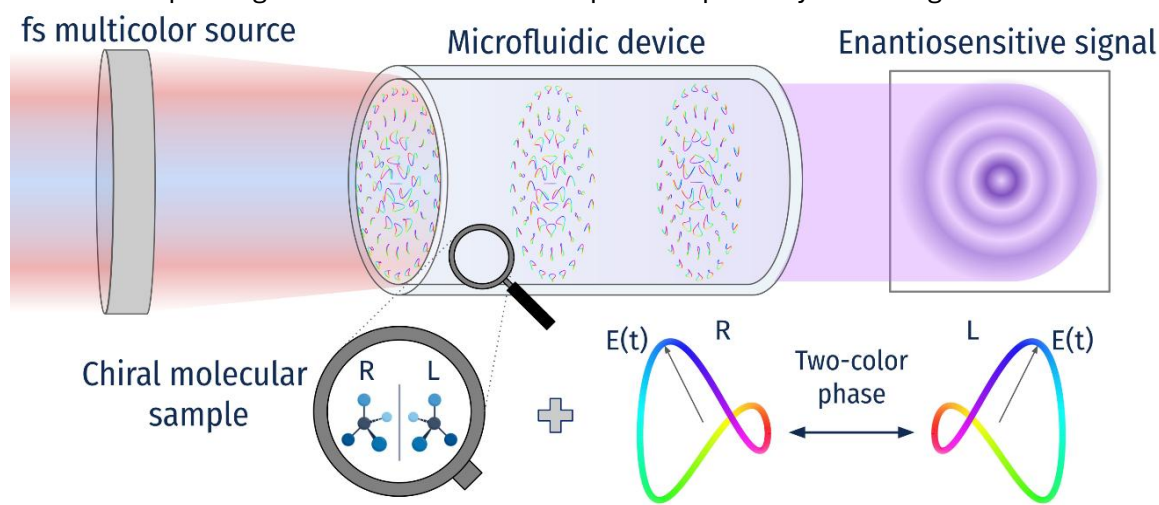

**Figure 1**. *Conceptual all-optical sensor for molecular chirality. A multicolor structured femtosecond beam is focused into a chiral-molecule-filled microfluidic waveguide, where the tailored input modes draw Lissajous field patterns driving chiral harmonic generation. The microfluidic capillary enables phase matching between different colours leading to the accumulated enantio-sensitive signal intensity at the output. The signal can be recorded as a function of two-color phase delay which controls the handedness of the locally chiral light.*

## Advances in science and technology to meet challenges

Unlocking the full potential of electric-dipole chiral coupling in all-optical setups requires both conceptual and technological advances. First, chiral signals can be enhanced by optimising phase-matching conditions, either through structured nanophotonic environments [13] or via quasi-phase-matching schemes. Second, using light with spatially structured polarization allows one to optimise the ratio of longitudinal to transverse fields at the sample, which directly controls the balance between chiral and achiral nonlinear responses. This can be achieved, for example, by guiding locally chiral light in optical fibres or by using microscopy configurations similar to the recently proposed chiral SFG setup [14] supplemented by spatial patterning via SLM. One may also exploit the different propagation constants of distinct fibre modes to maintain phase matching over macroscopic distances.

Conceptually, the chiral signal can be further refined by using structured light to imprint topological properties onto the chiral response. One natural route is to introduce local chirality into structured beams that carry orbital angular momentum, whose topological charge provides an additional handle on enantio-sensitive observables. Geometric effects also arise when circularly polarised light interacts with molecular electrons: the interaction term is highly sensitive to molecular orientation. This sensitivity couples rotational and electronic degrees of freedom, producing a new manifestation of Berry curvature linked to molecular chirality. By controlling the Lissajous figure of the light field or shaping molecular rotational wave packets, one can manipulate the associated Berry phases—and underlying enantio-sensitive observables. In gain–loss systems near an exceptional point, a locally chiral pulse could keep one enantiomer trivial while making the other topological, providing a purely topological handle for chiral discrimination.

Computationally, attosecond enantio-sensitive currents in randomly oriented chiral molecules remain largely unexplored. Recent advances in modelling charge-directed reactivity [15,16,17] following attosecond excitation provide solid foundation for understanding and quantifying chiral electronic response [3] and its coupling to nuclear motion. Prior theoretical work on electric-dipole electronic chiral response in non-linear photoelectron spectroscopy [3, 18,19] is an important milestone. Extending these approaches to all-optical detection of chiral electron dynamics in the strong-coupling regime is a key next step. A central challenge is the development of computationally efficient microscopic models that can be coupled self-consistently to Maxwell's equations describing light propagation in the medium. This difficulty is particularly pronounced for (i) dense media such as chiral solutions, (ii) structured light fields with spatially varying polarization, and (iii) structured nanophotonic environments, including optical fibres.

### Concluding remarks

Attosecond chiral dynamics can enhance chiroptical signals by several orders of magnitude compared to standard methods, as the response arises without requiring a magnetic field. Fully exploiting this purely electric-dipole effect, however, demands a systematic approach. The key limitations—and the order in which they must be addressed—are outlined below.

The first stage involves technological advances that are, in principle, within reach of current capabilities: structuring light in time and space, engineering environment to control light propagation, and using attosecond control of two-colour phases. The second stage concerns the ability to tailor and stabilise chiral signals by embedding topological features into the response. The temporal structure of chiral currents and chiral light fields—and their global organisation in parameter space—provide natural platforms where such topological properties can emerge. Taken together, these two steps could ultimately lead to compact, efficient, and potentially topologically robust chiral sensors on a chip, as well as new concepts for enantio-sensitive microscopy.

Although we have focused here on table-top setups, ultrafast chiral currents can also be excited and probed at FELs, where phase locking has been achieved through attosecond-pulse characterization that records the relative two-color phase and allows it to be sorted out after the measurement. Crucially, FEL facilities now provide the polarization control required to drive chiral currents [20].

## Acknowledgements

A.G.L. and O.S. acknowledge ERC-2021-AdG project ULISSES, grant agreement No 101054696. Views and opinions expressed are however those of the author(s) only and do not necessarily reflect those of the European Union or the European Research Council. Neither the European Union nor the granting authority can be held responsible for them. D. A. acknowledges funding from the Royal Society URF\R\251036.

## References

[1]   Ayuso D, Ordonez A and Smirnova O 2022 Ultrafast chirality:  a road to efficient chiral measurments *PCCP* **37** 074203

[2]   Beaulieu S *et al* 2018 Photoexcitation Circular Dichroism, *Nature Photonics* **23** 544

[3]   Wanie V, Bloch E, Månsson E P, Colaizzi L, Ryabchuk S, Saraswathula S, Ordonez A F, Ayuso D, Smirnova O, Trabattoni A, Blanchet V, Ben Amor N, Heitz M-C, Mairesse Y, Pons B, Calegari F,  2024 Capturing electron-driven chiral dynamics in UV-excited molecules *Nature* 630(8015), pp.109-115

[4]   Nahon, L., Garcia, G.A. and Powis, I., 2015. Valence shell one-photon photoelectron circular dichroism in chiral systems. *Journal of Electron Spectroscopy and Related Phenomena*, 204, pp.322-334.

[5]   Kastner, A., Lux, C., Ring, T., Züllighoven, S., Sarpe, C., Senftleben, A. and Baumert, T., 2016. Enantiomeric excess sensitivity to below one percent by using femtosecond photoelectron circular dichroism. *ChemPhysChem*, 17(8), pp.1119-1122..

[6]   Comby, A., Descamps, D., Petit, S., Valzer, E., Wloch, M., Pouységu, L., Quideau, S., Bocková, J., Meinert, C., Blanchet, V. and Fabre, B., 2023. Fast and precise chiroptical spectroscopy by photoelectron elliptical dichroism. *Physical Chemistry Chemical Physics*, 25(24), pp.16246-16263.

[7]   Ayuso, D., Neufeld, O., Ordonez, A.F., Decleva, P., Lerner, G., Cohen, O., Ivanov, M. and Smirnova, O., 2019. Synthetic chiral light for efficient control of chiral light–matter interaction. *Nature Photonics*, 13(12), pp.866-871

[8]   Král, P. and Shapiro, M., 2001. Cyclic population transfer in quantum systems with broken symmetry. *Physical Review Letters*, 87(18), p.183002.

[9]   Giordmaine, J.A., 1965. Nonlinear optical properties of liquids. *Physical Review*, 138(6A), p.A1599

[10]  Belkin, M.A., Kulakov, T.A., Ernst, K.H., Yan, L. and Shen, Y.R., 2000. Sum-frequency vibrational spectroscopy on chiral liquids: a novel technique to probe molecular chirality. *Physical review letters*, 85(21), p.4474

[11]  Fischer, P., Wiersma, D.S., Righini, R., Champagne, B. and Buckingham, A.D., 2000. Three-wave mixing in chiral liquids. Physical review letters, 85(20), p.4253

[12]  Fischer, P., Wise, F.W. and Albrecht, A.C., 2003. Chiral and achiral contributions to sum-frequency generation from optically active solutions of binaphthol. The Journal of Physical Chemistry A, 107(40), pp.8232-8238.

[13]  Ciriolo, A.G., Vázquez, R.M., Tosa, V., Frezzotti, A., Crippa, G., Devetta, M., Faccialà, D., Frassetto, F., Poletto, L., Pusala, A. and Vozzi, C., 2020. High-order harmonic generation in a microfluidic glass device. Journal of Physics: Photonics, 2(2), p.024005

[14]  Ji, Z., Yu, W., Dong, D., Yang, H., Liu, K., Xiao, Y.F., Gong, Q., Song, Q. and Shi, K., 2024. High spatial resolution collinear chiral sum-frequency generation microscopy. *Advanced Photonics Nexus*, 3(2), pp.026006-026006

[15]  Cardosa-Gutierrez, M., Levine, R.D. and Remacle, F., 2024. Electronic coherences built by an attopulse control the forces on the nuclei. *Journal of Physics B: Atomic, Molecular and Optical Physics*, 57(13), p.133501.

[16]  Alexander, O.G., Marangos, J.P., Ruberti, M. and Vacher, M., 2023. Attosecond electron dynamics in molecular systems. In *Advances In Atomic, Molecular, and Optical Physics* (Vol. 72, pp. 183-251). Academic Press.

[17] Grell, G., González-Vázquez, J., Fernández-Villoria, F., Palacios, A. and Martín, F., 2025. Modeling the Evolution of Laser-Induced Electronic Coherences with Trajectory Surface Hopping. *Journal of Chemical Theory and Computation*.

[18] Artemyev, A.N., Müller, A.D., Hochstuhl, D. and Demekhin, P.V., 2015. Photoelectron circular dichroism in the multiphoton ionization by short laser pulses. I. Propagation of single-active-electron wave packets in chiral pseudo-potentials. The Journal of chemical physics, 142(24)

[19] Goetz, R.E., Blech, A., Allison, C., Koch, C.P. and Greenman, L., 2025. Continuum-electron interferometry for enhancement of photoelectron circular dichroism and measurement of bound, free, and mixed contributions to chiral response. Physical Review Research, 7(3), p.L032036

[20] Ilchen, M., Allaria, E., Rebernik Ribič, P., Nuhn, H.D., Lutman, A., Schneidmiller, E., Tischer, M., Yurkov, M., Calvi, M., Prat, E. and Reiche, S., 2025. Opportunities for gas-phase science at short-wavelength free-electron lasers with undulator-based polarization control. *Physical Review Research*, 7(1), p.011001.

## 19. Attosecond-resolved coherent control

**Carlo Callegari[1]\*, Kevin C. Prince[1,2]\* and Giuseppe Sansone[3]\***

[1] Elettra-Sincrotrone Trieste S.C.p.A., Basovizza (Trieste), 34149, Italy
[2] Department of Surface and Plasma Science, Charles University, 18000, Czech Republic
[3] Institute of Physics, University of Freiburg, 79104, Freiburg, Germany.

carlo.callegari@elettra.eu, kevin.prince@elettra.eu, giuseppe.sansone@physik.uni-freiburg.de

**Status**

We define "attosecond-resolved coherent control" as the utilisation of the coherence of short light pulses to control the interaction of that light with matter. Strictly speaking, a sub-fs period implies light wavelength shorter than 300 nm, but when discussing "phase controlled" interferometric experiments, the resolution is determined by the fraction of the optical period that can be resolved, which can be sub-as [1, 2, 3]. We do not discuss experiments with laboratory or synchrotron light sources, except as inspiring the experiments that are the focus of our discussion: longitudinally coherent light generated by seeded Free-Electron Lasers (FELs), so far based on the high-gain harmonic generation (HGHG) scheme. Experiments described include some form of tailored light: two commensurate phase-locked colours, attosecond pulse trains, chirped pulses, time-varying polarisation; the control parameters may be relative phase, chirp, time delay between pulses, or polarisation. We focus attention on results from FERMI, as this has been the sole active source until recently, and note that other externally seeded FEL facilities are operating (Dalian, China) or about to begin operation with users (FLASH, Germany).

This is a fledgling field, and experiments have taken inspiration from well-established laser science; so far, the pulse envelopes are long and experimental protocols are interferometric. We can distinguish

**a) Schemes for the characterisation of the coherence properties of the FEL source.**

These are particularly relevant as the ultimate goal is to imprint the desirable and flexible properties of the seed laser field onto the FEL field [4], but the nature of the HGHG process, the bandwidth of the amplifier, and the properties of the electron bunch set a number of strong limitations on this goal [5, and references therein]. An early example of characterisation at FERMI was the demonstration that two commensurate wavelengths could be generated with a controllable phase relationship [2].

b) **Coherent-control schemes based exclusively on FEL radiation**

These experiments aim at replicating (with shorter-wavelength, ionising radiation) pivotal achievements at optical wavelengths. Notably, coherent control of photoelectron angular distributions or cross sections, and pump-dump, whose schemes are labelled after their optical namesakes: Yin-Chen-Elliot, Brumer-Shapiro, Tannor-Rice [5]. A variant of phase control is control of the chirp within a single pulse, and this has been applied to the investigation of Rabi dynamics [3].

c) **Coherent-control schemes based on the combination of FEL and optical pulses**

These experiments extend the results pioneered with table-top High Harmonic Generation (HHG) sources. Notably, two-colour photoionisation schemes such as the Reconstruction of Attosecond Beating By Two-photon Transitions (RABBITT) [6] have been widely used with extreme ultraviolet (XUV) table-top sources for the temporal characterisation of attosecond pulse trains and for the determination of the time taken by a photoelectron to leave the ionic potential (for brevity, Eisenbud-Wigner-Smith (EWS) time).

While HHG is making independent progress in generating short-wavelength, intense pulses, a breakthrough occurred with the advent of FERMI: not only are its pulses coherent and intense (pulse energies of few to hundreds of μJ), the flexibility of the light source makes it possible to manipulate many parameters of the light produced, and to develop innovative experimental schemes. For example, XUV energy tunability enables the selection of specific atomic and molecular resonances. Furthermore, the ability to generate different harmonics independently permits the synthesis of exotic harmonic combs, for example, consisting of non-consecutive harmonics [7]. Finally, standard pulses at FERMI are linearly or elliptically polarised, but pulses can be created with time-varying polarisation [8].

**Current and future challenges**

Scientific challenges can be grouped into a number of classes.

a) **Investigation and control of charge dynamics using non-linear techniques.**

Coherent control techniques are poised to make further major contributions to our knowledge of electron and nuclear dynamics. A substantial experimental effort is being devoted to the realisation of coherent Raman techniques, such as the theoretically described Transient redistribution of ultrafast electronic coherences in attosecond Raman signals (TRUECARS) experiment [9], which provide methods of monitoring transient electronic coherences in a photochemical reaction. Other areas of molecular dynamics, such as bond-breaking and -making, as well as roaming mechanisms, are currently being investigated intensively on the femtosecond scale and will be probed more deeply on shorter time scales. The use of wide-bandwidth attosecond pulses to tailor electronic wave packets and engineer charge-directed reactivity is debated and still controversial [10]. At present, two of the serious limitations that FERMI can address are the required strong intensity and the need for independent control of phases and amplitudes of a sufficiently large set of harmonics.

b) **Investigation and control of the photoionisation mechanism.**

The EWS time delay in photoionisation has been studied using the RABBITT technique [11], as well as interferometric methods with FEL light [12]. The investigation of EWS time delays continues to be a challenging and growing field of research, as they provide important information about the shape of the atomic or molecular potential as a photoelectron exits the potential. Future challenges in this area will include investigating photoemission delays for specific core resonances, which has been demonstrated at the LCLS [13]. This will require the generation of attosecond pulses in the soft X-ray region, which can be synchronised or reconstructed *a posteriori* with respect to a second optical/near-infrared pulse.

c) **Investigation and control of entanglement on the attosecond timescale.**

While the photoionisation process naturally leads to the creation of an entangled ion-photoelectron system, investigations have only recently begun to consider the consequences and scientific opportunities opened up by the control of the photoionisation process in single and two-colour schemes [14]. Using table-top sources it has been shown that phase-locked pairs of isolated attosecond pulses can be used to control the degree of entanglement in photoionisation [15]. Using intense XUV pulses it has been shown that the degree of entanglement can be controlled within the pulse duration [16]. FERMI combines high intensity in the XUV range and the tunability to address specific resonances in the photoionisation process, which will offer additional knobs to control the degree of entanglement of the bipartite ion-electron system on timescales as short as a few tens of attoseconds.

**Advances in science and technology to meet challenges**

Further machine development and new concepts in the generation of ultrashort, intense optical pulses will be required for tackling the scientific challenges outlined above.

Lower time-slice energy spread of the electron beam is an important goal as a cold beam allows the generation of higher pulse energies at high harmonics. Echo enabled operation has been implemented at FERMI [17], and produces higher stability and higher pulse energy, as well as an extended spectral range with respect to that allowed by the HGHG configuration.

An increase in repetition rate would enable better statistics and coincidence measurements, as well as electron beam multiplexing [18], which allows the number of beamlines on a single accelerator to be increased. Plasma-based compact accelerators [19] may offer a cheap and efficient means to obtain electron beams of the required energy, although there are still issues with achieving the required beam quality and stability.

Rapid polarisation switching will greatly improve the signal-to-noise ratio in polarisation-sensitive measurements. This has been implemented at synchrotrons using pairs of undulators separated by an electromagnetic phase shifter, and is currently under consideration at FERMI.

Among innovative machine operation modes, a high priority is to provide shorter pulses temporally locked to a second pulse, and a scheme has been proposed to generate twin attosecond pulses [20], based on a special undulator design. Currently the method of super radiant emission produces the shortest pulses at FERMI with a duration down to 4.7 fs, at 14.7 nm, and 1.7 fs at 7 nm, so far only in single-pulse mode. As well, more intense, fully coherent sources will provide light at shorter wavelengths.

In the femtosecond regime, seeded FELs can readily generate temporally overlapping, coherent pulses at different wavelengths; a key challenge ahead is to broaden the accessible range of seed wavelengths, i.e, harmonic separation; this could allow TRUECARS-like schemes. A currently unresolved challenge is to produce temporally well-separated pulses (of different and commensurate wavelengths) while maintaining coherence. Using machine techniques, short delays can be introduced with some loss of phase stability, but at longer delays, phase locking is lost.

Regarding end-station detectors and instrumentation, currently the most widely used electron spectrometers are Velocity Map Imaging, magnetic bottle, time-of-flight and (less commonly) hemispherical electron energy analysers. COLTRIMS/Reaction Microscope instruments are also widely used, but require sources at high repetition rates; each of these spectrometers has its advantages and disadvantages. For example, COLTRIMS provides the greatest amount of information, but at the price of stringent operating conditions. New approaches, seeking a different compromise, will enable users to optimally address the needs of each experiment. The development of new approaches will benefit the quality and quantity of data acquired during an experiment. Photon spectrometers are also vital in some experiments and further development will lead to gains in efficiency.

**Concluding remarks**

The field of attosecond coherent control is flourishing and developing rapidly. We can expect many new breakthroughs and discoveries in the coming years, as further applications of interferometric techniques are devised. X-ray FELs such as the LCLS are currently routinely investigating dynamics with attosecond pulses. A significant challenge is to generate phase-locked pulses, to enable coherent control in this wavelength range. Hard X-ray machine technology is developing rapidly, so such pulses may soon be available.

### Acknowledgements

G.S. acknowledges the financial support by FRIAS, by the Deutsche Forschungsgemeinschaft grant 46852490 (Project SA3470/14-1), Research Training Group DynCAM (RTG2717), and the European Union's Horizon Europe research and innovation programme under the Marie Skłodowska-Curie grant agreement No 101168628 (QU-ATTO). We thank Luca Giannessi for a careful reading of the manuscript and helpful comments.

### References

[1]  Prince K C, Allaria E, Callegari C, Cucini R, Ninno G, Di Mitri S, Diviacco B, Ferrari E, Finetti P, Gauthier D, Giannessi L, Mahne N, Penco G, Plekan O, Raimondi L, Rebernik P, Roussel E, Svetina C, Trovò M, Zangrando M, Negro M, Carpeggiani P, Reduzzi M, Sansone G, Grum-Grzhimailo A N, Gryzlova E V, Strakhova S I, Bartschat K, Douguet N, Venzke J, Iablonskyi D, Kumagai Y, Takanashi T, Ueda K, Fischer A, Coreno M, Stienkemeier F, Ovcharenko Y, Mazza T and Meyer M 2016 Coherent control with a short-wavelength free-electron laser *Nat. Photonics* **10** 176–179

[2]  Maroju P K, Grazioli C, Di Fraia M, Moioli M, Ertel D, Ahmadi H, Plekan O, Finetti P, Allaria E, Giannessi L, Ninno G, Spezzani C, Penco G, Spampinati S, Demidovich A, Danailov M B, Borghes R, Kourousias G, Sanches Dos Reis C E, Billé F, Lutman A A, Squibb R J, Feifel R, Carpeggiani P, Reduzzi M, Mazza T, Meyer M, Bengtsson S, Ibrakovic N, Simpson E R, Mauritsson J, Csizmadia T, Dumergue M, Kühn S, Nandiga Gopalakrishna H, You D, Ueda K, Labeye M, Bækhøj J E, Schafer K J, Gryzlova E V, Grum-Grzhimailo A N, Prince K C, Callegari C and Sansone G 2020 Attosecond pulse shaping using a seeded free-electron laser *Nature* **578** 386–391

[3]  Richter F, Saalmann U, Allaria E, Wollenhaupt M, Ardini B, Brynes A, Callegari C, Cerullo G, Danailov M, Demidovich A, Dulitz K, Feifel R, Fraia M D, Ganeshamandiram S D, Giannessi L, Gölz N, Hartweg S, von Issendorff B, Laarmann T, Landmesser F, Li Y, Manfredda M, Manzoni C, Michelbach M, Morlok A, Mudrich M, Ngai A, Nikolov I, Pal N, Pannek F, Penco G, Plekan O, Prince K C, Sansone G, Simoncig A, Stienkemeier F, Squibb R J, Susnjar P, Trovo M, Uhl D, Wouterlood B, Zangrando M and Bruder L 2024 Strong-field quantum control in the extreme ultraviolet domain using pulse shaping *Nature* **636** 337–341

[4]  Gauthier D, Rebernik Ribič P R, De Ninno G, Allaria E, Cinquegrana P, Danailov M B, Demidovich A, Ferrari E, Giannessi L, Mahieu B and Penco G 2015 Spectrotemporal Shaping of Seeded Free-Electron Laser Pulses *Phys. Rev. Lett.* **115** 114801

[5]  Callegari C, Grum-Grzhimailo A N, Ishikawa K L, Prince K C, Sansone G and Ueda K 2021 Atomic, molecular and optical physics applications of longitudinally coherent and narrow bandwidth Free-Electron Lasers Phys. Rep. **904** 1–59

[6]  Paul P M, Toma E S, Breger P, Mullot G, Augé F, Balcou Ph, Muller H G and Agostini P 2001 Observation of a Train of Attosecond Pulses from High Harmonic Generation *Science* **292** 1689–1692

[7]  Maroju P K, Benito de Lama M, Di Fraia M, Plekan O, Bonanomi M, Merzuk B, Busto D, Makos I, Schmoll M, Shah R, Rebernik Ribič P, Giannessi L, Allaria E, Penco G, Zangrando M, Simoncig A, Manfredda M, De Ninno G, Spezzani C, Demidovich A, Danailov M, Coreno M, Squibb R J, Feifel R, Bengtsson S, Simpson E R, Csizmadia T, Dumergue M, Kühn S, Ueda K, Zeni G, Frassetto F, Poletto L, Prince K C, Mauritsson J, Feist J, Palacios A, Callegari C and Sansone G 2025 Attosecond temporal structure of nonconsecutive harmonic combs revealed by multiple near-infrared photon transitions in two-color photoionisation *Comm. Phys.* **8** 20

[8]  Perosa G, Wätzel J, Garzella D, Allaria E, Bonanomi M, Danailov M B, Brynes A, Callegari C, De Ninno G, Demidovich A, Di Fraia M, Di Mitri S, Giannessi L, Manfredda M, Novinec L, Pal N, Penco G, Plekan O, Prince K C, Simoncig A, Spampinati S, Spezzani C, Zangrando M, Berakdar J, Feifel R, Squibb R J, Coffee R, Hemsing E, Roussel E, Sansone G, McNeil B W J and Rebernik Ribič P 2023 Femtosecond Polarization Shaping of Free-Electron Laser Pulses *Phys. Rev. Lett.* **131** 045001

[9]  Kowalewski M, Bennett K, Dorfman K E and Mukamel S 2015 Catching Conical Intersections in the Act: Monitoring Transient Electronic Coherences by Attosecond Stimulated X-Ray Raman Signals *Phys. Rev. Lett.* **115** 193003

[10]  Lépine F, Ivanov M, Vrakking M J J 2014 Attosecond molecular dynamics: fact or fiction? *Nat. Photonics* **8** 195–204

[11]  Dahlström J M, L'Huillier A and Maquet A 2012 Introduction to attosecond delays in photoionization *J. Phys. B: At., Mol. Opt. Phys.* **45** 18300

[12]  You D, Ueda K, Gryzlova E V, Grum-Grzhimailo A N, Popova M M, Staroselskaya E I, Tugs O, Orimo Y, Sato T, Ishikawa K L, Carpeggiani P A, Csizmadia T, Füle M, Sansone G, Maroju P K, D'Elia A, Mazza T, Meyer M, Callegari C, Di Fraia M, Plekan O, Richter R, Giannessi L, Allaria E, De Ninno G, Trovò M, Badano L, Diviacco B, Gaio G, Gauthier D, Mirian N, Penco G, Rebernik Ribič P, Spampinati S, Spezzani C and Prince K C 2020 New Method for Measuring Angle-Resolved Phases in Photoemission *Phys. Rev. X* **10** 031070

[13] Driver T, Mountney M, Wang J, Ortmann L, Al-Haddad A, Berrah N, Bostedt C, Champenois EG, DiMauro LF, Duris J, Garratt D, Glownia JM, Guo Z, Haxton D, Isele E, Ivanov I, Ji J, Kamalov A, Li S, Lin MF, Marangos JP, Obaid R, O'Neal JT, Rosenberger P, Shivaram NH, Wang AL, Walter P, Wolf TJA, Wörner HJ, Zhang Z, Bucksbaum PH, Kling MF, Landsman AS, Lucchese RR, Emmanouilidou A, Marinelli A, Cryan JP 2024 Attosecond delays in X-ray molecular ionization *Nature* **632** 762–7

[14] Lewenstein M, Baldelli N, Bhattacharya U, Biegert J, Ciappina M F, Grass T, Grochowski P T, Johnson A S, Lamprou Th, Maxwell A S, Ordonez A, Pisanty E, Rivera-Dean J, Stammer P and Tzallas P 2024 Attosecond Physics and Quantum Information Science *Proceedings of the 8th International Conference on Attosecond Science and Technology* ed Argenti L, Chini M and Fang L (Cham: Springer International Publishing) pp 27–44 ISBN 978-3-031-47938-0

[15] Koll L M, Maikowski L, Drescher L, Witting T and Vrakking M J J 2022 Experimental Control of Quantum-Mechanical Entanglement in an Attosecond Pump-Probe Experiment *Phys. Rev. Lett.* **128** 043201

[16] Nandi S, Stenquist A, Papoulia A, Olofsson E, Badano L, Bertolino M, Busto D, Callegari C, Carlström S, Danailov M B, Demekhin P V, Di Fraia M, Eng-Johnsson P, Feifel R, Gallician G, Giannessi L, Gisselbrecht M, Manfredda M, Meyer M, Miron C, Peschel J, Plekan O, Prince K C, Squibb R J, Zangrando M, Zapata F, Zhong S and Dahlström 2024 J M Generation of entanglement using a short-wavelength seeded free-electron laser *Sci. Adv.* **10** eado0668

[17] Rebernik Ribič P, Abrami A, Badano L, Bossi M, Braun H H, Bruchon N, Capotondi F, Castronovo D, Cautero M, Cinquegrana P, Coreno M, Couprie M E, Cudin I, Boyanov Danailov M, De Ninno G, Demidovich A, Di Mitri S, Diviacco B, Fawley W M, Feng C, Ferianis M, Ferrari E, Foglia L, Frassetto F, Gaio G, Garzella D, Ghaith A, Giacuzzo F, Giannessi L, Grattoni V, Grulja S, Hemsing E, Iazzourene F, Kurdi G, Lonza M, Mahne N, Malvestuto M, Manfredda M, Masciovecchio C, Miotti P, Mirian N S, Petrov Nikolov I, Penco G M, Penn G, Poletto L, Pop M, Prat E, Principi E, Raimondi L, Reiche S, Roussel E, Sauro R, Scafuri C, Sigalotti P, Spampinati S, Spezzani C, Sturari L, Svandrlik M, Tanikawa T, Trovò M, Veronese M, Vivoda D, Xiang D, Zaccaria M, Zangrando D, Zangrando M and Allaria E M 2019 Coherent soft X-ray pulses from an echo-enabled harmonic generation free-electron laser *Nat. Photonics* **13** 555–561

[18] Li S, Zhang Z, Alverson S, Cesar D, Driver T, Franz P, Isele E, Duris J P, Larsen K, Lin M F, Obaid R, O'Neal J T, Robles R, Sudar N, Guo Z, Vetter S, Walter P, Wang A L, Xu J, Carbajo S, Cryan J P and Marinelli A 2024 "Beam à la carte": Laser heater shaping for attosecond pulses in a multiplexed x-ray free-electron laser *Appl. Phys. Lett.* **125** 191101

[19] Galletti M, Assmann R, Couprie M E, Ferrario M, Giannessi L, Irman A, Pompili R and Wang W 2024 Prospects for free-electron lasers powered by plasma-wakefield-accelerated beams *Nat. Photonics* **18** 780–791

[20] Tanaka T and Rebernik Ribič P 2022 Proposal to generate a pair of intense independently tunable attosecond pulses from undulator radiation *Opt. Lett.* **47** 1411–1414

## 20. Attosecond spin dynamics in atoms and molecules

**Stefanos Carlström[1], Philip Flores[1], Serguei Patchkovskii[1], Felipe Morales[1], Olga Smirnova[1,2,4]\*, Misha Ivanov[1,4,5]\***

[1] Theory Department, Max-Born Institute, Berlin, Germany
[2] Institute of Physics, Technical University, Berlin, Germany
[3] Department of Chemistry, Imperial College, London, UK
[4] Solid state institute, Technion, Israeli Institute of Technology, Haifa, Israel
[5] Humboldt University Berlin, Berlin, Germany

olga.smirnova@mbi-berlin.de, mikhail.ivanov@mbi-berlin.de

**Status**

Attosecond spin dynamics may sound like an oxymoron since the spin-orbit splitting in atoms and molecules scales with the fine structure constant⍺ often leaving us with characteristic times for valence spin dynamics on the scale of tens of femtoseconds in light atoms to several femtoseconds in the heavy ones. In contrast, strong field ionization, also known as optical tunnelling, is the sub-cycle phenomenon responsible for all pre-FELs attosecond physics. That is why it came as a surprise that optical tunnelling can filter the electron spin [1,2]. Likewise, one can also control spins on the attosecond time scale as the freed electron is driven back by the laser field in the process known as recollision [3, 4] or in angular streaking by few-cycle pulses with controlled carrier envelope phase (CEP) [5]. Thus, strong field ionization can create attosecond bursts of spin-polarised electrons.

First, these electron bursts can be used in ultrafast electron diffraction [6]. Second, they can be injected into Plasma-based compact particle accelerators and subsequently accelerated to relativistic speeds [7]. High ionization yields typical for strong field ionization and its non-resonant nature are among the practical benefits of this approach, which can be used for a wide range of atomic species. Third, strong laser fields can also drive enantio-sensitive dynamics in chiral molecules [8, 9].

Recent discovery of Chiral Induced Spin Selectivity (CISS) [10] suggests chiral medium as a spin filter, with high spin polarization (tens of percent) reversing for opposite chirality. Combining these two advances – strong-field dynamics and chirality – can lead to attosecond molecular spintronics, an uncharted field where attosecond physics can take a lead. Specifically, attosecond physics brings opportunities to control spin via controlling electron or hole dynamics on the sub-laser-cycle scale using pulses shaped in time and space. Fourth, attosecond chemistry and charge directed reactivity triggered by charge migration [Lenz] are, in fact, spin dependent. How does charge migration couple to spin? Remarkably, spin migration can occur even without coherent charge migration, as spin–orbit–mediated coherence transfer is less affected by the dynamics of the nuclei than the coherence underlying charge migration .

Current spin-resolved attosecond physics has seen limited number of experiments [2,11]. First numerical models [2] ignored coupling between different spin-resolved channels and used different pseudopotentials to reproduce the respective spectra in each channel. While the approach worked well for IR drivers (800 nm) [2] it failed to reproduce the experimental spin-resolved ATI spectrum for visible light (400 nm) [11]. Modern theoretical approaches involve a range of solvers with different degree of complexity, which include coupled single active electron two-channel codes [12], time-dependent configuration interaction with spin (TDCIS) codes [13,20], fully relativistic Dirac equation based codes [14,19], single active electron codes which use pseudopotentials to describe spin resolved dynamics [15,16]. Coupling between different spin-resolved channels proved important for quantitative simulation of experiments [12,13,15], successfully describing previously unexplained features indicating that they are related to coupling of different spin channels, not necessarily to multielectron effects.

**Current and future challenges**

Several future challenges can be addressed with already developed tools. First, in wake field particle acceleration, accurate simulations of spin-polarization in multiple ionization can be used to optimize injection of spin polarised electrons into the plasma wake. Second, photoelectron spin polarization can be optimized by controlling laser frequency, intensity, and using phase-locked multicolour fields. Third, addressing the interplay of the angular momentum of the driving fields and the resulting spin -resolved bound or photoelectron currents can provide unexplored ways of controlling spin-orbit interaction on ultrafast time scale via multiple interfering multiphoton pathways.

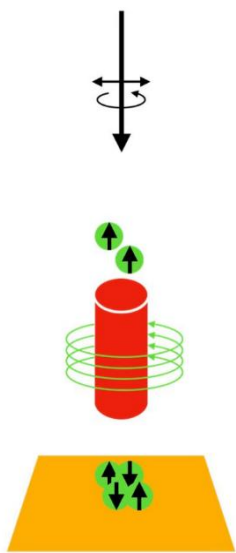

*Figure 1. Concept for combining chirality with ultrafast techniques for ultrafast control of chirality-induced spin selectivity. Spin-unpolarised electrons photoemitted from the surface travel through a chiral molecule, where an electronic ring current has been excited by an ultrafast pulse, modulating the chirality induced spin selectivity through the effective magnetic field generated by the current.*

Attosecond chemistry strives to image and control chemical reactions at the level of electrons. However, many elementary chemical steps are spin-dependent — electrons, radicals, and transition states obey spin selection rules. If one can control or bias electron spin, one effectively controls which reaction pathways are allowed or favoured. Typical analysis of spin-selective photodynamics is usually based on selection rules defined by light. However, there are deeper pathways of spin selectivity than the ones controlled by light or external magnetic fields. Recent results suggest that excited or photoionized chiral molecules possess intrinsic axis that orients the spin of the excited or photoelectron [17]. This axis realises an enantio-sensitive molecular compass. Its internal "compass axis" is locked to the molecular geometry itself — not to any external field. Remarkably, this compass activates even under isotropic illumination, where light provides no preferred direction. Just as a traditional compass needle aligns with Earth's magnetic field, the molecular compass aligns the electron spin with a built-in geometric direction inside the molecule — a direction defined by its handedness. In this way, the molecule generates its own "chiral north," guiding the electron spin without any magnetic interaction. In a single chiral molecule fixed in space, this compass causes the emitted or excited electron's spin to orient differently for left- and right-handed forms of the same molecule — exactly the kind of enantio-sensitive spin polarization observed in the CISS effect. This effect addresses the core of molecular spintronics and it quantification and analysis requires significant advances in computational methods.

**Advances in science and technology to meet challenges**

Exploring new mechanisms of ultrafast control of spin dynamics in atoms and molecules and taking full advantage of its function in chemical reactions requires both technological and conceptual advances.

In ultrafast spin-resolved photoelectron spectroscopy, these challenges are related, for example, to integrating spin detectors in standard attosecond set-ups in a way that both photoelectron energy and angular distributions and spin could be detected.  In solids, table-top HHG sources have already been implemented for spin- time- energy- and angle-resolved detection of photoelectrons [18]. Similar detector capabilities are desired for standard table-top attosecond set-ups involving phase locked IR-XUV pulses or two-color IR-UV pulses.

Conceptually, the framework for analysis of the interplay of ultrafast electron currents and electron spin has to be extended from one photon processes to multiphoton processes driven by multicolor fields. The discovery of CISS makes it clear that chiral molecular structure can couple to spin. The main geometric quantities mediating this process in one photon ionization such as the Bloch vector and spin-torque vector generated by the Berry curvature in chiral molecules will have to be extended to multiphoton process, and the extent to which they can control the electron spin has to be quantified.

Computationally, challenges are associated with extending current computational approaches mainly developed for atoms to molecules.  To accurately simulate spin-resolved ultrafast photodynamics in the gas and liquid phase molecules and molecules at interfaces first principles spin-resolved methods currently available for atoms should be developed to complement more generic TDDFT approaches. One needs to treat the energy structure of both the neutral molecule and its cation, generally beyond the Hartree–Fock approximation, including relativistic effects such as spin–orbit coupling at least at the two-component level. This has to be combined with including visible or near-infrared fields triggering ultrafast excitation, which can also appreciably accelerate the electron. The latter necessitates grid treatment that allows for large electron excursions and high angular momenta, both computationally expensive. An efficient method for computing spin-resolved photoelectron spectra beyond single ionization presents another important challenge. Notably, accurate description of the energy structure typically requires a basis localized at the molecular nuclei, requiring combination of grids with a localized basis set.

Longer term, one also needs conceptual advances in the theory of strongly-correlated systems. The existing techniques seem inadequate for modelling fast dynamics of complex spin systems -- which would be essential for developing e.g. practical ultrafast molecular spintronics devices, where multiple spins must interact, robustly and in a controlled way. Density-functional theories, especially without explicit spin-current dependence, have a tendency to artificially break spin-related symmetries. *Ab initio* theories tend to scale unfavourably with the number of unpaired spins, which need to be treated; and time-dependent density-matrix renormalization theories are at present difficult to apply to non-perturbative dynamics. Here new ideas are definitely welcome and needed.

**Concluding remarks**

The somewhat unexpected interface of attosecond physics and spin dynamics triggered by intense laser fields, discovered in atoms over a decade ago [1,2], is now poised to explore its potential in molecules. The combination of chiral molecular structures and polarization-tailored laser fields, including fields with conrolled time-dependent polarizaton, can be used to excite, control and use ultrafast electronic currents. Such currents generate geometric fields acting on electron spin, opening an exciting opportunity for attosecond molecular spintronics [17] and an attosecond perpective on CISS. Another related emerging new topic is ultrafast spin migration, an effect complementary to ultrafast charge migration in molecules that may prove a lot more resilient to nuclear motion: thanks to the local nature of spin-orbit interaction, spin-orbit split potential energy curves may remain parallel for changing internuclear distances (e.g between a heavy and a light atom.)

**Acknowledgements**

M.I. acknowledges CRC 1477 "Light-Matter Interactions at Interfaces" (ID: 441234705). O.S. and P.F. acknowledge ERC-2021-AdG project ULISSES, grant agreement No 101054696. Views and opinions expressed are however those of the author(s) only and do not necessarily reflect those of the European Union or the European Research Council. Neither the European Union nor the granting authority can be held responsible for them.

**References**

[1]  Barth, I. and Smirnova, O., 2013. Spin-polarized electrons produced by strong-field ionization, *Physical Review A*, 88, 013401.

[2]  Hartung, A., Morales, F., Kunitski, M., Henrichs, K., Laucke, A., Richter, M., Janke, T.,Kalinin, A., Schoffler, M., Schmidt, L., Ivanov,M., Smirnova, O., and Dörner, R., 2016. Electron spin polarization in strong-field ionization of xenon atoms. *Nature Photonics*, 10, p. 526.

[3]  Milošević, D. B., 2016. Possibility of introducing spin into attoscience with spin-polarized electrons produced by a bichromatic circularly polarized laser field, *Physical Review A* 93 (5), p. 051402.

[4]  Ayuso, D., Jiménez-Galán, A., Morales, F., Ivanov, M., & Smirnova, O., 2017. Attosecond control of spin polarization in electron ion recollision driven by intense tailored fields. *New Journal of Physics*, 19(7), 073007.

[5]  Kaushal, J., and Smirnova, O., 2018. Looking inside the tunnelling barrier III: spin polarisation in strong field ionisation from orbitals with high angular momentum. *Journal of Physics B: Atomic, Molecular and Optical Physics,* 51 (17), p. 174003.

[6]  Morimoto, Y., & Baum, P. (2018). Diffraction and microscopy with attosecond electron pulse trains. *Nature Physics,* 14(3), p. 252-256.

[7]  Nie, Z., Li, F., Morales, F., Patchkovskii, S., Smirnova, O., An, W., Nambu, N., Matteo, D., Marsh, K.A., Tsung, F. , Mori, W.B., and Joshi, Ch., 2021. In situ generation of high-energy spin-polarized electrons in a beam-driven plasma wakefield accelerator. *Physical Review Letters*, 126(5), p.054801.

[8]  Beaulieu, S., Comby, A., Clergerie, A., Caillat, J., Descamps, D., Dudovich, N., Fabre, B., Géneaux, R., Légaré, F., Petit, S. and Pons, B., 2017. Attosecond-resolved photoionization of chiral molecules. *Science*, 358(6368), pp.1288-1294.

[9]  Rozen, S., Comby, A., Bloch, E., Beauvarlet, S., Descamps, D., Fabre, B., Petit, S., Blanchet, V., Pons, B., Dudovich, N. and Mairesse,Y., 2019. Controlling subcycle optical chirality in the photoionization of chiral molecules. *Physical Review X*, 9(3), p.031004.

[10] Göhler, B., Hamelbeck, V., Markus, T.Z., Kettner, M., Hanne, G.F., Vager, Z., Naaman, R. and Zacharias, H., 2011. Spin selectivity in electron transmission through self-assembled monolayers of double-stranded DNA. *Science*, 331(6019), pp.894-897.

[11] Trabert, D., Hartung, A., Eckart, S., Trinter, F., Kalinin, A., Schöffler, M., Schmidt, L.P.H., Jahnke, T., Kunitski, M. and Dörner, R., 2018. Spin and angular momentum in strong-field ionization. *Physical review letters*, 120(4), p.043202.

[12] Zhang, L., Carlström, S., Smirnova, O., Ivanov, M. and Ye, D., 2025. Spin Polarization in Strong-Field Ionization as Sensor of Trapped Electron Orbits. Physical Review Letters, 135(19), p.193201.

[13] Carlström, S., Dahlström, J.M., Ivanov, M.Y., Smirnova, O. and Patchkovskii, S., 2023. Control of spin polarization through recollisions. Physical Review A, 108(4), p.043104.

[14] Zapata, F., Vinbladh, J., Ljungdahl, A., Lindroth, E. and Dahlström, J.M., 2022. Relativistic time-dependent configuration-interaction singles method. Physical Review A, 105(1), p.012802.

[15] Artemyev, A.N., Kutscher, E., Lagutin, B.M. and Demekhin, P.V., 2023. Theoretical study of spin polarization in multiphoton ionization of Xe. *The Journal of Chemical Physics*, 158(15) p. 154115.

[16] Artemyev, A.N., Kutscher, E., Lagutin, B.M. and Demekhin, P.V., 2025. Theoretical study of spin polarization in multiphoton ionization of the HI molecule. *Physical Review A*, 111(4), p.043113.

[17] Flores, P.C.M., Carlström, S., Patchkovskii, S., Ivanov, M., Mujica, V., Ordonez, A.F. and Smirnova, O., 2025. Enantiosensitive locking of photoelectron spin and cation orientation. *arXiv preprint arXiv*:2505.22433.

[18] Fanciulli, M., Schusser, J., Lee, M.I., Youbi, Z.E., Heckmann, O., Richter, M.C., Cacho, C., Spezzani, C., Bresteau, D., Hergott, J.F., d'Oliveira, P., Tcherbakoff, O., Ruchon, Th., Minár, J., and Hricovini, K., 2020. Spin, time, and angle resolved photoemission spectroscopy on WTe$_2$. *Physical Review Research*, 2(1), 013261.

[19] Tahouri, R., Papoulia, A., Carlström, S., Zapata, F. and Dahlström, J.M., 2024. Relativistic treatment of hole alignment in noble gas atoms. *Communications Physics*, 7(1), p.344.

[20] Carlström, S., Spanner, M. and Patchkovskii, S., 2022. General time-dependent configuration-interaction singles. I. Molecular case. *Physical Review A*, 106(4), p.043104.

## 21. Attosecond science in semiconductor metrology

**Peter M. Kraus[1,2]\*, Stefan Witte[3]\***

[1] Advanced Research Center for Nanolithography (ARCNL), Science Park 106, 1098 XG Amsterdam, The Netherlands

[2] Department of Physics and Astronomy, Vrije Universiteit, De Boelelaan 1081, 1081 HV Amsterdam, The Netherlands

[3] Imaging Physics Department, Faculty of Applied Sciences, Delft University of Technology, Lorentzweg 1, 2628 CJ Delft, The Netherlands

kraus@arcnl.nl, s.m.witte@tudelft.nl

**Status**

Nanostructures drive the world around us. Technology has developed to a point where highly sophisticated nanostructures, known as integrated circuits (ICs), provide amazing functionality that is at the core of smartphones and household devices, but also drive progress in virtually any form of advanced technology. Present-day ICs have evolved into complex multilayer architectures, containing many different materials, and having sub-10-nm feature sizes. The continued development of IC technology, and in general the ability to 'see' such nanostructures, requires advanced inspection tools that can reach extremely high spatial resolution and distinguish different materials, and must also be able to visualize delicate 3D structures without destroying them [1]. Modern ICs consist of many layers and often contain materials that are opaque for visible light, providing a further challenge to present-day metrology tools. Microscopy with extreme ultraviolet (EUV) radiation has the potential to address these challenges: its short wavelength ($\lambda$ =2-40 nm) enables high spatial resolution, penetration depth into most materials can reach several hundred nanometers, and many elements have well-defined spectral signatures that provide contrast. As such, metrology tools need to be located in a 'fab' cleanroom where space constraints are tight, the availability of compact EUV imaging systems is a crucial ingredient to enable EUV metrology in semiconductor applications.

High-harmonic generation (HHG) sources are compact, yet provide fully coherent and broadband EUV radiation, across technologically important wavelength ranges such as the water window ($\lambda$ = 2-5 nm) and actinic inspection wavelengths for EUV lithography masks ($\lambda$ = 13.5 nm) [2-4]. The broad bandwidth of HHG provides specific opportunities for nanoscale metrology, as the spectrally dependent reflectivity and scattering encode both material properties and 3D structure.

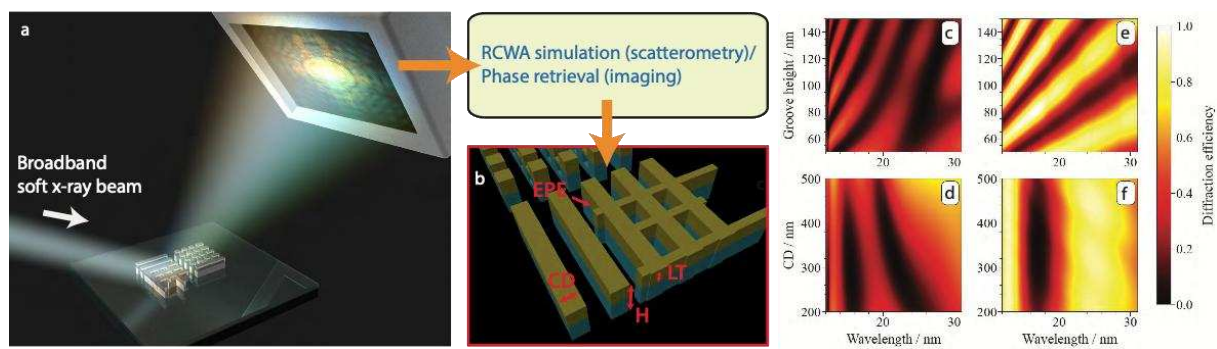

**Figure 1. a)** *Concept of HHG-based scatterometry and imaging. Coherent diffraction using broadband or monochromatized harmonics enables characterization of substrate-based nanostructures:* **b)** *either through full image reconstruction using phase retrieval methods, or model-based inference of key parameters such as line thickness (LT), height (H), critical dimension (CD) and edge-placement error (EPE). c-f) RCWA simulations of the 0th order diffraction efficiency for H and CD of a grating as a function of EUV wavelength (adapted from [7]).*

**Current and future challenges**

The most immediate opportunities for attosecond and high-harmonic generation (HHG) sources in semiconductor metrology are post-development overlay and profilometry [5], and actinic mask inspection [6] (Fig 1). Overlay metrology determines how accurately patterned layers align after lithography and etching, while profilometry characterizes three-dimensional topography of nanoscale structures [8,9]. Both rely on optical scatterometry to infer geometry and composition. Extending this principle to the EUV and soft x-ray range through HHG [10] offers clear advantages: first and foremost a short enough wavelength for at-resolution measurements on either markers with representative critical dimensions or directly on the device, coherence for interferometrically sensitive measurements, spectral tunability, and the ability to probe buried features via their wavelength-dependent response. Actinic mask inspection, operating near the lithography wavelength of 13.5 nm, benefits from the intrinsic coherence and narrow spectral bandwidth of isolated high-harmonic orders for coherent-diffraction or ptychographic imaging [11,12]. These two applications illustrate complementary operating modes: broadband HHG for overlay and material-specific analysis, and quasi-monochromatic HHG for actinic, high-resolution imaging.

Both modes confront similar physical and computational barriers. As semiconductor architectures become more complex and three-dimensional, multiple scattering within multilayer stacks can invalidate the single-scattering (first Born) approximation that underlies conventional reflectometry and diffractive imaging [13] (Fig. 1a). Retrieving quantitatively accurate structural parameters then requires more advanced computational approaches, based on more complete light-matter interaction models and advanced parameter inference techniques (Fig. 1b). A second bottleneck lies in the source. Present tabletop HHG systems deliver tens of microwatts of EUV or soft-x-ray power, far below the milliwatt-level flux desirable for quantitative metrology. Scaling the photon yield while preserving coherence and focusability is complicated by the phase matching requirements and HHG dynamics, leading to complications such as wavefront aberrations and spatiotemporal coupling [14-17]. However, accurate knowledge of the beam properties from either calibration measurements or probe-reconstructions in ptychography can provide a solution and enable accurate quantitative metrology.

Beyond static imaging, new directions are emerging. Nonlinear schemes such as harmonic deactivation microscopy (HADES) exploit control over the emission process itself to surpass the diffraction limit [18], while functional pump–probe metrology uses broadband HHG pulses to trace carrier or lattice dynamics across multilayer stacks [19]. These approaches link nanoscale structure to femtosecond function, but demand higher flux, precise focusing, and advanced reconstruction tools.

**Advances in science and technology to meet challenges**

Progress toward attosecond metrology will depend on parallel advances in source technology, beam control, and data analysis. On the source side, increasing photon flux to the milliwatt level will require improved phase matching and gas handling, high-repetition-rate mid-infrared drivers, and better thermal and plasma management. Two-color or multicolor driving fields can enhance conversion efficiency while favoring short-trajectory emission, thereby reducing divergence [20] (Fig. 2a,c). Adaptive optics implemented directly in the generation beamline will correct astigmatism and maintain diffraction-limited focusing across the HHG spectrum.

Equally important is precise characterization of the emitted wavefront. Multi-wavelength ptychographic diagnostics and Hartmann sensors [17] can quantify wavelength-dependent aberrations and feed them back to the generation optics in real time (Fig. 2b). Advances in multilayer and grazing-incidence optics—particularly deformable mirrors with tunable spectral response—will enable broadband focusing without introducing chromatic shift. Together, these measures will stabilize the HHG focus and ensure that increased power translates into usable, coherent flux.

The reconstruction problem calls for a new computational framework. Rigorous electromagnetic solvers combined with stochastic or gradient-based optimization can account for multiple scattering in realistic three-dimensional stacks (Fig. 1b-f). Embedding these solvers in maximum-likelihood or variational Bayesian formulations allows consistent treatment of noise and prior information such as design-layout constraints. Machine-learning models trained on large libraries of simulated patterns can further accelerate inversion and recover structure from incomplete or noisy data.

Integration of attosecond pulses into pump–probe schemes will extend HHG metrology from static imaging to functional probing by transient absorption and reflection studies through complex multilayers [19]. In combination with element-specific absorption edges in the soft-x-ray range, this approach allows layer-resolved tracking of carrier motion, defect formation, and ultrafast phase transitions.

Nonlinear imaging concepts such as HADES [18] will benefit from shaped control beams and high-speed modulation. Improved spatial light modulators and donut-mode generation will confine harmonic emission below the diffraction limit. In principle, methods like HADES have no intrinsic resolution limit [18]. Together, these developments—high-flux coherent sources, adaptive beam control, and physics-aware inversion—will transform HHG from a research tool into a quantitative method for structural and functional metrology of semiconductor devices.

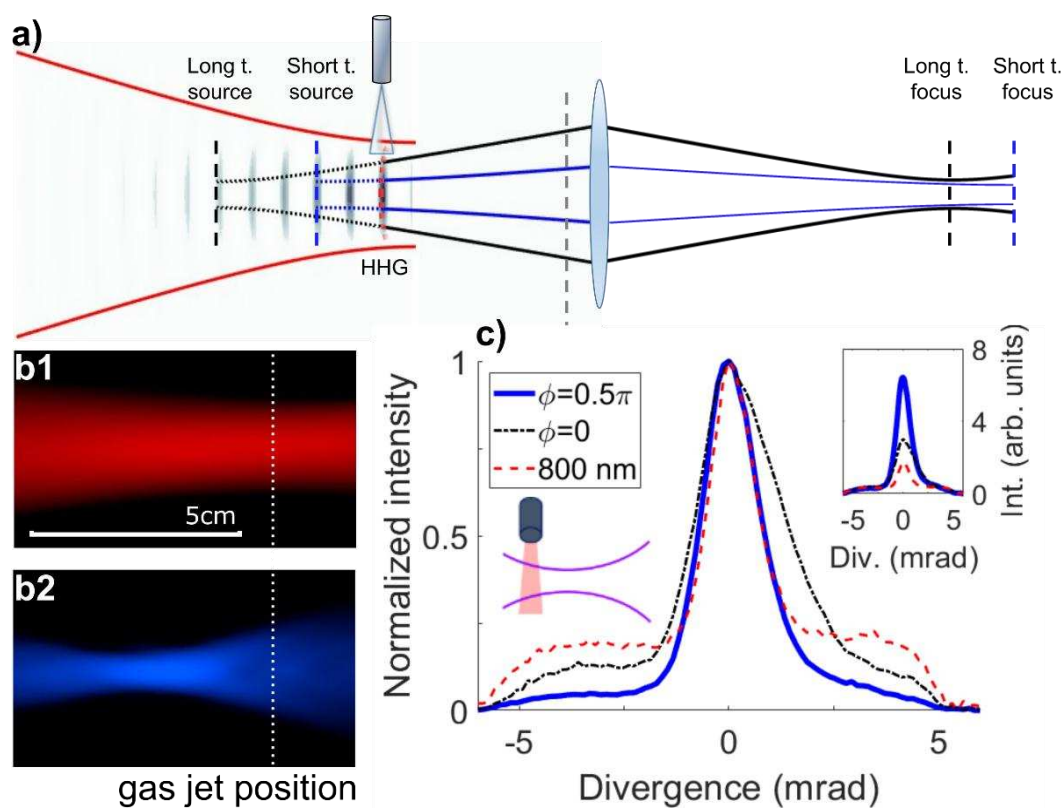

***Figure 2. (a)*** Schematic representation of an HHG source and refocusing (lens shown for simplicity). Different focus positions are caused by both chromatic aberrations and long/short-trajectory contributions. ***(b1,b2)*** Experimentally reconstructed virtual source positions for harmonics 15 (53 nm), 800 nm driver) and 25 (23 nm) via ptychography, showing strong chromatic aberrations. ***(c)*** Focus smearing due to long and short trajectories can be addressed in two-color HHG. The panel shows a spectrally integrated normalized and absolute (inset) beam profile for harmonic 13 (62 nm), for two different relative two-color phases and 800 nm (single color) only. Panels ***(a,c)*** are adapted from [20], ***(b)*** is adapted from [17].

## Concluding remarks

Reaching the next level in attosecond-based metrology will depend on making HHG sources brighter, and more stable. Average powers at or above the milliwatt level, combined with diffraction-limited focusing and minimal chromatic distortion, are needed to turn these sources into quantitative metrology tools rather than laboratory prototypes. A sharper, more coherent focus will directly translate into higher spatial precision and signal fidelity, enabling measurements on the relevant length scales of semiconductor devices.

The development of models capturing 3D light-matter interactions would allow quantitative recovery of shape and material information for realistic device geometries, wafer-bonding interfaces, and resists. Including line-edge and line-width roughness directly in the reconstruction—currently impossible below the nanometer scale—would turn scatterometry into a genuine dimensional metrology method rather than an imaging proxy.

Ultimately, a future academic challenge and opportunity for industry lie in combining structural and functional metrology: coupling 3D scatterometry with attosecond (or femtosecond) pump–probe spectroscopy to simultaneously resolve form, composition, and local electronic response—measuring not only how devices are built, but how they work.

## Acknowledgements

Part of this work was conducted at the Advanced Research Center for Nanolithography, a public-private partnership between the University of Amsterdam (UVA), Vrije Universiteit Amsterdam (VU), Rijksuniversiteit Groningen (RUG), the Netherlands Organization for Scientific Research (NWO), and the semiconductor equipment manufacturer ASML. PMK was supported by the European Research Council (101041819, ERC Starting Grant ANACONDA) and the NWO VIDI research programme (vi.vidi.223.133, HIMALAYA). SW acknowledges funding from the ERC (864016, Consolidator grant 3D-VIEW), and the NWO VICI programme (19405).

## References

[1] Den Boef AJ 2013 Optical metrology of semiconductor wafers in lithography *Proc. SPIE* **8769** 876907

[2] Chen M-C et al 2010 Bright, Coherent, Ultrafast Soft X-Ray Harmonics Spanning the Water Window from a Tabletop Light Source *Phys. Rev. Lett.* **105** 173901

[3] Popmintchev T et al 2012 Bright Coherent Ultrahigh Harmonics in the keV X-ray Regime from Mid-Infrared Femtosecond Lasers *Science* **336** 1287

[4] Teichmann SM et al 2016 0.5-keV Soft X-ray attosecond continua, *Nat. Commun.* **7** 11493

[5] Orji NG et al 2018 Metrology for the next generation of semiconductor devices *Nat. Electron.* **1** 532

[6] Mochi I et al 2020 Quantitative characterization of absorber and phase defects on EUV reticles using coherent diffraction imaging *J. Micro/Nanolith. MEMS MOEMS* **19** 014002

[7] Corrazza F et al 2025 Broadband extreme ultraviolet zero-order Scatterometry for nanostructure metrology *Research Square preprint: https://doi.org/10.21203/rs.3.rs-7215343/v1*

[8] Tanksalvala M et al 2021 Nondestructive, high-resolution, chemically specific 3D nanostructure characterization using phase-sensitive EUV imaging reflectometry *Sci. Adv.* **7** eabd9667

[9] Eschen W et al 2022 Material-specific high-resolution table-top extreme ultraviolet microscopy *Light: Sci. Appl.* **11** 117

[10] Porter CL et al 2023 Soft x-ray: novel metrology for 3D profilometry and device pitch overlay, *Metrology, Inspection, and Process Control XXXVII, SPIE* 124961I (2023)

[11] Nagata Y et al 2019 At wavelength coherent scatterometry microscope using high-order harmonics for EUV mask inspection *Int. J. Extrem. Manuf.* **1** 032001

[12] Porter CL et al 2017 General-purpose, wide field-of-view reflection imaging with a tabletop 13 nm light source *Optica* **4** 1552

[13] Loetgering L, Witte S, Rothhardt J 2022 Advanced in laboratory-scale ptychography using high-harmonic sources *Opt. Express* **30** 4133

[14] Wikmark H et al 2019 Spatiotemporal coupling of attosecond pulses *Proc. Natl. Acad. Sci. USA* **116** 4779

[15] Hoflund M et al 2021 Focusing properties of high-order harmonics *Ultraf. Sci.* **2021** 9797453

[16] Du M et al 2023 High-resolution wavefront sensing and aberration analysis of multi-spectral extreme ultraviolet beams *Optica* **10** 255

[17] Liu X et al 2023 Observation of chromatic effects in high-order harmonic generation *Phys. Rev. Res.* **5** 043100

[18] Murzyn K et al 2024 Breaking Abbe's diffraction limit with harmonic deactivation microscopy *Science Advances* **10** eadp3056

[19] Cushing SK 2020 Layer-resolved ultrafast extreme ultravioletmeasurement of hole transport in aNi-TiO2-Si photoanode et al *Science Advances* **6** eaay6650

[20] Roscam Abbing SDC et al 2020 Divergence Control of High-Harmonic Generation *Physical Review Applied* **13** 054029

## 22. Theory of high-harmonic generation in strongly correlated systems

**Lars Bojer Madsen**

Department of Physics and Astronomy, Aarhus University, 8000 Aarhus, Denmark

bojer@phys.au.dk

**Status**

For about a decade, high-order harmonic generation (HHG) at near-infrared frequencies has been pursued as an ultrafast spectroscopy technique aiming at mapping out electron motion in strongly correlated systems (SCSs) [1,2]. These efforts followed work on uncorrelated solids, where the generation is rationalized in terms of a three-step model involving promotion of an electron from the valence to the conduction band, its propagation in the conduction band and a final recombination step when the electron and hole collide. This picture has been instrumental, and analysis in terms of the semiconductor Bloch equations (SBE) has offered qualitative and quantitative insights [3].

In the case of SCS the band-structure picture ceases to exist. Many insights have been obtained by a consideration of the Fermi-Hubbard model [1,2]: In the Mott insulating phase, characterized by an onsite electron repulsion, $U$, much larger than the hopping term between sites, $t_0$, analysis of HHG can be made in terms of quasiparticles, doublons and holons, that describe doubly occupied and unoccupied sites in a crystal. A three-step model can be formulated in terms of these quasiparticles with the following steps: (i) excitation of a holon-doublon pair, (ii) propagation of the pair in Hubbard sub-bands and (iii) its recombination.

In recent efforts excitonic effects were explored using a SBE model with electron-electron interaction included at the Fock level [4], with time-dependent Hartree Fock [5] and with the extended Hubbard model including nearest-neighbour interaction $V$ [6]. The work with HHG and SCSs has also explored effects associated with finite lattices, doping and lattice imperfections (see, e.g., the discussion in [7]). A consideration of the 2-leg ladder model allowed an identification of contributions to HHG from polarons and magnons and a discussion of spin-charge coupling [8]. Combined experimental and theoretical work shows that HHG discriminates between phases in superconductors [9]. Such advances highlight that correlated electron dynamics is elucidated by HHG spectroscopy. Quite recently, a series of studies show that the photonic degrees of freedom (DOF) in some cases need be treated quantum mechanically [10–14]. This finding contrasts standard methodology for attosecond science treating only the electronic DOF quantum mechanically and describing the photonic DOF by classical electromagnetic fields. Work show that the light generated from a SCS is squeezed [15], i.e., nonclassical, and that the generation of this squeezing can be controlled by optically active Mott excitons [16].

**Current and future challenges**

To gain insights into electron motion in SCSs, we need theoretical models with predictive power. We can divide approaches into density-based and wave-function-based methodologies. The density-based approach allows the handling of large systems, see, e.g., a recent time-dependent density-functional theory (TDDFT) study of HHG including nondipole effects in crystalline Si [17]. However, even when augmented with a Hubbard-$U$ term as in the TDDFT+$U$ approach (see, e.g. ,[18]) such density-based approaches are challenged in describing SCSs.

With wavefunction-based methodologies, we quickly face the 'curse of dimensionality' and it is difficult to account for electron-electron interaction beyond the mean-field level of theory – there are simply too many DOF to allow for application of the wavefunction-based ab initio approaches that work for atoms, molecules and in cold-atom physics [19]. To manage computational complexity and to isolate physical mechanisms, we are therefore led to consider effective-Hamiltonian models such as the Fermi-Hubbard model, and typically in reduced dimensionality. To proceed further with such models, we need work on extending the description to higher dimensions as well as more bands under strong nonlinear drive. Effects of the couplings between charge, spin and orbitals would be more readily mapped out in such descriptions.

One particularly promising direction for HHG spectroscopy in SCS is the consideration of quantum light generated in the process [15,16]. We now know that SCSs generate quantum light with nonvanishing squeezing (see figure 1, [15,16]). Put in simple terms, the high degree of correlation between the electrons in SCSs leads to a correspondingly high degree of correlation between the generated photons, inducing nonclassical, quantum properties in the generated light. Progress in elucidating the interplay between quantum mechanical electronic and photonic DOF is currently challenged by the theoretical formulations that necessitate access to all field-free eigenstates of the electronic system [10,15]. In practice this demand means that at maximum Fermi-Hubbard chains of length around 10 sites at half-filling can be considered with exact diagonalisation techniques. Moreover, the description of the photonic DOF leads to a coupled set of differential equations for the photonic wave packets associated with each field-free electronic state, where all modes are coupled. This again leads to an explosion the Hilbert space size. Out of necessity and based on physical arguments, it has been proposed to decouple de photonic DOF, such that one mode of the electromagnetic field is considered at a time [10,15].

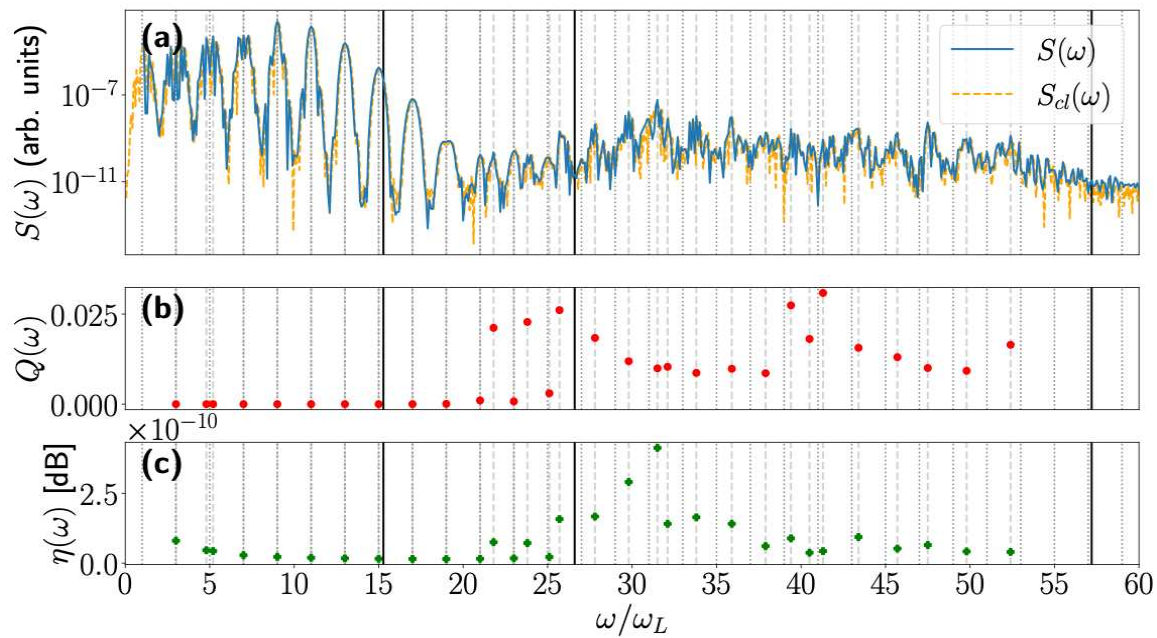

**Figure 1**. *(a) HHG spectrum with harmonic orders measured in units of driving laser frequency $\omega_L$ for a Hubbard chain (see [15] for details on the system and laser parameters). The classical $S_{cl}(\omega)$ and quantum $S(\omega)$ models for the HHG spectra give almost identical results. (b) A nonvanishing value for the Mandel $Q(\omega)$ parameter illustrates super-Poissonian statistics of the generated light. (c) A nonvanishing squeezing, illustrated here by the squeezing parameter $\eta(\omega)$ , shows that the generated light is nonclassical. Reprinted figure with permission from [15], © by the American Physical Society 2024.*

**Advances in science and technology to meet challenges**

While simulations of SCSs under strong-field driving in two and three dimensions still seem out of reach,  infinite time evolving Block dissemination calculations combined with matrix product state techniques  have very recently been applied to 2-leg ladder Fermi-Hubbard model systems [8]. This extension is promising in terms of perspectives for revealing dynamics that cannot be captured fully in one-dimensional simulation. Investigations along these avenues will allow us to gain more insight into spin-charge and orbital couplings under highly nonlinear conditions.

The perspective of generating HHG light with nonclassical properties in SCS imposes high demands on theory development for two reasons. (i) With the presently available theory, we must simulate the time-dependent Schrödinger equation (TDSE) starting in all the field-free electronic eigenstates. This demand imposes limitations on the number of DOF that can be accounted for in the electron system. (ii) A full account of all coupled modes in the photonic DOF leads to a very large Hilbert space, that often will make practical simulations unfeasible.  At present, a decoupling of photonic modes is assumed to be a reasonable accurate approximation. While the issue with the electronic DOF seems to persist in the theoretical formulations, the problem with the practically required decoupling of photonic modes in the TDSE-based approach for the theory, can possibly be dealt with in an alternative formulation based on the Heisenberg picture [20]. In this case the HHG spectra and other observables, such as the second-order coherence function or the squeezing link to dipole correlations and these can be evaluated without performing the decoupling procedure. In connection with the generation of nonclassical light, it is noted that theory suggests that the HHG spectra are accurately described by semiclassical theory and therefore focusing on measuring observables other than the HHG spectrum, e.g., the Mandel $Q$, the second-order coherence function or the degree of squeezing, may reveal more clearly insights into the properties of quantum light and the associated correlated quantum mechanical electron motion. We have situations where the quantum properties of the generated light are not clearly identifiable from the HHG spectra, i.e., the spectra look like those produced by semiclassical theory, as is also the case in figure 1, and, accordingly, other observables need be considered.

**Concluding remarks**

HHG spectroscopy provides unique new insights into electron dynamics, spin-charge and orbital couplings in SCSs. Theoretically, the nonperturbative dynamics of SC many-body systems is an area of research where new theory is continuously being developed. As an exciting new aspect of SCS and attoscience, the interplay between correlated electrons and photons is being investigated. Fully quantum treatments of both electronic and photonic DOF show that strong-field driving of SCSs lead to the generation of quantum light through the HHG process [15,16]. By studying the properties of the generated light, it is expected that several features of the electron dynamics, not accessible from the HHG spectrum alone, will be revealed. In this sense it is expected that insights into so far un-elucidated properties of the electron dynamics will be obtained by analysis of the 'quantumness' of the generated light.

**Acknowledgements**

This work is supported by the Independent Research Fund Denmark (Technology and Production Sciences 10.46540/4286-00053B) and the Novo Nordisk Foundation Project Grants in the Natural and Technical Sciences (0094623).

## References

[1] Silva R E F, Blinov I V, Rubtsov A N, Smirnova O and Ivanov M 2018 High-harmonic spectroscopy of ultrafast many-body dynamics in strongly correlated systems *Nature Photonics* **12** 266–70

[2] Murakami Y, Eckstein M and Werner P 2018 High-Harmonic Generation in Mott Insulators *Phys. Rev. Lett.* **121** 057405

[3] Vampa G, McDonald C R, Orlando G, Klug D D, Corkum P B and Brabec T 2014 Theoretical Analysis of High-Harmonic Generation in Solids *Phys. Rev. Lett.* **113** 073901

[4] Molinero E B, Amorim B, Malakhov M, Cistaro G, Jiménez-Galán Á, Picón A, San-José P, Ivanov M and Silva R E F 2024 Subcycle dynamics of excitons under strong laser fields *Science Advances* **10** eadn6985

[5] Jensen S V B, Madsen L B, Rubio A and Tancogne-Dejean N 2024 High-harmonic spectroscopy of strongly bound excitons in solids *Phys. Rev. A* **109** 063104

[6] Udono M, Sugimoto K, Kaneko T and Ohta Y 2022 Excitonic effects on high-harmonic generation in Mott insulators *Phys. Rev. B* **105** L241108

[7] Hansen T and Madsen L B 2024 Lattice imperfections and high-harmonic generation in correlated systems *New J. Phys.* **26** 063023

[8] Murakami Y, Hansen T, Takayoshi S, Madsen L B and Werner P 2025 Many-Body Effects on High-Harmonic Generation in Hubbard Ladders *Phys. Rev. Lett.* **134** 096504

[9] Alcalà J, Bhattacharya U, Biegert J, Ciappina M, Elu U, Graß T, Grochowski P T, Lewenstein M, Palau A, Sidiropoulos T P H, Steinle T and Tyulnev I 2022 High-harmonic spectroscopy of quantum phase transitions in a high-Tc superconductor *Proceedings of the National Academy of Sciences* **119** e2207766119

[10] Gorlach A, Neufeld O, Rivera N, Cohen O and Kaminer I 2020 The quantum-optical nature of high harmonic generation *Nat Commun* **11** 4598

[11] Lewenstein M, Ciappina M F, Pisanty E, Rivera-Dean J, Stammer P, Lamprou Th and Tzallas P 2021 Generation of optical Schrödinger cat states in intense laser–matter interactions *Nat. Phys.* **17** 1104–8

[12] Stammer P, Rivera-Dean J, Maxwell A S, Lamprou T, Argüello-Luengo J, Tzallas P, Ciappina M F and Lewenstein M 2024 Entanglement and Squeezing of the Optical Field Modes in High Harmonic Generation *Phys. Rev. Lett.* **132** 143603

[13] Yi S, Klimkin N D, Brown G G, Smirnova O, Patchkovskii S, Babushkin I and Ivanov M 2025 Generation of Massively Entangled Bright States of Light during Harmonic Generation in Resonant Media *Phys. Rev. X* **15** 011023

[14] Theidel D, Cotte V, Sondenheimer R, Shiriaeva V, Froidevaux M, Severin V, Merdji-Larue A, Mosel P, Fröhlich S, Weber K-A, Morgner U, Kovacev M, Biegert J and Merdji H 2024 Evidence of the Quantum Optical Nature of High-Harmonic Generation *PRX Quantum* **5** 040319

[15] Lange C S, Hansen T and Madsen L B 2024 Electron-correlation-induced nonclassicality of light from high-order harmonic generation *Phys. Rev. A* **109** 033110

[16] Lange C S, Hansen T and Madsen L B 2025 Excitonic Enhancement of Squeezed Light in Quantum-Optical High-Harmonic Generation from a Mott Insulator *Phys. Rev. Lett.* **135** 043603

[17] Jensen S V B, Tancogne-Dejean N, Rubio A and Madsen L B 2025 Beyond Electric-Dipole Treatment of Light-Matter Interactions in Materials: Nondipole Harmonic Generation in Bulk Si *Phys. Rev. Lett.* **134** 196902

[18] Tancogne-Dejean N, Oliveira M J T and Rubio A 2017 Self-consistent $\mathrm{DFT}+U$ method for real-space time-dependent density functional theory calculations *Phys. Rev. B* **96** 245133

[19] Lode A U J, Lévêque C, Madsen L B, Streltsov A I and Alon O E 2020 Colloquium: Multiconfigurational time-dependent Hartree approaches for indistinguishable particles *Rev. Mod. Phys.* **92** 011001

[20] Stammer P, Rivera-Dean J and Lewenstein M 2025 Theory of quantum optics and optical coherence in high harmonic generation

## 23. R-matrix approaches for atoms and molecules

**Zdeněk Mašín[1]\*, Jimena D. Gorfinkiel[2], Jakub Benda[1], Andrew C. Brown[3] and Hugo W. van der Hart[3]**

[1] Institute of Theoretical Physics, Faculty of Mathematics and Physics, Charles University, V Holešovickách 2, Prague 8, 180 00, Czech Republic
[2] School of Physical Sciences, The Open University, Milton Keynes, United Kingdom
[3] Centre for Light-Matter Interactions, Queen's University Belfast, Belfast, Northern Ireland, United Kingdom, BT7 1NN

zdenek.masin@matfyz.cuni.cz

**Status**

The R-matrix approach [1] is a highly accurate method for the investigation of light-particle (electron/positron) scattering and photoionization processes involving atoms, molecules and small clusters. It is a wavefunction-based, variational and multi-electron method which makes it advantageous for problems where the structure of the continuum plays the main role. Typical applications include studies of multi-electron effects (field-free and/or field-induced) and resonance formation in both scattering and photoionization.

Time-dependent and time-independent R-matrix approaches have been developed to study attosecond processes. The R-matrix with time-dependence (RMT) approach allows the description of general atoms and molecules driven by arbitrarily polarised laser pulses and can describe relativistic electron dynamics in the former [2]. The time-independent approach allows the determination of multi-photon above- and below-threshold ionization amplitudes of many-electron atoms and molecules. Two state-of-the-art computational codes implement the methods: RMT (see Chapter 24), which treats atoms and molecules, and the time-independent molecular UKRmol+ [3]. Other approaches and software implementations, which share some of their characteristics with UKRmol+ and RMT, can be applied to the study of attosecond molecular processes, for example XChem [4], tRecX-haCC [5], Tiresia [6], etc.

Over the past decade, the molecular R-matrix method has matured as a method for photoionization problems: UKRmol+ now delivers highly accurate wavefunctions for small to medium-sized molecules (typically including 1st – 3rd row atoms). Capabilities now include constructing molecular photoionization and Siegert states, evaluating arbitrary-order perturbative multiphoton amplitudes [7], and providing the stationary molecular data required for time-dependent simulations using the RMT codes.

Recent successful uses of both approaches include the evaluation of RABBITT time-delays for small molecules [8] and supporting experiments showing how the entanglement between the photoion and the photoelectron created in the photoionisation process manifests in attosecond delays [8]. In atoms, RMT was used to resolve individual partial wave contributions beyond the perturbative regime, extending the RABBITT technique into the few-photon regime [10].

The future development of the methods and software is aimed at enabling studies of the multi-electron effects in core-hole states, nuclear dynamics in resonant molecular photoionization and the modeling of relativistic effects in systems including heavy atoms.

**Current and future challenges**

Free electron lasers are quickly establishing themselves as an important tool for the study of attosecond dynamics, as their tunable frequency and intensity enable applications with element-specific resolution [11]. As a step towards this goal, recently, high-repetition rate attosecond pulses in the X-ray domain have been generated [12]. Accessing core-hole states in molecules by FELs allows the study of, for example, chemical reactions, highly correlated electronic states, coherent dynamics in core-excited states and the relativistic effects on ultrafast time-scales. While FELs will remain key facilities for this purpose, the development of table-top sources of soft X-rays would greatly speed up progress.

Specific quantum processes involving entanglement of the atomic and molecular states or coherence [12] can be controlled by a tunable external field. For example, resonant tuning of the photon energy in multi-color photoionization experiments would allow access to the entangled nature of the photoelectron wavefunction [8] which is not directly probed in traditional 1-photon experiments. Similarly, expanding the spectral width of two-colour experiments to span several electronic states would enable probing coherence and entanglement by directly measuring the photoelectron density matrix [13]. Significant progress has been made recently in generating VUV and UV ultrashort pulses for Time-Resolved Photoelectron Spectroscopy (TR-PES), see e.g. [14]. Continuing this effort to achieve applicability to a wide range of molecules, especially with biological relevance, would improve our understanding of chemical dynamics and non-radiative decay pathways, including conical intersections. Another emerging technique is the use tailored light fields to manipulate 3D electron dynamics and investigate chiral molecules. With the main techniques of ultrafast physics firmly established, application to other fields becomes possible. In particular time-domain studies of electron-molecule collisions could be performed through photodetachment experiments which access the neutral scattering continuum.

Providing theoretical support for the analysis and interpretation of these novel experiments requires continued development of the computational tools. Even for atomic systems, the description of core-vacancies or multiple ionisation stages made accessible by FEL sources, is a challenge. While there is no fundamental reason these processes cannot be described with R-matrix methods, the scale of the computations is prohibitive, especially if relativistic effects are to be included. Techniques such as attosecond transient absorption spectroscopy allow 'real-time' measurement of dynamics in strong fields [15]; but while describing atomic systems in such fields is already possible with RMT, extending molecular calculations into the strong-field regime will require significant code development.

**Advances in science and technology to meet challenges**

A particular challenge, still not successfully resolved, is the description of double continua in both atoms and molecules, relevant for photoionization and impact ionization problems. Solving this problem will allow explicit time-dependent calculations of double ionization processes, such as non-sequential double ionization, with RMT. A prototype has already been demonstrated for helium [16]. For non-sequential processes, each stage can be described with a single RMT calculation, and the evolution of the residual ion in between times can be computed using a density-matrix-based approach [17].

Specific key challenges and opportunities for molecular targets include treating highly correlated and/or non-symmetrical systems (chiral molecules), High-Harmonic Generation, inclusion of heavy atoms where relativistic effects must be included, coupled electron–nuclear dynamics and two-electron continua relevant to double ionization and Auger processes.

The currently implemented molecular R-matrix approach is non-relativistic, while novel experiments employing FEL sources are routinely accessing core molecular orbitals, where relativistic effects are ubiquitous. The recently implemented Effective Core Potentials (ECPs) describe the relativistic effects of the core on the valence electrons, but an explicit inclusion of relativistic effects among the active electrons is lacking. Much experience is to be drawn from the atomic relativistic R-matrix implementations originating in the 1980s and culminating in 2009 with the fully relativistic (Dirac) R-matrix approach, but they are not the only possible routes for inclusion of relativistic effects.

A major step which has enabled many recent applications was the implementation of an accurate description of the molecular continuum, required for the high energies and large angular momenta typically encountered in molecules and ultrafast experiments. UKRmol+ uses GBTOLib, a molecular integral library implementing a flexible hybrid B-spline/Gaussian continuum description. Nonetheless, computing molecular integrals remains a demanding component of many calculations, especially for larger molecules, and its optimization an active area of research.

Molecular R-matrix calculations use the fixed-nuclei approximation, which is mostly adequate except when transiently bound (resonant) states are formed and the electron–nuclear coupling can drive various reaction channels. This problem is not solvable in its full dimensionality using the current computational codes and resources. Instead, prior R-matrix work in electron–molecule collisions employed Feshbach partitioning and successfully parametrized the transient coupled electron-nuclear dynamics using fixed-nuclei scattering data [18]; extensions of this approach to molecular photoionization and photodetachment are underway.

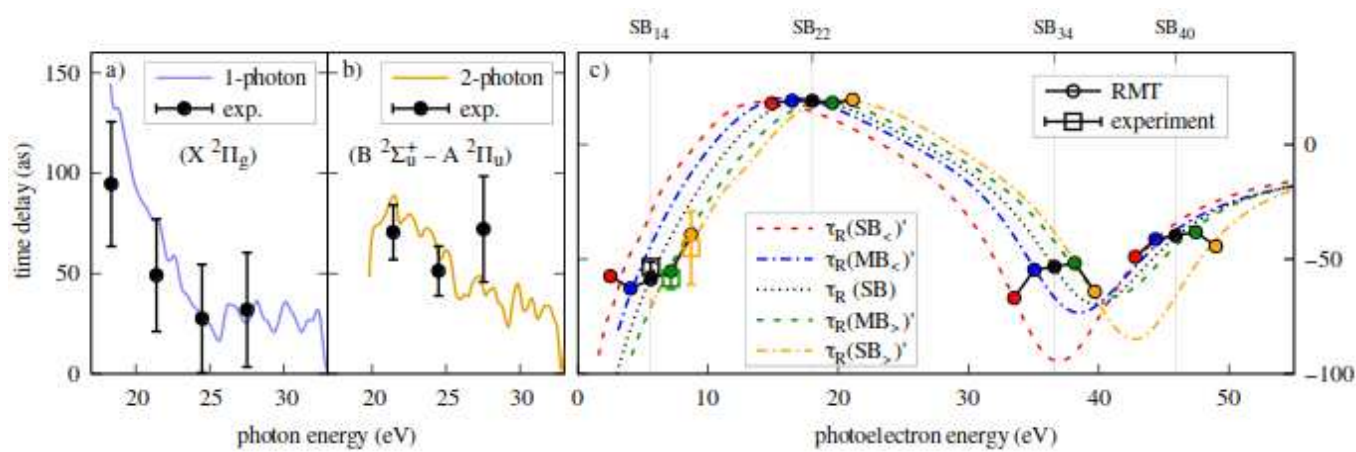

*Figure 1: Attosecond delays obtained using R-matrix methods. a) One-photon ionization delays for ground state of CO2+ calculated in UKRmol+. b) Relative RABBITT delays for ionization into the first two excited states of CO2+. c) RABBITT (sideband) delays and "higher-order RABBITT" (mainband and outer sideband) delays in argon in two-harmonic field configurations around four selected central sidebands calculated using the time-independent perturbative (dashed lines) and time-dependent (RMT) approaches. Panels a) and b) are adapted from [8]. Panel c) is adapted from [19].*

## Concluding remarks

The R-matrix approach to atoms and molecules continues to be actively developed by the international community. Traditionally, the development of the molecular R-matrix was driven mostly by applications to scattering processes. More recently, this drive has been replaced with the need to serve the ultrafast community. Despite additional challenges associated with molecules due to a lower symmetry and the nuclear motion, molecular developments are catching up with the more mature atomic codes.

The grand challenge for the future of both atomic and molecular R-matrix methods lies, in our opinion, in treating highly-correlated systems and two-electron continua. The key principle behind the R-matrix (Green's function) approach is the division of space into a multi-electron part, where ab initio techniques must be employed, and a much simpler 1-electron part, where analytical approaches can be used and help to formulate a physical interpretation. [7, 20]. Possible extensions of the analytical R-matrix approach [20] to two-electron continua may be a useful complement to the purely ab initio calculations.

## Acknowledgements

The authors recognize the indispensable contributions of many developers of the R-matrix codes: Gregory S.J. Armstrong, Daniel D.A. Clarke, Jack Wragg , Kathryn R. Hamilton, Alex Harvey, Michael Lysaght, Robert McGibbon, Laura Moore, Lampros Nikolopoulos, Jonathan Parker, Martin Plummer, Ken Taylor, Ahmed Al-Refaie, Rui Zhang, Daniel Darby-Lewis, Dermot Madden, Andrew Sunderland, Jo Carr, Paul Roberts, Vincent Graves, Cliff Noble, Lesley Morgan, Charles Gillan and Jonathan Tennyson. ZM acknowledges the support of the Czech Science Foundation (Grant no. 25-18015K) and the Charles University Research Center Grant No. UNCE/24/SCI/016. JDG acknowledges support by EPSRC under grants EP/P022146/1 and EP/R029342/1. ACB and HWvdH acknowledge funding from the EPSRC under Grants No.  EP/T019530/1, EP/V05208X/1, and EP/R029342/1.

## References

[1]  Burke  P G 2011  R-Matrix Theory of Atomic Collisions Springer Series on Atomic, Optical, and Plasma Physics 61 *Springer*

[2]  Wragg J, Balance  C and  van der Hart  H.  202 Breit–Pauli R-Matrix approach for the time-dependent investigation of ultrafast processes *Comput. Phys. Commun.* **254** 107274

[3]  Mašín Z, Benda J, Gorfinkiel  J D, Harvey A G and Tennyson J 2020 UKRmol+: A suite for modelling electronic processes in molecules interacting with electrons, positrons and photons using the R-matrix method. *Comput. Phys. Commun* **249** 107092

[4]  Borràs V J, Fernández-Milán P, Argenti L, González-Vázquez J and Martín F 2024  Photoionization cross sections and photoelectron angular distributions of molecules with xchem-2.0 *Comput. Phys. Commun.* **296**, 109033

[5]  Scrinzi A 2022  tRecX - An environment for solving time-dependent Schrödinger-like problems, *Comput. Phys. Commun.* **270** 108146

[6]  Toffoli D, Coriani S, Stener M, and Decleva P  2024 Tiresia: A code for molecular electronic continuum states and photoionization  *Comput. Phys. Commun.* **297** 109038

[7]  Benda J and Mašín Z 2021 *Sci. Rep.* **11** 11686

[8]  Benda J, Mašín Z and  Gorfinkiel  J D 2022 Analysis of RABITT time delays using the stationary multiphoton molecular *R*-matrix approach *Phys. Rev.* A **105** 053101

[9]  Makos I, Busto D, Benda J, Ertel D, Merzuk B, Steiner B,  Frassetto F,  Poletto L, Schröter C D, Pfeifer T, Moshammer R,  Patchkovskii S, Mašín Z and Sansone G 2025 Entanglement in photoionisation reveals the effect of ionic coupling in attosecond time delays *Nat Commun* **16** 8554

[10]  Jiang W, Roantree L, Han L, Ji J, Xu Y, Zuo Z., Wörner H J, Ueda K, Brown A C, Van Der Hart,  H W, Gong X and  Wu J 2025 Heterodyne analysis of high-order partial waves in attosecond photoionization of Helium *Nat Commun*  **16** 381

[11]  Chergui M, Beye M, Mukamel S, Svetina C and Masciovecchio C 2023 "Progress and prospects in nonlinear extreme-ultraviolet and X-ray optics and spectroscopy," *Nat Rev Phys* **5** 578–596

[12]  Wang J, Driver T , Franz P L , Kolorenč P, Thierstein E, Robles R R,  Isele E, Guo Z, Cesar D, Alexander O, Beauvarlet S, Borne K, Cheng K, DiMauro L F, Duris J, Glownia J M, Graßl M, Hockett P, Hoffman K, Kamalov A, Larsen K A, Li X, Lin M-F,  Obaid R, Rosenberger P, Walter, Wolf T J A, Marangos J P, Kling M F, Bucksbaum P H, Marinelli A and Cryan J P 2025 Probing Electronic Coherence between Core-Level Vacancies at Different Atomic Sites *Phys. Rev.* X **15** 011008

[13]  Laurel H, Finkelstein-Shapiro D, Dittel C, Guo C, Demjaha R, Ammitzböll M, Weissenbilder R, Neoričić L, Luo S, Gisselbrecht M, Arnold C L, Buchleitner A, Pullerits T, L'Huillier A, and  Busto D 2022 Continuous-variable quantum state tomography of photoelectron  *Phys. Rev. Res.* **4** 033220

[14]  Crego A, Severino S, Mai L,  Medeghini F,  Vismarra F,  Frassetto F,  Poletto L,  Lucchini M,  Reduzzi M,  Nisoli M and Borrego-Varillas B 2024 Sub-20-fs UV-XUV beamline for ultrafast molecular spectroscopy *Sci Rep* **14** 26016

[15]  Hutcheson L, Hartmann M, Borisova G D, Birk P, Hu S, Ott C, Pfeifer T, van der Hart H W and Brown A C 2025 Phase evolution of strong-field ionization *Phys. Rev. Res.* **7** L022074

[16]  Wragg J, Parker J S, and van der Hart H W  2015 Double ionization in R-matrix theory using a two-electron outer region *Phys. Rev.* A **92** 022504

[17] Lavery H, Brown A C and van der Hart H W   2025 Impact of electron correlation on photoelectron angular distributions in sequential double photoionization of Ne⁺ *Phys. Rev.* A accepted

[18] Zawadzk M, Čížek M, Houfek K, Čurík R, Ferus M, Civiš S, Kočišek J, and  Fedor J 2018 Resonances and Dissociative Electron Attachment in HNCO *Phys. Rev. Lett*. **121** 143402

[19] Benda J, Mašín Z, Palakkal S, Lépine F, Nandi S, Loriot V 2025 Angular momentum dependence in multiphoton ionization and attosecond time delays *Phys. Rev. A* **111** 013110

[20] Torlina L,  Ivanov M,  Walters Z B and  Smirnova O 2012  Time-dependent analytical R-matrix approach for strong-field dynamics. II. Many-electron systems *Phys. Rev.* A **86** 043409

## 24. Attosecond Transient Absorption Spectroscopy using R-Matrix with Time-Dependence Methods

**Andrew C. Brown[1]\*, Lynda R. Hutcheson[2], Sean Marshallsay[1] and Hugo W. van der Hart[1]**

[1] School of Mathematics and Physics, Queen's University Belfast, Belfast, BT7 1NN, United Kingdom
[2] School of Physics, University College Dublin, Belfield, Dublin 4, D04 P7W1, Ireland

andrew.brown@qub.ac.uk

### Status

Attosecond transient absorption spectroscopy (ATAS) has emerged as a premier technique for probing electronic dynamics on their natural timescale[1]. Unlike traditional spectroscopy that measures ejected electrons, ATAS analyzes transmitted light, providing time-resolved measurements of bound state dynamics, and ionization processes[2]. The field has evolved from early ionization yield measurements[3] to sophisticated phase-resolved spectroscopy[4]. Advances in spectrometer resolution allow analysis of absorption line shapes, providing access to both amplitude and phase information of the dipole response.

Simulating ATAS presents unique challenges due to the need to describe electronic structure and strong-field dynamics simultaneously. The R-Matrix with Time-dependence (RMT) method has proven particularly successful, solving the time-dependent Schrödinger equation (TDSE) for multielectron systems in intense laser fields[5]. RMT calculations have achieved remarkable agreement with experimental observations, correctly predicting complex phase evolution and interference patterns in atomic systems[6]. Importantly, RMT provides direct access to the dipole, which will become increasingly important as complex interplay between multiple processes modifies absorption line shapes, making them difficult to fit, and confounding dipole reconstruction.

### Current and future challenges

The greatest challenge lies in accurately modeling many-body dynamics during strong-field processes. Computational demands limit RMT calculations to molecules in relatively weak IR fields[7] or atoms. In strong, long wavelength fields, tens or hundreds of photons may be absorbed, necessitating prohibitively large angular momentum expansions. While this is currently possible for atoms, approximations will be required for even small molecules. Extending to larger molecules will require alternative approaches to mitigate the exponential scaling of the many-electron problem.

The connection between microscopic dynamics and macroscopic observables requires deeper understanding. Focal volume averaging, carrier-envelope phase effects, and pulse propagation all influence measured spectra, yet are often treated phenomenologically. While ordinarily[8], ATAS is less sensitive to macroscopic effects than high-harmonic spectroscopy [9], the problem remains of reproducing pulse profiles used in experiment. While RMT has been used to calibrate experimental pulses, calculations are relatively expensive, especially as we move towards multidimensional spectroscopy or molecular targets where we may no longer exploit the symmetry of atomic systems in linearly polarized light.

As we push toward more intense fields, threshold dynamics, including double ionization, increase in importance, but are computationally expensive to include. Currently, RMT calculations neglect these processes, leading to discrepancies with experiment at higher intensities[10]. Double ionization capability is being developed for RMT but maintaining computational tractability for systems larger than helium is a significant challenge.

Phase retrieval represents another frontier challenge. While fitting procedures can extract phase information from line shapes, direct experimental measurement of quantum phases remains elusive. Developing new approaches providing direct phase access would eliminate model dependencies and enable more rigorous tests of theoretical predictions.

**Advances in science and technology to meet challenges**

While new attosecond sources, advanced phase-retrieval algorithms and better pulse metrology will enhance ATAS experiments, computational advances are equally crucial. Hybrid methods combining RMT with reduced-dimensionality approaches, or development of effective potential methods that capture correlation effects with reduced computational cost could extend calculations to larger systems. Graphical Processing Unit acceleration and quantum computing algorithms may eventually enable full many-body calculations for molecular systems, but the software development required to leverage these new technologies is non-trivial.

Novel experimental geometries offer new capabilities. Two-dimensional spectroscopy techniques in the attosecond domain[11] and coincidence measurements combining ATAS with photoelectron or ion detection could provide complementary information about the same physical processes.

Machine learning approaches are transforming both experimental design and theoretical modeling. Neural networks can optimize experimental parameters, while physics-informed machine learning may accelerate solution of the TDSE. These tools could eventually enable real-time feedback control of quantum dynamics, but may also be used to drive large-scale, *ab initio* simulations.

**Concluding remarks**

The field stands at an exciting crossroads where experimental capabilities are approaching fundamental limits while theoretical methods are becoming increasingly sophisticated. The excellent agreement between RMT calculations and experimental observations validates our understanding of the underlying physics and provides confidence in predictions for unexplored regimes.

Future developments will focus on extending capabilities to more complex systems, particularly molecules where vibronic coupling creates richer dynamics. The prospect of controlling entanglement between photoelectrons and residual ions[12] represents a particularly exciting frontier, with potential applications in quantum information processing.

Integrating attoscience with emerging technologies promises transformative applications. Real-time monitoring and control of chemical reactions[13], development of ultrafast electronic devices[14], and creation of novel quantum states[15] all depend on our ability to understand and manipulate attoscale electronic motion.

As computational power grows and experimental techniques become more sophisticated, ATAS will remain at the forefront of ultrafast science and RMT will occupy an important niche in supporting and guiding experimental efforts. The technique's combination of experimental accessibility and theoretical tractability makes it an ideal platform for exploring the quantum mechanical foundations of light-matter interaction and pushing the boundaries of our understanding of electronic dynamics. Success will require continued close collaboration between experimentalists and theorists, leveraging the complementary strengths of advanced measurement techniques and sophisticated computational methods.

## Acknowledgements

ACB and HWvdH acknowledge funding from the EPSRC under Grants No. EP/T019530/1, EP/V05208X/1, and EP/R029342/1. LRH acknowledges EPSRC studentship number 2442954. SM is funded by the Department for the Economy, Northern Ireland.

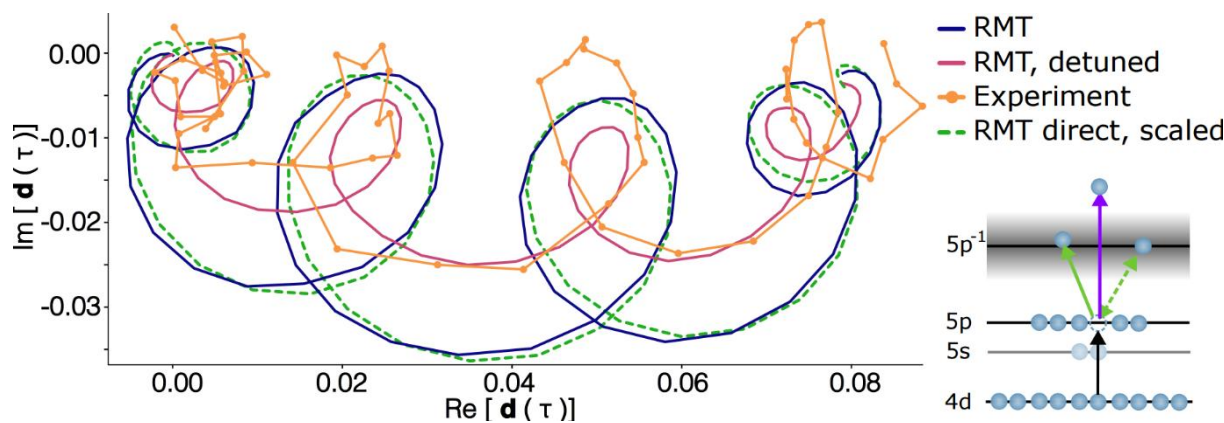

*Figure 1.* ATAS with strong-field-ionized xenon allows reconstruction of both the amplitude and phase of the dipole– here shown together in the complex plane– as extracted from the absorption spectrum for experiment (orange) and two separate RMT calculations: one with the central energy of the broadband XUV pulse tuned to the $4d \rightarrow 5p$ resonance (blue), and one with central energy detuned by 3 eV (pink). Also included is the dipole extracted directly from RMT calculations (green). Inset: The complex behaviour is caused by interference between direct, strong-field ionization (purple arrow) and excitation into several Rydberg states (green arrows). Adapted with permission from Ref. [6], published under CC BY 4.0 by the American Physical Society.

## References

[1] Wirth, A., Hassan, M. T., Grguraš, I., Gagnon, J., Moulet, A., Luu, T. T., Pabst, S., Santra, R., Alahmed, Z. A., Azzeer, A. M., Yakovlev, V. S., Pervak, V., Krausz, F., & Goulielmakis, E. 2011. Synthesized Light Transients. Science **334**, 195

[2] Gaarde, M. B., Buth, C., Tate, J. L., & Schafer, K. J. 2011. Transient absorption and reshaping of ultrafast XUV light by laser-dressed helium. Phys. Rev. A **83**, 013419

[3] Loh, Z.-H., Khalil, M., Correa, R. E., Santra, R., Buth, C., & Leone, S. R. 2007. Quantum State-Resolved Probing of Strong-Field-Ionized Xenon Atoms Using Femtosecond High-Order Harmonic Transient Absorption Spectroscopy. Phys. Rev. Lett. **98**, 143601

[4] Ott, C., Kaldun, A., Raith, P., Meyer, K., Laux, M., Evers, J., Keitel, C. H., Greene, C. H., & Pfeifer, T. 2013. Lorentz Meets Fano in Spectral Line Shapes: A Universal Phase and Its Laser Control. Science **340**, 716

[5] Brown, A. C., Armstrong, G. S. J., Benda, J., Clarke, D. D. A., Wragg, J., Hamilton, K. R., Mašín, Z., Gorfinkiel, J. D., & van der Hart, H. W. 2020. RMT: R-matrix with time-dependence. Solving the semi-relativistic, time-dependent Schrödinger equation for general, multielectron atoms and molecules in intense, ultrashort, arbitrarily polarized laser pulses. Comput. Phys. Commun. **250**, 107062

[6] Hutcheson, L., Hartmann, M., Borisova, G. D., Birk, P., Hu, S., Ott, C., Pfeifer, T., van der Hart, H. W., & Brown, A. C. 2025. Phase evolution of strong-field-ionization. Phys. Rev. Res. 7, L022074

[7] Benda, J., Gorfinkiel, J. D., Mašín, Z., Armstrong, G. S. J., Brown, A. C., Clarke, D. D. A., van der Hart, H. W., & Wragg, J. 2020. Perturbative and nonperturbative photoionization of H₂ and H₂O using the molecular R-matrix-with-time method. Phys. Rev. A **102**, 052826

[8] *Some studies exploit the macroscopic effects to manipulate the absorption: see e.g.* Phys. Rev. Res. 6 013103 (2024)

[9] Hutcheson, L., van der Hart, H. W. and Brown, A. C. 2023. Modelling intensity volume averaging in ab initio calculations of high harmonic generation J. Phys. B: At. Mol. Opt. Phys. **56**, 135402

[10] Hartmann, M., Hutcheson, L., Borisova, G. D., Birk, P., Hu, S., Brown, A. C., van der Hart, H. W., Ott, C., & Pfeifer, T. 2022. Core-resonance line-shape analysis of atoms undergoing strong-field ionization J. Phys. B: At. Mol. Opt. Phys. **55**, 245601

[11] Marroux, H. J. B., Fidler, A. P., Neumark, D. M., & Leone, S. R. 2018. Multidimensional spectroscopy with attosecond extreme ultraviolet and shaped near-infrared pulses Sci. Adv. 4 eaau3783

[12] Ishikawa, K., Prince, K. C. & Ueda, K. 2023. Control of Ion-Photoelectron Entanglement and Coherence Via Rabi Oscillations. J. Phys. Chem. A 127 10638

[13] Vismarra, F. *et al* 2024. Few-femtosecond electron transfer dynamics in photoionized donor–π–acceptor molecules. Nat. Chem. 16 2017

[14] de la Torre, A. *et al* 2021. Colloquium: Nonthermal pathways to ultrafast control in quantum materials. Rev. Mod. Phys. 93 041002

[15] Park, H., Park, N. & Lee, J. 2024. Novel Quantum States of Exciton–Floquet Composites: Electron–Hole Entanglement and Information. Nano. Lett. 24 13192

## 25. *Atto-photochemistry: a theoretical perspective*

**Morgane Vacher[1]**

[1] Nantes Université, CNRS, CEISAM, UMR 6230, F-44000 Nantes, France

morgane.vacher@univ-nantes.fr

**Status**

Photo-induced processes is one of the fundamental and widespread processes in chemistry and physics. As a result of light absorption and excitation, the distribution of electrons and thus the reactivity of a molecule can differ significantly from the ones in the ground state. Thanks to this conceptually simple process, photochemistry has broadened the spectrum of possible reactions, compared to thermal chemistry [1].

In 2001, the first attosecond (1 as = $10^{-18}$ s) domain pulses were generated [2]. Because of the time-energy uncertainty principle, pulses of extremely short duration have a large spectral bandwidth, potentially larger than the energy difference between electronic excited states. Attosecond pulses can coherently populate multiple excited electronic states, forming an electronic wavepacket. Because of interference phenomena between the populated electronic states, the electronic distribution of the wavepacket is not the simple average of the electronic distributions of the individual states: a coherent electronic wavepacket has a new electronic distribution and thus can be considered as a new type of initial electronic state, with a potentially new reactivity. This concept is called "charge-directed reactivity" [3] or "atto-photochemistry" and corresponds to photochemistry induced by electronic wavepackets excited by attosecond domain pulses. It has been successfully explored experimentally and theoretically in diatomic molecules [4,5]. Extending it to real chemical reactions in polyatomic molecules is a current ambitious goal which might revolutionize photochemistry, in the same way the latter has revolutionized chemistry [6].

**Current and future challenges**

In order to control the outcome of photochemical reactions in polyatomic molecules using electronic wavepackets, several challenges need to be faced. First of all, polyatomic molecules contain numerous nuclear coordinates known to lead to electronic decoherence [7,8], while electronic coherence is a key property for attosecond control. A primary challenge is whether electronic coherence, or its "legacy," survives long enough to affect a photochemical reaction. An encouraging result regarding this is the recent prediction of induced vibrational coherences lasting for at least 50 fs, despite electronic decoherence in less than one femtosecond upon ionization of ethylene, a prototype organic chromophore (Figure 1) [9].

Polyatomic molecules possess a large number of electronically excited states of varying character. Irradiating such systems with an attosecond pulse that populates many of these states in an arbitrary manner is unlikely to have a significant impact on the photochemical process of interest. Effective control instead requires the preparation of a well-defined, carefully tailored electronic wavepacket, typically composed of only a small number of selected electronic eigenstates. When too many states of diverse nature are coherently populated, their interference generally produces a hole density that is delocalized across the molecule and exhibits only weak temporal modulation. Under these conditions, the capacity to steer photochemical dynamics via electronic coherence is limited. The second challenge therefore lies in identifying which electronic wavepacket - namely, which combination of complex amplitudes, including both weights and relative phases in the superposition - can either trigger a chemical reaction that would otherwise be inaccessible or enhance the yield of a photochemical process that is intrinsically inefficient.

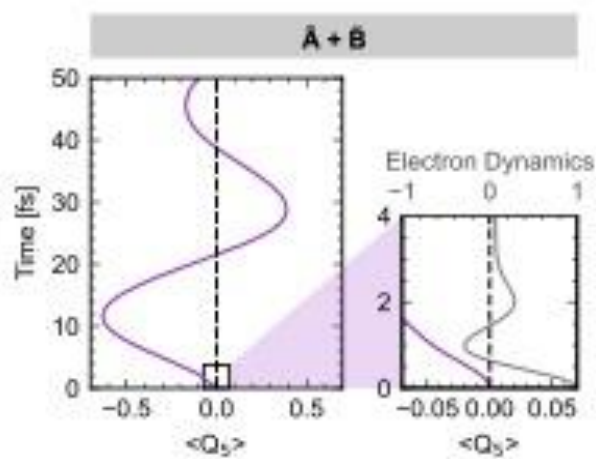

*Figure 1. Time evolution of nuclear position expectation value following excitation of equally weighted in-phase coherent superposition of A and B states along vibrational normal mode 5. The inset shows magnification of the early-time position expectation value (purple), together with the electron dynamics (grey). Adapted with permission from J. Phys. Chem. Lett. 2025, 16, 34, 8745-8751. Copyright 2025 ACS.*

**Advances in science and technology to meet challenges**

The theoretical study of atto-photochemical reactions requires simulations of coupled electron–nuclear dynamics [10–12]. Although the most accurate quantum dynamics methods are, in principle, preferable, they are often constrained by the exponential scaling of their computational cost with the number of nuclear degrees of freedom, which typically limits applications to reduced-dimensionality models [13]. However, the selection of a limited set of nuclear coordinates can bias the resulting dynamics. In addition, these approaches usually rely on model potential energy surfaces constructed within the harmonic approximation. While such potentials are well suited for describing photophysical processes, they become less adequate for chemical reactions involving large-amplitude motions, such as dissociation or cis–trans isomerization. "Direct" or "on-the-fly" dynamical methods overcome these limitations by propagating nuclear trajectories that explicitly include all nuclear degrees of freedom on the true potential energy surfaces. The most widely used variants are the semiclassical Ehrenfest [14] and surface-hopping [15] methods. A key limitation of these approaches, however, is their approximate treatment of electronic coherence, which is a critical aspect of atto-photochemistry [16]. A promising compromise is provided by the DD-vMCG method [17,18], which is based on the propagation of variationally coupled Gaussian wavepackets along quantum trajectories. In this framework, the time-dependent Schrödinger equation governs not only the evolution of the wavepacket expansion coefficients but also the mean positions and momenta of each Gaussian basis function.

**Concluding remarks**

The new paradigm of atto-photochemistry is to act directly on electrons and to use electronic coherences to create a new electronic density directing the nuclear motion in the desired direction. To achieve this both fundamental and potentially ground-breaking idea, the community needs to overcome several challenges.

## Acknowledgements

M.V. acknowledges the European Union for the ERC Starting Grant No. 101040356 - ATTOP. Views and opinions expressed are those of the author only and do not necessarily reflect those of the European Union or the European Research Council Executive Agency. Neither the European Union nor the granting authority can be held responsible for them.

## References

[1]  Hoffman N, 2008 Photochemical reactions as key steps in organic synthesis Chem. Rev. 108 1052-1103
[2]  Hentschel M, Kienberger R, Spielmann C, Reider GA, Milosevic N, Brabec T, Corkum P, Heinzmann U, Drescher M, Krausz R, 2001 Attosecond Metrology Nature 414 509
[3]  Weinkauf R, Schanen P, Metsala A, Schlag EW, Burgle M, Kessler H, 1996 Highly efficient charge transfer in peptide cations in the gas phase: threshold effects and mechanism. J. Phys. Chem. 100, 18567-18585.
[4]  Roudnev V, Esry EB, Ben-Itzhak, 2004 Controling HD+ and H2+ dissociation with the carrier-envelope phase difference of an intense ultrashort laser pulse. Phys. Rev. Lett. 93, 163601
[5]  Kling MF, von den Hoff P, Znakovskaya I, de Vivie-Riedle R, 2013 Sub-femtosecond control of molecular reactions via tailoring the electric field of light Phys. Chem. Chem. Phys. 15 9448-9467
[6]  Merritt ICD, Jacquemin D, Vacher M, 2021 Attochemistry : is controlling electrons the future of photochemistry? J. Phys. Chem. Lett. 12 8404-8415
[7]  Vacher M, Steinberg L, Jenkins AJ, Bearpark MJ, Robb MA, 2015 Electron dynamics following photoionization: decoherence due to the nuclear wave packet width Phys. Rev. A 92 040502
[8]  Vacher M, Bearpark MJ, Robb MA, Malhado JP, 2017 Electron dynamics upon ionization of polyatomic moelcules: coupling to quantum nuclear motion and decoherence Phys. Rev. Lett. 118 083001
[9]  Fransén L, Gomez S, Vacher M, 2025 Attochemical control of nuclear motion despite fast electronic decoherence J. Phys. Chem. Lett. 16 34 8745-8751
[10] Vacher M, Bearpark MJ, Robb MA, 2016 Direct methods for non adiabatic dynamics: connecting the single-set variational multi-configurational Gaussian (vMCG) and Ehrenfest perspectives Theo. Chem. Acc. 135 1-11
[11] Crespo-Otero R, Barbatti M, 2018 Recent advances and perspectives on non-adibatic mixed quantum-classical dynamics. Chem. Rev. 118, 7026-7068.
[12] Curchod B, Martinez TJ, 2018 Ab initio nonadiabatic quantum molecular dynamics. Chem. Rev. 118 3305-3336.
[13] Meyer H-D, Manthe U, Cederbaum L. 1990 The multi-configurational time-dependent Hartree approach. Chem. Phys. Lett. 165, 73 – 78
[14] Ehrenfest P, 1927 Bemerkung über die angenäherte Gültigkeit der klassischen Mechanik innerhalb der Quantenmechanik Z. Phys. 45 455-457
[15] Tully JC, 1990 Molecular dynamics with electronic transitions J. Chem. Phys. 93 1061-1071
[16] Tran T, Ferté A, Vacher M, 2024 Simulating attochemistry : which dynamics method to use? J. Phys. Chem. Lett. 15 13 3646-3652
[17] Burghardt I, Meyer HD, Cederbaum L, 1999 Approaches to the approximate treatment of complex molecular systems by the multiconfiguration time-dependent Hartree method. J. Chem. Phys. 111, 2927.
[18] Worth GA, Robb MA, Lasorne B, 2008 Solving the time-dependent Schrödinger equation for nuclear motion in one step: direct dynamics of non-adiabatic systems. Mol. Phys. 106, 2077-2091.

## 26. Attosecond electron delays in photoionized molecules

**Sreelakshmi Palakkal and Franck Lépine\***

Université Claude Bernard Lyon 1, CNRS, Institut Lumière Matière, UMR5306, F-69100, Villeurbanne, France

franck.lepine@univ-lyon1.fr

**Status**

One of the most fascinating aspects of attosecond technology is its ability to shed new light on traditional concepts of quantum mechanics. In this context, photoionization certainly plays a special role due to the historical paradigm shift created by Einstein's photoelectric law. In attosecond science, energetic photons contained in attosecond XUV pulses typically induce the ionization of matter. Since ionization is a quantum scattering process, it produces an outgoing electronic wave-packet that carries information about its interaction with its surrounding in its phase. This scattering phase can be accessed experimentally using a photoelectron interferometric technique and attosecond technology[1]. While the relevant microscopic information is the quantum scattering phase, the link with the semiclassical scattering time delay in photoemission is often discussed in terms of the Wigner-Smith-Eisenbud theory[2]. The transition from energetic considerations of photoelectric effects to measuring how electron ejection time or phase varies with different initial electron angular momentum was successfully demonstrated in atoms in the pioneering work of Schultze et al.[3] and Klunder et al.[4]. Similar investigations were soon pursued in the case of molecules, for which the complexity of electronic structure combined with the additional role of nuclear degrees of freedom, posed additional challenges. As with atoms, properties such as electron energy, angular momentum, correlation and resonances are expected to affect the quantum phase. But the question of how sensitive the phase is and whether it is experimentally accessible with nowadays technics has attracted the interest of many researchers. The first demonstration that the scattering phase is indeed sensitive to the presence of a resonance in a molecule was presented in Ref.[5] for the $N_2$ molecule. The concept was further developed in Ref.[6], where the effect of a shape resonance was measured in $N_2O$, showing a delay of 160 as, associated with the electron's trapping time. In Ref.[7], a shape resonance was investigated in $N_2$, showing that very small variations in nuclear bond lengths can correspond to a large change in time delay of about 200 attoseconds, which calls into question the definition of the Frank-Condon principle. Since electron scattering in molecules is a three-dimensional problem influenced by molecular structure, molecular frame measurement is important approach that have been applied to highlight the role of the three-dimensional nature of molecular potentials. In Ref.[8], space- and phase-resolved photoionization dynamics were measured in the molecular frame near a shape resonance of $CF_4$. In Ref.[9], the orientational dependence of the delay provided direct information about the initial localization of the electron wave packet within the molecular potential of the CO molecule.

**Current and future challenges**

Following these first demonstrations of measuring electron quantum scattering phase and associated ionization delays in molecules, new research avenues have emerged, making more advanced experiments accessible. In this regard, on example among others concerned the role of symmetry breaking in delay measurements that was studied in Ref. [10], which showed that delay differences arise from the Renner-Teller effect-induced shape resonance in $CO_2$. One of the most fascinating recent developments in the study of attosecond delays is related to the emergence of attosecond free electron lasers, which made it possible to study core-level photoionization with attosecond precision. In Ref [11], the authors used X-ray photons to perform phase measurements in core-level photoemission in NO molecules. This illuminated the effects of shape resonances and multi-electron scattering processes. Alongside the development of more advanced experimental techniques, the exploration of more complex molecular species, including large molecules or molecular clusters, has also begun. The study of large, several tens of atoms, molecules was successfully addressed in Ref. [12], in which the effects of molecular size and symmetry were explored in a series of aromatic molecules. The study demonstrated that attosecond delays do not simply scale with molecular size, it can vary with the symmetry, as in the case of 2D systems. In this case, the measured delay is directly connected to the dimension of the delocalized hole created upon ionization. Delays in the 3D fullerene carbon cage were also investigated in Refs. [13] and [14], which demonstrated for the first time the influence of plasmon resonances on the delay and the role of electron correlation that shapes the electronic response of the molecule in the XUV domain. The increase in complexity was also studied through the influence of chemical groups in Ref. [15], in which ethyl iodide was studied using a giant dipole resonance in the I atoms as a reference. A change in delay was observed due to the presence of the functional ethyl group. Opening a route to the study of the role of chemical environment. Molecular clusters, which are an intermediate between the isolated molecule and its liquid phase, have been studied. In reference [16], the case of water clusters showed that the delay increased with the number of water molecules in the cluster, reaching saturation due to structural disorder in larger clusters. In Ref. [17], extending attosecond spectroscopy to the liquid phase showed an increase of a few tens of attoseconds from liquid to gaseous water due to the effect of the local environment, opening the domain to the condensed molecular phase.

**Advances in science and technology to meet challenges**

Although striking results have been obtained with increasingly complex polyatomic molecules, the large number of degrees of freedom and photoreaction pathways encountered in molecules that interact with attosecond pulses, poses a significant challenge to attosecond physics. In the context of attosecond ionization delay measurements, improving spectroscopic techniques is essential to increasing the flexibility of experiments. Coincidence approaches, such as COLTRIMS, to decipher pathways and connect electron and ion properties are very important and already demonstrate promising performances. Precise measurements of attosecond delays also challenged the capabilities of our light sources. It is important to note that delay measurements do not require attosecond pulses but rather attosecond precision of the interferometric technique. Therefore, improving light sources in terms of the accessible spectral range from the terahertz (THz) to ultraviolet (UV) and X-ray domains opens new areas. Obviously, advances in attosecond light sources in the X-ray domain, using table-top or free-electron lasers, offer striking opportunities in terms of site selectivity. Improved performances in terms of reliability, tunability, stability, and high repetition and flux rates remain a crucial aspect. However, having adequate light sources is only one side of the problem. Making the method more general and relevant to other scientific communities will also relies on our ability to address a broader range of molecular systems. For this, more diverse sample delivery technologies are required that have to be combined and adapted to our attosecond technologies. A recent example is the implementation of molecular ion electrospray sources that have been coupled to attosecond technology[19] opening the possibility to investigate systems as complex as an entire protein or DNA strand. Currently, the accurate quantum mechanical description of a delay measurement is still challenging, even for small molecules. Needless to say, investigating more complex objects will require innovative theoretical approaches which should stimulate on active dialog between theory and experiment.

**Concluding remarks**

Attosecond electron interferometry provides accurate information on the quantum phase of electrons liberated in the ionization process. Having detailed, accurate information about many-electron atoms is challenging enough, but targeting molecular species is even more intriguing because of the role played by additional degrees of freedom. Indeed, the variety of quantum phenomena associated to symmetry, correlation, electro-nuclear couplings and the complexity of the potential energy landscape experienced by electrons in molecules make such investigations very demanding. However, recent improvements to these techniques are making this possible. These developments raise the question of the link between quantum and classical behavior in a system of increasing number of degrees of freedom and how quantum properties can be experimentally captured. Ultimately, this would allow accessing a complete reconstruction of the quantum states of the electrons and of their dynamics. At the same time, simple considerations based on general rules of quantum scattering show that this technique could be used more broadly to measure hole dimensions, electrostatic landscape, and electron dynamics both in isolated molecules and molecules in condensed phase.

**Acknowledgements**

We acknowledge financial support from CNRS and ANR-DFG FAUST (ANR-21-CE30-0052)

**References**

[1] Paul P M, Toma E S, Breger P, Mullot G, Augé F, Balcou Ph, Muller H G, and Agostini P 2001 Observation of a train of attosecond pulses from high harmonic generation. Science 292 1689-1692 (doi:10.1126/science.1059413)

[2] Maquet A, Caillat J and Taieb R 2014 Attosecond delays in photoionization: time and quantum mechanics. Journal of Physics B: Atomic, Molecular and Optical Physics 47 204004 (doi :10.1088/0953-4075/47/20/204004)

[3] Schultze M, Fies M, Karpowicz N, Gagnon J, Korbman M, Hofstetter M, Neppl S, Cavalieri A L, Komninos Y, Mercouris TH, Nicolaides C A, Pazourek R, Nagele S, Feist J, Burgdörfer J, Azzeer A M, Ernstorfer R, Kienberger R, Kleineberg U, Goulielmakis E, Krausz F, and Yakovlev V.S 2010 Delay in photoemission. Science 328, 5986 (doi: 10.1126/science.1189401)

[4] Klünder K, Dahlström J M, Gisselbrecht M, Fordell T, Swoboda M, Guénot D, Johnsson P, Caillat J, Mauritsson J, Maquet A, Taïeb R, and L'Huillier A 2011 Probing single-photon ionization on the attosecond time scale. Phys. Rev. Lett. 106, 169904 (doi: 10.1103/PhysRevLett.106.143002)

[5] Haessler S, Fabre B, Higuet J, Caillat J, Ruchon T, Breger P, Carré B, Constant E, Maquet A, Mével E, Salières P, Taïeb R, and Mairesse Y 2009 Phase-resolved attosecond near-threshold photoionization of molecular nitrogen. Phys. Rev. A 80, 011404 (doi: 10.1103/PhysRevA.80.011404)

[6] Huppert M, Jordan I, Baykusheva D, Conta A and Wörner H. J 2016 Attosecond delays in molecular photoionization. Phys. Rev. Lett. 117, 093001 (doi: 10.1103/PhysRevLett.117.093001)

[7]    Nandi S, Plésiat E, Zhong S, Palacios A, Busto D, Isinger M, Neoričić L, Arnold C L , Squibb R J , Feifel R, Decleva P, L'Huillier A, Martín F and Gisselbrecht M 2020 Attosecond timing of electron emission from a molecular shape resonance. Sci.Adv. 6,eaba7762 (doi:10.1126/sciadv.aba7762)

[8]    Heck S, Baykusheva D, Han M, Ji J B, Perry C, Gong X, Wörner H J 2021 Attosecond interferometry of shape resonances in the recoil frame of CF4. Sci. Adv. 7, eabj8121 (doi: 10.1126/sciadv.abj8121)

[9]    Vos J, Cattaneo L, Patchkovskii S, Zimmermann T, Cirelli C, Lucchini M, Kheifets A, Landsman A S, Keller U 2018 Orientation-dependent stereo Wigner time delay and electron localization in a small molecule.  Science 360, 1326–1330 (doi: 10.1126/science.aao4731)

[10]    Li M, Zhao L, Wang H, Li J, Wang W, Cai J, Hong X, Shi X, Zhang M, Zhao X, Weissenbilder R, Busto D, Gisselbrecht M, Ueda K, Luo S, Li Z, Ding D 2025 Attosecond spectroscopy reveals spontaneous symmetry breaking in molecular photoionization Sci. Adv, 11, 38 (doi: 10.1126/sciadv.adw5415)

[11]    Driver T, Mountney M, Wang J, Ortmann L, Al-Haddad A, Berrah N, Bostedt C, Champenois E G, DiMauro L F, Duris J, Garratt D, Glownia J M, Guo Z, Haxton D, Isele E, Ivanov I, Ji J, Kamalov A, Li S, Lin M F, Marangos J P, Obaid R, O'Neal J T, Rosenberger P, Shivaram N H, Wang 4 A L, Walter P, Wolf T A J, Wörner H J, Zhang Z, Bucksbaum P H, Kling M F, Landsman A S, Lucchese R, Emmanouilidou A, Marinelli A, Cryan J P 2024 Attosecond delays in X-ray molecular ionization.  Nature 632, 762–767 (doi: 10.1038/s41586-024-07771-9)

[12]    Loriot V, Boyer A, Nandi S, González-Collado C M, Plésiat E, Marciniak A, Garcia C L, Hu Y, Lara-Astiaso M, Palacios A, Decleva P, Martín F and Lépine F 2024 Attosecond metrology of the two-dimensional charge distribution in molecule. Nature Physics 20 765 (doi: 10.1038/s41567-024-02406-2)

[13]    Barillot T, Cauchy C, Hervieux P A, Gisselbrecht M, Canton S E, Johnsson P, Laksman J, Mansson E P, DahlströmJ M, Magrakvelidze M, Dixit G, Madjet M E,  Chakraborty H S, Suraud E, Dinh P M, Wopperer P, Hansen K, Loriot V, Bordas C, Sorensen S and Lépine F 2015 Angular asymmetry and attosecond time delay from the giant plasmon resonance in $C_{60}$ photoionization. *Phys. Rev. A* **91**, 033413 (doi : 10.1103/PhysRevA.91.033413)

[14]    Biswas S, Trabattoni A, Rupp P, Magrakvelidze M, Madjet M E A, De Giovannini U, Castrovilli M C, Galli M, Liu Q, Månsson E P, Schötz J, Wanie V, Wnuk P, Colaizzi L, Mocci D, Reduzzi M, Lucchini M, Nisoli M, Rubio A, Chakraborty H S, Kling M S, Calegari F 2025 Correlation-driven attosecond photoemission delay in the plasmonic excitation of C60 fullerene science advances 11, 7(doi: 10.1126/sciadv.ads0494)

[15]    Biswas S, Förg B, Ortmann L, Schötz J, Schweinberger W, Zimmermann T, Pi L, Baykusheva D, Masood H A, Liontos I, Kamal A M, Kling N G, Alharbi A F, Alharbi, M, Azzeer A M, Hartmann G, Wörner H J, Landsman A S and Kling M F 2020 Probing molecular environment through photoemission delays. Nature Physics 16, 778 (doi : 10.1038/s41567-020-0887-8)

[16]    Gong X, Heck S, Jelovina D, Perry C, Zinchenko K, Lucchese R, Wörner H J 2022 Attosecond spectroscopy of size-resolved water clusters.  Nature 609, 507 doi : 10.1038/s41586-022-05039-8

[17]    Jordan I, Huppert M, Rattenbacher D, Peper M, Jelovina D, Perry C, von Conta A, Schild A and Wörner H J, 2020 Attosecond spectroscopy of liquid water Science 369, 974–979 (doi: 10.1126/science.abb0979)

[18]    Constant E, Nandi S, Picot C, Prost E, Palakkal S, Lépine F, Lorio V 2025 High order harmonic generation-based attosecond light sources and applications to quantum phenomena APL Photonics 10, 010907 (doi : 10.1063/5.0235171)

[19]    Hervé M, Boyer A, Brédy R, Allouche A R, Compagnon I and Lépine F 2022 On-the-fly investigation of XUV excited large molecular ions using a high harmonic generation light source Scientific Reports 12, 13191 (doi: 10.1038/s41598-022-17416-4)

## 27. Chemistry at the attosecond time scale

**Vincent Wanie[1]\*, Andrea Trabattoni[1,2,3]\* and Francesca Calegari[1,4,5]\***

[1] Center for Free-Electron Laser Science CFEL, Deutsches Elektronen-Synchrotron DESY, Hamburg, Germany
[2] Institute of Quantum Optics, Leibniz Universität Hannover, Hannover, Germany
[3] Cluster of Excellence PhoenixD (Photonics, Optics, and Engineering-Innovation Across Disciplines), Hannover, Germany
[4] Department of Physics, Universität Hamburg, Hamburg, Germany
[5] The Hamburg Centre for Ultrafast Imaging, Universität Hamburg, Hamburg, Germany

vincent.wanie@desy.de, andrea.trabattoni@desy.de, francesca.calegari@desy.de

**Status**

The exploration of matter on ultrafast timescales has transformed our understanding of microscopic phenomena. Advances in femtosecond (fs) and attosecond (as) light sources have enabled real-time observation and control of quantum processes. Femtosecond light pulses have been used to reveal rapid mechanisms such as charge transfer, energy flow, and molecular conformational changes, key to biological functions like photosynthesis, vision, and DNA repair. Nobel laureate Ahmed Zewail was awarded in 1999 for pioneering work in the field of femtochemistry [1]. The field has rapidly shifted from merely observing to actively controlling these processes, inspiring the development of artificial light-harvesting devices and molecular machines. However, some natural processes, like electron motion, occur on attosecond timescales, previously inaccessible until technological advances broke the so-called "femtosecond barrier." About twenty-five years ago, High-order Harmonic Generation (HHG) [2], an extremely nonlinear optical phenomenon, enabled the generation of attosecond pulses [3, 4]. Recognized with the 2023 Nobel Prize in Physics to Anne L'Huillier, Ferenc Krausz, and Pierre Agostini, HHG has opened new avenues for probing and controlling the earliest moments of the light-matter interaction.

Attosecond light sources provide exceptional temporal resolution and enable instantaneous, broadband excitation of molecular systems. These unique capabilities can be harnessed to activate a purely electronic mechanism known as charge migration [5]. Upon sudden excitation, a coherent superposition of electronic states can be prepared within a frozen nuclei configuration, often resulting in a non-static charge density distribution that periodically migrates along the molecular backbone. Initially predicted through theoretical studies [6], this phenomenon has subsequently been explored using various experimental techniques [7, 8]. Despite advancements in controlling and observing electronic processes like charge migration, there remains a lack of conclusive experimental evidence demonstrating that these electronic coherences can fundamentally influence the photochemistry of the molecule. Notably, it is predicted that nuclear motion may induce rapid decoherence [9], thereby limiting the ability to utilize this charge motion for selective bond formation or cleavage. However, recent theoretical studies suggest that, due to the quantum correlation between electrons and atomic nuclei, electronic coherences could influence the nuclear dynamics occurring at later time scales by affecting the total force field acting on the nuclei [10, 11]. Manipulating the force field through coherent electronic excitation could potentially enable unconventional reaction pathways, in line with the proposed mechanism of charge-directed reactivity. Demonstrating this ability would open new and significant opportunities for the emerging field of attosecond chemistry, which is still in its infancy.

## Current and future challenges

Despite the tremendous importance of studies on charge migration in a variety of isolated gas-phase molecules of increasing complexity [7, 8], effort must be made for investigating attosecond electron dynamics in more realistic chemical environments.

A crucial step is the investigation of solvation effects over the coherent electron dynamics. Recent experiments performed in liquid water - probing photoemission delays [12], electron wave packet propagation [13], and emission bands [14] - have established a pathway toward the systematic implementation of attosecond techniques in the liquid phase. Extending attosecond molecular studies to solvated systems will allow direct access to ultrafast electronic processes occurring in real chemical environments, where solvation, hydrogen-bond networks, and charge transport play key roles.

With the recent demonstration of time-resolved spectroscopy of electron dynamics in chiral molecules (15), a new challenge is opening for attosecond science. In a chiral backbone, coherent electronic excitation can lead to the creation of chiral currents that induce dynamical changes in the properties of the chiral system. Engineering such currents by chiral light excitation can be used to enantioselectively control chemical interaction and search for spin-selective interactions via the chirality-induced spin selectivity (CISS) effect, which is essential for spintronics applications. Another ambitious challenge lies in probing the fundamental mechanisms of chiral recognition with ultrafast temporal resolution. Achieving this goal requires moving beyond single-molecule targets and developing new protocols for investigating transient intermolecular chiral responses.

In recent years, attosecond studies have entered the realm of solid-state physics. Another crucial challenge is to merge condensed-phase and molecular studies and establish attosecond spectroscopy of molecular interfaces. Ultrafast charge transport through interfaces, for example, is a key mechanism underlying all molecular optoelectronic devices such as organic molecular films or monomolecular layers. Moreover, studying electron dynamics at molecular interfaces is essential for understanding adsorption and catalysis, as well as macromolecular biological systems and membranes. Although the slow ($>10^{-12}$ s) relaxation dynamics of these structures have already been extensively characterized, the ultrafast interrogation of interfacial phenomena is still in its infancy - despite remarkable examples of photo-control of molecular switches on the femtosecond timescale [16] - and clearly represents a future challenge for attosecond chemistry.

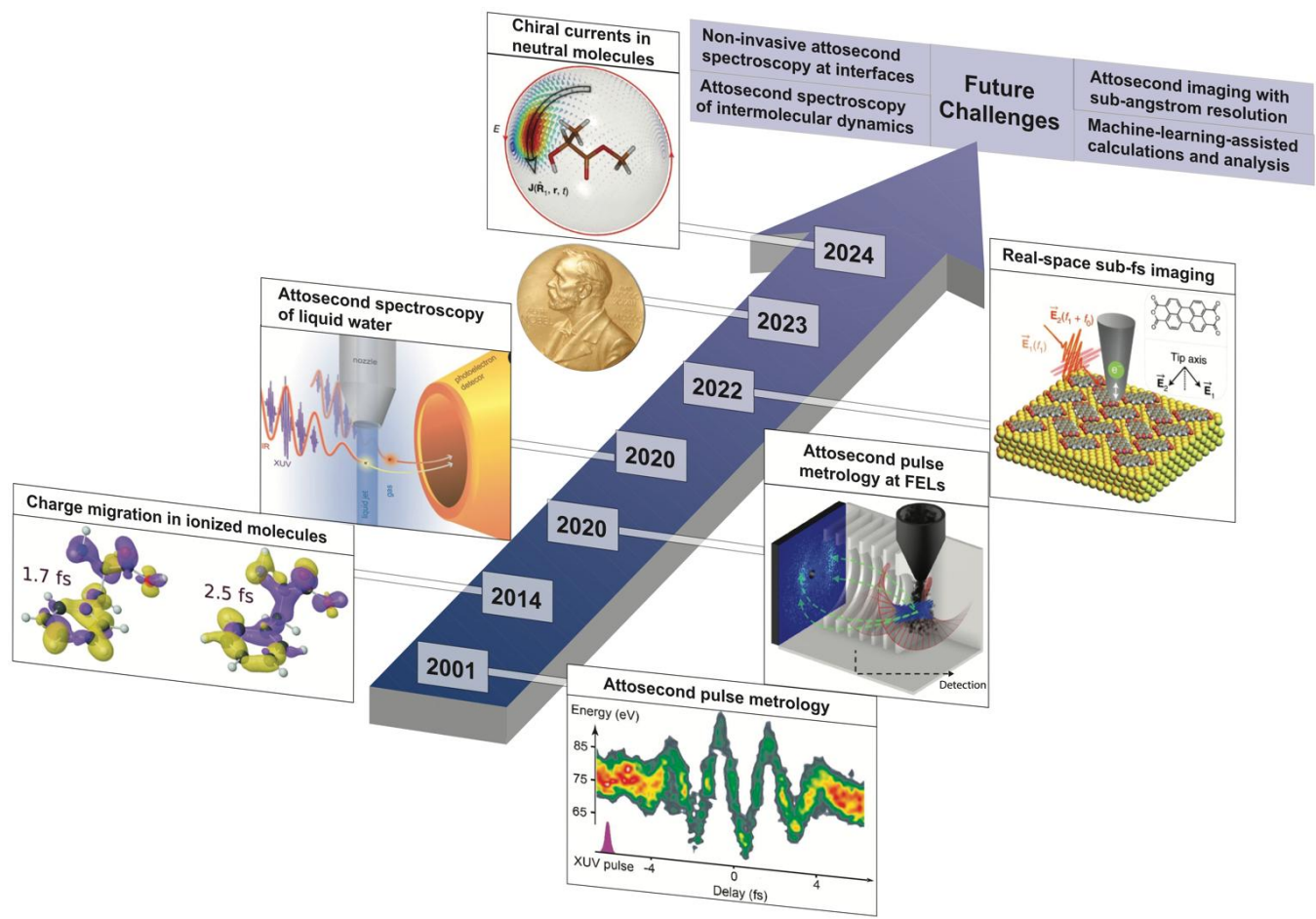

**Figure 1. Roadmap of attosecond chemistry**. *Important milestones along the path toward attosecond chemistry - including attosecond technology developments and advances in tracing charge dynamics in molecules - are reported together with a vision of the future challenges.*

## Advances in science and technology to meet challenges

Progress in attosecond chemistry will rely on a concerted evolution of light sources, methodologies and theoretical frameworks that bridge fundamental electron-driven dynamics and charge transfer processes with functional molecular environments, enabling technologically relevant scenarios. Achieving this requires extending attosecond concepts beyond isolated molecules towards the condensed phase and interfacial systems, where solvation and surface effects govern functionality. Emerging interdisciplinary studies on small aggregates of molecules, droplets, nanoparticles and molecular crystals already illustrate the trend to meet this major challenge and anticipate attosecond/electronic control for catalysis, atmospheric and bio-chemistry as well as water and material sciences. Understanding the role of electron dynamics in nuclear transition is another example within reach for advances in metrology thanks to the continuous advancements in light sources [17].

Equally crucial is the development of non-invasive spectroscopic approaches that access electron and nuclear motion under natural, weak-field conditions. Generating attosecond-to-few-femtosecond pulses in the ultraviolet and vacuum-ultraviolet range enables excitation below the ionization threshold, thus avoiding strong-field distortion in experiments while preserving high temporal resolution [15]. Such light sources open the path to pump-probe techniques capable of following charge migration and transfer phenomena with minimal perturbation. Among promising directions is a variant of laser-induced electron diffraction (LIED), where a weak-field excitation pulse is followed by gentle ionization and rescattering, retaining structural sensitivity without the disruptive fields used in conventional strong-field LIED [18].

Since very recently, a few X-ray Free-Electron laser (XFEL) facilities, including the Linac Coherent Light Source (LCLS) and the European XFEL, can deliver attosecond pulses, providing new perspectives for attosecond measurements [19, 20]. Notably, the high photon energy and the high photon flux of the attosecond XFELs enable nonlinear light matter interactions at specific atomic sites. Methods such as Impulsive X-ray stimulated Raman scattering (IXSRS) [13] and transient redistribution of ultrafast electronic coherences in the attosecond Raman signals (TRUECARS) [21] have been proposed to dynamically drive the coherent electronic excitation from a specific atomic site or to measure electronic coherences emerging across conical intersections with attosecond resolution. Early demonstrations using pairs of attosecond X-ray pulses to monitor core-ionized dynamics already indicate the feasibility of tracking electronic and structural evolution on their natural timescale (22). Combining attosecond temporal with Ångström spatial resolution represents another transformative goal. The advent of hard X-ray attosecond sources enables, in combination with ad-hoc designed X-ray lenses, diffraction experiments on molecular crystals with unprecedented spatio-temporal precision [23].

Finally, new approaches for theoretical modelling and data processing must evolve in parallel. The modelling and data analysis for complex, correlated molecular systems requires both enhanced computational capabilities and machine-learning-assisted approaches for efficient sampling and interpretation of high-dimensional data. Beyond coherence-based control, molecular manipulation via quantum entanglement has emerged as a novel paradigm, demanding new theoretical formalisms to describe and exploit such effects [24, 25]. In summary, close feedback between experiment, simulation, and data science will be vital to transform attosecond spectroscopy into a predictive and design-oriented tool for chemistry and materials science.

**Concluding remarks**

In the past two decades, attosecond technology has transformed our ability to observe, understand, and control the earliest moments of light-molecule interactions central to photochemistry. Building on the achievements of femtochemistry, which offered detailed insights into the molecular machinery, the field has advanced to the attosecond regime. Here, sudden, broadband excitation opens opportunities for a new coherent control, selectively activating vibrational modes via the creation of electronic coherences. Ultrafast light sources, spanning from ultraviolet to x-ray, now allow real-time study of charge migration and electronic coherences. These advances enrich our knowledge of molecular dynamics and enable new ways to manipulate chemical reactivity at its inception. Major challenges remain as the field moves from isolated molecules to complex environments, such as solvated, chiral, or condensed-phase systems. Meeting these challenges requires better light sources, precise synchronization of attosecond x-ray pulses from XFELs, and novel spectroscopic techniques for high spatio-temporal resolution with site-selectivity and minimal perturbation. Progress in theoretical models and data analysis, driven by high-performance computing and machine learning, is also essential. As attosecond sources become widely available, their use will expand across catalysis, spintronics, and materials science. Looking ahead, the coming years promise remarkable progress in the ambition of engineering chemical processes with attosecond temporal precision and atomic spatial resolution.

## Acknowledgements

V.W. acknowledges funding from the German Research Foundation (DFG) - project ID 545611997. A.T. acknowledges support from the Helmholtz Association under the Helmholtz Young Investigator Group VH-NG-1603, and financial support from the European Research Council under the ERC SoftMeter Grant No. 101076500. F.C. acknowledges financial support from the Cluster of Excellence "CUI: Advanced Imaging of Matter" of the Deutsche Forschungsgemeinschaft (DFG)-EXC 2056–Project No. 390715994. Views and opinions expressed are however those of the author(s) only and do not necessarily reflect those of the European Union or the European Research Council Executive Agency.

## References

[1] A. H. Zewail, Femtochemistry: Atomic-scale dynamics of the chemical bond using ultrafast lasers (Nobel lecture). Angewandte Chemie - International Edition 39, 2586–2631 (2000).

[2] P. B. Corkum, Plasma Perspective on Strong-Field Multiphoton Ionization. Phys Rev Lett 71, 1994–1997 (1993).

[3] P. M. Paul, E. S. Toma, P. Breger, G. Mullot, F. Augé, P. Balcou, H. G. Muller, P. Agostini, Observation of a train of attosecond pulses from high harmonic generation. Science (1979) 292, 1689–1692 (2001).

[4] M. Hentschel, R. Kienberger, C. Spielmann, G. a Reider, N. Milosevic, T. Brabec, P. Corkum, U. Heinzmann, M. Drescher, F. Krausz, Attosecond metrology. Nature 414, 509–513 (2001).

[5] M. Nisoli, P. Decleva, F. Calegari, A. Palacios, F. Martín, Attosecond Electron Dynamics in Molecules. Chem Rev 117, 10760–10825 (2017).

[6] L. S. Cederbaum, J. Zobeley, Ultrafast charge migration by electron correlation. Chem Phys Lett 307, 205–210 (1999).

[7] F. Calegari, D. Ayuso, A. Trabattoni, L. Belshaw, S. De Camillis, S. Anumula, F. Frassetto, L. Poletto, A. Palacios, P. Decleva, J. B. Greenwood, F. Martín, M. Nisoli, Ultrafast electron dynamics in phenylalanine initiated by attosecond pulses. Science (1979) 346, 336–339 (2014).

[8] E. P. Månsson, S. Latini, F. Covito, V. Wanie, M. Galli, E. Perfetto, G. Stefanucci, H. Hübener, U. De Giovannini, M. C. Castrovilli, A. Trabattoni, F. Frassetto, L. Poletto, J. B. Greenwood, F. Légaré, M. Nisoli, A. Rubio, F. Calegari, Real-time observation of a correlation-driven sub 3 fs charge migration in ionised adenine. Communications Chemistry 4 (2021).

[9] M. Vacher, M. J. Bearpark, M. A. Robb, J. P. Malhado, Electron Dynamics upon Ionization of Polyatomic Molecules: Coupling to Quantum Nuclear Motion and Decoherence. Physical Review Letters 118, 083001 (2017).

[10] M. Cardosa-Gutierrez, R. D. Levine, F. Remacle, Electronic coherences built by an attopulse control the forces on the nuclei. Journal of Physics B: Atomic, Molecular and Optical Physics 57 (2024).

[11] L. Fransén, S. Gómez, M. Vacher, Attochemical Control of Nuclear Motion despite Fast Electronic Decoherence. J Phys Chem Lett 16, 8745–8751 (2025).

[12] I. Jordan, M. Huppert, D. Rattenbacher, M. Peper, D. Jelovina, C. Perry, A. Von Conta, A. Schild, H. J. Wörner, Attosecond spectroscopy of liquid water. Science (1979) 369, 974–979 (2020).

[13] O. Alexander, F. Egun, L. Rego, A. Martinez Gutierrez, D. Garratt, G. A. Cárdenas, J. J. Nogueira, J. P. Lee, K. Zhao, R.-P. Wang, D. Ayuso, J. C. T. Barnard, S. Beauvarlet, P. H. Bucksbaum, D. Cesar, R. Coffee, J. Duris, L. J. Frasinski, N. Huse, K. M. Kowalczyk, K. A. Larsen, M. Matthews, S. Mukamel, J. T. O'neal, T. Penfold, E. Thierstein, J. W. G. Tisch, J. R. Turner, T. Driver, N. Berrah, M.-F. Lin, G. L. Dakovski, S. P. Moeller, J. P. Cryan, A. Marinelli,

A. Picón, J. P. Marangos, Attosecond impulsive stimulated X-ray Raman scattering in liquid water. Sci Adv 10, eadp0841 (2024).

[14] S. Li, L. Lu, S. Bhattacharyya, C. Pearce, K. Li, E. T. Nienhuis, G. Doumy, R. D. Schaller, S. Moeller, M. Lin, G. Dakovski, D. J. Hoffman, D. Garratt, K. A. Larsen, J. D. Koralek, C. Y. Hampton, D. Cesar, J. Duris, Z. Zhang, N. Sudar, J. P. Cryan, A. Marinelli, X. Li, L. Inhester, R. Santra, L. Young, Attosecond-pump attosecond-probe x-ray spectroscopy of liquid water. Science (1979) 383, 1118–1222 (2024).

[15] V. Wanie, E. Bloch, E. P. Mânsson, L. Colaizzi, S. Ryabchuk, K. Saraswathula, A. F. Ordonez, D. Ayuso, O. Smirnova, A. Trabattoni, V. Blanchet, N. Ben Amor, M. C. Heitz, Y. Mairesse, B. Pons, F. Calegari, Capturing electron-driven chiral dynamics in UV-excited molecules. Nature 630, 109–115 (2024).

[16] D. Peller, L. Z. Kastner, T. Buchner, C. Roelcke, F. Albrecht, N. Moll, R. Huber, J. Repp, Sub-cycle atomic-scale forces coherently control a single-molecule switch. Nature 585, 58–62 (2020).

[17] M. Seitz, F. Calegari, P. G. Thirolf, A. Trabattoni, Towards the time-resolved spectroscopy of photoinduced electron dynamics in nuclear transitions. Phys Rev A (Coll Park) 112, 040101 (2025).

[18] U. De Giovannini, J. Küpper, A. Trabattoni, New perspectives in time-resolved laser-induced electron diffraction. Journal of Physics B: Atomic, Molecular and Optical Physics 56 (2023).

[19] J. Yan, W. Qin, Y. Chen, W. Decking, P. Dijkstal, M. Guetg, I. Inoue, N. Kujala, S. Liu, T. Long, N. Mirian, G. Geloni, Terawatt-attosecond hard X-ray free-electron laser at high repetition rate. Nat Photonics 18, 1293–1298 (2024).

[20] Z. Guo, T. Driver, S. Beauvarlet, D. Cesar, J. Duris, P. L. Franz, O. Alexander, D. Bohler, C. Bostedt, V. Averbukh, X. Cheng, L. F. DiMauro, G. Doumy, R. Forbes, O. Gessner, J. M. Glownia, E. Isele, A. Kamalov, K. A. Larsen, S. Li, X. Li, M. F. Lin, G. A. McCracken, R. Obaid, J. T. O'Neal, R. R. Robles, D. Rolles, M. Ruberti, A. Rudenko, D. S. Slaughter, N. S. Sudar, E. Thierstein, D. Tuthill, K. Ueda, E. Wang, A. L. Wang, J. Wang, T. Weber, T. J. A. Wolf, L. Young, Z. Zhang, P. H. Bucksbaum, J. P. Marangos, M. F. Kling, Z. Huang, P. Walter, L. Inhester, N. Berrah, J. P. Cryan, A. Marinelli, Experimental demonstration of attosecond pump–probe spectroscopy with an X-ray free-electron laser. Nat Photonics 18, 691–697 (2024).

[21] M. Kowalewski, K. Bennett, K. E. Dorfman, S. Mukamel, Catching Conical Intersections in the Act: Monitoring Transient Electronic Coherences by Attosecond Stimulated X-Ray Raman Signals. Phys Rev Lett 115 (2015).

[22] T. Driver, M. Mountney, J. Wang, L. Ortmann, A. Al-Haddad, N. Berrah, C. Bostedt, E. G. Champenois, L. F. DiMauro, J. Duris, D. Garratt, J. M. Glownia, Z. Guo, D. Haxton, E. Isele, I. Ivanov, J. Ji, A. Kamalov, S. Li, M. F. Lin, J. P. Marangos, R. Obaid, J. T. O'Neal, P. Rosenberger, N. H. Shivaram, A. L. Wang, P. Walter, T. J. A. Wolf, H. J. Wörner, Z. Zhang, P. H. Bucksbaum, M. F. Kling, A. S. Landsman, R. R. Lucchese, A. Emmanouilidou, A. Marinelli, J. P. Cryan, Attosecond delays in X-ray molecular ionization. Nature 632, 762–767 (2024).

[23] H. N. Chapman, C. Li, S. Bajt, M. Butola, J. L. Dresselhaus, D. Egorov, H. Fleckenstein, N. Ivanov, A. Kiene, B. Klopprogge, V. Kremling, P. Middendorf, D. Oberthuer, M. Prasciolu, T. E. S. Scheer, J. Sprenger, J. C. Wong, O. Yefanov, M. Zakharova, W. Zhang, Convergent-beam attosecond x-ray crystallography. Structural Dynamics 12 (2025).

[24] M. J. J. Vrakking, Control of Attosecond Entanglement and Coherence. Phys Rev Lett 126, 113203 (2021).

[25] H. Laurell, S. Luo, R. Weissenbilder, M. Ammitzböll, S. Ahmed, H. Söderberg, C. L. M. Petersson, V. Poulain, C. Guo, C. Dittel, D. Finkelstein-Shapiro, R. J. Squibb, R. Feifel, M. Gisselbrecht, C. L. Arnold, A. Buchleitner, E. Lindroth, A. Frisk Kockum, A. L'Huillier, D. Busto, Measuring the quantum state of photoelectrons. Nat Photonics 19, 352–357 (2025).

## 28. Attosecond spectroscopy in liquids

**Gabriele Crippa[1,2], Hugo J. B. Marroux[2*]**

[1] Sorbonne Université, CNRS, Laboratoire de Chimie Physique-Matière et Rayonnement, LCPMR, F-75005 Paris, France
[2] Université Paris-Saclay, CEA, LIDYL, Gif-sur-Yvette 91191, France

hugo.marroux@cea.fr

**Status**

Electron dynamics in liquids, in particular in water, plays a paramount role in a large range of scientific disciplines. For instance, low-energy electrons in water (i.e., with kinetic energies (KE) below 20 eV) are key reactants in radiochemical processes that drive the formation of cytotoxic radicals [1].

However, as of now there is no general consensus on the values of cross sections and mean free paths of electron scattering in water, especially in the low energy region (<20 eV). In order to pinpoint the scattering dynamics, seminal studies recorded the angular dependence of photoemission in water at the oxygen K-edge [2] or from the valence shell [3], and attributed the observed divergence from the gas phase Photoelectron Angular Distribution (PAD) to elastic and inelastic events occurring on the photoelectron trajectory towards the detector. However, it has been recognized that due to the complexity of the process a large discrepancy exists with theoretical models [3,4]. Taking into account results of subsequent studies [5,6], the reported values of mean free paths below 100 eV of kinetic energies extend up to a few nanometres. Converted to **mean free times between collisions, these correspond to a scattering timescale from a few tens of femtoseconds down to the hundreds of attoseconds, which is accessible to attosecond spectroscopy.**

Toward this end, two approaches have been taken so far. In high harmonic spectroscopy (HHS), the cut-off extension has been shown to be independent from the driving wavelength due to decoherence induced by elastic scattering of the electron wavepacket (EWP) over the first solvation shell [7]. This study demonstrated that HHS is a viable tool for the study of electron scattering in the low kinetic energy region, which is particularly difficult to access with standard photoelectron spectroscopy [8].

On the other hand, attosecond photoemission spectroscopy grants access to the ionisation and transport dynamics in the medium. In a seminal study, Reconstruction of Attosecond Beating By interference of two-photon Transitions (RABBIT) was used to compare photoionization time delays from water molecules in the liquid and gas phase, highlighting the effect of the local solvation environment on ionization time delays [9,10].

**Current and future challenges**

Although attosecond spectroscopy holds great promise for uncovering remarkable new dynamics in liquids, it faces significant experimental and computational challenges that currently hinder its broader application.
The photoemission spectra of liquid species are particularly broad (typically 8-10 eV) and present multiple ionization channels with large contributions of both homogeneous and inhomogeneous broadening. Several procedures have been proposed to deal with spectral congestion [11,12], but their assumptions allow to access only partial information as state-specific time delay remains difficult to resolve.

Investigation of the low kinetic energy region is particularly interesting as the scattering electron interacts with various degrees of freedom of the encountered molecules (vibration, libration etc..). However, this region is dominated by background from secondary electrons inducing lineshape distortion and thus its study is experimentally challenging [8,13,14].

Another current limitation of attosecond spectroscopy is its poor sensitivity towards solvated species. This is due to their much weaker contribution compared to the solvent signal and spectral congestion. Lifting this limitation will greatly broaden the scope of attosecond spectroscopy in liquids and open the route towards controlling the scattering process by chemically tuning the escape depth of electrons [15], for example.

Finally, a lively debate in the attosecond community is the role and predominance of quantum effects in the process of photoionization and a few experimental and theoretical approaches have been developed for isolated systems [16,17]. In the liquid phase, this picture is complicated as the EWP is no longer isolated but is in interaction with a bath, and should therefore be treated as an open quantum system. Moreover, in liquids various regimes have to be taken into account depending on the De Broglie wavelength of the outgoing electron compared to the intermolecular distances, with different contributions from localized excitations and delocalized entangled states [3].

**Advances in science and technology to meet challenges**

More detailed and accurate theoretical models are needed to understand the impact of different scattering channels on the measured RABBIT traces. This will allow the separation of the scattering delays, encoding the EWP propagation in the liquid, from the initial photoionization delays which carry information on the local solvation environment.

The increased efficiency of high harmonic generation sources in the soft X-ray is nowadays giving access to atomic and molecular core-levels with table-top experiments. The photoemission lines of core-levels are significantly simplified compared to their valence counterparts. The shifted binding energy compared to the solvent band also allows for the spectral isolation of the solute features, simplifying data interpretation.

Finally, the landmark of electron scattering studies has been the recording of PAD in liquids, but current experimental approaches require scanning the incident light polarisation while detecting only a small portion of the photoemission sphere. These schemes are hardly compatible with time-resolved spectroscopy, which requires the scanning of an additional delay dimension. The anticipated development of velocity map imaging in liquid-phase photoemission would lift this limitation and permit the recording of the angularly resolved amplitude and phase of the EWP using RABBIT spectroscopy, potentially revealing, for example, the spatial dependence of decoherence.

**Concluding remarks**

The description of electron scattering in liquids is a topic of pivotal interest in physics and chemistry, to which photoelectron spectroscopy contributed extensively through critical investigations such as the recording of PADs in water. Nowadays attosecond spectroscopy techniques stand as mature tools to provide further insight into these processes as they probe the EWP dynamics in the time domain and access the complementary phase information. Several seminal studies have already highlighted the role of elastic scattering on the EWP decoherence and the effect of the local solvation environment on the measured photoionization delays.

These approaches, however, have only scratched the surface of the complex and intricate behaviour of electrons in liquids. To achieve a complete understanding, many challenges lie

ahead — from disentangling congested spectral features and modelling open quantum systems, to resolving low-energy scattering events and solute-specific dynamics. Yet, promising solutions exist. Advances in theoretical frameworks, more efficient soft X-ray sources, and experimental schemes capable of recording the full photoemission sphere will together push the frontier further. These developments will gradually bridge the gap toward a comprehensive spatiotemporal and energetic description of electron dynamics in liquids.

## Acknowledgments

Gabriele Crippa acknowledges the ANR-24-CE29-0141-EPAD and PALM. Hugo Marroux and Gabriele Crippa acknowledge support from the European Research Council (ERC) under the European Union's [starting grant SATTOC (101078595)]

## References

[1] Garrett B C ,Dixon D A, Camaioni D M, Chipman D M, Johnson M A, Jonah C D, Kimmel G A, Miller J H, Rescigno T N, Rossky P J, Xantheas S S, Colson S D, Laufer A H, Ray D, Barbara P F, Bartels D M, Becker K H, Bowen Jr. K H, Bradforth S E, Carmichael I, Coe J V, Corrales L R, Cowin J P, Dupuis M, Eisenthal K B, Franz J A, Gutowski M S, Jordan K D, Kay B D, LaVerne J A, Lymar S V, Madey T E, McCurdy C W, Meisel D, Mukamel S, Nilsson A R, Orlando T M, Petrik N G, Pimblott S M, Rustad J R, Schenter G K, Schenter S J, Tokmakoff A, Wang L-S and Zwier T S 2004 Role of Water in Electron-Initiated Processes and Radical Chemistry: Issues and Scientific Advances *Chem. Rev.* 105 355-390

[2] Thürmer S, Seidel R, Faubel M, Eberhardt W, Hemminger J C, Bradforth S E and Winter B 2013 Photoelectron Angular Distributions from Liquid Water: Effects of Electron Scattering *Phys. Rev. Lett.* 111 173005

[3] Gozem S, Seidel R, Hergenhahn U, Lugovoy E, Winter B, Krylov A I and Bradforth S E 2020 Probing the Electronic Structure of Bulk Water at the Molecular Length Scale with Angle-Resolved Photoelectron Spectroscopy *J. Phys. Chem. Lett.* 11 5162–5170

[4] Seidel R, Winter B and Bradforth S E 2016 Valence Electronic Structure of Aqueous Solutions: Insights from Photoelectron Spectroscopy *Annu. Rev. Phys. Chem.* 67 283–305

[5] Signorell R. 2020 Electron Scattering in Liquid Water and Amorphous Ice: A Striking Resemblance *Phys. Rev. Lett.* 124 205501

[6] Schild A, Peper M, Perry C, Rattenbacher D, Wörner H J 2020 Alternative Approach for the Determination of Mean Free Paths of Electron Scattering in Liquid Water Based on Experimental Data *J. Phys. Chem. Lett.* 11 1128–1134

[7] Mondal A, Neufeld O, Yin Z, Nourbakhsh Z, Svoboda V, Rubio A, Tancogne-Dejean N and Wörner H J 2023 High-harmonic spectroscopy of low-energy electron-scattering dynamics in liquids *Nat. Phys.* 19 1813–1820

[8] Malerz S, Trinter F, Hergenhahn U, Ghrist A, Ali H, Nicolas C, Saak C-M, Richter C, Hartweg S, Nahon L, Lee C, Goy C, Neumark D M, Meijer G, Wilkinson I, Winter B and Thürmer S 2021 Low-energy constraints on photoelectron spectra measured from liquid water and aqueous solutions *Phys. Chem. Chem. Phys.* 23 8246-8260

[9] Jordan I, Huppert M, Rattenbacher D, Peper M, Jelovina D, Perry C, von Conta A, Schild A and Wörner H J 2020 Science 369 974-979

[10] Rattenbacher D, Jordan I, Schild A and Wörner H J 2018 Nonlocal mechanisms of attosecond interferometry and implications for condensed-phase experiments *Phys. Rev. A* 97 063415

[11] Jordan I and Wörner H J 2018 Extracting attosecond delays from spectrally overlapping interferograms *J. Opt.* 20 024013

[12] Gebauer A, Schabbehard T, Maschmann L and Pfeiffer W 2025 Disentanglement of overlapping spectrograms in attosecond time-resolved photoelectron spectroscopy of solids *Phys. Rev. B* **111** 125146

[13]     Dupuy R, Buttersack T, Trinter F, Richter C, Gholami S, Björneholm O, Hergenhahn U, Winter B and Bluhm H 2024 The solvation shell probed by resonant intermolecular Coulombic decay *Nat. Commun.* **15** 6926

[14]     Signorell R and Winter B 2022 Photoionization of the aqueous phase: clusters, droplets and liquid jets *Phys. Chem. Chem. Phys.* **24** 13438-13460

[15]     Dupuy R, Filser J, Richter C, Buttersack T, Trinter F, Gholami S, Seidel R, Nicolas C, Bozek J, Egger D, Oberhofer H, Thürmer S, Hergenhahn U, Reuter K, Winter B and Bluhm H 2023 Ångstrom-Depth Resolution with Chemical Specificity at the Liquid-Vapor Interface *Phys. Rev. Lett.* **130** 156901

[16]     Bourassin-Bouchet C, Barreau L, Gruson V, Hergott J-F, Quéré F, Salières P and Ruchon T 2020 Quantifying Decoherence in Attosecond Metrology *Phys. Rev. X* **10** 031048

[17]     Laurell H, Luo S, Weissenbilder R, Ammitzböll M, Ahmed S, Söderberg H, Petersson C L M , Poulain V, Guo C, Dittel C, Finkelstein-Shapiro D, Squibb R J, Feifel R, Gisselbrecht M, Arnold C L, Buchleitner A, Lindroth E, Kockum A F, L'Huillier A and Busto D 2025 Measuring the quantum state of photoelectrons *Nature Photonics* **19** 352–357

# 29. Attosecond spectroscopy of solids and molecules of opto-electronic interest

**Rocío Borrego-Varillas[1]\*, Maurizio Reduzzi[2], Matteo Lucchini[1,2] and Mauro NIsoli[1,2]**

[1] Istituto di Fotonica e Nanotecnologie, Consiglio Nazionale delle Ricerche, Milano, Italy
[2] Dipartimento di Fisica, Politecnico di Milano, Milano, Italy

rocio.borregovarillas@cnr.it

**Status**

One of the major challenges of modern physics is the understanding of fundamental physical and chemical processes at the nanoscale, where the behaviour of individual electrons determines the outcome of chemical reactions and the full electro-optical response of materials. Gaining insight into how electrons move and interact within matter is essential not only for basic science but also for the development of next-generation technologies in fields such as photovoltaics, molecular computing, and quantum information. A full understanding of these mechanisms requires real-time imaging of electronic motion. Attosecond technologies [1–3] represent a major conceptual and experimental advance, offering new opportunities to explore and manipulate electronic processes in matter with a level of detail that was previously unattainable [4].

Various experimental techniques have been demonstrated in which attosecond sources serve either as pump or probe pulses in time-resolved spectroscopy. For instance, attosecond transient absorption (ATAS) and attosecond transient reflection (ATRS) spectroscopies [5], in which attosecond pulses probe the dynamics with elemental specificity, have been applied to study processes such as molecular dynamics driven by conical intersections [6,7], or the response of solid-state materials to electric fields [8–10]. Another effective approach is attosecond photoelectron/ion spectroscopy, which, by using attosecond pulses to ionise the system, has been employed to explore phenomena such as charge migration in aromatic amino acids [11], and charge transfer in donor-acceptor molecules [12].

**Current and future challenges**

Despite their potential, attosecond techniques have largely been limited to studying photoionized molecules in the gas phase, whereas most natural photochemical processes involve neutral molecules excited by sunlight in solution or on surfaces. Initiating and probing electronic coherences in such systems requires ultrashort UV–visible pulses (< 3 fs), due to the need to excite simultaneously multiple electronic states separated by ~0.1–1 eV and to outpace rapid decoherence occurring within a few femtoseconds. A key challenge is therefore the generation of few-cycle pulses in the UV–visible spectral region.

For opto-electronic applications, molecules are typically integrated into devices via deposition on solid substrates with engineered electrodes. A key challenge lies in sample preparation: both device integration and attosecond-scale investigations require precise control of molecular orientation and conformation on the surface. Additionally, ATAS experiments demand ultrathin samples due to the high absorption of extreme UV (EUV) radiation and soft X-rays in solids.

The third challenge is related to the sensitivity of ATAS/ATRS experiments. In solid-state experiments, the low efficiency of detectors, strong background absorption, sample-induced scattering, and pulse-to-pulse noise fluctuations typically limit the minimum detectable signal to ΔOD levels of approximately 0.1-1 mOD [13]. This sensitivity constraint has, for instance, hindered the investigation of monolayer materials.

The fourth challenge is the employment of circularly polarised and structured attosecond pulses for spectroscopic applications. Its use has, for example, very recently enabled attosecond coherent control over photoelectron circular dichroism [14] and would open the door to the investigation of magnetic processes on attosecond time scales.

The last challenge concerns the development of theoretical methods that can accurately describe the dynamics and reproduce the observables measured in attosecond experiments, including electron correlation, many-body effects, and the coupling between electronic and nuclear motion, all on ultrafast time scales.

**Advances in science and technology to meet challenges**

In recent years, significant progress has been made in the generation of ultrashort UV pulses [15,16]. These advancements are pushing the temporal resolution of UV-EUV experiments below the 3-fs threshold [17,18] and are very promising for ATAS experiments [19].

The emergence of thulium fibre lasers and Fe:ZnSe and $ZnGeP_2$ CPA systems [20] have led to substantial improvements towards higher photon energies and photon flux in high-order harmonic generation. These advances are enabling attosecond light sources to access deeper core-level transitions across a wider range of elements, particularly in the soft X-ray regime. At the same time, the resulting increase in photon flux will support higher signal fidelity and faster data acquisition, as well as the generation of more intense circularly polarised attosecond pulses.

Finally, advances in high-repetition-rate laser systems (most notably Ytterbium lasers and OPCPAs), coupled with the development of more efficient optics (such as reflection zone plates) and low-noise detectors (for instance, based on sCMOS technology) compatible with spectrometer setups in the EUV and soft-X-ray spectral regions, can significantly enhance the signal-to-noise ratio in ATAS and ATRS experiments. These improvements are particularly critical for studying weakly absorbing systems, such as monolayer materials or diluted molecular systems, where the detectable changes in optical density are often close to the current sensitivity limits. Higher repetition rates allow for better statistical averaging without increasing sample damage, while improved detection efficiency extends the applicability of attosecond techniques to more complex and realistic environments.

**Concluding remarks**

Attosecond spectroscopy holds the promise of directly observing, and ultimately controlling, electronic motion in opto-electronic molecules and materials, representing a potentially transformative step toward next-generation technologies. However, realising this potential requires overcoming several challenges, including the development of advanced light sources, highly sensitive detection schemes, and precise control over material properties and sample preparation.

## Acknowledgements

We thank financial support from the European Research Council (ERC) under the European Union's Horizon 2020 research and innovation programme for grants ERC SynG no. 951224 TOMATTO, and Ministero dell'Università della Ricerca (grant no. 202239HFZN, grant no. 2022WZ8LME, grant no. 2022PX279E and grant no. R209LXZRSL).

## References

[1] Alexander O, Ayuso D, Matthews M, Rego L, Tisch J W G, Weaver B and Marangos J P 2025 Attosecond physics and technology *Appl. Phys. Lett.* **126** 170501

[2] Biegert J, Calegari F, Dudovich N, Quéré F and Vrakking M 2021 Attosecond technology(ies) and science *J. Phys. B At. Mol. Opt. Phys.* **54** 070201

[3] Midorikawa K 2022 Progress on table-top isolated attosecond light sources *Nat. Photonics* **16** 267–78

[4] Borrego-Varillas R, Lucchini M and Nisoli M 2022 Attosecond spectroscopy for the investigation of ultrafast dynamics in atomic, molecular and solid-state physics *Reports Prog. Phys.* **85** 066401

[5] Kraus P M, Zürch M, Cushing S K, Neumark D M and Leone S R 2018 The ultrafast X-ray spectroscopic revolution in chemical dynamics *Nat. Rev. Chem.* **2** 82–94

[6] Kobayashi Y, Chang K F, Zeng T, Neumark D M and Leone S R 2019 Direct mapping of curve-crossing dynamics in IBr by attosecond transient absorption spectroscopy *Science (80-. ).* **365** 79–83

[7] Zinchenko K S, Ardana-Lamas F, Seidu I, Neville S P, van der Veen J, Lanfaloni V U, Schuurman M S and Wörner H J 2021 Sub-7-femtosecond conical-intersection dynamics probed at the carbon K-edge *Science (80-. ).* **371** 489–94

[8] Schultze M, Bothschafter E M, Sommer A, Holzner S, Schweinberger W, Fiess M, Hofstetter M, Kienberger R, Apalkov V, Yakovlev V S, Stockman M I and Krausz F 2013 Controlling dielectrics with the electric field of light *Nature* **493** 75–8

[9] Inzani G, Adamska L, Eskandari-asl A, Di Palo N, Dolso G L, Moio B, D'Onofrio L J, Lamperti A, Molle A, Borrego-Varillas R, Nisoli M, Pittalis S, Rozzi C A, Avella A and Lucchini M 2023 Field-driven attosecond charge dynamics in germanium *Nat. Photonics* **17** 1059–65

[10] Dolso G L, Sato S A, Inzani G, Di Palo N, Moio B, Borrego-Varillas R, Nisoli M and Lucchini M 2025 Attosecond virtual charge dynamics in dielectrics *Nat. Photonics* **19** 999–1005

[11] Calegari F, Ayuso D, Trabattoni A, Belshaw L, De Camillis S, Anumula S, Frassetto F, Poletto L, Palacios A, Decleva P, Greenwood J B, Martín F, Nisoli M, Camillis S De, Anumula S, Frassetto F, Poletto L, Palacios A, Decleva P, Greenwood J B and Nisoli M 2014 Ultrafast electron dynamics in phenylalanine initiated by attosecond pulses *Science (80-. ).* **346** 336–9

[12] Vismarra F, Fernández-Villoria F, Mocci D, González-Vázquez J, Wu Y, Colaizzi L, Holzmeier F, Delgado J, Santos J, Bañares L, Carlini L, Castrovilli M C, Bolognesi P, Richter R, Avaldi L, Palacios A, Lucchini M, Reduzzi M, Borrego-Varillas R, Martín N, Martín F and Nisoli M 2024 Few-femtosecond electron transfer dynamics in photoionized donor–π–acceptor molecules *Nat. Chem.* **16** 2017–24

[13] Di Palo N, Inzani G, Dolso G L, Talarico M, Bonetti S and Lucchini M 2024 Attosecond absorption and reflection spectroscopy of solids *APL Photonics* **9** 020901

[14] Han M, Ji J-B, Blech A, Goetz R E, Allison C, Greenman L, Koch C P and Wörner H J 2025 Attosecond control and measurement of chiral photoionization dynamics *Nature* **645** 95–100

[15] Galli M, Wanie V, Lopes D P, Månsson E P, Trabattoni A, Colaizzi L, Saraswathula K, Cartella A, Frassetto F, Poletto L, Légaré F, Stagira S, Nisoli M, Martínez Vázquez R, Osellame R and Calegari F 2019 Generation of deep ultraviolet sub-2-fs pulses *Opt. Lett.* **44** 1308

[16] Travers J C, Grigorova T F, Brahms C and Belli F 2019 High-energy pulse self-compression and ultraviolet generation through soliton dynamics in hollow capillary fibres *Nat. Photonics* **13** 547–54

[17] Colaizzi L, Mocci D, Pini M, Kotsina N, Nordmann J, Brahms C, Travers J, Lucchini M, Borrego-Varillas R, Reduzzi M and Nisoli M 2024 A UV-XUV attosecond beamline with few-femtosecond tunable ultraviolet pump pulses ed L De Stefano, R Velotta and E Descrovi *EPJ Web Conf.* **309** 07004

[18] Wanie V, Ryabchuk S, Colaizzi L, Galli M, Månsson E P, Trabattoni A, Wahid A B, Hahne J, Cartella A, Saraswathula K, Frassetto F, Lopes D P, Martínez Vázquez R, Osellame R, Poletto L, Légaré F, Nisoli M and Calegari F 2024 A flexible beamline combining XUV attosecond pulses with few-femtosecond UV and near-infrared pulses for time-resolved experiments *Rev. Sci. Instrum.* **95** 083004

[19] Lee J, Avni T, Alexander O, Maimaris M, Ning H, Bakulin A, Burden P, Moutoulas E, Georgiadou D, Brahms C, Travers J, Marangos J and Ferchaud C 2024 Few-femtosecond soft X-ray transient absorption spectroscopy with tuneable DUV-Vis pump pulses *Optica* **11** 1320–3

[20] Marra Z A, Wu Y, Zhou F and Chang Z 2023 Cryogenically cooled Fe:ZnSe-based chirped pulse amplifier at 4.07 μm *Opt. Express* **31** 13447

## 30. Attosecond Spectroscopy: A New Window on Many-Body Physics

**Jens Biegert[1,2]\***, **Igor Tyulnev[1]**, **Julita Poborska[1]** and **Fernando Ardana-Lamas[1]**

[1] ICFO - Institut de Ciencies Fotoniques, The Barcelona Institute of Science and Technology, 08860 Castelldefels (Barcelona), Spain
[2] ICREA, Pg. Lluís Companys 23, 08010 Barcelona, Spain.

jens.biegert@icfo.eu

**Status**

The central challenge in many-body physics is understanding how electrons, holes, and nuclei interact across vastly different energy and time scales to grasp, and hopefully control, the emergence of material properties or the outcome of chemical reactions [1]. In modern condensed matter physics [2], key areas include emergent phases such as superconductivity, charge density waves, and excitonic insulators, or in chemistry, the fundamental dynamics of molecular pathways like isomerisation. Whether ultrafast or incoherent, traditional probes provide invaluable snapshots through many measurement techniques, such as absorption, emission, or scattering. However, they largely lack the combined temporal and state selectivity to resolve real-time multibody interactions or their onset [3]. As a result, many questions remain, for example, in Mott physics, high-$T_c$ superconductivity, the role of long- or short-range order, phase transitions, and topology, as well as in non-adiabatic charge dynamics or isomerisation. Recent progress in attosecond soft X-ray absorption spectroscopy (AXAS) and high-harmonic spectroscopy (HHS), enabled by high-flux tabletop coherent sources, now offers a transformative new approach. Combining attosecond temporal resolution with spectroscopic coherence and elemental and orbital selectivity allows these methods to access microscopic many-body dynamics, scattering, collective ordering, and vibronic interactions with unprecedented clarity.

For example, in solids, AXAS examined ultrafast charge dynamics and band dynamics in Silicon [12] and electron–hole asymmetries in graphite, where electrons dephase through impact excitation while holes quickly switch to Auger heating [13]. It has also identified strongly coupled optical phonons as the primary channels for decoherence. Simultaneously, transport studies on layered materials near van Hove singularities have shown anomalous conductivity due to flat-band occupation [14]. Complementing this microscopic perspective, polarisation-resolved HHS in $TiSe_2$ has sensitively detected chiral charge-density-wave formation [15] and allowed control over valley polarisation in $MoS_2$ [16]. Beyond crystalline materials, AXAS has been extended early on to molecular systems, exemplified by vibronic [5,6,17] and ring-opening dynamics [18–20]. The method directly tracked sequential conical intersections, revealed site-specific quantum beats across different carbon atoms, and captured the transition from electronic to vibrational coherence. These findings underscore the universality of attosecond spectroscopy, applicable from molecular gases to strongly correlated solids. These advances establish AXAS and HHS as unifying frameworks for many-body physics.

**Current and future challenges**

Attosecond spectroscopy has opened a transformative window on many-body physics, but several challenges remain for its application across materials and chemical sciences.

A first challenge lies in photon energy reach and flux. Tabletop high-harmonic generation (HHG) sources provide access to absorption edges of light elements, but the K-edges of heavier elements remain out of reach. Semi-valence edges are accessible but can mask the correlated dynamics of interest, complicating interpretation. At the same time, HHG sources remain flux-limited, preventing systematic studies of weak signals, complex samples, or cases where systems do not return immediately to their initial state.

A second challenge concerns detection and efficiency. Present soft X-ray spectrometers use diffraction gratings with low efficiency, and most experiments rely on standard CCD or sCMOS cameras. These detectors were not designed for attosecond applications and limit sensitivity and speed. This inefficiency means that many experiments remain photon-starved, even when bright sources are available.

A third challenge is theory and interpretation. Attosecond methods probe nonlinear dynamics where multiple interactions overlap: electronic correlations, scattering, vibronic couplings, and symmetry-breaking effects. Conventional theoretical tools rarely capture all these processes simultaneously. Green's function approaches and dynamical mean-field theory (DMFT) hold promise, but remain computationally expensive and have yet to routinely connect microscopic interactions with macroscopic observables such as conductivity or order parameters.

Finally, there are cross-disciplinary challenges. Attoscience grew from ultrafast, atomic and strong-field physics, where simplified few-body models dominate. However, condensed matter physics and chemistry deal with disorder, collective phenomena, and reaction coordinates that require different languages and frameworks. Bridging these cultural and conceptual gaps — ensuring that attoscience engages with the intricacies of materials and molecules, while condensed matter and chemistry embrace new attosecond probes — is imperative.

In short, the field must overcome source reach and flux limitations, detection efficiency, theoretical description, and cultural integration to fulfil its promise as a unifying framework for many-body physics.

### Advances in science and technology to meet challenges

Several scientific and technological advances are emerging that can address these limitations.

On the source side, HHG is steadily advancing toward higher photon energies, improved phase matching, and higher repetition rates. These developments will broaden access to elemental edges and deliver higher photon flux. Complementarily, free-electron lasers (FELs) operate naturally in the multi-keV regime, where they can probe heavy elements with high brightness. Rather than competing, HHG and FELs form a synergistic toolkit: laboratory-scale HHG for soft X-ray studies with attosecond resolution, and FELs for hard X-ray access to heavier elements.

Significant gains are possible on the detection side. Zone-plate spectrometers already offer an order-of-magnitude improvement over conventional gratings, and purpose-built detectors with higher quantum efficiency, lower noise, and faster readout are on the horizon. Such advances could shift the bottleneck from detection to photon production, making low-signal experiments feasible.

Hybrid frameworks are beginning to emerge for theory and analysis. Combining ab initio electronic-structure methods, time-dependent density functional theory, and many-body approaches such as DMFT makes it possible to simulate correlated, nonlinear, and ultrafast phenomena in a unified manner. These efforts promise a closer connection between microscopic interactions and macroscopic observables, enabling direct comparison to experiment. A further aspect is machine learning techniques, which may aid in detecting multiple subtle changes in measurements to extract observables in a better way than currently possible.

Finally, cross-disciplinary integration is accelerating. Strong-field and atomic physics concepts, such as recollision and tunnelling, were extended to solids. In contrast, condensed matter notions such as quasiparticles, Mott physics, and order parameters are being imported into attoscience. In chemistry, attosecond probes map conical intersections and vibronic coherence, complementing established ultrafast techniques. Such dialogue creates a shared conceptual framework that connects atomic, condensed matter, and molecular sciences.

These advances in sources, detectors, theory, and cultural integration point to a future where attosecond spectroscopy is no longer a niche technique but a core tool for tackling some of the most profound questions in physics and chemistry.

## Concluding remarks

In just over two decades, attoscience has grown from its roots in atomic and strong-field physics into a versatile probe of condensed matter and chemistry. Its unique ability to combine attosecond temporal resolution with elemental and orbital specificity has already revealed electron–hole asymmetries, phonon-driven decoherence, hidden order parameters, and site-specific vibronic dynamics. These breakthroughs demonstrate that attoscience can provide a unifying lens on many-body physics across scales, from molecules to correlated solids.

The challenges ahead — extending photon reach, improving flux and detection, advancing theory, and bridging disciplinary boundaries — are substantial, but so are the opportunities. Progress in tabletop HHG sources, synergy with FEL facilities, new spectrometers and detectors, and hybrid theoretical frameworks all point toward a future where attosecond methods become routine. Perhaps the most vital step will be cultural: fostering dialogue between atomic physicists, condensed matter researchers, and chemists to translate capabilities into insight. If these challenges are met, attosecond spectroscopy will not only allow us to observe correlated dynamics in real time. It may also open the door to controlling them, transforming how we understand and design quantum materials, chemical pathways, and emergent phases of matter.

## Acknowledgements

J.B. and group acknowledges financial support from the European Research Council for ERC Advanced Grant "TRANSFORMER" (788218), ERC Proof of Concept Grant "miniX" (840010), FET-OPEN "PETACom" (829153), FET-OPEN "OPTOlogic" (899794), FET-OPEN "TwistedNano" (101046424), MINECO for Plan Nacional PID2024-162757NB-I00; QU-ATTO, 101168628; AGAUR for 2021 SGR 01449, MINECO for "Severo Ochoa" (CEX2019-000910-S), Fundació Cellex Barcelona, the CERCA Programme/Generalitat de Catalunya, and the Alexander von Humboldt Foundation for the Friedrich Wilhelm Bessel Prize. JB also acknowledges Lasers4EU, which is funded by the European Union funds under HEU-GA 101131771.

## References

[1] Leone S R 2024 Reinvented: An Attosecond Chemist *Annual Review of Physical Chemistry* **75** 1–19
[2] Tsymbal E Y and Dowben P A 2013 Grand challenges in condensed matter physics: from knowledge to innovation *Front. Phys.* **1**
[3] Biegert J 2024 Attosecond science: a new era for many-body physics *Europhysics News* **55** 12–5
[4] Cousin S L, Silva F, Teichmann S, Hemmer M, Buades B and Biegert J 2014 High-flux table-top soft x-ray source driven by sub-2-cycle, CEP stable, 1.85-mu m 1-kHz pulses for carbon K-edge spectroscopy *Optics Letters* **39** 5383–6
[5] Pertot Y, Schmidt C, Matthews M, Chauvet A, Huppert M, Svoboda V, von Conta A, Tehlar A, Baykusheva D, Wolf J-P and Wörner H J 2017 Time-resolved x-ray absorption spectroscopy with a water window high-harmonic source *Science* **355** 264–7

[6]   Loh Z-H and Leone S R 2013 Capturing Ultrafast Quantum Dynamics with Femtosecond and Attosecond X-ray Core-Level Absorption Spectroscopy *J. Phys. Chem. Lett.* **4** 292–302

[7]   Goulielmakis E, Loh Z H, Wirth A, Santra R, Rohringer N, Yakovlev V S, Zherebtsov S, Pfeifer T, Azzeer A M, Kling M F, Leone S R and Krausz F 2010 Real-time observation of valence electron motion *Nature* **466** 739-U7

[8]   Kienberger R, Hentschel M, Uiberacker M, Spielmann C, Kitzler M, Scrinzi A, Wieland M, Westerwalbesloh T, Kleineberg U, Heinzmann U, Drescher M and Krausz F 2002 Steering attosecond electron wave packets with light *Science* **297** 1144–8

[9]   Biegert J, Heinrich A, Hauri C P, Kornelis W, Schlup P, Anscombe M, Schafer K J, Gaarde M B and Keller U 2005 Enhancement of high-order harmonic emission using attosecond pulse trains *Laser Physics* **15** 899–902

[10]  Mairesse Y, Higuet J, Dudovich N, Shafir D, Fabre B, Mével E, Constant E, Patchkovskii S, Walters Z, Ivanov M Y and Smirnova O 2010 High harmonic spectroscopy of multichannel dynamics in strong-field ionization *Physical Review Letters* **104** 213601–213601

[11]  Boutu W, Haessler S, Merdji H, Breger P, Waters G, Stankiewicz M, Frasinski L J, Taieb R, Caillat J, Maquet A, Monchicourt P, Carre B and Salieres P 2008 Coherent control of attosecond emission from aligned molecules *Nature Phys* **4** 545–9

[12]  Schultze M, Ramasesha K, Pemmaraju C D, Sato S A, Whitmore D, Gandman A, Prell J S, Borja L J, Prendergast D, Yabana K, Neumark D M and Leone S R 2014 Attosecond band-gap dynamics in silicon *Science* **346** 1348–52

[13]  Sidiropoulos T P H, Di Palo N, Rivas D E, Severino S, Reduzzi M, Nandy B, Bauerhenne B, Krylow S, Vasileiadis T, Danz T, Elliott P, Sharma S, Dewhurst K, Ropers C, Joly Y, Garcia M E, Wolf M, Ernstorfer R and Biegert J 2021 Probing the Energy Conversion Pathways between Light, Carriers, and Lattice in Real Time with Attosecond Core-Level Spectroscopy *Phys. Rev. X* **11** 041060

[14]  Sidiropoulos T P H, Di Palo N, Rivas D E, Summers A, Severino S, Reduzzi M and Biegert J 2023 Enhanced optical conductivity and many-body effects in strongly-driven photo-excited semi-metallic graphite *Nat Commun* **14** 7407

[15]  Tyulnev I, Zhang L, Vamos L, Poborska J, Bhattacharya U, Chhajlany R W, Grass T, Mañas-Valero S, Coronado E, Lewenstein M and Biegert J 2025 High harmonic spectroscopy reveals anisotropy of the charge-density-wave phase transition in TiSe2 *Commun Mater* **6** 152

[16]  Tyulnev I, Jimenez-Galan A, Poborska J, Vamos L, Silva R E F, Russell P St J, Tani F, Smirnova O, Ivanov M and Biegert J 2024 Valleytronics in bulk MoS2 with a topologic optical field *Nature* **626** 746–51

[17]  Saito N, Sannohe H, Ishii N, Kanai T, Kosugi N, Wu Y, Chew A, Han S, Chang Z and Itatani J 2019 Real-time observation of electronic, vibrational, and rotational dynamics in nitric oxide with attosecond soft x-ray pulses at 400 eV *Optica* **6** 1542–1542

[18]  Attar A R, Bhattacherjee A, Pemmaraju C D, Schnorr K, Closser K D, Prendergast D and Leone S R 2017 Femtosecond x-ray spectroscopy of an electrocyclic ring-opening reaction *Science* **356** 54–8

[19]  Wörner H J, Bertrand J B, Fabre B, Higuet J, Ruf H, Dubrouil A, Patchkovskii S, Spanner M, Mairesse Y, Blanchet V, Mével E, Constant E, Corkum P B and Villeneuve D M 2011 Conical intersection dynamics in NO2 probed by homodyne high-harmonic spectroscopy *Science* **334** 208–12

[20]  Severino S, Ziems K M, Reduzzi M, Summers A, Sun H-W, Chien Y-H, Gräfe S and Biegert J 2024 Non-Adiabatic Electronic and Vibrational Ring-Opening Dynamics resolved with Attosecond Core-Level Spectroscopy *Nature Photonics* **18** 731–7

## *31. High Harmonics as a Universal Response of Matter to High Intensity Light*

**Shima Gholam-Mirzaei and P.B. Corkum***

Joint Attosecond Science Laboratory, University of Ottawa and National Research Council of Canada, Ottawa, Canada

pcorkum@uottawa.ca

**Status**

In a short pulse of light, atoms [1], molecules [2], transparent solids [3], and even liquids [4] all produce high harmonics as they approach breakdown. Is this a universal response of matter? Metals have long been thought to be exception due to their strong reflectivity at infrared wavelengths. Recent observation of harmonics from titanium nitride (TiN) — a semi-metal with metallic conductivity and plasmonic behavior — suggests that strong-field emission can occur even in systems dominated by free electrons [5]. However, the observation of harmonics from classical metals and the responsible mechanism remains unclear.

Metals are characterized by free electrons that dominate many of the material properties including how infrared light couples to the solid [6]. To first order, light interacts primarily with these electrons, rejecting most of the incident field — nearly 99% in noble metals — until the onset of damage. The lattice acts as a scaffold within which the electrons move. To second order new electrons may be generated, but their influence is largely diluted by the sea of existing electrons. Even under strong-field excitation, metals retain high conductivity and remain far from transparency.

However, for high harmonic generation, newly formed electrons are unique: The electron-hole pair enables a coherent emission pathway that is fundamentally different from the response of the pre-existing free electrons. While metals are dominated by conduction electrons, the bound electrons — though fewer in number — can play a critical role when released in the strong field. Thus, harmonics from metals might resemble those in other condensed media, involving coherent electron-hole pairs [7] and possibly even plasmons [8] driven by the optical field.

If we are to search for harmonics from classical metals, atomic tunnelling [9] suggests that the driving pulse should be as short as possible. Tunneling rates increase rapidly near the damage threshold, and shorter pulses compress the ionization window into a single optical cycle. As rates rise, the distinction between different ionization potentials diminishes — electrons from various bound states can be freed with similar probability. This favors conditions where coherent electron–hole dynamics, including possible recollision, may emerge.

But is it likely that we will see recollision harmonics from metals? In ZnO, calculations show that interband emission from free carriers exceeds intraband (recombination) emission by a factor of $\sim 10^4$ [10]. Near damage intensities, an even larger fraction of bound electrons can be liberated from valence bands — enough to potentially support recollision harmonics in metals. While in TiN the damage threshold restricts harmonic generation to intraband dynamics, the possibility of interband contributions in classical metals with higher thresholds remains open — particularly near the cutoff, where recombination could benefit from both kinetic and interband energy.

Although the ratio is sensitive to the band structure of the substrate, $10^4$ is a large number and it is prudent to see if recollision harmonics are within reach near cut-off.

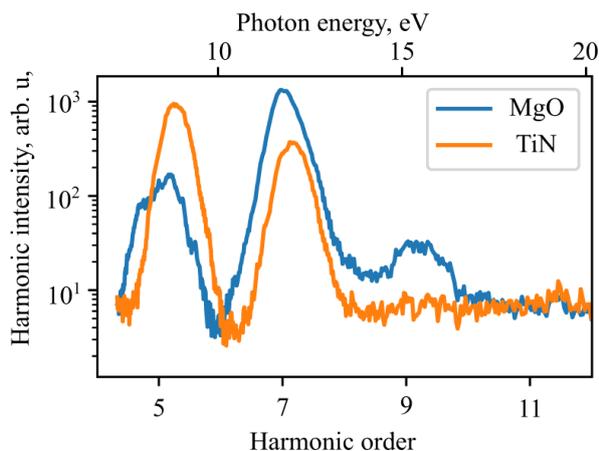

*Figure 1: Harmonic yield in reflection for TiN irradiated with a few cycles of 800 nm light [5]*

The first experimental evidence of harmonics from semi-metal titanium nitride (TiN), epitaxially grown on MgO suggests that the intraband transition enables nonlinear emission. Driven with 800 nm light, TiN produces harmonics up to ~15 eV at peak intensities near $1.3 \times 10^{13}$ W/cm$^2$. Remarkably, the harmonic yield was comparable to that of MgO at the same intensity. MgO, however, survives higher fields, and so its harmonic brightness and cutoff can be extended further. In TiN, the emission mechanism is explained by Bloch oscillations of conduction electrons.

**Current and future challenges**

Solid-state harmonics are often divided into single-band (intraband) motion and two-band (interband) recombination. However, the real picture in a classical metal might be more sophisticated since the electrons explore many bands. The key question is whether the liberated electrons maintain correlation with a hole. If they do, interband trajectories can extend across multiple bands and still recombine coherently. If they do not, only intraband motion remains. Exploring multiple bands imprints a distinct chirp in the recollision harmonics compared to intraband emission. Distinguishing these signatures in experiment is therefore central to identifying the mechanisms at work.

**Advances in science and technology to meet challenges**

Long-wavelength, few-cycle infrared pulses are especially advantageous for solid-state harmonics. They extend the cutoff, and generate more harmonic lines, which makes phase measurements between harmonics more accessible — a critical tool for distinguishing interband from intraband mechanisms.

**Concluding remarks**

In metals, light penetrates only a very short distance. For silver, with a free-carrier density of about $5.8 \times 10^{22}$ electrons/cm$^3$ [6], the collisionless plasma model gives a penetration depth of roughly $(c/\omega_p)^{-1}$, where $\omega_p$ is the plasma frequency. As a result, only a surface layer on the order of 20 nm contributes to the harmonic emission. This confinement makes HHG in metals naturally sensitive to surfaces and interfaces, offering non-destructive in-situ spectroscopy of ultrathin layers.

Metals also provide robust platforms for hybrid structures. Their high reflectivity under strong fields creates standing waves [11], with intensity maxima forming one quarter of a wavelength inside the adjacent dielectric. Harmonics generated in such conditions can be exceptionally strong. Recent findings [12] suggest that even quintessential metals may support such dynamics, pointing to rich opportunities at surfaces and interfaces.

## Acknowledgements

We acknowledge financial support from the U.S. ARO (Award no. FA9550-16-1-0109) and the National Research Council of Canada. We also acknowledge discussions with A. Jimenez-Galan, A. Korobenko, G. Vampa and A. Staudte.

## References

[1]  Ferray M, L'Huillier A., Li X. F., Lompre L. A., Mainfray G. and Manus C. 1988 Multiple-harmonic conversion of 1064 nm radiation in rare gases *Journal of Physics B: Atomic, Molecular and Optical Physics* **21** L31

[2]  J. Itatani J., Levesque J., Zeidler D., Niikura H., Pépin H., Kieffer J-C., Corkum P. B. and Villeneuve D. M. 2004 Tomographic imaging of molecular orbitals *Nature* **432** 867

[3]  Ghimire S., DiChiara A. D., Sistrunk E. and Agostini P. Observation of high-order harmonic generation in a bulk crystal 2011 *Nature Physics* **7** 138

[4]  Luu T. T., Yin Z., Jain A., Gaumnitz T., Pertot Y., Ma J. and Wörner, H. J. 2018 Extreme–ultraviolet high–harmonic generation in liquids *Nature Communications* **9** 3723

[5]  Korobenko A., Saha S., Godfrey A. T. K., Gertsvolf M., Naumov A. Y., Villeneuve D. M., Boltasseva A., Shalaev V. M. and Corkum P. B. 2021 High-harmonic generation in metallic titanium nitride, *Nature Communications* **12**, 4981

[6]  N. W. Ashcroft N. W. and Mermin N. 1976 Solid state physics holt. *Rinehart and Winston, New York* Appendix C

[7]  Vampa G., Hammond T. J., Thiré N., Schmidt B. E., Légaré F., McDonald C. R., Brabec T. and Corkum, P. B. 2015 Linking high harmonics from gases and solids *Nature* **522** 462-464

[8]  Vampa G., Ghamsari B. G., Siadat Mousavi S., Hammond T. J., Olivieri A., Lisicka-Skrek E., Naumov A. Y. Villeneuve D. M., Staudte A., Berini P. and Corkum, P. B. 2017. Plasmon-enhanced high-harmonic generation from silicon. *Nature Physics* **13** 659-662.

[9]  Ammosov M. V., Delone N. B. and Krainov, V. P. 1986 Tunnel ionization of complex atoms and atomic ions in electromagnetic field *High intensity laser processes* **664** 138-141

[10]  Vampa G., McDonald C. R., Orlando G., Klug D. D., Corkum P. B. and Brabec, T. 2014 Theoretical analysis of high-harmonic generation in solids *Physical Review Letters* **113** 073901

[11]  Korobenko A., Hammond T. J., Zhang C., Naumov A. Yu., Villeneuve D. M. and Corkum P. B. 2018 High-harmonic generation in solids driven by counter-propagating pulses. *Optics Express* **27** 32630

[12]  Gholam-Mirzaei, S., Korobenko, A., Haram, N., Purschke, D. N., Saha, S., Naumov, A. Y., Vampa, G., Villeneuve, D.M., Silva, R.E., Staudte, A. and Boltasseva, A. Shalaev V. M., Galán A. J-. and Corkum, P. B. 2025 High Harmonic Generation from a Noble Metal. *arXiv preprint* **arXiv:2503.05073**

## 32. Attosecond Electronics

**Marcus Ossiander[1]\* and Martin Schultze[1]\***

[1] Institute of Experimental Physics, Graz University of Technology, Graz, Austria

schultze@tugraz.at, marcus.ossiander@tugraz.at

**Status**

Twenty-five years after the first attosecond experiments, it appears safe to conclude that today's extremely accurate and sophisticated quantum mechanical description of atomic physics derives from experiments that can be grouped into three influential categories. Seeking to harness nuclear power, scattering experiments revealed the atomic structure. The invention of the laser and its application as a spectroscopy tool rendered an atom's energy levels, i.e., the eigen-energy spectrum of its electrons, an easily detectable observable. However, the temporal evolution of electron wave functions in atoms remained elusive to experiments, and the widely accepted spell "too fast to be seen" was cast on many dynamic processes in atoms - including the seemingly simple conversion of a photon into an electronic excitation. Attosecond physics finally answered *when* this happens, made the temporal phase of electrons observable, and stimulated theory development that, in the case of atomic physics, has reached an impressive level of maturity [1–5].

Experimental physics has then decamped to explore wave function dynamics also in the condensed phase, and different experimental attosecond spectroscopy techniques have been pursued to explore the dynamics of electronic processes in solids.

The spectrally resolved high-order harmonic radiation (HHG) emitted by solids, when they are irradiated by intense optical waveforms, has shown that oscillatory carrier motion in a solid can act out at Petahertz frequencies, i.e., considerably faster than the oscillation period of visible light [6–8]. Recording the high-frequency radiation has underpinned that such reciprocating currents are microscopically phase coherent and reveal intricate details of the solid's band structure [9,10].

Attosecond transient absorption experiments performed in dielectric, semiconducting, and metallic solids have resolved the timing of photodoping & bandgap renormalization [11,12] and dynamical band-structure modifications [13–15]. Finally, inspecting minute changes of an ultrafast laser pulse's electric field upon passage through a solid has determined the temporal evolution of energy exchange between the electric field of light and a solid's electronic system [16].

**Current and future challenges**

From the start of those experiments, it was speculated that the unification of ultrafast optics and electronic circuitry would pave the way towards a new hyper-fast information technology platform. The most direct link between these fields is currents: in electronics, applied voltages accelerate electrons in wires and change the conductivity of field-effect transistors. In optics, the oscillating currents induced by light electric fields yield a dynamic polarization that manifest as the refractive index. Carrier-envelope phase stabilized light fields and attosecond pulses expanded light-based control to direct currents: Fig. 1 and its caption introduce this capability and highlight differences between gas-phase and solid-state attosecond experiments.

In gas-phase samples, optically induced direct currents are linearly proportional to (i.e. they measure) the electric field of light waves [17] due to the free electron dispersion relation's infinite extent, its parabolicity, and its bijective momentum-energy mapping. A solid's dispersion relation does not possess these properties: its periodicity with the Brillouin zone excludes infinite extent and parabolicity, and multiple valence and conduction bands exclude a one-to-one momentum-energy mapping. Moreover, bands are often spaced closely in energy, enabling visible light-driven linear & nonlinear transitions and other rich dynamics [18].

The magnitude of a current depends on the number of carriers and their velocity. In a solid, the latter depends on the band- and momentum-dependent band slope (see Fig. 1 caption). Consequently, experimentally measured optically induced currents reveal a material's dispersion relation, its band-resolved population, and transitions within with sub-femtosecond resolution, especially when paired with time-dependent density functional theory and semiconductor Bloch equation modelling. Current is a well-defined observable in modelling and the experiment, facilitating their unambiguous comparison.

**Advances in science and technology to meet challenges**

The technological application of optical currents interfaced with electronics will require a multi-fold advance of the attosecond toolbox: spatial resolution competitive to the structure size of modern electronic devices, accurate modelling capturing many-body dynamics, and millimeter-wave electronics technology:

Optical current control before decoherence will require device dimensions on the order of the excited state wave function or the inelastic mean free path. Attosecond spectroscopy of one-dimensional heterostructures in the optical propagation direction conquers such dimensions and demonstrates that engineering coherent currents is feasible [19]. The next steps, examining two- and three-dimensional devices, require observing minuscule functional areas also in the transverse dimensions, and thus demand microscopy concepts capable of unifying extreme spatial and temporal resolution (we have given an outlook on the prospects of attosecond microscopy elsewhere [20]).

Carrier interaction, thermalization, and cooling rapidly decohere excited wave packets and have major consequences on the connection between incident light fields and generated currents. However, predicting these effects from first principles is hard, and electron-electron and electron-phonon scattering are often treated phenomenologically. Therefore, numerical techniques accurately predicting solid-state many-body dynamics and their dependence on the excitation density, the device material, and the device schematic will be invaluable for experiment and device design.

Today, optical currents are often read out with kilohertz to megahertz bandwidths. I.e., even parallel readout would require unrealistically high channel numbers to extract their full, multi-terahertz rate information content in real time. Here, collaborative device miniaturization and design with microwave engineers promises to push electronic single-channel data rates beyond 100 GHz and to achieve manageable channel numbers.

**Concluding remarks**

Attosecond technology's capacity to track and control electronic motion has reached a maturity that suggests first applications are becoming feasible (although many details of light-matter interaction dynamics in the condensed phase are still under examination). Examples include ultrafast reincarnations of the Auston switch sampling the electric field of visible light using conventional electronics, ultrafast large-current-density optoelectronic switches testing next-generation high-frequency electronics, and solid-state-based optical frequency conversion schemes synthesizing short ultraviolet waveforms.

In parallel, experimental sophistication and enhanced theoretical dexterity gradually permit investigating the relations between electronic correlations, charge and lattice motion couplings, and topological material properties. These efforts fuel the hope to identify material systems & light-waveform combinations that permit coherently controlling carriers in solids (and, as an extension, in conventional microelectronics) – a prospect that could overcome dissipation, the limiting factor for the processing speed of current technology.

**Acknowledgements**

M.O. acknowledges funding from the European Union (grant agreement 101076933EUVORAM). The views and opinions expressed are, however, those of the author(s) only and do not necessarily reflect those of the European Union or the European Research Council Executive Agency. Neither the European Union nor the granting authority can be held responsible for them.

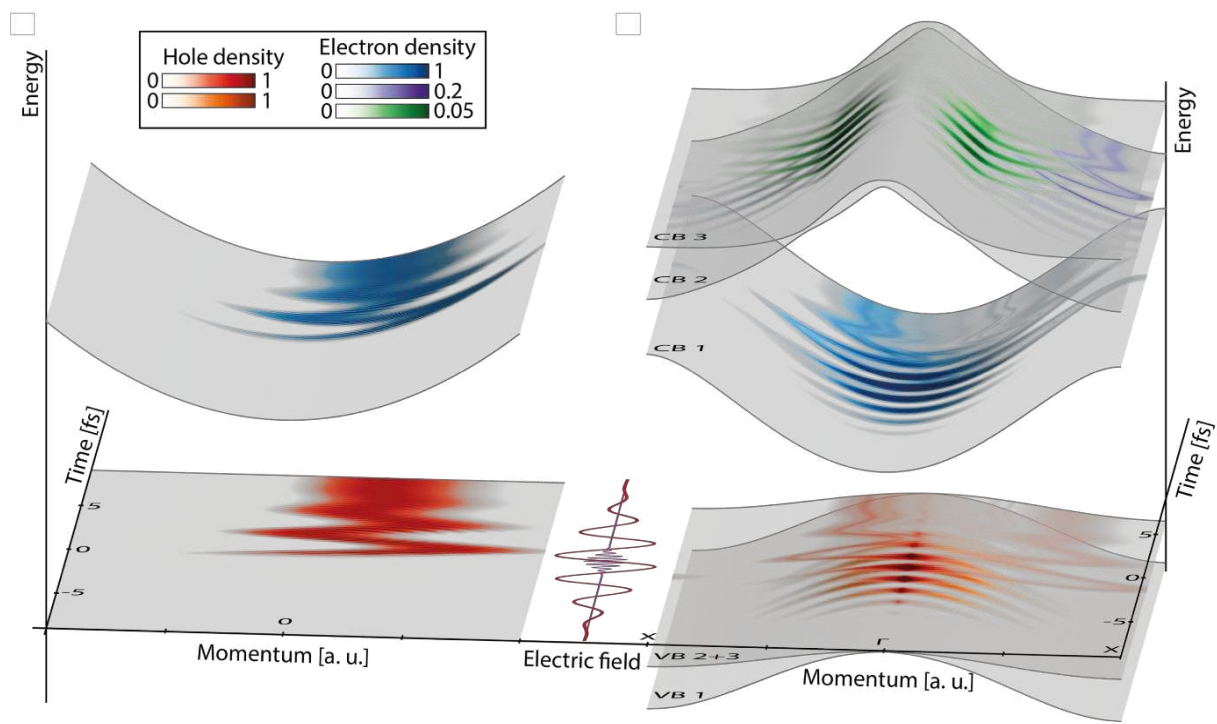

*Figure 1 Time-resolved coherent wave packet dynamics in the electronic dispersion relation of atoms and solids. High-frequency photons of a low-intensity attosecond pulse (center panel, purple) prepare an excited electronic wavepacket through single photon absorption that is subsequently accelerated by the intense light field of a low-frequency laser pulse (center panel, red). Left Panel: During the photoionization of a noble gas, an electron is excited from a bound core level to the continuum of free states. From a solid-state perspective, the initial state must have a flat momentum distribution because it is spatially localized, and a hole (red) describes the electron's absence. The final, free electron can be described as a carrier wave packet (blue) in an infinitely extended, parabolic, conduction band. Because the hole and the carrier are charged, the incident laser light field modulates their crystal momenta. As a wave packet's group velocity is given by its dispersion relation's momentum derivative ∂E/∂k, the bound hole's velocity vanishes for all crystal momenta. The carrier's velocity changes linearly with the carrier momentum (as expected from classical mechanics), resulting in a detectable direct current after both light pulses vanish. Right panel: During the photoexcitation of a semiconductor or dielectric, an electron is elevated from one of multiple occupied valence bands to one of multiple empty conduction bands. As soon as a band is partially filled, the incident light field modulates the holes' and carriers' crystal momenta and group velocities. Each band's dispersion relation is only approximated by a parabola close to high-symmetry points and contains regions where hole and carrier wave packets possess opposite-than-intuitive group velocities (e.g., close to the Brillouin zone boundaries) and population coupling to adjacent bands can occur. The right panel is partially reproduced from [18] under a CC BY 4.0 License (http://creativecommons.org/licenses/by/4.0).*

## References

[1]   Eckle P, Pfeiffer A N, Cirelli C, Staudte A, Dörner R, Muller H G, Büttiker M and Keller U 2008 Attosecond Ionization and Tunneling Delay Time Measurements in Helium *Science* **322** 1525-29

[2]   Schultze M, Fieß M, Karpowicz N, Gagnon J, Korbman M, Hofstetter M, Neppl S, Cavalieri A L, Komninos Y, Mercouris Th, Nicolaides C A, Pazourek R, Nagele S, Feist J, Burgdörfer J, Azzeer A M, Ernstorfer R, Kienberger R, Kleineberg U, Goulielmakis E, Krausz F and Yakovlev V S 2010 Delay in Photoemission *Science* **328** 1658–62

[3]   Klünder K, Dahlström J M, Gisselbrecht M, Fordell T, Swoboda M, Guénot D, Johnsson P, Caillat J, Mauritsson J, Maquet A, Taïeb R and L'Huillier A 2011 Probing Single-Photon Ionization on the Attosecond Time Scale *Phys. Rev. Lett.* **106** 143002

[4]   Calegari F, Sansone G, Stagira S, Vozzi C and Nisoli M 2016 Advances in attosecond science *J. Phys. B At. Mol. Opt. Phys.* **49** 062001

[5]   Ott C, Kaldun A, Raith P, Meyer K, Laux M, Evers J, Keitel C H, Greene C H and Pfeifer T 2013 Lorentz Meets Fano in Spectral Line Shapes: A Universal Phase and Its Laser Control *Science* **340** 716–20

[6]   Ghimire S, DiChiara A D, Sistrunk E, Agostini P, DiMauro L F and Reis D A 2010 Observation of high-order harmonic generation in a bulk crystal *Nat. Phys.* **7** 138–41

[7]   Gertsvolf M, Spanner M, Rayner D M and Corkum P B 2010 Demonstration of attosecond ionization dynamics inside transparent solids. *J. Phys. B At. Mol. Opt. Phys.* **43** 131002

[8]   Vampa G and Villeneuve D M 2015 High-harmonic generation: To the extreme *Nat. Phys.* **11** 529–30

[9]   Garg M, Zhan M, Luu T T, Lakhotia H, Klostermann T, Guggenmos A and Goulielmakis E 2016 Multi-petahertz electronic metrology *Nature* **538** 359–63

[10]  Uzan-Narovlansky A J, Faeyrman L, Brown G G, Shames S, Narovlansky V, Xiao J, Arusi-Parpar T, Kneller O, Bruner B D, Smirnova O, Silva R E F, Yan B, Jiménez-Galán Á, Ivanov M and Dudovich N 2024 Observation of interband Berry phase in laser-driven crystals *Nature* **626** 66–71

[11]  Schultze M, Ramasesha K, Pemmaraju C D, Sato S A, Whitmore D, Gandman A, Prell J S, Borja L J, Prendergast D, Yabana K, Neumark D M and Leone S R 2014 Attosecond band-gap dynamics in silicon *Science* **346** 1348–52

[12]  Schlaepfer F, Lucchini M, Sato S A, Volkov M, Kasmi L, Hartmann N, Rubio A, Gallmann L and Keller U 2018 Attosecond optical-field-enhanced carrier injection into the GaAs conduction band *Nat. Phys.* **14** 560–4

[13]  Mashiko H, Oguri K, Yamaguchi T, Suda A and Gotoh H 2016 Petahertz optical drive with wide-bandgap semiconductor *Nat. Phys.* **12** 741–5

[14]  Lucchini M, Sato S A, Ludwig A, Herrmann J, Volkov M, Kasmi L, Shinohara Y, Yabana K, Gallmann L and Keller U 2016 Attosecond dynamical Franz-Keldysh effect in polycrystalline diamond *Science* **353** 916–9

[15]  Schultze M, Bothschafter E M, Sommer A, Holzner S, Schweinberger W, Fiess M, Hofstetter M, Kienberger R, Apalkov V, Yakovlev V S, Stockman M I and Krausz F 2013 Controlling dielectrics with the electric field of light *Nature* **493** 75–8

[16]  Sommer A, Bothschafter E M, Sato S A, Jakubeit C, Latka T, Razskazovskaya O, Fattahi H, Jobst M, Schweinberger W, Shirvanyan V, Yakovlev V S, Kienberger R, Yabana K, Karpowicz N, Schultze M and Krausz F 2016 Attosecond nonlinear polarization and light–matter energy transfer in solids *Nature* **534** 86–90

[17]  Goulielmakis E, Uiberacker M, Kienberger R, Baltuska A, Yakovlev V, Scrinzi A, Westerwalbesloh Th, Kleineberg U, Heinzmann U, Drescher M and Krausz F 2004 Direct measurement of light waves. *Science* **305** 1267–9

[18]  Ossiander M, Golyari K, Scharl K, Lehnert L, Siegrist F, Bürger J P, Zimin D, Gessner J A, Weidman M, Floss I, Smejkal V, Donsa S, Lemell C, Libisch F, Karpowicz N, Burgdörfer J, Krausz F and Schultze M 2022 The speed limit of optoelectronics *Nat. Commun.* **13** 1620

[19]  Siegrist F, Gessner J A, Ossiander M, Denker C, Chang Y-P, Schröder M C, Guggenmos A, Cui Y, Walowski J, Martens U, Dewhurst J K, Kleineberg U, Münzenberg M, Sharma S and Schultze M 2019 Light-wave dynamic control of magnetism *Nature* **571** 240–4

[20]  Vogelsang J, Mikkelsen A, Ropers C, Gaida J H, Garg M, Kern K, Miao J, Schultze M and Ossiander M 2025 Attosecond microscopy —Advances and outlook *Europhys. Lett.* **149** 36001

## *33. Attomicroscopy: Attosecond Electron Motion Imaging in Real Time and Space*

### Dandan Hui[1,2]*, Mohammed Hassan[3,4]*

[1] State Key Laboratory of Ultrafast Optical Science and Technology, 710119, Shaanxi, China.
[2] Xi'an Institute of Optics and Precision Mechanics, Chinese Academy of Sciences (XIOPM-CAS), 710119, Shaanxi, China.
[3] Department of Physics, University of Arizona, Tucson, 85721, USA.
[4] James C. Wyant College of Optical Sciences, University of Arizona, Tucson, 85721, USA.

huidandan@opt.ac.cn, mohammedhassan@arizona.edu

### Introduction

At the dawn of the new millennium, the rise of attosecond science opened a path to explore electron motion in matter [1]. The generation of extreme ultraviolet (XUV) pulses via high-harmonic generation (HHG), [2] which naturally possesses attosecond resolution, provided a crucial tool for probing electron dynamics on ultrafast timescales [3,4]. This advancement enabled the development of attosecond spectroscopic methods such as HHG spectroscopy, the attoclock, and XUV transient absorption spectroscopy [5-7]. These attosecond techniques have made it possible to monitor electron dynamics in real time. As the field has matured, extensive efforts have shifted toward extending attosecond science to real-world applications [8]. These include the development of petahertz optoelectronics [9-11] and the possibility of controlling chemical reactions using light [12]. Consequently, understanding how electron motion correlates with the morphology of complex systems—particularly in solid-state materials—has become increasingly essential [13]. This need has accelerated the development of attosecond electron microscopy and real-time electron imaging approaches[14].

### Attosecond electron imaging

Alongside the progress of attosecond science over the past three decades, researchers have advanced ultrafast transmission electron microscopy (UTEM), and ultrafast scanning electron microscopy (USEM) [15] and ultrafast electron diffraction (UED) to visualize ultrafast dynamics in matter[16]. The temporal resolution of UTEM has improved by several orders of magnitude—from the nanosecond to the femtosecond regime (see Fig. 1a). These developments have enabled imaging of various ultrafast processes, including melting, phase changes, and atomic or molecular motion in spatiotemporal domains. Recently, we achieved attosecond resolution in electron microscopy using an optical gating technique, culminating in the first demonstration of "*Attomicroscopy*", *which refers to the imaging of electron motion dynamics using attosecond electron microscopy, which achieves attosecond temporal resolution through single attosecond electron pulses.*"[14] as shown in Fig. 1b. Remarkably, the attomicroscopy is different than the light interference-based (CW light electron coupled interference [17] or self-laser pulse inference imprinted on modulated electron[18]) electron microscopy which uses CW and few hundred femtosecond pulses. In that case the generated train of attosecond electron pulses is limited to image the light propagation dynamics on nanostructure symmetrically repeated every half-cycle therefore the ultrafast matter dynamics imaging stays beyond the reach by these methodologies [17-19]. Hence, based on the basic principle of the time-resolve approach, the claimed attosecond resolution is untenable and the realistic temporal resolution of this modulated-electron approach remain hundreds of femtoseconds or infinity [13].

This breakthrough technique allows us to image attosecond electron motion in materials like graphene using time-resolved attosecond electron diffraction. In addition, USEM is also being employed to introduce attosecond optical pulses, aiming to push towards attosecond temporal resolution in surface dynamics studies. Electron motion underlies fundamental phenomena in nature, shaping material properties, chemical mechanisms, and biological behaviour. Emerging electron imaging techniques—such as attosecond electron microscopy, 4D-STEM , diffraction, spectroscopy, elemental mapping and cryo-imaging—enable a broad spectrum of applications across physics, materials science, chemistry, and biology [13].

In materials science, for instance, atomicroscopy of electron motion in graphene has enabled us to manipulate photocurrents in graphene-based phototransistors using light fields, demonstrating attosecond scale current switching. This breakthrough offers a path to building petahertz transistors [20]. In quantum physics, attosecond electron microscopy provides a new platform for probing real-time wavefunction dynamics at the atomic scale. In chemistry and biochemistry, the technology allows us to monitor redox reactions in cable bacteria in both spatial and temporal domains. Additionally, cryo-atomicroscopy could transform biological and medical research, for example, by enabling studies of cardiac sarcomere dynamics—the molecular basis of heart muscle contraction—and exploring laser-based modulation of these processes. These applications showcase the vast potential of atomicroscopy to drive the next wave of scientific exploration. In addition to attosecond electron microscopy, major efforts across the scientific community have focused on developing attosecond XUV microscopy and high-energy ultrafast electron diffraction tools—such as MeV UED [21]—aimed at resolving electron dynamics in matter. A notable milestone includes the demonstration of attosecond free-electron pulses at LCLE at SLAC-Stanford, which have already been used to measure photoemission delays of core-level electrons in molecules [22].

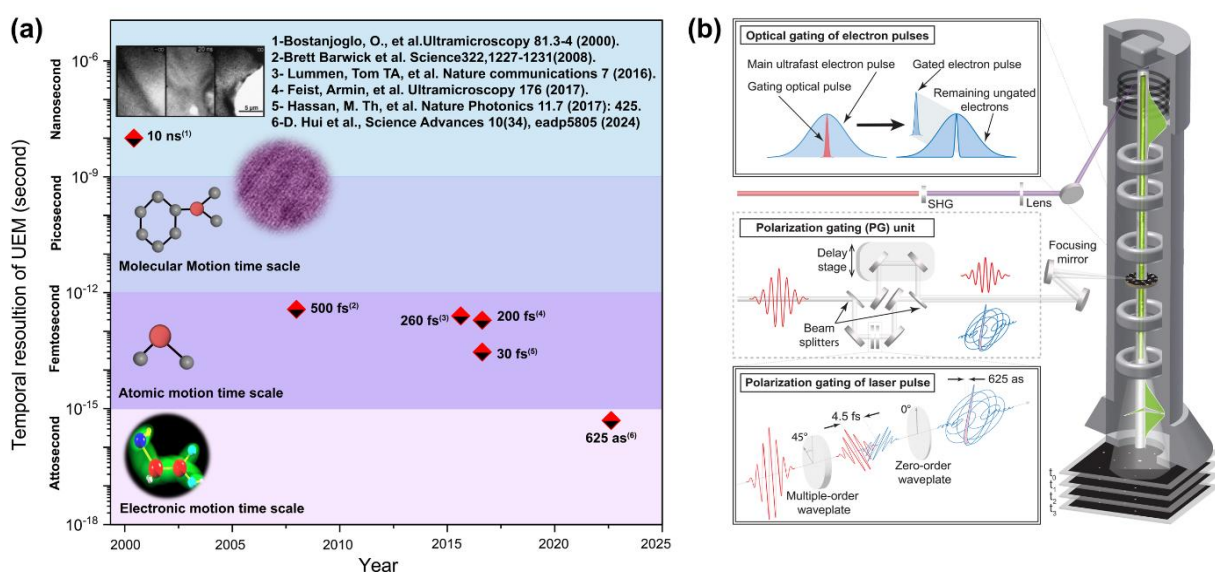

**Figure 1. (a)** *The UEM temporal resolution enhancement over the last 25 years.* **(b)** *Attomciroscopy setup for imaging the electron motion based on the optical and polarization gating approaches.*

**Attomicroscopy limitation and challenging**

Despite the promise of attomicroscopy, technical challenges remain. One major limitation is the low number of ultrafast or attosecond electron pulses, caused by the broadening of the initial electron pulse generated from the photocathode. As the electron wavepacket propagates through the microscope, Coulomb interactions—known as the space-charge effect—cause some electrons to move faster than others, elongating the pulse duration. To reduce space-charge effects, recent strategies have employed radiofrequency (RF) compression techniques, which allow the electron pulses to be temporally compressed to shorter durations at the sample plane. Another method involves using a deflection cavity inside the transmission electron microscope (TEM), which chops a continuous electron beam into pulses lasting a few hundred femtoseconds, while retaining a high electron count.

Another significant challenge lies in spatial resolution. Although attosecond time resolution has been achieved, current electron imaging systems are often limited to reciprocal space. To realize the ultimate vision of recording real-space movies of electron motion, imaging tools with sub-atomic spatial resolution are required. Scanning tunnelling microscopy (STM) offers this resolution. Over the last few decades, many groups have attempted to integrate time-resolved pump-probe methods with STM [23]. Early demonstrations achieved picosecond resolution using a single laser pulse as a pump. However, these efforts were limited by the timing jitter of synchronization circuits and the duration of the laser pulses used.

Significant progress has since been made, notably by Huber's group, which used terahertz pulses to generate tunnelling currents, pushing STM's temporal resolution into the few hundred femtosecond range [24]. This breakthrough enabled the study of laser-induced ultrafast carrier dynamics in solid-state systems. Furthermore, the spectroscopic capabilities of ultrafast STM were leveraged to visualize real-time and real-space dynamics in quantum materials. In a landmark study, Garg and Kern demonstrated STM imaging with 6-femtosecond temporal resolution using two visible laser pulses [25]. They successfully captured charge migration oscillations in perylene tetracarboxylic dianhydride molecules [26]. In principle, this resolution could be further pushed into the sub-femtosecond regime by confining the tunnelling current signal to a single half-cycle of a visible driving pulse.

**Attosecond scanning tunnelling microscope (Quantum attomicroscopy)**

In addition to attomicroscopy and cryo-attomicroscopy, we are developing the attosecond scanning tunnelling microscope. This system generates tunnelling currents from the STM tip through light-induced quantum tunnelling, driven by attosecond laser pulses (see Fig. 2a). We refer to this technique as quantum attomicroscopy (Q-attomicroscopy). The resulting current pulse lasts only a few hundred attoseconds, enabling electron motion imaging with both attosecond temporal and angstrom-scale spatial resolution.

In this setup, STM images are collected as a function of time delay between the tunnelling current and a pump pulse. The sample is scanned across spatial dimensions as well. This technique will yield both a static image and a dynamic "movie" of electron motion, directly visualized in real space and time.

These new attomicroscopy imaging tools—central to the core facilities of our Attomicroscopy Quantum Imaging Centre (see Fig. 2b)—open exciting opportunities for global, multidisciplinary collaboration. They enable a wide range of attosecond imaging applications across diverse scientific fields, including:

**(i) Quantum Physics:** Visualize laser-induced quantum motion in electron wave packets, address fundamental questions about their evolution, and explore the transition between classical and quantum behaviour. These capabilities offer unprecedented insights into the dynamic processes of quantum systems.

**(ii) Quantum Biology:** Observe quantum tunnelling in DNA, deepening our understanding of DNA-protein interactions, drug mechanisms, and DNA repair processes. Investigate charge transfer dynamics in biological molecules, advancing research in molecular electronics and bioinformatics. Image optogenetic processes to gain a deeper understanding of neuronal activity, potentially aiding in the development of treatments for neurological disorders.

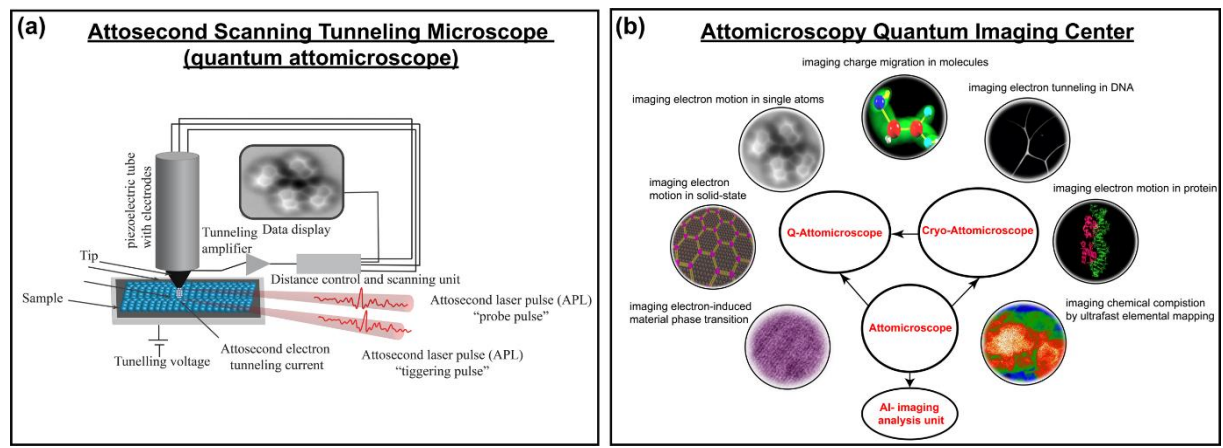

**Figure 2. (a)** Schematic of attosecond Scanning tunnelling electron microscope (Q-attomicroscope). **(b)** the structure and the potential applications at the attomicroscopy quantum imaging center.

**(iii) Quantum Chemistry:** Capture chemical interactions with unmatched spatial and temporal resolution. Study electron motion in organic solar cells and solid-state batteries, facilitating the development of efficient energy solutions. Manipulate and image charge migration in photochemical reactions, supporting the design of novel pharmaceuticals.

**(iv) Quantum Materials:** Examine electronic structures under varying conditions to advance our knowledge of quantum phase transitions, superconductivity, and electron spin dynamics. These insights will drive innovation in ultrafast optoelectronics, which are critical for the future of information technology, artificial intelligence, and quantum computing.

**Conclusion**

Imaging electron motion and electronic structure dynamics on their native attosecond timescale is crucial for uncovering deeper insights into quantum phenomena, material properties, and chemical reactivity. Real-time visualization of ultrafast processes brings science closer to fulfilling a long-standing dream: the ability to control electron motion—and thereby, matter—at will. This level of control could revolutionize our capacity to engineer chemical reactions, tailor material properties, generate ultrafast currents, investigate quantum effects, and resolve the 3D electronic structure of biological systems such as DNA and proteins.

In recent years, major strides have been made in developing attosecond imaging technologies, including attosecond X-ray free-electron laser imaging, attomicroscopy, and ultrafast scanning electron microscopy. When these tools are paired with next-generation machine learning and artificial intelligence technologies—like those being developed at the Ultrafast Quantum

Imaging Centre—their potential expands significantly. They could enable petahertz-speed electronics, light-triggered superconductivity, quantum process control in biological systems, and the long-sought ability to watch and direct chemical reactions in real time. Attosecond science is not only revealing the quantum world—it is beginning to shape it.

### Acknowledgements

M. Th. Hassan would like to thank the Gordon and Betty Moore Foundation Grant (GBMF 11476) and the Air Force Office of Scientific Research (award number FA9550-22-1-0494) for funding the attomicroscopy quantum imaging centre facilities.

### References

[1]   P. Corkum and F. Krausz, Nat. Phys. **3**, 381 (2007).
[2]   M. Lewenstein, P. Balcou, M. Y. Ivanov, A. L'huillier, and P. B. Corkum, Phys. Rev. A **49**, 2117 (1994).
[3]   M. Nisoli and G. Sansone, Prog. Quantum. Electron. **33**, 17 (2009).
[4]   J. Li, J. Lu, A. Chew, S. Han, J. Li, Y. Wu, H. Wang, S. Ghimire, and Z. Chang, Nature Communications **11**, 1 (2020).
[5]   A. L. Cavalieri *et al.*, Nature **449**, 1029 (2007).
[6]   P. Eckle, M. Smolarski, P. Schlup, J. Biegert, A. Staudte, M. Schöffler, H. G. Muller, R. Dörner, and U. Keller, Nat. Phys. **4**, 565 (2008).
[7]   E. Goulielmakis *et al.*, Nature **466**, 739 (2010).
[8]   F. Calegari, G. Sansone, S. Stagira, C. Vozzi, and M. Nisoli, J Phys. B At. Mol. Opt. Phys. **49**, 062001 (2016).
[9]   M. T. Hassan, ACS Photonics **11**, 334 (2024).
[10]  D. Hui, H. Alqattan, S. Zhang, V. Pervak, E. Chowdhury, and M. T. Hassan, Science Advances **9**, eadf1015 (2023).
[11]  C. Heide, P. D. Keathley, and M. F. Kling, Nature Reviews Physics  (2024).
[12]  F. Calegari and F. Martin, Communications Chemistry **6**, 184 (2023).
[13]  M. T. Hassan, Physics Today **77** 38 (2024).
[14]  D. Hui, H. Alqattan, M. Sennary, N. V. Golubev, and M. T. Hassan, Science Advances **10**, eadp5805 (2024).
[15]  O. F. Mohammed, D.-S. Yang, S. K. Pal, and A. H. Zewail, Journal of the American Chemical Society **133**, 7708 (2011).
[16]  A. H. Zewail, Science **328**, 187 (2010).
[17]  D. Nabben, J. Kuttruff, L. Stolz, A. Ryabov, and P. Baum, Nature, 1 (2023).
[18]  J. H. Gaida, H. Lourenço-Martins, M. Sivis, T. Rittmann, A. Feist, F. J. García de Abajo, and C. Ropers, Nat. Photon., 1 (2024).
[19]  D. Hui, H. Alqattan, M. Sennary, N. V. Golubev, and M. T. Hassan, arXiv preprint arXiv:2502.06592  (2025).
[20]  M. Sennary, J. Shah, M. Yuan, A. Mahjoub, V. Pervak, N. Golubev, and M. Hassan, 2024.
[21]  S. Weathersby *et al.*, Review of Scientific Instruments **86** (2015).
[22]  T. Driver *et al.*, Nature **632**, 762 (2024).
[23]  A. van Houselt and H. J. Zandvliet, Rev. Mod. Phys. **82**, 1593 (2010).
[24]  T. Siday *et al.*, Nature **629**, 329 (2024).
[25]  M. Garg and K. Kern, Science **367**, 411 (2020).
[26]  M. Garg, A. Martin-Jimenez, M. Pisarra, Y. Luo, F. Martín, and K. Kern, Nat. Photon., 1 (2021)

## 34. Attosecond Quantum Optics

**Giulio Vampa[1]\*, David N. Purschke[1,2], Thomas Brabec[3]**

[1]Joint Attosecond Science Laboratory, University of Ottawa & National Research Council of Canada, Ottawa K1A0R6, Ontario, Canada
[2]Laboratory for Laser Energetics, University of Rochester, Rochester 14623, New York, USA
[3]Department of Physics, University of Ottawa, Ottawa K1N 6N5, Ontario, Canada

Giulio.Vampa@nrc-cnrc.gc.ca

**Status**

Attosecond science is rooted in the development of high-intensity lasers, where billions of photons work in unison to create oscillating electric fields with a well-defined amplitude and phase, as illustrated in Figure 1a. While the matter system has been considered as a quantum entity, and exquisitely quantum-mechanical features such as entanglement between electrons and ions have been measured on attosecond timescale [1-3], the non-classical aspects of light have been all but forgotten. Nevertheless, in recent years, driven both by curiosity and by the potential for combining attosecond precision with the so-called "quantum advantage" in measurement, the quantum-optical nature of intense light-matter interaction has come into focus. This new field of attosecond quantum optics (AQO) is poised to bring a *time-centered* perspective on quantum electrodynamics, a perspective that is – at first sight – estranged from the more typical *photon* picture. A similar motivation applied to multi-terahertz waves has recently pioneered the time-domain measurement of quantum states of light [4].

Early experiments by Tsatrafyllis et al. [5] demonstrated that an infrared driver upon emission of a high-harmonic photon from a Xe gas exhibited a nearly discrete distribution of the pulse energy that resembled that of the high-harmonic spectrum. Such distribution is unexpected because any modifications of the pump were always assumed to elicit classical changes to the light state. Adding quantum state tomography to this conditional – or *projective* – measurement seemingly suggested that the spent infrared driver is in a quantum superposition of two coherent states—a Schrödinger cat state—composed of the initial state of the laser and that attenuated due to the emission of high harmonics [6]. For the first time, quintessential quantum-optical metrology inspired an attosecond experiment.

Following these first ripples, researchers at Technion developed a theoretical framework for high-harmonic emission with quantized fields [7]. They predicted that emission from gases driven by an intense coherent laser can have mildly super-Poissonian statistics and show squeezing of the field fluctuations below the shot noise, which indicate non-classical correlations within the emitted harmonics. These results were later extended using a *phase-space* approach to AQO (see Fig. 1) based on the P- or Husimi distributions. This approach showed that photon shot-noise of non-coherent drivers, such as thermal or squeezed states, dresses the accelerated electrons in the continuum and results in a dramatic extension of the harmonic cut-off [8] and added an effective *photon-statistics* force that changes the timing of harmonic emission [9]. Fang et al. studied strong-field ionization by a squeezed vacuum field with a similar approach [10]. Experimental work in Erlangen utilized bright squeezed vacuum (BSV) to trigger multi-photon electron emission from metallic nanotips [11] and drive high-harmonic generation in a dielectric [12]. Post-selection of particular BSV energies collapses the photoelectron energy distribution, validating theoretical predictions based on the phase-space approach.

The role of bipartite entanglement came into sharper focus thanks to experiments that measured intensity cross-correlations between low-order harmonics from semiconductors. Intensity cross-correlations, i.e. the Hanbury-Brown-Twiss experiment, measure whether the photons arrive randomly distributed in time or whether their time of arrival is correlated, e.g. they prefer to arrive together (bunched) or separated (anti-bunched). Thiedel et al. [13] compared the intensity cross-correlation between harmonics 3 and 5 with the cross-correlation from each harmonic separately. Cauchy-Schwarz inequalities relating these quantities can be violated in the presence of multi-partite entanglement, and such violation was measured in the experiment.

Yet another approach to AQO involves utilizing high-harmonic generation driven by a classical coherent field to transduce quantum optical properties of a perturbing beam, such as a BSV, to the high-order harmonics [14]. The frequency mixing of the two fields results in the emission of high-harmonic sidebands with super-Poissonian statistics. A theoretical derivation of the experiment indicates that multi-partite entanglement is created between the sidebands and the spent perturbation, which can be used in a projective measurement to generate Schrödinger cat states at the high harmonics or photon-added squeezed vacuum states at the perturbation [15]. This approach is potentially useful to transduce any truly quantum-optical beam with proven quantum correlations, which tend to have only few photons, or anyways not of sufficient intensity to drive strong-field dynamics.

Theory also points to new ways of controlling quantum optical aspects of high-harmonic generation. For example, the role of *electronic correlations* in AQO has been explored in the Dicke-state responsible for super-radiant emission [16] and in a Mott-insulator [17]. In both cases, the matter correlations manifest as non-classical features in the emitted high-order harmonics, pointing towards new ways to probe electronic correlations with AQO. Finally, it has been suggested that intense *resonant* nonlinear interactions, coupled with cavities to enhance the light-matter interaction, intrinsically leads to quantum correlations between harmonics [18].

**Current and future challenges**

At the heart of quantum optics are non-classical correlations, such as squeezing and entanglement, between various sub-systems involved in the interaction, be it different optical modes such as harmonics and perturbing and driving fields, or electrons in matter. The truly quantum optical features measured in experiments so far [6, 13] have yet to be confirmed by independent groups, and other approaches have traced a path towards achieving quantum-valid results but have yet to achieve that goal. In addition, most of the theoretical work is yet to be tested in the laboratory.

The purity of the quantum-optical state, and therefore ability to measure and exploit entanglement, can be compromised if incomplete information about the system is collected. This is an experimental challenge, as entanglement is created across vastly different frequencies. Moreover, the large number of photons in the driving field is at odds with the small outcome of the nonlinear interaction, such as weak ionization levels or low number of high-harmonic photons. This ever-present imbalance is both the intriguing oxymoron within AQO and a challenge that must be overcome. For example, small quantum-optical modifications to the intense laser field are extremely difficult to measure without introducing loss, which washes out any quantum feature. Upconverting a quantum-optical perturbation with high quantum efficiency would be a possible solution.

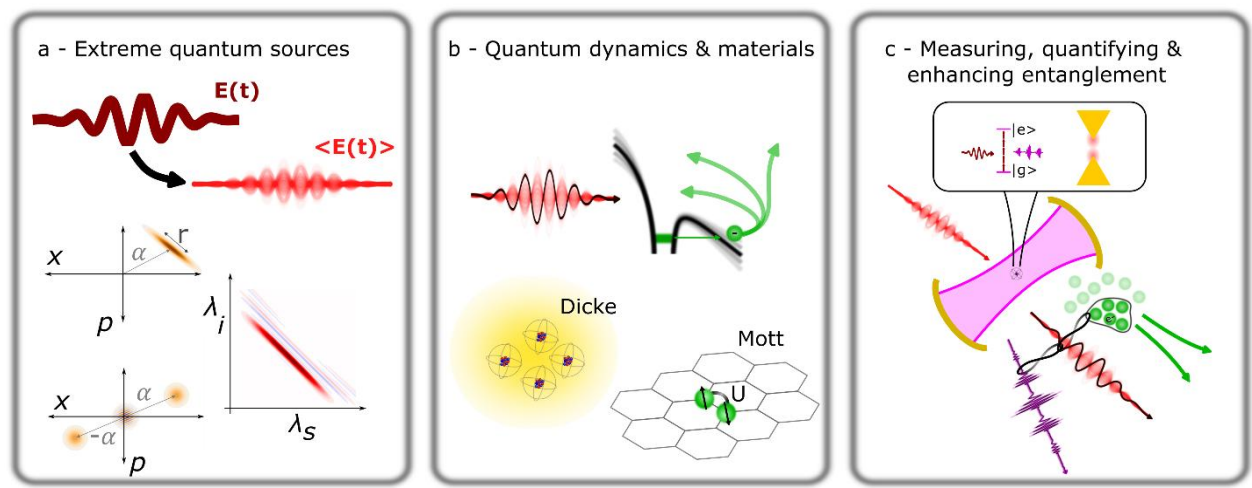

**Figure 1.** *Three areas of development in AQO. a – Extreme quantum sources. Top: The familiar classical description of light used in attosecond science, a coherent state (dark red), with a well-defined amplitude and phase. In quantum optics, more exotic states of light, such as bright squeezed vacuum (bright red), can be constructed and generated in the lab. Here, the field is visualized as an ensemble of coherent states, which is most appropriate for understanding the dynamics of intense light-matter interaction. Bottom: Phase-space picture of states commonly encountered in AQO: top left, Wigner functions of a generalized squeezed state with displacement α and squeeze parameter r (squeezed vacuum is the special case with vanishing α), a Schrodinger cat state (bottom left), and a two-mode squeezed state. b – Quantum dynamics & materials. Top: Illustration of a quantum optical strong field process, where uncertainty in the energy landscape (tunnel barrier) and electron dynamics is correlated to quantum fluctuations in the laser field. Bottom: Correlated matter systems that are predicted to yield non-classical high-harmonic generation, a Dicke state (left) and a Mott insulator (right). c – Measuring, quantifying and enhancing entanglement. The observables of an AQO experiment lie, i.e., the generated harmonics (purple), electrons (green) as well as the spent driver & perturbing beams (red), pictured bottom right, are predicted to be entangled in some cases. Bringing in and adapting new tools from quantum optics will be key to measuring entanglement and combining AQO with tools such as resonance and plasmonics from strong-field photonics will enable enhancing and sculpting non-classical light generated by HHG.*

**Advances in science and technology to meet challenges.**

In Figure 1, we highlight three inter-related focus areas required for advancement in AQO. First, technological advances in extreme quantum sources. Although advances in the generation of bright squeezed vacuum were instrumental for these early demonstrations, new innovations are required for maintaining and verifying their quantum properties at extreme brightness [19]. For the latter, attosecond metrology could provide a fresh approach to quantum measurement, for example, by revealing temporal mode structure and measuring squeezing in the time domain. Furthermore, new approaches to generating extreme quantum light [20] will enable more flexibility and help drive innovation.

Second, the study of the *fundamental physics* of AQO. Theory and experiments should focus on elucidating uniquely non-classical effects as opposed to aspects of AQO that could be consistent with an ensemble of coherent states. Simultaneously, experimentalists must verify predictions that correlations in matter are transferred onto the emitted high-order harmonics, which could render high-harmonic spectroscopy an invaluable tool for materials scientists.

Third, theory predicts that the observables of an AQO experiment (Fig. 1c) are entangled. While some pioneering experiments suggest that this entanglement has already been verified, further development of the methodologies and technologies to quantify non-classicality are needed. Meeting these challenges requires, more broadly, familiarizing ourselves with the intricacies of understanding and measuring correlated photons and electrons, a skill that – we predict – will become front and central in the future of attosecond science. Furthermore, emerging tools such as photon number-resolving detectors can be adapted to the ultraviolet range and are poised to facilitate measurement of entanglement across many modes in AQO. Finally, researchers must find new ways to enhance the non-classical aspects of light generated via high-harmonic generation, for example, by enhancing the efficiency of harmonic up-conversion or using cavity enhancement and resonances to tailor the state of emitted light.

## Concluding remarks

Thanks to the pioneering works demonstrated to date, attosecond quantum optics is developing into a truly exciting cross-discipline research field with many open questions and opportunities. The payout of such exploration could be major: generating quantum-optical states at the short high-harmonic wavelengths, creation of entanglement across vastly different spectral ranges and between optical and matter systems, to name a few, could enable element and chemically specific molecular and solid-state spectroscopy that beats the shot noise limit. Furthermore, the innate randomness of quantum light could be integrated in Lightwave electronics.

## References

[1]   H. Niikura, F. Légaré, R. Hasbani, M. Y. Ivanov, D. M. Villeneuve, and P. B. Corkum 2003 Probing molecular dynamics with attosecond resolution using correlated wave packet pairs, Nature *421*(6925) 826-829.

[2]   L. M. Koll, L. Maikowski, L. Drescher, T. Witting, and M. J. Vrakking 2022 Experimental control of quantum-mechanical entanglement in an attosecond pump-probe experiment, Physical Review Letters *128*(4), 043201.

[3]   H. Laurell, et al. 2025. Measuring the quantum state of photoelectrons, Nature Photonics *19*(4), 352-357.

[4]   C. Riek, D. V. Seletskiy, A. S. Moskalenko, J. F. Schmidt, P. Krauspe, S. Eckart, S. Eggert, G. Burkard, and A. Leitenstorfer 2015 Direct sampling of electric-field vacuum fluctuations, Science 350 6259 420-423

[5]   N. Tsatrafyllis, I. K. Kominis, I. A. Gonoskov & P. Tzallas 2017 High-order harmonics measured by the photon statistics of the infrared driving-field exiting the atomic medium Nature Communications 6479 15170

[6]   M. Lewenstein, M. F. Ciappina, E. Pisanty, J. Rivera-Dean, P. Stammer, Th. Lamprou, P. Tzallas 2021 Generation of optical Schrodinger cat states in intense laser-matter interactions Nature Physics 17 1104-1108

[7]   A. Gorlach, O. Neufeld, N. Rivera, O. Cohen, I. Kaminer 2020 The quantum-optical nature of high harmonic generation Nature Communications 11 4598

[8]   A. Gorlach, M. E. Tzur, M. Birk, M. Kruger, N. Rivera, O. Cohen, I. Kaminer 2023 High-harmonic generation driven by quantum light Nature Physics 19 1689-1696

[9]   M. E. Tzur, M. Birk, A. Gorlach, M. Kruger, I. Kaminer, O. Cohen, 2023 Photon-statistics force in ultrafast electron dynamics Nature Photonics 17 501-509

[10]  Y. Fang, F.-X. Sun, Q. He, Y. Liu 2023 Strong-field ionization of hydrogen atoms with quantum light Physical Review Letters 130 253201

[11]  J. Heimerl, A. Mikhaylov, S. Meier, H. Hollerer, I. Kaminer, M. Chekhova, P. Hommelhoff 2024 Multiphoton electron emission with non-classical light Nature Physics 20 945-950

[12]  A. Rasputnyi, Z. Chen, M. Birk, O. Cohen, I. Kaminer, M. Kruger, D. Seletskiy, M. Chekhova, F. Tani 2024 High-harmonic generation by a bright squeezed vacuum

[13]  D. Theidel, V. Cotte, R. Sondenheimer, V. Shiriaeva, M. Froidevaux, V. Severin, A. Merdji-Larue, P. Mosel, S. Froehlich, K.-A. Weber, U. Morgner, M. Kovacev, J. Biegert, H. Merdji 2024 Evidence of the quantum optical nature of high-harmonic generation PRX Quantum 5 040319

[14]  S. Lemieux, S. A., Jalil, D. N. Purschke, N. Boroumand, TJ Hammond, D. Villeneuve, A. Naumov, T. Brabec, G. Vampa 2025 Photon bunching in high-harmonic emission controlled by quantum light Nature Photonics 19 767-771

[15]  N. Boroumand, A. Thorpe, G. Bart, L. Wang, D. N. Purschke, G. Vampa, T. Brabec 2025 Quantum engineering of high harmonic generation arXiv:2505.22536

[16]  A. Pizzi, A. Gorlach, N. Rivera, A. Nunnenkamp, I. Kaminer 2022 Light emission from strongly driven many-body systems Nature Physics 19 551-561

[17]  C. S. Lange, T. Hansen, L. B. Madsen 2024 Electron-correlation-induced nonclassicality of light from high-order harmonic generation

[18]  S. Yi, N. D. Klimkin, G. G. Brown, O. Smirnova, S. Patchkovskii, I. Babushkin, M. Ivanov 2025 Generation of massively entangled bright states of light during harmonic generation in resonant media Physical Review X 15 1

[19]  S. Panahiyan, C. S. Munoz, M. Chekhova, F. Schlawin 2023 Nonlinear interferometry for quantum-enhanced measurements of multiphoton absorption

[20]  Th. Lamprou, J. Rivera-Dean, P. Stammer, M. Lewenstein, P. Tzallas 2025 Nonlinear optics using intense optical coherent state supersoitions Physical Review Letters 124 013601

## 35. Attosecond dynamics of quantum entanglement and decoherence in photoemission

**Charles Bourassin-Bouchet[1], David Busto[2], Jérémie Caillat[3] and Pascal Salières[4]\***

[1] Université Paris-Saclay, CNRS, Institut d'Optique Graduate School, Laboratoire Charles Fabry, 91127 Palaiseau, France
[2] Department of Physics, Lund University, Lund, Sweden
[3] Sorbonne Université, CNRS, Laboratoire de Chimie Physique-Matière et Rayonnement, LCPMR, 75005 Paris, France
[4] Université Paris-Saclay, CEA, LIDYL, 91191 Gif-sur-Yvette, France

pascal.salieres@cea.fr

### Status

Photoemission, which produces a multi-partite ion-electron quantum system, is an efficient playground for preparing, studying and controlling entangled states of massive particles. Entanglement between, e.g., the electron and/or ion degrees of freedom (DoFs) induces decoherence in the measurement of one sub-system if the other one is not simultaneously measured. For instance, static experiments using synchrotron radiation have shown how electron-electron entanglement in double ionization can smear out interference fringes in the angular distributions [1].

Yet, entanglement -as well as decoherence- have long been disregarded, or at least overlooked, when investigating photoemission from a time-dependent perspective with the tools of attosecond science. Historically, this field built upon the excitation and control of *coherent* electron/ion wavepackets, and an assumption of full coherence was –and still is– underlying most interpretations of interferometric attosecond experiments.

While some coherence issues were investigated in the early days of attosecond physics [2], they later became central in theoretical studies of attosecond charge migration. The sudden removal of an electron from a molecule launches a hole wavepacket that may migrate in just a few 100's attoseconds, opening the prospect of charge-directed reactivity [3]. However, electron-electron interactions [4] and coupling of the electronic and nuclear DoFs [5] lead to an enhanced entanglement between the photoelectron and the parent ion, resulting in a reduced coherence of the ionic hole wavepacket.

In experiments, indirect signatures of entanglement — loss of visibility in spectral interferences — were recently investigated and attributed to couplings between the electronic and vibrational DoFs [6, 7], the radial and angular electronic DoFs [8], or the electronic and dressed ionic DoFs [9]. No complete characterization of the corresponding quantum state, i.e., no density matrix *measurement*, was performed, except for an early pioneering study [10] that accessed the *ion* Reduced Density Matrix (RDM) of a spin-orbit wavepacket via attosecond transient absorption measurements of strong-field ionized krypton.

Much later, the first measurements of *photoelectron* RDMs were carried out by quantum tomography *via* two different schemes, both using infrared (IR) probe fields. They are reminiscent of the standard interferometric schemes of attosecond metrology, but they explicitly take into account and exploit decoherence as a source of physical information rather than a technical limitation. The mixed-FROG protocol [11] has enabled RDM measurement over a broad 10-eV bandwidth compatible with attosecond resolution, cf. figure1(b). It evidenced high instrumental decoherence in standard conditions**.** The KRAKEN protocol [12] has provided direct access to RDMs over a ~0.2-eV bandwidth compatible with a ~10-fs temporal resolution, cf. Figure 1(a). It was applied to the characterization of quantum decoherence due to partially resolved spin-orbit splitting in argon photoemission.

In *all* the above-mentioned experimental studies, only the final -asymptotic- state was observed/measured, i.e., the dynamics of entanglement/decoherence is not accessed. Yet, the experimental reconstruction of these dynamics is of high fundamental and applied interest. This would open a new window on matter, as entanglement constitutes an 'observable' of the interactions and correlations intrinsic to the systems under study, from the physical interactions (electronic and vibronic couplings [13]) to the fundamental laws (conservation of energy and angular momentum) governing its structure and dynamics. Moreover, measuring and controlling the speed of entanglement/decoherence could open up prospects for technological applications in quantum information and detection.

**Current and future challenges**

**Challenge 1: Measurement of the attosecond dynamics of entanglement**. Obviously, this is not a straightforward task. In quantum optics, the fastest dynamics that have been measured are on the ps timescale in weakly-coupled Rydberg-excited atoms. Since the strength of the interaction that creates the entanglement determines the speed of the latter ('quantum speed' limit), the strong electromagnetic forces between electrons and nucleus correspond to attosecond entanglement dynamics. How can this be accessed? A first strategy is to rely on time-domain analyses of the asymptotic quantum state characterized in the spectral domain (amplitude and phase of the RDM) [14], see figure 1(c). This involves decomposing the final quantum state in coherent vectors (e.g., Schmidt vectors) [15] that can be submitted to Fourier-based algorithms initially established for fully coherent dynamics (see, e.g., [16]). A second strategy is to directly probe in time the entanglement dynamics through pump-probe experiments. This will require advanced protocols able to capture the transient density matrix. This could be further generalized to perform the complete characterization of the quantum dynamics of the system via quantum process tomography (QPT) [17]. Instead of focusing on the temporal evolution of a specific initial quantum state, QPT aims at characterizing the process matrix that governs the evolution of an arbitrary initial state, providing the most general description of the process. This would require preparing different initial states and measuring their temporal evolutions.

**Challenge 2: Control of the attosecond dynamics of entanglement**. The coupling with the ionizing field during photoemission is a direct route to temporal control of coherence and entanglement in the photoionized systems. A straightforward possibility is to control the intensity and spectro-temporal properties of the ionizing light [6, 9, 12, 18]. Another way of influencing entanglement and its evolution is provided by adequate preparation of the system prior to photoemission. For instance, the effectiveness of control by prior coherent electronic, vibrational or rotational excitation will reveal the influence of the relative phase of the different states on entanglement dynamics.

**Challenge 3: Resolving photoemission induced by non-classical light.** The recent development of ultrafast sources of non-classical light, from massively-entangled IR fields to multimode squeezed XUV light [19], raises a number of burning questions. How do the properties of non-classical light influence/transfer to the ion-electron system and with which dynamics? Can the photoemission quantum tomography schemes be used to characterize the ionizing non-classical light? Can we produce novel ultrafast light-matter entangled states through photoemission?

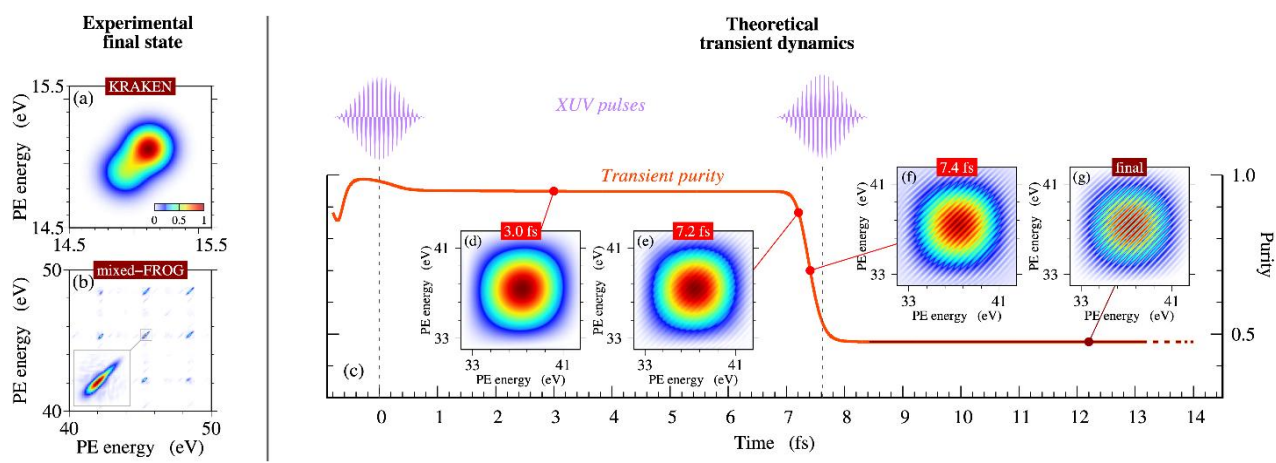

**Figure 1.** *Entangled photoemission dynamics from the photoelectron (PE) perspective. Left panels: Final PE RDMs measured with: (a) the KRAKEN [12] protocol in experimental narrowband ionization of Ar overlapping the 1/2 and 3/2 spin-orbit channels; and (b) the mixed-FROG [11] protocol in broadband ionization of Ne. The inset in (b) is a close-up of the RDM's central structure (45.0-45.8 eV) evidencing instrumental decoherence. Right panel: Transient PE dynamics in simulated broadband molecular photoemission of a model H2 by a composite XUV pulse, with a ~10-eV bandwidth concealing the open vibrational channels [14]. Two identical time-shifted pulses (violet) induce channel-dependent Ramsay-like interferences, which results in delay-dependent PE-ion entanglement [6]. The field-driven entanglement build-up is evidenced by the evolution of the transient purity computed at each time of the simulation (c) and comprehensively revealed by the partial loss of coherence appearing in the corresponding transient RDM shown at four representative times (d-g). Only the moduli of the normalized complex-valued RDMs are shown, see colorbar in panel (a).*

**Advances in science and technology to meet challenges**

**Development of a universal attosecond quantum state tomography scheme**. This scheme should go beyond the technical and conceptual limitations of the current protocols. This involves at least three requirements:

i) To identify and overcome the various sources of instrumental decoherence. Such decoherence should be minimized or at least fully characterized in order to take it into account in the data analysis (calibration).

ii) To resolve the maximum number of DoFs of the system under study, i.e., to perform simultaneous measurements of the ion and/or electron DoFs. For instance, combining electron momentum imaging with attosecond spectroscopy allows to identify couplings between radial and angular electronic DoFs and the corresponding decoherence [8]. The recent development of high rep-rate post-compressed Ytterbium lasers also opens wide prospects for electron-ion coincidences.

iii) To develop reconstruction algorithms as general as possible. In Mixed-FROG, the assumptions underlying the inversion algorithm (strong-field and central momentum approximations) prevent its use on energies close to the ionization threshold (E < 10 eV) or over too wide spectral ranges (DE/E >0.5). In KRAKEN, the distortions induced by the continuum-continuum transitions need to be accounted for in the analysis.

**Certification of entanglement**. Current approaches rely on indirect signatures such as reduced interference contrast or mixedness in one subsystem, assuming a pure global state. However, robust certification of entanglement requires measurements on both subsystems in multiple, non-commuting bases, without assuming purity. Bell tests offer a method to reveal non-local correlations and thus entanglement. A recent proposal [20], applies this to spin correlations between ion and photoelectron, showing that a Bell parameter exceeding 2 could be measured, thus certifying both entanglement and nonlocality. Yet, not all entangled states violate Bell inequalities. In such cases, entanglement witnesses can be designed to detect specific entangled states without requiring a full Bell test [15].

## Concluding remarks

The paradigm shift in attosecond science now opens the prospect of studying fundamental quantum properties such as entanglement and decoherence on the natural timescale of electron dynamics in matter. In photoemission, it provides a new 'observable' of the dynamics of correlations intrinsic to the ionized systems, from vibronic couplings to charge migration processes. The generalization of the quantum tomography protocols developed on isolated atomic and molecular systems to 'open' systems such as clusters and liquid jets promises new insights into attosecond dynamics of elastic/inelastic electron scattering and dissipation in matter. Finally, the dynamic control of quantum entanglement and decoherence, in particular of their respective speed, is another important direction as it may open new prospects for technological applications in quantum information and detection.

## Acknowledgements

This research received the financial support of the French National Research Agency through Grants No. ANR-20-CE30-0007-DECAP, ANR-22-EXLU-0002 Ultrafast-LUMA and ANR-24-RRII-0004 CEA-Audace, and of the European Union through HORIZON-MSCA-2023-DN-QU-ATTO-101168628. D.B. acknowledges support from the Knut and Alice Wallenberg Foundation through the Wallenberg Centre for Quantum Technology.

## References

[1]   Akoury D, Kreidi K, Jahnke T, Weber T, Staudte A, Schöffler M, Neumann N, Titze J, Schmidt L P H, Czasch A, Jagutzki O, Costa Fraga R A, Grisenti R E, Díez Muiño R, Cherepkov N A, Semenov S K, Ranitovic P, Cocke C L, Thompson J C, Prior M H, Belkacem A, Landers A L, Schmidt-Böcking H and Dörner R 2007 The simplest double slit: interference and entanglement in double photoionization of $H_2$ *Science* **318** 949
[2]   Smirnova O, Yakovlev V S and Scrinzi A 2003 Quantum Coherence in the Time-Resolved Auger Measurement *Phys. Rev. Lett.* **91** 253001
[3]   Breidbach J and Cederbaum L S 2003 Migration of holes: Formalism, mechanisms, and illustrative applications *J. Chem. Phys.* **118** 3983
[4]   Pabst S, Greenman L, Ho P J, Mazziotti D A and Santra R 2011 Decoherence in Attosecond Photoionization *Phys. Rev. Lett.* **106** 053003
[5]   Vacher M, Bearpark M J, Robb M A and Malhado J P 2017 Electron Dynamics upon Ionization of Polyatomic Molecules: Coupling to Quantum Nuclear Motion and Decoherence *Phys. Rev. Lett.* **118** 083001
[6]   Vrakking M J J 2021 Control of Attosecond Entanglement and Coherence *Phys. Rev. Lett*. **126** 113203
[7]   Koll L-M, Maikowski L, Drescher L, Witting T and Vrakking M J J 2022 Experimental control of quantum-mechanical entanglement in an attosecond pump-probe experiment *Phys. Rev. Lett.* **128** 043201
[8]   Busto D, Laurell H, Finkelstein-Shapiro D, Alexandridi C, Isinger M, Nandi S, Squibb R J, Turconi M, Zhong S, Arnold C L, Feifel R, Gisselbrecht M, Salières P, Pullerits T, Martín F, Argenti L and L'Huillier A 2022 Probing electronic decoherence with high-resolution attosecond photoelectron interferometry *Eur. Phys. J. D* **76** 112
[9]   Nandi S, Stenquist A, Papoulia A, Olofsson E, Badano L, Bertolino M, Busto D, Callegari C, Carlström S, Danailov M B, Demekhin P V, Di Fraia M, Eng-Johnsson P, Feifel R, Gallician G, Giannessi L, Gisselbrecht M, Manfredda M, Meyer M, Miron C, Peschel J, Plekan O, Prince K C, Squibb R J, Zangrando M, Zapata F, Zhong S and Dahlström J M. 2024 Generation of entanglement using a short-wavelength seeded free-electron laser *Science Advances* **10** eado0668
[10]  Goulielmakis E, Loh Z-H, Wirth A, Santra R, Rohringer N, Yakovlev V S, Zherebtsov S, Pfeifer T, Azzeer A M, Kling M F, Leone S R and Krausz F 2010 Real-time observation of valence electron motion *Nature* **466** 739
[11]  Bourassin-Bouchet C, Barreau L, Gruson V, Hergott J-F, Quéré F, Salières P and Ruchon T 2020 Quantifying decoherence in attosecond metrology *Phys. Rev. X* **10** 031048

[12] Laurell H, Luo S, Weissenbilder R, Ammitzböll M, Ahmed S, Söderberg H, Petersson C L M, Poulain V, Guo C, Dittel C, Finkelstein-Shapiro D, Squibb R J, Feifel R, Gisselbrecht M, Arnold C L, Buchleitner A, Lindroth E, Kockum A F, L'Huillier A and Busto D 2025 Measuring the quantum state of photoelectrons *Nat. Photon.* **19** 352

[13] Blavier M, Levine R D and Remacle F 2022 Time evolution of entanglement of electrons and nuclei and partial traces in ultrafast photochemistry *Phys. Chem. Chem. Phys.* **24** 17516

[14] Berkane M, Taïeb R, Granveau G, Salières P, Bourassin-Bouchet C, Lévêque C and Caillat J 2025 Complete retrieval of attosecond photoelectron dynamics from partially coherent states in entangled photoemission *Phys. Rev. A* **111** L041101

[15] Tichy J, Mintert F and Buchleitner A 2011 *J. Phys. B: At. Mol. Opt. Phys.* **44** 192001

[16] Gruson V, Barreau L, Jiménez-Galán Á, Risoud F, Caillat J, Maquet A, Carré B, Lepetit F, Hergott J-F, Ruchon T, Argenti L, Taïeb R, Martín F and Salières P 2016 Attosecond dynamics through a Fano resonance: Monitoring the birth of a photoelectron *Science* **354** 734

[17] Yuen-Zhou J, Krich J J, Mohseni M and Aspuru-Guzik A 2011 Quantum state and process tomography of energy transfer systems via ultrafast spectroscopy *Proc. Natl. Acad. Sci. U.S.A.* **108** 17615

[18] Carlstrom S, Mauritsson J, Schafer K J, L'Huillier A and Gisselbrecht M 2018 Quantum coherence in photo-ionisation with tailored XUV pulses *J. Phys. B: At. Mol. Opt. Phys.* **51** 015201

[19] Cruz-Dominguez L, Dey D, Freibert A and Stammer P 2024 Quantum phenomena in attosecond science *Nat. Rev. Phys.* **6** 691

[20] Ruberti M, Averbukh V and Mintert F 2024 Bell test of quantum entanglement in attosecond photoionization *Phys. Rev. X* **14** 041042

## 36. Entanglement in Attosecond Science

**Dong Hyuk Ko[1] and Paul B Corkum[1]\***

[1] Joint Attosecond Science Laboratory, University of Ottawa and National Research Council of Canada, Ottawa, Canada

pcorkum@uottawa.ca

**The background**

In 1964 Keldish published his famous paper [1] on tunnel ionization and the subsequent motion of the electron — later known as above-threshold ionization (ATI) [2]. Influenced by plasma physics [3], I recalculated the ATI spectrum as the first two steps of the semiclassical three-step model of high-harmonic generation [4, 5]. Of course, scientists used many-electron atoms and even excited molecules [6]. However, the three-step model and the single-active electron approximation together with phase matching [7] had become the fundamental principle for generating attosecond pulses using high harmonic radiations [8]. Adding a quantum source as a perturbation on the signal's nonlinear growth during propagation seemed like the best path to quantum optics in the VUV [9] although post-selection was always silently present.

One of the most unusual aspects of modern quantum mechanics is entanglement. The first serious problem with single-active electron approximation for attosecond science appeared for transparent solids [10, 11] where the motion of both electrons and holes (i.e. the representative of all other electrons) is tracked. In fact, holes are as responsible for the cut-off frequency of the solid-state harmonics as electrons [12], implying that multiple active electrons were automatically taken into account for solid-state theory.

Although the underlying process of recollision provided sufficient time resolution to generate attosecond pulses, for several decades, multi-electron dynamics was only addressed, using already generated attosecond pulses. Xenon, used for high harmonic generation as early as 1990 [13], provides a chance to use recollision directly. Two forms of multi-electron dynamics lead to entangled wave packets in xenon. In one, ionizing the $5p$ electron leads to spin-dynamic (no spatial charge) during the hole's interaction with the core [14]. As shown in Fig. 1 for the experimental result, the recollision electron finds recombination blocked by the Pauli principle at the key time of 3.4 fs. The spin wave packet is measured by the recollision wave packet. In the second, the ionized electron simultaneously comes from two ion states, thereby creating a moving wave packet. This allows one side of the $5p$ wave function to ionize while leaving the other intact [15].

Thus, in xenon any recollision is entangled with a hole wave packet that must make its way in the core. In xenon the core contains a plasmon. We assume that (1) a plasmon has all its energy in the field at one time, (2) the volume of the plasmon is the volume of wave function of a $4d$ electron and (3) the energy of a single plasmon is 80 eV. With these assumptions, the field is ~480 V/Å (for only 1 quantum! This is approximately 10 times the voltage of 1 electron at r=1 atomic unit.) Two types of hole wave packets are responsible for the modulation in the Fig. 1. They co-exist in post selected xenon and can be followed by time-dependent density functional theory [16] as stationary or moving holes. Most interesting is the spin-orbit hole. It has a well-known spatial, temporal and intensity structure. It splits over the core while hugging it. It covers ½ of its orbit in ~2.3 fs from tunnelling to recollision and it produces the high-frequency component of the spectrum. A 1.8-mm laser pulse reaches its peak field just before the wave packet reaches ½ way around the core. The laser field allows xenon to break the 3.17 $U_p$ + IP cut-off low at this time.

**Current and future challenges**

Attosecond pulses are used to observe time-resolved ultrafast electronic motion in many molecular systems [17]. Entangled wave packets can lead us to a different perspective. Tunnelling often forms a recollision electron that is entangled with a hole wave packet. Applying the perturbed trajectory method [18], the entangled recollision wave packet measures the hole dynamics in the presence of the field!

Why, you might ask, do we care? It has been suggested that all small molecules have a very characteristic fast response caused by other electrons rushing to take place of the departing (or returning) electron [19]. This is like the plasmon that we have just seen where electrons respond to internal fields. Therefore, both general and molecule specific information might hide in the attosecond optical signal that we measure.

But there are practical reasons to care as well. With two electrons to contribute, the 3.17 $U_p$ + IP limit on the emission frequency will be broken. In fact, it already has been!

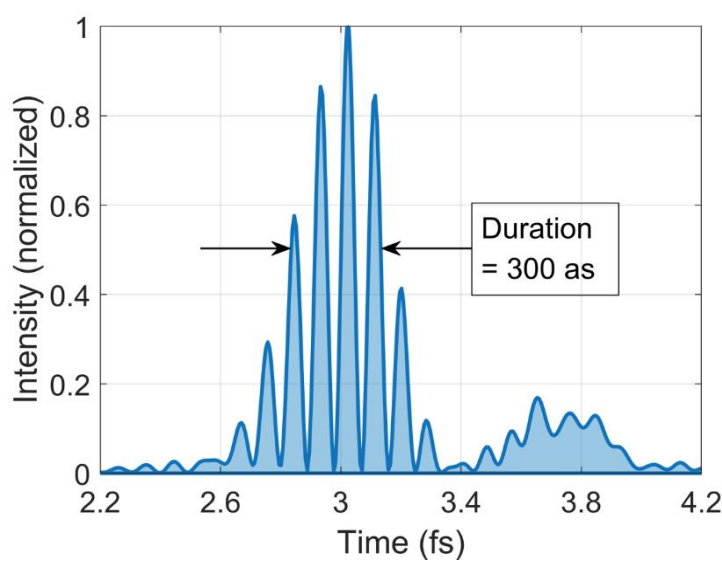

*Figure 1. Time-dependent attosecond pulse produced from xenon using 1.8-μm laser fields.*

**Advances needed to meet the challenges**

The greatest challenge is the limited time-window of observation that high-harmonic emission allows. If we are restricted to short trajectories, then the effective observation range is ~⅔ of an optical period of the driving laser. Xenon required ~1.8 mm. For molecular systems we should use as long a wavelength as feasible. Readers, see the section by Professor Z. Chang for the progress on infrared drivers [20].

**The implications**

Nonlinear optics through parametric down conversion and post selection is the basis of almost all quantum optics. Recollision physics is nonlinear optics at its extreme and attosecond pulse generation is a superb high bandwidth filter through which to post select dynamics that returns the system to its initial state. Like parametric down conversion, other pathways exist but are not selected. Furthermore, a recollision electron cannot lose its coherence in a low-density atomic gas, nor can dynamics in the core destroy coherence. Even the strong laser field will not destroy coherence.

With the underlying wave packet dynamics entangled and selected, where will quantum technology take us? It allows us to break the $3U_p$ limit. For the same intensity, our simulations show that it will give us a ~50% augmentation of the cut-off. With much shorter trajectories, it may also work for better phase matching. Proton transfer is like a spin-orbit wave packet. Therefore, time-resolving proton transfer may become feasible using attosecond pulse generation just as time-resolved spin-orbit wave packets are.

## Acknowledgements

We acknowledge financial support from the US: AFOSR: (Award no. FA9550-16-1-0109) and from the Canada Foundation for Innovation, Canada's NSERC, and NRC. We also acknowledge discussions with C. Zhang, G. G. Brown, and K. Jana.

## References

[1] Keldysh L V 1965 Ionization in the Field of a Strong Electromagnetic Wave *Sov. Phys. JETP* **20** 1307-1314

[2] Eberly J H and Javanainen J 1988 Above-Threshold Ionisation *Eur. J. Phys.* **9** 265-275

[3] Corkum P B, Burnett N H and Brunel F 1989 Above-Threshold Ionization in the Long-Wavelength Limit *Phys. Rev. Lett.* **62** 1259

[4] Corkum P B 1993 Plasma Perspective on Strong Field Multiphoton Ionization *Phys. Rev. Lett.* **71** 1994

[5] Lewenstein M, Balcou Ph, Ivanov M Y, L'Huillier A and Corkum P B 1994 Theory of High-Harmonic Generation by Low-Frequency Laser Fields *Phys. Rev. A* **49** 2117

[6] Niikura H, Villeneuve D M and Corkum P B 2005 Mapping Attosecond Electron Wave Packet Motion *Phys. Rev. Lett.* **94** 083003

[7] Popmintchev T, Chen M-C, Bahabad A, Gerrity M, Sidorenko P, Cohen O, Christov I P, Murnane M M and Kapteyn H C 2009 Phase Matching of High-Harmonic Generation in the Soft and Hard X-ray Regions of the Spectrum *Proc Natl Acad Sci USA* **106** 10516-10521

[8] Corkum P B, Burnett N H and Ivanov M Y 1994 Subfemtosecond Pulses *Opt. Lett.* **19** 1870-1872

[9] Gorlach A, Tzur M E, Birk M, Krüger M, Rivera N, Cohen O and Kaminer I 2023 High-harmonic generation driven by quantum light Nat. Phys. 19 1689-1696

[10] Ghimire S, DiChiara A D, Sistrunk E, Agostini P, DiMauro L F and Reis D A 2011 Observation of High-Order Harmonic Generation in a Bulk Crystal *Nat. Phys.* **7** 138-141

[11] Vampa G, McDonald C R, Orlando G, Klug D D, Corkum P B and Brabec T 2014 Theoretical Analysis of High-Harmonic Generation in Solids *Phys. Rev. Lett.* **113** 073901

[12] Ashcroft N W and Mermin N D 1976 *Solid State Physics* (Holt, Rinehart and Winston, New York)

[13] Lompré L A, L'Huillier A, Ferray M, Monot P, Mainfray G and Manus C 1990 High-Order Harmonic Generation in Xenon: Intensity and Propagation Effects *J. Opt. Soc. Am. B* **7** 754-761

[14] Blume M and Watson R E 1962 Theory of Spin-Orbit Coupling in Atoms I. Derivation of the Spin-Orbit Coupling Constant *Proc. R. Soc. Lond. A* **270** 127-143

[15] Stewart G A, Hoerner P, Debrah D A, Lee S K, Schlegel H B and Li W 2023 Attosecond Imaging of Electronic Wave Packets *Phys. Rev. Lett.* **130** 083202

[16] Burke K, Werschnik J and Gross E K U 2005 Time-Dependent Density Functional Theory: Past, Present, and Future *J. Chem. Phys.* **123** 062206

[17] Cheng K T and Johnson W R 1983 Orbital Collapse and the Photoionization of the Inner $4d$ Shells for Xe-Like Ions *Phys. Rev. A* **28** 2820

[18] Ko D H, Brown G G, Zhang C and Corkum P B 2021 Near-Field Imaging of Dipole Emission Modulated by an Optical Grating *Optica* **8** 1632-1637

[19] Breidbach J and Cederbaum L S 2005 Universal Attosecond Response to the Removal of an Electron *Phys. Rev. Lett.* **94** 033901

[20] Marra Z A, Wu Y, Zhou F and Chang Z 2023 Cryogenically Cooled Fe:ZnSe-Based Chirped Pulse Amplifier at 4.07 mm *Opt. Express* **31** 13447-13454

## *37. Attosecond control of quantum trajectories*

**Pierre Agostini[1,2] and Louis F. DiMauro[1,2]\***

[1] Department of Physics, The Ohio State University, Columbus, OH, 43210 USA
[2] Institute for Optical Science, The Ohio State University, Columbus, OH 43210 USA

dimauro.6@osu.edu

**Status**

Attosecond Pulse Trains (APT) and Isolated Attosecond Pulses (IAP) generated from high harmonics have been reported 25 years ago [1,2]. Studies using attosecond pulses synchronized to the fundamental laser have produced the first "picture of an electron" [3], control of the High Harmonic Generation (HHG) process [4-9] and Non-Sequential Double Ionization (NSDI) [10]. In contrast, control is diminished for single color strong field ionization since the laser phases leading to a free electron wave packet (EWP) are dictated by the laser intensity through the tunneling probability (Fig.1). Consequently, attosecond selective control of quantum trajectories can shed a new light on these processes and explore physics beyond the strong field semi-classical model [11,12]. Recently, a new method dubbed Quantum Trajectory Selector (QTS) replaces tunneling by single-photon XUV photoionization induced by the shaped APT at precise phases of the dressing field thus selecting quantum trajectories, as shown in Fig. 1 and contrasted against tunneling. QTS results in superior control of the initial conditions defining the EWP trajectories which can further amplify applications. Initial measurements were performed in argon, but the QTS method that combines NIR pulses and spectral sculpted XUV APT possessing a few eV bandwidth near the atomic ionization threshold is quite universal: it is possible to generate EWPs like the one produced by tunneling with zero kinetic energy [10]. During the experiment both the photoelectron energy spectrum and ion charge state are measured and the relative delay between the APT and NIR fields is varied with attosecond precision. A typical QTS spectrogram is shown in Fig. 2 where a clear delay dependence of the electron and double ion yield is evident, each oscillating at twice the NIR period. In addition, by extracting the $\pi/2$ phase jump (~35 eV in Fig.2) in the electron spectrum it is possible to calibrate the phase corresponding to the zero of the electric field thus allowing the direct clocking of the recollision processes [13]. For example, the maximum in the NSDI oscillation has a phase that is near the semi-classical expectation but unexpectedly diverges from this prediction as the NIR intensity is increased. The results demonstrate control of strong field recollision in the time domain with attosecond precision for both NSDI and elastic rescattering thus allowing an unprecedented comparison with theoretical calculations. Control of trajectories in space with elliptical polarization, two-color fields, and time-dependent polarization gating, can allow a complete space-time control of the recollision process and offer a route to the optimization of attosecond sources by trajectory selection at the single atom level.

**Current and future challenges**

Among the current challenges, several questions need to be addressed. First, the QTS double ion yield shows a 20% dc-component that is independent of the initial phase but wavelength dependent. Currently this signal is attributed to a sequential process whose detailed understanding must be clarified. The second is a long-standing unresolved issue in strong-field ionization physics: NSDI is observed at an intensity such that the recolliding energy is significantly below the ion binding energy or even below the excitation energy, *i.e.* non-classical NSDI. How can the energy deficit be overcome remains an open question. Another challenge is the time resolution of the APT pulses. The temporal profile of the APT is reconstructed using the standard RABBITT method which yields a FWHM of ~ 900 attoseconds compared to the half cycle of 4.1 fs of the 2.4 mm laser, thus determining the precision of the measured EWP emission phase. It is therefore important to engineer these pulses as short as possible and perhaps, remove the phase irregularities. Consequently, a compromise must be found between the bandwidth and the sculpting of the APT spectrum to mimic a zero-kinetic energy, tunnelling wave packet.

However, these challenges also present a unique opportunity to explore physics whose origins deviate from the classical behaviour. In principle, the QTS method can be extended to more complex species, different EWP initial conditions and core ionization, allowing exploration of the physics of recollision involving atomic relaxations. Another future direction would be to extend the method to solid state samples and control the role of intraband and interband processes in crystalline solids HHG.

Furthermore, the current QTS scheme ignores the phase of the EWP. Determining the phase in the time domain is challenging due to the electron dynamics occur on an attosecond time scale. In the spectral domain however, it has been shown that EWP created through a Fano autoionizing resonance can measure the amplitude and phase by spectrally resolved attosecond electron interferometry [14].

**Advances in science and technology to meet challenges**

Future improvements in the QTS method include pushing the laser wavelength further into the mid-infrared and implementing greater control of the spectral filtering and sculpting of the APT. New laser materials can provide drivers beyond 3 μm wavelength while progress in spectral filters and multilayer mirrors are expected to improve the latter. To help identify sequential processes responsible for the offset, it is perhaps necessary to modify the RABBITT setup to detect excited states of the target atom. The growing availability of CPA laser systems with repetition rates as high as MHz will greatly improve the QTS method and offer higher precision in determining the rescattered EWP energy.

**Concluding remarks**

For over 30 years, the semiclassical recollision model of strong-field physics has provided a unified understanding of the properties of energetic electrons, photons, and highly charged ions created by an intense IR laser field interacting with atoms or molecules. By sculpting the bandwidth of photon energies in an XUV APT to a few eV around the atomic ionization potential, it is possible to produce an EWP that mimics the zero kinetic energy EWP produced by tunneling in an intense NIR field. Therefore, one can study in greater detail the dynamics of strong-field ionization and recollision with precise control over the electron trajectories, ascertaining the QTS method as illustrated in Fig. 1. The QTS method, which includes the preparation of an initial state, the time evolution of the state under a comparable Hamiltonian, and measurements of relevant quantities, form an ideal strong-field simulator.

**Acknowledgements**
The authors acknowledge funding from the U.S. Department of Energy, Office of Science, Basic Energy Sciences, under Awards No. DE-FG02-04ER15614. LFD acknowledges support from the Edward & Sylvia Hagenlocker endowment.

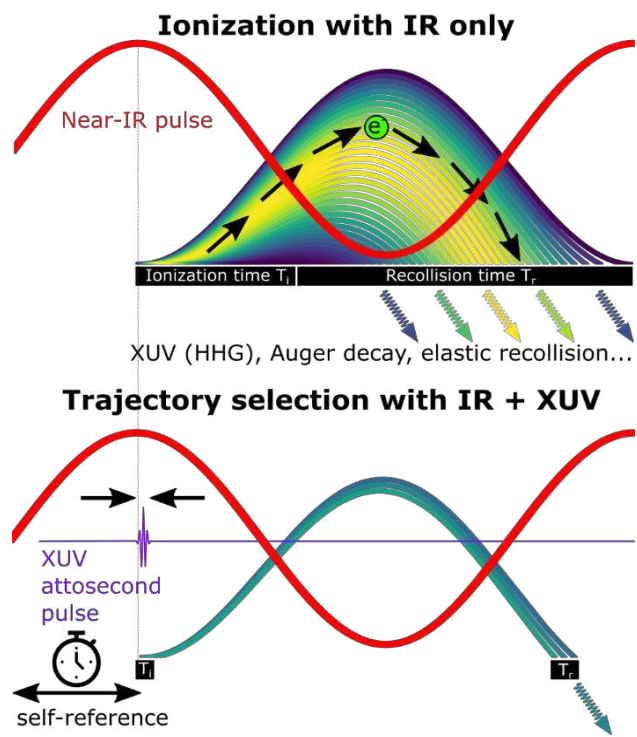

**Figure 1.** Comparison between the strong field ionization process and the quantum trajectory selector approach.

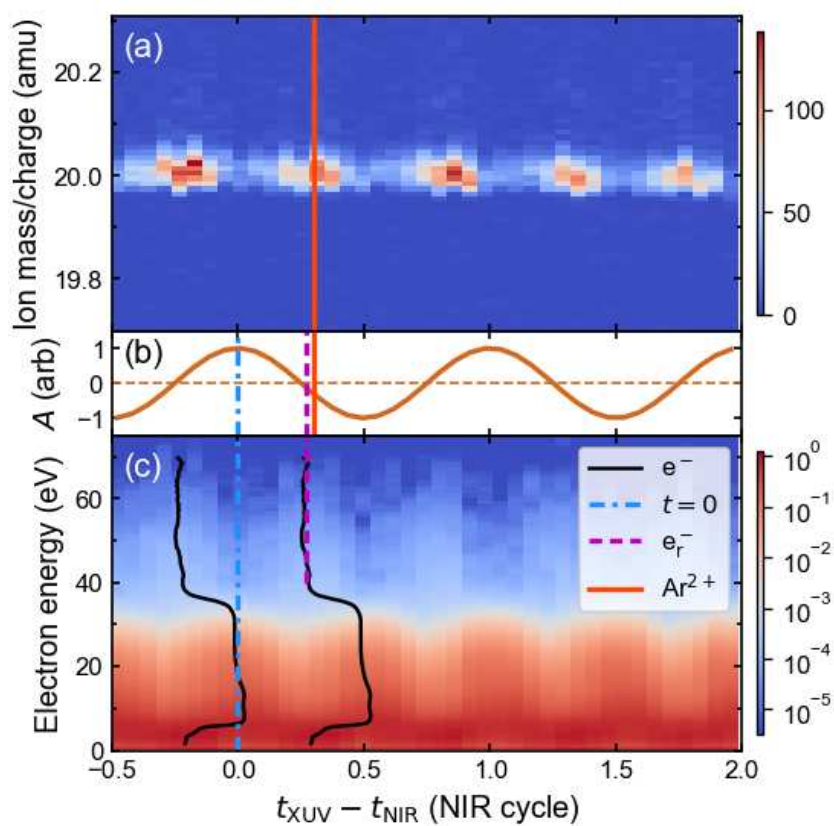

**Figure 2.** QTS Spectrogram from measurements in argon. In (**a**) the subcycle character of the argon double ion signal is clearly observed. (**b**) illustrates the NIR vector potential and (**c**) shows the calibrations of the time origin determined through the analysis of the photoelectron spectrum.

*(Reproduced from Piper et al., Phys. Rev. Lett.* **134***, 073201).*

## References

[1]   P. M. Paul et al. 2001 Observation of a Train of Attosecond Pulses from High Harmonic Generation, *Science* **292** 1289.

[2]   M. Hentschel, R. Kienberger et al. 2001 Attosecond metrology *Nature* **414**: 509-513.

[3]   J. Mauritsson et al. 2008 Coherent Electron Scattering Captured by an Attosecond Quantum Stroboscope *Phys. Rev. Lett.* **100** 073003.

[4]   K. J. Schafer, M. B. Gaarde, A. Heinrich, J. Biegert, and U. Keller 2004 Strong field quantum path control using attosecond pulse trains, *Phys. Rev. Lett*. **92**, 023003.

[5]   J. Biegert, A. Heinrich, C. P. Hauri, W. Kornelis, P. Schlup, M. P. Anscombe, M. B. Gaarde, K. J. Schafer, and U. Keller 2006 Control of high-order harmonic emission using attosecond pulse trains *J. Mod. Opt.* **53**, 87.

[6]   M. B. Gaarde, K. J. Schafer, A. Heinrich, J. Biegert, and U. Keller 2005 Large enhancement of macroscopic yield in attosecond pulse train-assisted harmonic generation *Phys. Rev. A* **72** 013411.

[7]   G. Gademann, F. Kelkensberg, W. K. Siu, P. Johnsson, M. B. Gaarde, K. J. Schafer, and M. J. Vrakking 2011 Attosecond control of electron-ion recollision in high harmonic generation *New J. Phys.* **13**, 033002.

[8]   D. Azoury, M. Krüger, G. Orenstein, H. R. Larsson, S. Bauch, B. D. Bruner, and N. Dudovich 2017 Self-probing spectroscopy of XUV photo-ionization dynamics in atoms subjected to a strong-field environment *Nat. Commun.* **8**, 1453.

[9]   T. Heldt, J. Dubois, P. Birk, G. D. Borisova, G. M. Lando, C. Ott, and T. Pfeifer 2023 Attosecond real-time observation of recolliding electron trajectories in helium at low laser intensities *Phys. Rev. Lett.* **130** 183201.

[10]  A J Piper et al. 2025 Attosecond Clocking and Control of Strong Field Quantum Trajectories *Phys. Rev Lett.* **134** 073201.

[11]  P. Corkum 1993 Plasma perspective on strong field multiphoton ionization, *Phys. Rev. Lett.* **71**, 1994.

[12]  K. J. Schafer, et al. 1993 Above threshold ionization beyond the high harmonic cutoff Phys. Rev. Lett. **70,** 1599.

[13]  B. Walker et al. 1996 Elastic Rescattering in the Strong Field Tunneling Limit *Phys. Rev. Lett.* **77**, 5031.

[14]  Gruson, V. et al. 2016 Attosecond dynamics through a Fano resonance: monitoring the birth of a photoelectron *Science* **354**, 734–738.